\documentclass[runningheads]{cl2emult}

\usepackage{graphicx} % standard LaTeX graphics tool
\def\RRe{I\!\!R}
\def\const{{\rm constant}}
\begin{document}
\pagenumbering{roman}
\setcounter{tocdepth}{3}
\tableofcontents
\newpage
\setcounter{page}{1}
\pagenumbering{arabic}

\title*{Selected solutions of Einstein's field equations: their role in
general relativity and astrophysics}

\toctitle{The role of exact solutions of Einstein's equations
\protect\newline in the developments of general relativity
\protect\newline and astrophysics selected themes}
% allows explicit linebreak for the table of content
%
%
\titlerunning{The role of exact solutions}
% allows abbreviation of title, if the full title is too long
% to fit in the running head
%
\author{Ji\v r\' \i\ Bi\v c\' ak}
\authorrunning{Ji\v r\' \i\ Bi\v c\' ak}
% if there are more than two authors,
% please abbreviate author list for running head
%
%
\institute{Institute of Theoretical Physics,\\
Charles University, Prague
%, Czech Republic
}

\maketitle              % typesets the title of the contribution

%\begin{abstract}
%The abstract\index{abstract} should summarize the contents of the paper
%in at least 70 and at most 150 words; neither too long
%nor too short but to the point!
%\end{abstract}

%%%%%%%%%%%%%%%%%%%%%%%%%%%%%%%%%%%%%%%%%%%%%%%%%%%%%%%%%%%%%%%%%%%%%
%%%%%%%%\input intro.tex
%%%%%%%%%%%%%%%%%%%%%%%%%%%%%%%%%%%%%%%%%%%%%%%%%%%%%%%%%%%%%%%%%%%%%
\section{Introduction and a few excursions}

The primary purpose of all physical theory is rooted in reality,
and most relativists pretend to be physicists. We may often be
members of departments of mathematics and our work oriented
towards the mathematical aspects of Einstein's theory, but even
those of us who hold a permanent position on ``scri'', are
primarily looking there for gravitational waves. Of course, the
builder of this theory and its field equations was {\it the}
physicist. J\"urgen Ehlers has always been very much interested
in the conceptual and axiomatic foundations of physical theories
and their rigorous, mathematically elegant formulation; but he
has also developed and emphasized the importance of such
areas of relativity as kinetic theory, the mechanics of continuous
media, thermodynamics and, more recently, gravitational lensing.
Feynman expressed his view on the relation of physics to
mathematics as follows \cite{FEM}:

``The physicist is always interested in the special case; he is
never interested in the general case. He is talking about
something; he is not talking abstractly about anything. He wants
to discuss the gravity law in three dimensions; he never wants
the arbitrary force case in $n$ dimensions. So a certain amount
of reducing is necessary, because the mathematicians have
prepared these things for a wide range of problems. This is very
useful, and later on it always turns out that the poor physicist
has to come back and say, `Excuse me, when you wanted to tell me
about four dimensions...' '' Of course, this is Feynman, and from
1965...

However, physicists are still rightly impressed by special
explicit formulae. Explicit solutions enable us to discriminate
more easily between a ``physical'' and ``pathological'' feature.
Where are there singularities? What is their character? How do
test particles and fields behave in given background spacetimes?
What are their global structures? Is a solution stable and, in
some sense, generic? Clearly, such questions have been asked not
only within general relativity.

By studying a {\it special} explicit solution one acquires an
intuition which, in turn, stimulates further questions
relevant to more {\it general} situations. Consider, for example,
charged black holes as described by the Reissner-Nordstr\"om
solution. We have learned that in their interior a Cauchy horizon
exists and that the singularities are timelike. We shall discuss
this in greater detail in Section 3.1. The singularities can be
seen by, and thus exert an influence on, an observer travelling in
their neighborhood. However, will this violation of the (strong)
cosmic censorship persist when the black hole is perturbed by
weak (``linear'') or even strong (``nonlinear'') perturbations?
We shall see that, remarkably, this question can also be studied
by explicit exact special model solutions. Still more
surprisingly, perhaps, a similar question can be addressed and
analyzed by means of explicit solutions describing completely
diverse situations -- the collisions of plane waves. As we shall
note in Section 8.3, such collisions may develop Cauchy horizons
and subsequent timelike singularities. The theory of black holes
and the theory of colliding waves have intriguing structural
similarities which, first of all, stem from the circumstance that
in both cases there exist two symmetries, i.e. two Killing
fields. What, however, about more general situations? This is a
natural question inspired by the explicit solutions. Then ``the
poor physicists have to come back'' to a mathematician, or today
alternatively, to a numerical relativist, and hope that somehow
they will firmly learn whether the cosmic censorship is the
``truth'', or that it has been a very inspirational, but in general
false conjecture. However, even {\it after} the formulation of a
conjecture about a general situation inspired by particular exact
solutions, {\it newly} discovered exact solutions can play an
important role in verifying, clarifying, modifying, or ruling out
the conjecture. And also ``old'' solutions may turn out to act as
asymptotic states of general classes of models, and so become
still more significant.

Exact explicit solutions have played a crucial role in the
development of many areas of physics and astrophysics. Later on
in this Introduction we will take note of some general features which
are specific to the solutions of Einstein's equations. Before
that, however, for illustration and comparison we shall indicate
briefly with a few examples what influence exact explicit
solutions have had in other physical theories.
Our next introductory excursion, in Section 1.2, describes in
some detail the (especially early) history of Einstein's route to
the gravitational field equations for which his short stay in
Prague was of great significance. The role of Ernst Mach (who
spent 28 years in Prague before Einstein) in the construction of
the first modern cosmological model, the Einstein static
universe, is also touched upon. Section 1.3 is devoted to a few
remarks on some old and new impacts of the other simplest
``cosmological'' solutions of Einstein's equations -- the
Minkowski, the de Sitter, and the anti de Sitter spacetimes. Some
specific features of solutions in Einstein's theory, such as
the observability and interpretation of metrics, the role of
general covariance, the problem of the equivalence of two
metrics, and of geometrical characterization of solutions are
mentioned in Section 1.4. Finally, in the last (sub)sections of
the ``Introduction'' we give some reasons why we consider our
choice of solutions to be ``a natural selection'', and we briefly
outline the main body of the article.
\newpage

\subsection{A word on the role of explicit solutions in other
parts of physics and astrophysics}

Even in a linear theory like Maxwell's electrodynamics one needs
a good sample, a useful kit, of exact fields like the homogeneous
field, the Coulomb field, the dipole, the quadrupole and other
simple solutions, in order to gain a physical intuition and
understanding of the theory. Similarly, of course, with the
linearized theory of gravity. Going over to the Schr\"odinger
equation of standard quantum mechanics, again a linear theory,
consider what we have learned from simple, explicitly soluble
problems like the linear and the three-dimensional harmonic
oscillator, or particles in potential wells of various shapes. We
have acquired, for example, a transparent insight into such basic
quantum phenomena as the existence of minimum energy states
whose energy is not zero, and their associated wave functions which
have a certain spatial extent, in contrast to classical
mechanics. The three-dimensional problems have taught us, among
other things, about the degeneracy of the energy levels. The case
of the harmonic oscillator is, of course, very exceptional since
Hamiltonians of the same type appear in all problems involving
quantized oscillations. One encounters them in quantum
electrodynamics, quantum field theory, and likewise in the theory
of molecular and crystalline vibrations. It is thus perhaps not
so surprising that the Hamiltonian and the wave functions of the
harmonic oscillator arise even in the minisuperspace models
associated with the Hartle-Hawking no-boundary proposal for the
wave function of the universe \cite{HHT}, and in the
minisuperspace model of homogeneous spherically symmetric dust filled
universes \cite{KP}.

In {\it nonlinear} problems explicit solutions play still a greater
role since to gain an intuition of nonlinear phenomena is hard.
Landau and Lifshitz in their Fluid Mechanics (Volume 6 of their
course) devote a whole section to the exact solutions of the
nonlinear Navier-Stokes equations for a viscous fluid (including
Landau's own solution for a jet emerging from the end of a narrow
tube into an infinite space filled with fluid).

Although Poisson's equation for the gravitational potential in
the classical theory of gravity is linear, the combined system of
equations describing both the field and its {\it fluid} sources
(not rigid bodies, these are simple!) characterized by Euler's
equations and an equation of state are nonlinear. In classical
astrophysical fluid dynamics perhaps the most distinct
and fortunate example of the role of explicit solutions is given
by the exact descriptions of 
ellipsoidal, uniform density masses of self-gravitating fluids.
These ``ellipsoidal figures of equilibrium'' \cite{CHDR} 
include the familiar Maclaurin
spheroids and triaxial Jacobi ellipsoids, which are characterized
by rigid rotation, and a wider class discovered by Dedekind and
Riemann, in which a motion of uniform vorticity exists, even in a
frame in which the ellipsoidal surface is at rest. The 
solutions representing the rotating ellipsoids did not
only play an inspirational role in developing basic concepts of
the theory of rigidly rotating stars, but quite unexpectedly in 
the study of inviscid, differentially rotating polytropes.
These closely resemble Maclaurin spheroids, although they do
not maintain rigid rotation. As noted in the well-known monograph on
rotating stars \cite{TAS}, ``the classical work on uniformly
rotating, homogeneous spheroids has a range of validity much
greater than was usually anticipated''. It also influenced
galactic dynamics \cite{BT}: the existence of Jacobi ellipsoids
suggested that a rapidly rotating galaxy may not remain
axisymmetric, and the Riemann ellipsoids demonstrated that there
is a distinction between the rate at which the matter in a
triaxial rotating body streams and the rate at which the figure
of the body rotates. Since rotating incompressible ellipsoids
adequately illustrate the general feature of rotating
axisymmetric bodies, they are also used in the studies of double
stars whose components are close to each other. The disturbances
caused by a neighbouring component are treated as first order
perturbations. Relativistic effects on the rotating
incompressible ellipsoids have been investigated in the
post-Newtonian approximation by various authors, recently with
a motivation to understand the coalescence of binary neutron stars
near their innermost stable circular orbit (see \cite{TI} for the
latest work and a number of references).

\vskip -0.4pt
As for the last subject, which has a more direct connection with
exact explicit solutions of Einstein's equations, we want to
say a few words about integrable systems and their soliton
solutions. Soliton theory has been one of the most interesting
developments in the past decades both in physics and mathematics,
and gravity has played a role both in its birth and recent
developments. It has been known from the end of the last century
that the celebrated Korteweg-de Vries nonlinear evolution
equation, which governs one dimensional surface gravity waves
propagating in a shallow channel of water, admits solitary wave
solutions. However, it was not until Zabusky and Kruskal (the
Kruskal of section 2.4 below) did extensive {\it
numerical} studies of this equation in 1965 that the remarkable
properties of the solitary waves were discovered: the nonlinear
solitary waves, named solitons by Zabusky and Kruskal, can
interact and then continue, preserving their shapes and
velocities. This discovery has stimulated extensive studies of
other nonlinear equations, the inverse scattering methods of
their solution, the proof of the existence of an infinite number
of conservation laws associated with such equations, and  the
construction of explicit solutions (see \cite{ABL} for a recent
comprehensive treatise). Various other nonlinear equations, similar to
the sine-Gordon equation or the nonlinear Schr\"odinger equation,
arising for example in plasma physics, solid state physics, and
nonlinear optics, have also been successfully tackled by these
methods. At the end of the 1970s several authors discovered that
Einstein's vacuum equations for axisymmetric stationary systems
can be solved by means of the inverse scattering methods, and it
soon became clear that one can employ them also in situations
when both Killing vectors are spacelike (producing, for example,
soliton-type cosmological gravitational waves). Dieter Maison,
one of the pioneers in applying these techniques in general
relativity, describes the subject thoroughly in this volume. We
shall briefly meet the soliton methods when we discuss the
uniformly rotating disk solution of Neugebauer and Meinel
(Section 6.3), colliding plane waves (Section 8.3), and 
inhomogeneous cosmological models (Section 12.2).
Our aim, however, is to understand the meaning of solutions,
rather than generation techniques of finding them. From this
viewpoint it is perhaps first worth noting the interplay between
numerical and analytic studies of the soliton solutions --
hopefully, a good example of an interaction for numerical and
mathematical relativists. However, the explicit solutions of
integrable models have played important roles in various other
contexts. The most interesting {\it multi}-dimensional integrable
equations are the four-dimensional self-dual Yang-Mills equations
arising in field theory. Their solutions, discovered by R. Ward
using twistor theory, on one hand stimulated Donaldson's most
remarkable work on inequivalent differential structures on
four-manifolds. On the other hand, Ward indicated that many of
the known integrable systems can be obtained by dimensional
reduction from the self-dual Yang-Mills equations. Very recently
this view has been substantiated in the monograph by Mason and
Woodhouse \cite{MAW}. The words by which these authors finely
express the significance of exact solutions in integrable systems
can be equally well used for solutions of Einstein's equations:
``they combine tractability with nonlinearity, so they make it
possible to explore nonlinear phenomena while working with
explicit solutions''.\footnote{In 1998, in 
the discussion after his Prague lecture on
the present role of physics in mathematics, Prof. Michael Atiyah
expressed a similar view that even with more powerful
supercomputers and with a growing body of general mathematical
results on the existence and uniqueness of solutions of
differential equations, the exact, explicit solutions of
nonlinear equations will not cease to play a significant role. (As
it is well known, Sir Michael Atiyah has made fundamental
contributions to various branches of mathematics and mathematical
physics, among others, to the theory of solitons,
instantons, and to the twistor theory of Sir Roger Penrose, with
whom he has been interacting ``under the same roof'' in Oxford
for 17 years \cite{ATI}.)}

\subsection{Einstein's field equations}

Since J\"urgen Ehlers has always been, among other things,
interested in the history
of science, he will hopefully tolerate a few remarks on the early
history of Einstein's equations to which not much attention has
been paid in the literature. It was during his stay in  Prague 
in 1911 and 1912 that Einstein's intensive interest in quantum
theory diminished, and his systematic effort in constructing a
relativistic theory of gravitation began. In his first ``Prague
theory of gravity'' he assumed that gravity can be described by a
single function -- the local velocity of light. This assumption
led to insurmountable difficulties. However, Einstein learned
much in Prague on his way to general relativity \cite{GEB}: he
understood the local significance of the principle of
equivalence; he realized that the equations describing the
gravitational field must be nonlinear and have a form invariant
with respect to a larger group of transformations than the
Lorentz group; and he found that ``{\it spacetime coordinates
lose their simple physical meaning}'', i.e. they do not determine
directly the distances between spacetime 
points.\footnote{At that time Einstein's view on the future theory of
gravity are best summarized in his reply to M. Abraham
\cite{EAB}, written just before departure from Prague.}
In his ``Autobiographical Notes'' Einstein says: ``Why were seven
years ... required for the construction of general relativity? The
main reason lies in the fact that it is not easy to free oneself
from the idea that coordinates must have an immediate metrical
meaning''... Either from Georg Pick while still in Prague, or
from Marcel Grossmann during the autumn of 1912 
after his return to Zurich (cf. \cite{GEB}), 
Einstein learned that an appropriate
mathematical formalism for his new theory of gravity was
available in the work of Riemann, Ricci, and Levi-Civita. Several
months after his departure from Prague and his collaboration with
Grossmann, Einstein had general relativity almost in hand. Their
work \cite{EG} was already based on the generally invariant line
element
\renewcommand{\theequation}{\Roman{equation}}
\begin{eqnarray}
ds^2 = g_{\mu \nu} dx^{\mu} dx^{\nu}
\end{eqnarray}
in which the spacetime metric tensor $g_{\mu \nu}(x^\rho),
\mu , \nu , \rho = 0,1,2,3,$ plays a dual role: on the one hand it
determines the spacetime geometry, on the other it represents the
(ten components of the) gravitational potential and is thus a
dynamical variable. The disparity between geometry and physics,
criticized notably by Ernst Mach,\footnote{
Mach spent 28 years as Professor of Experimental Physics
in Prague, until 1895, when he took the History and
Theory of Inductive Natural Sciences chair in Vienna.}
had thus been removed. When searching for the field equations
for the metric tensor, Einstein and Grossmann {\it had} already
realized that a natural candidate for generally covariant field
equations would be the equations relating -- in present-day
terminology -- the Ricci tensor and the energy-momentum tensor
of matter. However, they erroneously concluded that such
equations would not yield the Poisson equation of Newton's theory
of gravitation as a first approximation for weak gravitational
fields (see both \S 5 in the ``Physical part'' in \cite{EG}
written by Einstein and \S 4, below equation (\ref{Equ46}), in the
``Mathematical part'' by M. Grossmann). Einstein then rejected
the general covariance. In a subsequent paper with Grossmann
\cite{EG1}, they supported this mis-step by a well-known ``hole''
meta-argument and obtained (in today's terminology) four gauge
conditions such that the field equations were covariant only with
respect to transformations of coordinates permitted by the gauge
conditions. We refer to, for example, \cite{Pai} for more detailed information
on the further developments leading to the final
version of the field equations. Let us only summarize that in
late 1915 Einstein first readopted the generally covariant field
equations from 1913, in which the Ricci tensor $R_{\mu \nu}$
was, up to the gravitational coupling constant, equal to the
energy-momentum tensor $T_{\mu \nu}$ (paper submitted to the
Prussian Academy on November 4). From his {\it vacuum field
equations}
\begin{eqnarray}
R_{\mu \nu} = 0, %\nonumber
\end{eqnarray}
where $R_{\mu \nu}$ depends nonlinearly on $g_{\alpha \beta}$ and
its first derivatives, and linearly on its second derivatives, he
was able to explain the anomalous part of the perihelion
precession of Mercury -- in the note presented to the Academy on November
18. And finally, in the paper \cite{EFE} submitted on November 25
(published on December 2, 1915), the final version of the {\it
gravitational field equations}, or {\it Einstein's field equations}
appeared:\footnote{
David Hilbert submitted his paper on these field equations 
five days before Einstein, though it was published only on March 31, 1916.
Recent analysis \cite{CRS} of archival materials has revealed that
Hilbert made significant changes in the proofs. The originally submitted 
version of his paper contained the theory which is not generally covariant,
and the paper did not include equations (III).}
\begin{equation}
R_{\mu \nu} - \frac{1}{2} g_{\mu \nu} R = \frac{8\pi G}{c^4}
T_{\mu \nu}, %\\ \nonumber
\end{equation}
where the scalar curvature $R = g^{\mu \nu} R_{\mu \nu}$. 
Newton's gravitational constant $G$ and the velocity of light $c$ are the
(only) fundamental constants appearing in the theory. If not
stated otherwise, in this article we use the geometrized units in
which $G=c=1$, and the same conventions as in \cite{MTW} and
\cite{Wa}.

Now it is well known that Einstein further generalized his field
equations by adding a cosmological term $+\Lambda g_{\mu \nu}$ on
the left side of the field equations (III). The cosmological
constant $\Lambda$ appeared first in Einstein's work
``Cosmological considerations in the General Theory of Relativity''
\cite{ECO} submitted on February 8, 1917 and published on February 15, 1917,
which contained the {\it closed static model of the Universe}
(the Einstein static universe) -- an exact solution of equations
(III) with $\Lambda > 0$ and an energy-momentum tensor of
incoherent matter (``dust''). This solution marked the birth of
modern cosmology.

We do not wish to embark upon the question of the role that
Mach's principle played in Einstein's thinking when constructing
general relativity, or upon the intriguing issues relating to
aspects of Mach's principle in present-day relativity and
cosmology\footnote{It was primarily Einstein's recognition of the role of
Mach's ideas in his route towards general relativity, and in his
christening them by the name ``Mach's principle'' (though Schlick
used this term in a vague sense three years before Einstein), that
makes Mach's Principle influential even today. After the 1988
Prague conference on Ernst Mach and his influence on the
development of physics \cite{PRG}, the 1993 conference
devoted exclusively to Mach's principle was held in T\"ubingen,
from which a remarkably thorough volume was prepared \cite{TUB},
covering all aspects of Mach's principle and recording carefully
all discussion. The clarity of ideas and insights of J\"urgen
Ehlers contributed much to both conferences and their proceedings.
For a brief more recent survey of various aspects of Mach's
principle in general relativity, see the introductory section in
the work \cite{DJJ}, in which Mach's principle is analyzed in the
context of perturbed Robertson-Walker universes. Most recently,
Mach's principle seems to enter even into M theory \cite{HORA}.}
--  a problem which in any event would far exceed the scope of this
article. Although it would not be inappropriate to include it
here since exact solutions (such as G\"odel's universe or Ozsv\'ath's
and Sch\"ucking's closed model) have played a prominent role in
this context. However, it should be at least stated that Einstein
originally invented the idea of a closed space in order to
eliminate boundary conditions at spatial infinity. The boundary
conditions ``flat at infinity'' bring with them an inertial frame
unrelated to the mass-energy content of the space, and Einstein,
in accordance with Mach's views, believed that merely mass-energy
can influence inertia. Field equations (III) are not inconsistent
with this idea, but they admit as the simplest solution an empty
flat {\it Minkowski space} $(T_{\mu \nu}=0,\,\,\, g_{\mu \nu}=
\eta _{\mu
\nu}=$ diag $(-1, +1, +1, +1))$, so some restrictive boundary conditions
are essential if the idea is to be maintained. Hence, Einstein
introduced the cosmological constant $\Lambda$, hoping that with
this space will always be closed, and the boundary conditions
eliminated. But it was also in 1917 when {\it de Sitter}
discovered the {\it solution} \cite{dS} of the vacuum field
equations (II) with added cosmological term $(\Lambda > 0)$ which
demonstrated that a nonvanishing $\Lambda$ does not necessarily
imply a 
%closed universe, and extra conditions for closure are still needed.
nonvanishing mass-energy content of the universe.

\subsection{``Just so'' notes on the simplest solutions: 
the Minkowski, de Sitter, and anti de Sitter spacetimes}

Our brief intermezzo on the cosmological constant brought up
three explicit simple exact solutions of Einstein's field
equations -- the Minkowski, Einstein, and de Sitter spacetimes. To
these also belongs the anti de Sitter spacetime, corresponding to
a negative $\Lambda$. The de Sitter spacetime has the topology
$R^1\times S^3$ (with $R^1$
corresponding to the time) and is best represented geometrically
as the 4-dimensional hyperboloid $-v^2+w^2+x^2+y^2+z^2=(3/\Lambda )$
in 5-dimensional flat space with metric 
$ds^2=-dv^2+dw^2+dx^2+dy^2+dz^2$. The anti de Sitter spacetime has
the topology $S^1\times R^3$, and can be visualized as the
4-dimensional hyperboloid $-U^2-V^2+X^2+Y^2+Z^2=(-3/\Lambda)$,
$\Lambda < 0$, in flat 5-dimensional space with
metric $ds^2 = - dU^2 - dV^2 + dX^2 + dY^2 + dZ^2$. As is usual
(cf. e.g. \cite{HE,Pen}), we mean by ``anti de Sitter spacetime'' the
universal covering space which contains no closed timelike lines;
this is obtained by unwrapping the circle $S^1$.

These spacetimes will not be discussed in the following sections.
Ocassionally, for instance, in Sections 5 and 10, we shall
consider spacetimes which become asymptotically de Sitter.
However, since these solutions have played a crucial
role in many issues in general relativity and cosmology, and most
recently, they have become important prerequisites on the stage
of the theoretical physics of the ``new age'', including string
theory and string cosmology, we shall make a few comments on
these solutions here, and give some references to recent
literature.

The basic geometrical properties of these spaces are analyzed in
the Battelle Recontres lectures by Penrose \cite{Pen}, and in the
monograph by Hawking and Ellis \cite{HE}, where also references to
older literature can be found. The important role of the de Sitter
solution in the theory of the expanding universe is finely described
in the book by Peebles \cite{PEE}, and in much greater detail in
the proceedings of the Bologna 1988 meeting on the history of
modern cosmology \cite{BLG}.

\vskip -0.2pt
The Minkowski, de Sitter and anti de Sitter spacetimes are the
simplest solutions in the sense that their {\it metrics} are {\it
of constant} (zero, positive, and negative) {\it curvature}. They admit
the same number (ten) of independent Killing vectors,
but the interpretations of corresponding symmetries differ
for each spacetime. Together with the Einstein static universe,
they all are conformally flat, and can be represented as portions
of the Einstein static universe \cite{HE,Pen}. However, their
{\it conformal structure} is globally different. In Minkowski
spacetime one can go to infinity along timelike geodesics and
arrive to the future (or past) {\it timelike infinity} $i^+$
(or $i^-$); along null geodesics one reaches the future (past) {\it
null infinity} ${\cal{J}}^+ ({\cal{J}}^-)$; and spacelike geodesics
lead to {\it spatial infinity} $i^0$. Minkowski spacetime
can be compactified and mapped onto a finite region by an
appropriate conformal rescaling of the metric. One thus obtains the
well-known Penrose diagram in which the three types of infinities
are mapped onto the boundaries of the compactified spacetime --
see for example the boundaries on the ``right side'' in the Penrose
diagram of the Schwarzschild-Kruskal spacetime in Fig. 3,
Section 2.4, or the Penrose compactified diagram of
boost-rotation symmetric spacetimes in Fig. 13,  Section 11.
(The details of the conformal rescaling of the metric and resulting diagrams
are given in \cite{HE,Pen} and in standard textbooks,
for example \cite{MTW,Wa,DIN}.) In the de Sitter
spacetime there are only past and future conformal infinities
$\cal J^-, \cal J^+$, both being {\it spacelike} (cf. the
Penrose diagram of the ``cosmological'' Robinson-Trautman 
solutions in Fig. 11, Section 10); 
the conformal infinity in anti de Sitter spacetime
is {\it timelike}.

\vskip -0.2pt
These three spacetimes of constant curvature offer many basic
insights which have played a most important role elsewhere in
relativity. To give just a few examples 
(see e.g. \cite{HE,Pen}): both the particle
(cosmological) horizons and the event horizons for geodesic observers 
are well illustrated in the de Sitter 
spacetime; the Cauchy horizons in the anti de Sitter space; 
and the simplest acceleration horizons in Minkowski space 
(hypersurfaces $t^2 = z^2$ in Fig. 12, Section 11). With the de
Sitter spacetime one learns (by considering different cuts
through the 4-dimensional hyperboloid) that the concept of
an ``open'' or
``closed'' universe depends upon the choice of a spacelike slice
through the spacetime. 
There is perhaps no simpler way to
understand that Einstein's field equations are of local nature,
and that the spacetime topology is thus not given a priori, than by
considering the following construction in Minkowski spacetime.
Take the region given in the usual coordinates by
$|x| \leq 1, |y| \leq 1, |z| \leq 1$, remove the rest
and identify pairs of boundary points of the form $(t, 1, y, z)$
and $(t, -1, y, z)$, and similarly for $y$ and $z$. In this way
the spatial sections are identified to obtain a 3-torus -- a flat
but closed manifold.\footnote{This 
very simple point was apparently unknown to
Einstein in 1917, although soon after the publication of his
cosmological paper, E. Freundlich and F. Klein pointed out to him
that an elliptical topology (arising from the identification of
antipodal points) could have been chosen instead of the spherical
one considered by Einstein. Although topological questions have
been followed with a great interest in recent decades, the
chapter by Geroch and Horowitz in ``An Einstein Centenary Survey''
\cite{GHO} remains the classic; for more recent texts, see for example
\cite{Jo} and references therein.}

The {\it spacetimes of constant curvature} have been resurrected
as basic {\it arenas of new physical theories} since their
first appearance. After the role of the de Sitter universe
decreased with the refutation of the steady-state cosmology, it
has inflated again enormously in connection with the theory of
early quasi-exponential phase of expansion of the universe,
due to the false-vacuum state of a hypothetical scalar
(inflaton) field(s) (see e.g. \cite{PEE}). We shall mention the
de Sitter space as the asymptotic state of cosmological models
with a nonvanishing $\Lambda$ (so verifying the ``cosmic no-hair
conjecture'') in Section 10 on Robinson-Trautman spacetimes.
Motivated by its importance in inflationary cosmologies, several
new useful papers reviewing the properties of de Sitter spacetime
have appeared \cite{HJS,GRN}; they also contain many
references to older literature. For the most recent work on the
quantum structure of de Sitter space, see \cite{BOU}.

In the last two years, anti de Sitter spacetime has come to the
fore in light of Maldacena's conjecture \cite{MDC} relating
string theory in (asymptotically) anti de Sitter space to a
non-gravitational conformal field theory on the boundary at
spatial infinity, which is timelike as mentioned above (see,
e.g. \cite{BLB}, where among others, in the Appendix various
coordinate systems describing anti de Sitter spaces in arbitrary
dimensions are discussed).

Amazingly, the Minkowski spacetime has recently entered the
active new area of so called pre-big bang string cosmology
\cite{VEN}. String theory is here applied to the problem of
the big bang. The idea is to start from a simple Minkowski space
(as an ``asymptotic past triviality'') and to show that it is in
an unstable false-vacuum state, which leads to a long {\it
pre}-big bangian inflationary phase. This, at later times, should
provide a hot big bang. Although such a scenario has been
criticized on various grounds, it has attractive features, and
most importantly, can be probed through its observable relics
\cite{VEN}.

Since it is hard to forecast how the roles of these three
spacetimes of constant curvature 
will develop in new and exciting theories
in the next millennium, let us better conclude our
``just so'' notes by stating three ``stable'' results of
complicated, rigorous mathematical analyses of (the classical)
Einstein's equations.

In their recent treatise \cite{CHK}, Christodoulou and
Klainerman prove that any  smooth, asymptotically flat initial
data set which is ``near flat, Minkowski data'' leads to a unique,
smooth and geodesically complete solution of Einstein's vacuum
equations with vanishing cosmological constant. This demonstrates
the {\it stability of the Minkowski space} with respect to
nonlinear (vacuum) perturbations, and the existence of
singularity-free, asymptotically flat radiative vacuum
spacetimes. Christodoulou and Klainerman, however, are able to
show only a somewhat weaker decay of the field at null infinity
than is expected from the usual assumption of a sufficient
smoothness at null infinity in the framework of Penrose (see
e.g. \cite{KVB} for a brief account).

Curiously enough, in the case of the vacuum Einstein equations
with a {\it nonvanishing cosmological constant}, a more complete
picture has been known for some time. By using his
regular conformal field equations, Friedrich \cite{HF1}
demonstrated that initial data sufficiently close to de Sitter
data develop into solutions of Einstein's equations with a
positive cosmological constant, which are ``asymptotically
simple'' (with a smooth conformal infinity), as required in
Penrose's framework. More recently, Friedrich \cite{HF2} has
shown the existence of asymptotically simple solutions to the
Einstein vacuum equations with a negative cosmological constant.
For the latest review of Friedrich's thorough work on
asymptotics, see \cite{HF3}.

Summarizing, thanks to these profound mathematical achievements
we know that the {\it Minkowski, de Sitter}, and {\it anti de Sitter
spacetimes} are the solutions of Einstein's field equations which
are {\it stable with respect to general, nonlinear} (though
``weak'' in a functional sense) {\it vacuum perturbations}.
A result of this type is not known for any other solution of
Einstein's equations.

\subsection{On the interpretation and characterization of metrics}

Suppose that a metric satisfying Einstein's field equations is
known in some region of spacetime and  in a given coordinate
(reference) system $x^{\mu}$. A fundamental question, frequently
``forgotten'' to be addressed in modern theories which extend
upon general relativity, is whether the {\it metric tensor}
$g_{\alpha \beta}(x^{\mu})$ {\it is a measurable quantity}.
Classical general relativity offers (at least) three ways of
giving a positive answer, depending on what objects are considered
as ``primitive tools'' to perform the measurements. The first,
elaborated and emphasized primarily by M{\o}ller \cite{MOL},
employs standard rigid rods in the measurements. However, a ``rigid
rod'' is not really a simple primitive concept. The second procedure,
due to Synge \cite{SG}, accepts as the basic concepts a
``particle'' and a ``standard clock''. If $x^{\mu}$ and $x^\mu +
dx^\mu$ are two nearby events contained in the worldline of a
clock, then the separation (the spacetime interval) between the
events is equal to the interval measured by the clock. The main
drawback of this approach appears to lie in the fact that it does
not explain why the same  functions $g_{\alpha \beta}(x^{\mu})$
describe the behavior of the clock as well as paths of free
particles, as explained in more detail by Ehlers, Pirani and
Schild \cite{EPS}, in the motivation for their own axiomatic but
constructive procedure for setting up the spacetime geometry.
Their method, inspired by the work of Weyl and others, uses
neither rods nor clocks, but instead, light rays and freely
falling test particles, which are considered as basic tools for
measuring the metric and determining the spacetime geometry. (For a
simple description of how this can be performed, see exercise
13.7 in \cite{MTW}; for some new developments which build upon, among
others, the Ehlers-Pirani-Sachs approach, see \cite{MAS}.)
After indicating that the metric tensor is a measurable quantity
let us briefly turn to the {\it role of spacetime
coordinates}.

\vskip -0.3pt
In special relativity there are infinitely many global inertial
coordinate systems labelling events in the Minkowski manifold
$\RRe ^4$; they are related by elements of the Poincar\'e
group. The inertial coordinates labels $X^0, X^1, X^2, X^3$ of a
given event do not thus have intrinsic meaning. However, the spacetime
interval between two events, determined by the Minkowski metric
$\eta_{\mu \nu}$, represents an intrinsic property of spacetime.
Since the Minkowski metric is so simple, the differences between
inertial coordinates can have a metrical meaning (recall
Einstein's reply to Abraham mentioned in Section 1.2). In
principle, however, both in special and general relativity, it is
the metric, the line element, which exhibits intrinsically the
geometry, and gives all relevant information.
As Misner \cite{MER} puts it, if you write down for someone the
Schwarzschild metric in the ``canonical'' form (equation (\ref{Equ2}) in
Section 2.2) and receive the reaction ``that [it] tells me the
$g_{\mu \nu}$ gravitational potentials, now tell me in which $(t,
r, \theta, \varphi)$ coordinate system they have these values?'',
then there are two valid responses: (a) indicate that it is an
indelicate and unnecessary question, or (b) ignore it. Clearly, the
Schwarzschild metric describes the
geometrical properties of the coordinates used in (\ref{Equ2}). For
example, it implies that worldlines with fixed $r, \theta ,
\varphi$ are timelike at $r>2M$, orthogonal to the lines with $t
= \const$. It determines local null cones (given by $ds^2=0$),
i.e. the {\it causal structure} of the spacetime. In addition, in
Schwarzschild coordinates the metric (\ref{Equ2}) indicates how to
{\it measure} the radial coordinate of a given event, because the
proper area of the sphere going through the event is given
just by the Euclidean expression $4\pi r^2$ ($r$ is thus often
called ``the curvature coordinate''). On each sphere the angular
coordinates $\theta, \varphi$ have the same meaning as on a
sphere in Euclidean space. The Schwarzschild coordinate time
$t$, geometrically preferred by the timelike (for $r>2M$) Killing
vector, which is just equal to $\partial/\partial t$, can be
measured by radar signals sent out from spatial infinity
($r \gg 2M$) where $t$ is the proper time (see e.g. \cite{MTW}).
The coordinates used in (\ref{Equ2}) are in fact ``more unique'' than
the inertial coordinates in Minkowski spacetime, because the only
possible continuous transformations preserving the form (\ref{Equ2})
are rigid rotations of a sphere, and $t \rightarrow t + \const$.
Such a simple interpretation of coordinates is exceptional.
However, the simple case of the Schwarzschild metric clearly
demonstrates that all intrinsic information is contained in the
line element.

It is interesting, and for some purposes useful, to consider not
just one Schwarzschild metric with a given mass $M$ but the {\it
family} of such metrics for all possible $M$. In order to cover
also the future event horizon let us describe the metrics by
using Eddington-Finkelstein ingoing coordinates as in equation (\ref{Equ4}),
Section 2.3. This equation can be interpreted as a family of
metrics with various values of $M$ given on a {\it fixed
background manifold} ${\bar{\cal M}}_1$, with $v \in \RRe, r \in(0,
\infty)$, and $\theta \in [0, \pi], \varphi \in [0, 2\pi)$.
Alternatively, however, we may use, for example, the Kruskal null
coordinates $\tilde U, \tilde V$ in which the metric is given by
equation (\ref{Equ6}), Section 2.3, with $\tilde U = V-U, \tilde V = V +
U$. We may then consider metrics on a background manifold $\bar
{\cal {M}}_2$ given by $\tilde{U} \in \RRe, \tilde V \in(0,
\infty),\;\theta \in [0, \pi]$, and $\varphi \in [0, 2 \pi)$, which
corresponds to $\bar {\cal {M}}_1$. However, these two background
manifolds are {\it not} the same: the transformation
between the Eddington-Finkelstein coordinates and the Kruskal
coordinates is not a map from $\bar {\cal {M}}_1$  to 
$\bar {\cal{M}}_2$ because it
depends on the value of mass $M$. Therefore, the ``background
manifold'' used frequently in general relativity, for example in
problems of conservation of energy, or in quantum gravity, is not
defined in a natural, unique manner. The above simple
pedagogical observation has recently been made in connection with
gauge fixing in quantum gravity by H\'aj\'{\i}\v{c}ek \cite{HP}
in order to explain the old insight by Bergmann and Komar, that
the gauge group of general relativity is much larger 
than the diffeomorphism group of one manifold.
To identify points when working with backgrounds, one
usually fixes coordinates in all solution manifolds by some gauge
condition, and identifies those points of all these manifolds
which have the same value of the coordinates.

Returning back to a single solution $({\cal {M}}, g_{\alpha
\beta})$,
described by a manifold ${\cal {M}}$ and a metric $g_{\alpha \beta}$ in
some coordinates, a notorious (local) {\it ``equivalence problem''}
often arises. A given (not necessarily global) solution has the
variety of representations which equals the variety of choices of
a 4-dimensional coordinate system. Transitions from one choice to
another are isomorphic with the group of 4-dimensional
diffeomorphisms which expresses the general covariance of the
theory.\footnote{As pointed out by Kretschmann soon after the birth of
general relativity, one can always make a theory generally
covariant by taking more variables and inserting them as new
dynamical variables into the (enlarged) theory. Thus, standard
Yang-Mills theory is covariant with respect to the
transformations of Yang-Mills potentials, corresponding to a
particular group, say $SU(2)$. However, the theory is usually
formulated on a fixed background spacetime with a given metric.
The evolution of a dynamical Yang-Mills solution is thus
``painted'' on a given spacetime. When the metric -- the
gravitational field -- is incorporated as a dynamical variable in
the Einstein-Yang-Mills theory, the whole spacetime metric and
Yang-Mills field are ``built-up'' from given data (cf. the article by
Friedrich and Rendall in this volume). The resulting theory is
covariant with respect to a much larger group. The dual role of
the metric, determined only up to 4-dimensional diffeomorphisms,
makes the character of the solutions of Einstein's equations
unique among solutions of other field theories, which do not
consider spacetime as being dynamical.}
Given another set of functions $g'_{\alpha \beta}(x'^{\gamma})$ which
satisfy Einstein's equations, how do we learn that they are not
just transformed components of the metric $g_{\alpha
\beta}(x^{\gamma})$? In 1869 E. B. Christoffel raised a more general
question: under which conditions is it possible to transform a
quadratic form $g_{\alpha \beta}(x^\gamma )dx^\alpha dx^\beta$ in
$n$-dimensions into another such form $g'_{\alpha \beta}(x'^\gamma)dx'^\alpha
dx'^\beta$ by means of smooth transformation $x^\gamma
(x'^\kappa)$? As Ehlers emphasized in his paper \cite{ECH} on
the meaning of Christoffel's equivalence problem in modern field
theories, Christoffel's results apply to metrics of arbitrary
signature, and can be thus used directly in general relativity.
Without going into details let us say that today the solution
to the equivalence problem as presented by Cartan is most commonly
used. For both
metrics $g_{\alpha \beta}$ and $g'_{\alpha \beta}$
one has to find a frame (four 1-forms) in
which the frame metric is constant, and find the frame components
of the Riemann tensor and its covariant derivatives up to --
possibly -- the 10th order. The two metrics $g_{\alpha \beta}$ and
$g'_{\alpha \beta}$ are then
equivalent if and only if there exist coordinate and Lorentz
transformations under which one whole set of frame components
goes into the other. 
In a practical algorithm given by Karlhede \cite{KL},
recently summarized and used in \cite{PRM},
the number of derivations required is reduced. 

A natural first idea of how to solve the equivalence problem is
to employ the scalar invariants from the Riemann tensor and its
covariant derivatives. This, however, does not work. For example,
in all Petrov type {\it N} and {\it III} nonexpanding and nontwisting solutions all these
invariants vanish as shown recently (see Section 8.2), 
as they do in Minkowski spacetime.

However, even without regarding invariants, at present much can be
learnt about an exact solution (at least locally) in geometrical
terms, without reference to special coordinates. This is thanks to
the progress started in the late 1950s, in which the group of
Pascual Jordan in Hamburg has played the leading role, with
J\"urgen Ehlers as one of its most active members. Ehlers'
dissertation\footnote{The English translation of the title of the
dissertation reads: ``The construction and characterization of
the solutions of Einstein's gravitational field equations''. In
\cite{EK} the original German title is quoted, as in our citation
\cite{EHD}, but ``of the solutions'' is erroneously omitted. This
error then reemerges in the references in \cite{Wa}.}
\cite{EHD} from 1957 is devoted to the
characterization of exact solutions.

The problem of exact solutions also forms the content of his
contribution to the Royaumont GR-conference \cite{EHRO}, as well
as his plenary talk in the London GR-conference \cite{EHLO}. 
A detailed description of the results of the Hamburg group on
invariant geometrical characterization of exact solutions by
using and developing the Petrov classification of Weyl's tensors,
groups of isometries, and conformal transformations are contained in
the first paper \cite{JEK} in the (today ``golden oldies'')
series of articles published in the ``Abhandlungen der Akademie der
Wissenschaften in Mainz''. An English version, in a somewhat
shorter form, was published by Ehlers and Kundt \cite{EK} in the
``classic'' 1962 book ``Gravitation: An Introduction to Current
Research'' compiled by L. Witten. (We shall meet these references
in the following sections.) In the second paper of the
``Abhandlungen'' \cite{JEK2}, among others, algebraically special
vacuum solutions are studied, using the formalism of the
2-component spinors, and in particular, geometrical properties
of the congruences of null rays are analyzed in terms of their
expansion, twist, and shear.

These tools became essential for the discovery by Roy Kerr in
1963 of the solution which, when compared with all other solutions
of Einstein's equations found from the beginning of the
renaissance of general relativity in the late 1950s until today, has
played the most important role. As Chandrasekhar \cite{CHS}
eloquently expresses his wonder about the remarkable fact that
all stationary and isolated black holes are {\it exactly}
described by the Kerr solution: ``This is the only instance we
have of an exact description of a macroscopic object. Macroscopic
objects, as we see them all around us, are governed by a variety
of forces, derived from a variety of approximations to a variety
of physical theories. In contrast, the only elements in the
construction of black holes are our basic concepts of space and
time ...'' The Kerr solution can also serve as one of finest
examples in general relativity of ``the incredible fact that a
discovery motivated by a search after the beautiful
in mathematics should find its exact replica in Nature...'' \cite{SCH}.

The technology developed in the classical works 
\cite{EK,JEK}, and in a number of subsequent contributions, is mostly
concerned with the local geometrical characterization of exact
spacetime solutions. A well-known feature of the solutions of
Einstein's equations, not shared by solutions in other physical
theories, is that it is often very complicated to analyze their
global properties, such as their extensions, completeness, or topology.
If analyzed globally, almost any solution can tell us something
about the basic issues in general relativity, like the nature of
singularities, or cosmic censorship.

\subsection{The choice of solutions}

Since most solutions, when properly analyzed, can be of potential
interest, we are confronted with a richness of
material which puts us in danger of mentioning many of them, but
remaining on a general level, and just enumerating rather than
enlightening. In fact, because of lack of space (and of our
understanding) we shall have to adopt this attitude in many
places. However, we have selected some solutions, hopefully the
fittest ones, and when discussing their role, we have chosen
particular topics to be analyzed in some detail, and left other
issues to brief remarks and references.

Firstly, however, let us ask what do we understand by the term
``exact solution''. In the much used ``exact-solution-book'' \cite{KSH},
the authors ``do not intend to provide a definition'', or, rather,
they have decided that what they ``chose to include was, by
definition, an exact solution''. A mathematical relativist-purist
would perhaps consider solutions, the existence of which has been
demonstrated in the works of Friedrich or Christodoulou and
Klainerman, mentioned at the end of Section 1.3,
as ``good'' as the Schwarzschild metric. Most
recently, Penrose \cite{ROPE} presented a strong conjecture which
may lead to a general vacuum solution described in the complicated
(complex) formalism of his twistor theory. Although in this article
we do not mean by exact
solutions those just mentioned, we also do not consider as exact
solutions only those
explicit solutions which can be written in terms of elementary
functions on half of a page. We prefer, recalling Feynman,
simple ``special cases'', but we also discuss, for example, the
late-time behaviour of the Robinson-Trautman solutions for which
rigorously convergent series expansions can be obtained, which
provide sufficiently rich ``special information''.

\vskip -0.15pt
Concerning the selection of the solutions, the builder of
general relativity and the gravitational field equations (III)
himself indicates which solutions should be preferred \cite{AI}:
``The theory avoids all internal discrepancies which we have
charged against the basis of classical mechanics... But, it is
similar to a building, one wing of which is made of fine marble
(left part of the equation), but the other wing of which is built
of low grade wood (right side of equation). The phenomenological
representation
of matter is, in fact, only a crude substitute for a
representation which would correspond to all known properties of
matter. There is no difficulty in connecting Maxwell's theory... so
long as one restricts himself to space, free of ponderable matter
and free of electric density...''

\vskip -0.15pt
Of course, Einstein was not aware when he was writing this of
Yang-Mills-Higgs fields, or of the dilaton field, etc. However,
remaining on the level of field theories with a clear classical
meaning, his view has its strength and motivates us to prefer
(electro)vacuum solutions. A physical interpretation of the
vacuum solutions of Einstein's equations have been reviewed in
papers by Bonnor \cite{Bo}, and Bonnor, Griffiths and MacCallum
\cite{BoGM} five years ago. Our article, in particular in
emphasizing and
describing the role of solutions in giving rise to various
concepts, conjectures, and methods of solving problems in general
relativity, and in the astrophysical impacts of the solutions, is
oriented quite differently, and gives more detail. However, up to
some exceptions, like, for example, metrics for an infinite line-mass or
plane, which are discussed in \cite{Bo}, and new solutions which have been
discovered after the reviews \cite{Bo,BoGM} appeared as,
for example, the solution describing a rigidly rotating thin disk of
dust, our choice of solutions is similar to that of \cite{Bo,BoGM}.

\vskip -0.15pt
In selecting particular topics for a more detailed discussion we
will be led primarily by following overlapping aspects: (i) the
``commonly acknowledged'' significance of a solution -- we will
concentrate in particular on the Schwarzschild, the Kerr, the Taub-NUT, and
plane wave solutions, and (ii) the solutions and their
properties that I (and my colleagues) have been directly
interested in, such as the Reissner-Nordstr\"om metric, vacuum
solutions outside rotating disks, or radiative solutions such as
cylindrical waves, Robinson-Trautman solutions, and the
boost-rotation symmetric solutions.
Some of these have also been connected with the interests of
J\"urgen Ehlers, and we shall indicate whenever we are aware of
this fact.

\vskip -0.15pt
Vacuum cosmological solutions are discussed in less detail
than they deserve. A
possible excuse -- from the point of view of being a relativist, a
rather unfair one -- could be that a special recent issue of
Reviews of Modern Physics (Volume 71, 1999), marking the
Centennial of the American Physical Society, contains 
discussion of the Schwarzschild, the Reissner-Nordstr\"om and
other black hole solutions, and even remarks on the work of Bondi
et al. \cite{e} on radiative solutions, but among the cosmological solutions
only the standard models are mentioned. A real reason is the author's
lack of space, time, and energy. In the concluding remarks we will
try to list at least the most important solutions (not only the
Friedmann models!) which have not been ``selected'' and give
references to the literature in which more information can be
found.

\subsection{The outline}

Since the titles of the following sections
characterize the contents rather specifically, we restrict
ourselves to only a few explanatory remarks. In our discussion of
the Schwarschild metric, after mentioning its role in the solar
system, we indicate how the Schwarzschild solution gave rise to
such concepts as the event horizon, the trapped surface, and the
apparent horizon. We pay more attention to the concept of a bifurcate
Killing horizon, because this is usually not treated in textbooks,
and in addition, J\"urgen Ehlers played a role in its first
description in the literature. Another point which has not received
much attention is Penrose's nice presentation of evidence against
Lorentz-covariant field theoretical approaches to gravity, based on
analysis of the causal structure of the Schwarzschild
spacetime. Among various astrophysical implications of the
Schwarzschild solution we especially note recent
suggestions which indicate that we may have evidence of the existence
of event horizons, and of a black hole in the centre
of our Galaxy.

The main focus in our treatment of the Reissner-Nordstr\"om
metric is directed to the instability of the Cauchy horizon and
its relation to the cosmic censorship conjecture. We also briefly
discuss extreme black holes and their role in string theory.

About the same amount of space as that given to the Schwarzschild
solution is devoted to the Kerr metric. After explaining a few
new concepts the metric inspired, such as locally nonrotating
frames and ergoregions, we mention a number of physical processes
which can take place in the Kerr background, including the
Penrose energy extraction process, and the Blandford-Znajek
mechanism. In the section on the astrophysical evidence for a
Kerr metric, the main attention is paid to the broad iron line,
the character of which, as most recent observations indicate, is
best explained by assuming that it originates very close to a
maximally rotating black hole. The discussion of recent results
on black hole uniqueness and on multi-black hole solutions 
concludes our exposition of spacetimes representing black holes.
In the section on axisymmetric fields and relativistic disks a
brief survey of various static solutions is first given, then we
concentrate on relativistic disks as sources of the Kerr metric
and other stationary fields; in particular, we summarize briefly
the recent work on uniformly rotating disks.

An intriguing case of Taub-NUT space is introduced by a new
constructive derivation of the solution. Various pathological
features of this space are then briefly listed.

Going over to radiative spacetimes, we analyze in some detail
plane waves -- also in the light of the thorough study by Ehlers
and Kundt \cite{EK}. Some new developments are then noted, in
particular, impulsive waves generated by boosting various
``particles'', their symmetries, and recent use of the Colombeau
algebra of generalized functions in the analyses of impulsive
waves. A fairly detailed discussion is devoted to various effects
connected with colliding plane waves.

In our treatment of cylindrical waves we concentrate in particular
on two issues: on the proof that these waves provide explicitly
given spacetimes, which admit a smooth global null infinity,
even for strong initial data within a $(2+1)$-dimensional
framework; and on the role that cylindrical waves have played in the
first construction of a midisuperspace
model in quantum gravity. Various other developments concerning
cylindrical waves are then summarized only telegraphically.

A short section on Robinson-Trautman solutions points out how
these solutions with a nonvanishing cosmological constant can be
used to give an exact demonstration of the cosmic no-hair
conjecture under the presence of gravitational radiation, and
also of the existence of an event horizon which is smooth but not
analytic.

As the last class of radiative spacetimes we analyze the
boost-rotation symmetric solutions representing uniformly
accelerated objects. They play
a unique role among radiative spacetimes since they are
asymptotically flat, in the sense that they admit global smooth
sections of null infinity. And as the only known radiative
solutions describing finite sources they can provide expressions
for the Bondi mass, the news function, or the radiation patterns
in explicit forms. They have also been used as test-beds in
numerical relativity, and as the model spacetimes describing the
production of black hole pairs in strong fields.

Vacuum cosmological solutions such as the vacuum Bianchi 
models and Gowdy solutions
are mentioned, and their significance in the development of
general relativity is indicated in the last section.
Special attention is paid to their role in understanding 
the behaviour of a general model near an initial singularity.

In the concluding remarks, several important, in particular
{\it non}-vacuum solutions, which have not been included in the main
body of the paper, are at least listed, together with some
relevant references. A few remarks on the possible future
role of exact solutions ends the article.

Although we give over 360 references in the bibliography, we do
not at all pretend to give all relevant citations. When
discussing more basic facts and concepts, we quote primarily
textbooks and monographs. Only when mentioning more recent
developments do we refer to
journals. The complete titles of all listed
references will hopefully offer the reader a more complete idea
of the role the explicit solutions have played on the
relativistic stage and in the astrophysical sky.
\newpage

%%%%%%%%%%%%%%%%%%%%%%%%%%%%%%%%%%%%%%%%%%%%%%%%%%%%%%%%%%%%%%%%%%%%%
%%%%%%%%\input schwarz.tex
%%%%%%%%%%%%%%%%%%%%%%%%%%%%%%%%%%%%%%%%%%%%%%%%%%%%%%%%%%%%%%%%%%%%%
\section{The Schwarzschild solution}

In his thorough ``Survey of General Relativity Theory'' \cite{JE},
J\"{u}rgen Ehlers begins with an empirical motivation of the
theory, goes in depth and detail through his favourite topics
such as the axiomatic approach, kinetic theory, geometrical optics,
approximation methods, and only in the last section turns to
spherically symmetric spacetimes. As T. S. Eliot says, ``to make
an end is to
make a beginning -- the end is where we start from'', and so here
we start with a few remarks on spherical symmetry.

\subsection{Spherically symmetric spacetimes}

In the early days of general relativity spherical symmetry was
introduced in an intuitive manner. It is because of
the existence of exact solutions which are singular at their
centres (such as the Schwarzschild or the Reissner-Nordstr\"{o}m
solutions), and a realization that spherically symmetric,
topologically non-trivial smooth spacetimes without any centre
may exist \cite{Kun}, that today the group-theoretical definition of
spherical symmetry is preferred (for a detailed analysis,
see e.g. \cite{Wa,HE,JE}).

Following Ehlers \cite{JE}, we define a spacetime $({\cal {M}}, g_{\alpha
\beta})$ to be spherically symmetric if the rotation group $SO_{3}$
acts on $({\cal {M}}, g_{\alpha \beta})$ as an isometry group with simply
connected, complete, spacelike, 2-dimensional orbits. One can then
prove the theorem \cite{JE,Schm} that a spherically
symmetric spacetime is the direct product ${\cal {M}}=S^2\times N$, where $S^2$
is the 2-sphere manifold with the standard metric $g_S$ on the unit sphere;
and $N$ is a 2-dimensional manifold with a Lorentzian (indefinite) metric
$g_N$, and with a scalar $r$ such that the complete spacetime metric
$g_{\alpha \beta}$ is ``conformally decomposable'', i.e.
$r^{-2}g_{\alpha \beta}$ is the direct sum of the 2-dimensional parts
$g_N$ and $g_S$. Leaving further technicalities aside (see e.g.
\cite{HE,JE,Schm}) we write down the final spherically
symmetric line element in the form

\setcounter{equation}{0}
\renewcommand{\theequation}{\arabic{equation}}
\begin{equation}
\label{Equ1}
ds^2 = -e^{2\phi} dt^2 + e^{2\lambda}dr^2 + r^2(d \theta ^2 +
\sin^2 \theta ~d \varphi ^2),
\end{equation}
where (following \cite{JE}) we permit $\phi(r,t)$ and
$\lambda (r,t)$ to have an imaginary part $i \pi/2$ so that the
signs of $dt^2$ and $dr^2$ in (\ref{Equ1}), and thus the role of $r$
and $t$ as space- and time- coordinates may interchange (a
lesson learned from the vacuum Schwarzschild solutions -- see
below). The ``curvature coordinate'' $r$ is defined invariantly by
the area, $4\pi r^2$, of the 2-spheres $r = \const$, $t = \const$.
There is no a
priori relation between $r$ and the proper distance from the
centre (if there is one) to the spherical surface.

\subsection{The Schwarzschild metric and its role in the solar
system}

Starting from the line element (\ref{Equ1}) and imposing 
Einstein's {\it vacuum} field equations, but allowing spacetime to be
in general dynamical, we are led uniquely (cf. Birkhoff's theorem
discussed e.g. in \cite{MTW,HE}) to the Schwarzschild
metric
\begin{equation}
\label{Equ2}
ds^2 = - \left(1-\frac{2M}{r}\right) dt^2 +
{\left(1-\frac{2M}{r}\right)}^{-1} dr^2 +
r^2\left(d\theta ^2 + \sin ^2 \theta~d \varphi ^2\right),
\end{equation}
where $M = \const$ has to be interpreted as a mass, as test particle
orbits show. The resulting spacetime is static at $r>2M$ (no
spherically symmetric gravitational waves exist), and
asymptotically flat at $r \rightarrow \infty$.

Undoubtedly, the Schwarzschild solution, describing the exterior
gravitational field of an arbitrary -- static, oscillating,
collapsing or expanding -- spherically symmetric body of
(Schwarzschild) mass $M$, is among the most influential solutions
of the gravitational field equations, if not of any type of field
equations invented in the 20th century. It is the first
exact solution of Einstein's equations obtained -- by K.
Schwarzschild in December 1915, still before Einstein's theory
reached its definitive form and, independently, in May 1916, by J.
Droste, a Dutch student of H. A. Lorentz 
(see \cite{IS} for comprehensive survey).

However, in its exact form (involving regions near $r \approx
2M)$ the metric (\ref{Equ2}) has not yet been experimentally
tested (a more optimistic recent suggestion will be mentioned in
Section 2.6). When in 1915 Einstein explained the perihelion
advance of Mercury, he found and used only an approximate 
(to second order in the gravitational potential) spherically
symmetric solution. In order to find the value of the deflection
of light passing close to the surface of the Sun, in his famous
1911 Prague paper, Einstein used just the equivalence principle
within his ``Prague gravity theory'', based on the variable
velocity of light. Then, in 1915, he obtained this value to
be twice as big in general relativity, when, in addition to the
equivalence principle, the curvature of space (determined from
(\ref{Equ2}) to first order in $M/r$) was taken into account.

Despite the fact that for the purpose of solar-system observations
the Schwarzschild metric in the form (\ref{Equ2}) is, quoting \cite{MTW},
``too accurate'', it has played an important role in experimental
relativity. Eddington, Robertson and others introduced the method
of expanding the Schwarzschild metric at the order beyond Newtonian
theory, and then multiplying each post-Newtonian term by a
dimensionless parameter which should be determined by experiment.
These methods inspired the much more powerful {\it PPN (``Para\-metrized
post-Newtonian'') formalism} which was developed at the end of
the 1960s and the beginning of the 1970s for testing general relativity and
alternative theories of gravity. It has been very effectively used to
compare general relativity with observations (see e.g.
\cite{MTW,CW,Will} and references therein). In
order to gain at least some concrete idea, let us just write down
the simplest generalization of (\ref{Equ2}), namely the metric
\begin{eqnarray}
\label{Equ3}
\!\!\!\!\!ds^2 = -\left[1-\frac{2M}{r} \right. &+& \left. 2\left(\beta  - 
\gamma\right)\frac{M^2}{r^2}\right] dt^2 +
\left(1+2 \gamma \frac{M}{r}\right)dr^2 \nonumber \\ 
~ & \lefteqn{+ r^2\left(d\theta^2 + \sin^2\theta~d \varphi^2\right)~,} & ~
\end{eqnarray}
which is obtained by expanding the metric (\ref{Equ2}) in $M/r$ up
to one order beyond the Newtonian approximation, and
multiplying each post-Newtonian term by dimensionless parameters
which distinguish the post-Newtonian limits of different metric
theories of gravity, and should be determined experimentally. (In
general, one needs not just two but ten PPN parameters
\cite{MTW,CW,Will}.) In Einstein's theory: $\beta
= \gamma = 1$. Calculating from metric (\ref{Equ3}) the 
advance of the pericentre of a test particle 
orbiting a central mass $M$ on an
ellipse with semi-major axis $a$ and eccentricity $e$, one
finds $\Delta \phi = \frac{1}{3}(2+2 \gamma - \beta) 6 \pi M/
[a(1-e^2)]$, whereas the total deflection angle of electromagnetic
waves passing close to the surface of the body is $\Delta \psi =
2(1+\gamma)M/r_0 $, where $r_0$ is the radius of closest approach
of photons to the central body.

Measurements of the deflection of radio waves and microwaves by
the Sun (recently also of radio waves by Jupiter) at present
restrict $\gamma$ to $\frac{1}{2}(1+\gamma) = 1.0001\pm 0.001$
\cite{CW,Will}. Planetary radar rangings, mainly to
Mercury, give from the perihelion shift measurements the
result $(2\gamma + 2 - \beta)/3 = 1.00 \pm 0.002$, so that $\beta
= 1.000 \pm 0.003$, whereas the measurements of periastron advance
for the binary pulsar systems such as PSR 1913+16 implied 
agreement with Einstein's theory to better than about 1\% (see e.g.
\cite{CW,Will} for reviews). There are other solar-system
experiments verifying the leading orders of the Schwarzschild
solution to a high accuracy, such as gravitational redshift,
signal retardation, or lunar geodesic precession. A number
of advanced space missions have been proposed which could lead to
significant improvements in values of the PPN parameters, and even
to the measurements of post-post-Newtonian effects \cite{Will}.

Hence, though in an approximate form, the Schwarzschild solution
has had a great impact on {\it experimental relativity}.
In addition, the
observational effects of gravity on light propagation in
the solar system, and also today routine observations of
gravitational lenses in cosmological contexts \cite{SE},
have significantly increased
our confidence in taking seriously similar predictions of
general relativity in more extreme conditions.

\subsection{Schwarzschild metric outside a collapsing star}

I recall how Roger Penrose, at the beginning of his
lecture at the 1974 Erice Summer School on gravitational
collapse, placed two figures side by side. The first illustrated
schematically the bending of light rays by the Sun (surprisingly, Penrose
did not write
``Prague 1911'' below the figure). I do not
remember exactly his second figure but it was similar
to Fig. 1 below: the spacetime diagram showing spherical
gravitational collapse through the Schwarzschild radius into a
spherical black hole.

\begin{figure}
\centering
\includegraphics[width=.7\textwidth]{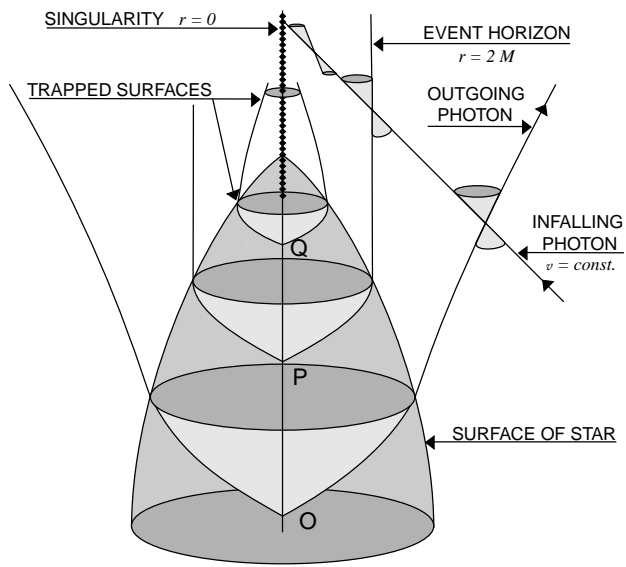}
\label{Figure 1}
\caption{
The gravitational collapse of a spherical star (the interior 
of the star is shaded). The light cones of the three events, 
$O$, $P$, $Q$, at the centre of the star, and of the three events
outside the star are illustrated. The event horizon, 
the trapped surfaces, and the singularity formed 
during the collapse are also shown. Although the singularity appears 
to lie in a ``time direction'', from the character of the 
light cone outside the star but inside
the event horizon it is seen that it has a spacelike character.
}
\end{figure}

It is in all modern books on general relativity that the
Schwarzschild radius at $R_s = 2M$ is the place where
Schwarzschild coordinates $t,r$ are unsuitable, and that metric (\ref{Equ2})
has a coordinate singularity but not a physical one. One has to
introduce other coordinates to extend the Schwarzschild metric
through $R_s$. In order to describe all spacetime outside a
collapsing spherical body it is advantageous to use ingoing
Eddington-Finkelstein coordinates $(v, r, \theta, \varphi)$ where
$ v = t + r + 2M \log (r/2M - 1)$. Metric (\ref{Equ2}) takes the form
\begin{equation}
\label{Equ4}
ds^2 = -\left(1-\frac{2M}{r}\right)dv^2 + 2dvdr + r^2\left(d \theta^2 +
\sin^2 \theta ~d \varphi^2\right),
\end{equation}
$(v, \theta, \varphi) = \const$ are ingoing radial null
geodesics. Fig. 1, plotted in these coordinates, demonstrates
well several basic concepts and facts which were introduced and
learned after the end of 1950s when a more complete understanding
of the Schwarzschild solution was gradually achieved. The
metric (\ref{Equ4}) holds only outside the star, there will be another
metric in its interior, for example the Oppenheimer-Snyder collapsing
dust solution (i.e. a portion of a collapsing Friedmann universe),
but the precise form of the interior solution is not
important at the moment. Consider a series of flashes of light
emitted from the centre of the star at events $O,P,\;Q$ (see Fig.
1) and assume that the stellar material is transparent. As the
Sun has a focusing effect on the light rays, so does matter during
collapse. As the matter density becomes higher and higher, the
focusing effect increases. At event $P$ a special wavefront will
start to propagate, the rays of which will emerge from the surface
of the star with zero divergence, i.e. the null vector $k^{\alpha} =
dx^{\alpha}/ dw, \,\,\, w$ being an affine parameter, tangent to null
geodesics, satisfies $k^{\alpha}_{; \alpha} = 0$. The wave\-front
then ``stays'' at the hypersurface $r=2M$ in metric (\ref{Equ4}), and the
area of its 2-dimensional cross-section remains constant. The
null hypersurface representing the history of this critical
wavefront is the (future) {\it event horizon}. Note that the light
cones turn more and more inwards as the event horizon is
approached. They become tangential to the horizon in such a way
that radial outgoing photons stay at $r = 2M$ whereas ingoing
photons fall inwards, and will eventually reach the curvature singularity
at $r=0$. As Fig. 1 indicates, wavefronts emitted still later
than the critical one, as for example that emitted from event $Q$, will
be focused so strongly that their rays will start to converge,
and will form (closed) {\it trapped surfaces}. The light cones at
trapped surfaces are so turned inwards that both ingoing and
outgoing radial rays converge, and their area decreases.

Consider a family of spacelike hypersurfaces $\Sigma (\tau)$
foliating spacetime ($\tau$ is a time coordinate, e.g. $v-r$).
The boundary of the region of $\Sigma (\tau)$ which contains
trapped surfaces lying in $\Sigma (\tau)$ is called the {\it apparent
horizon} in $\Sigma(\tau)$.

\begin{figure}[b]
\centering
\includegraphics[width=.71\textwidth]{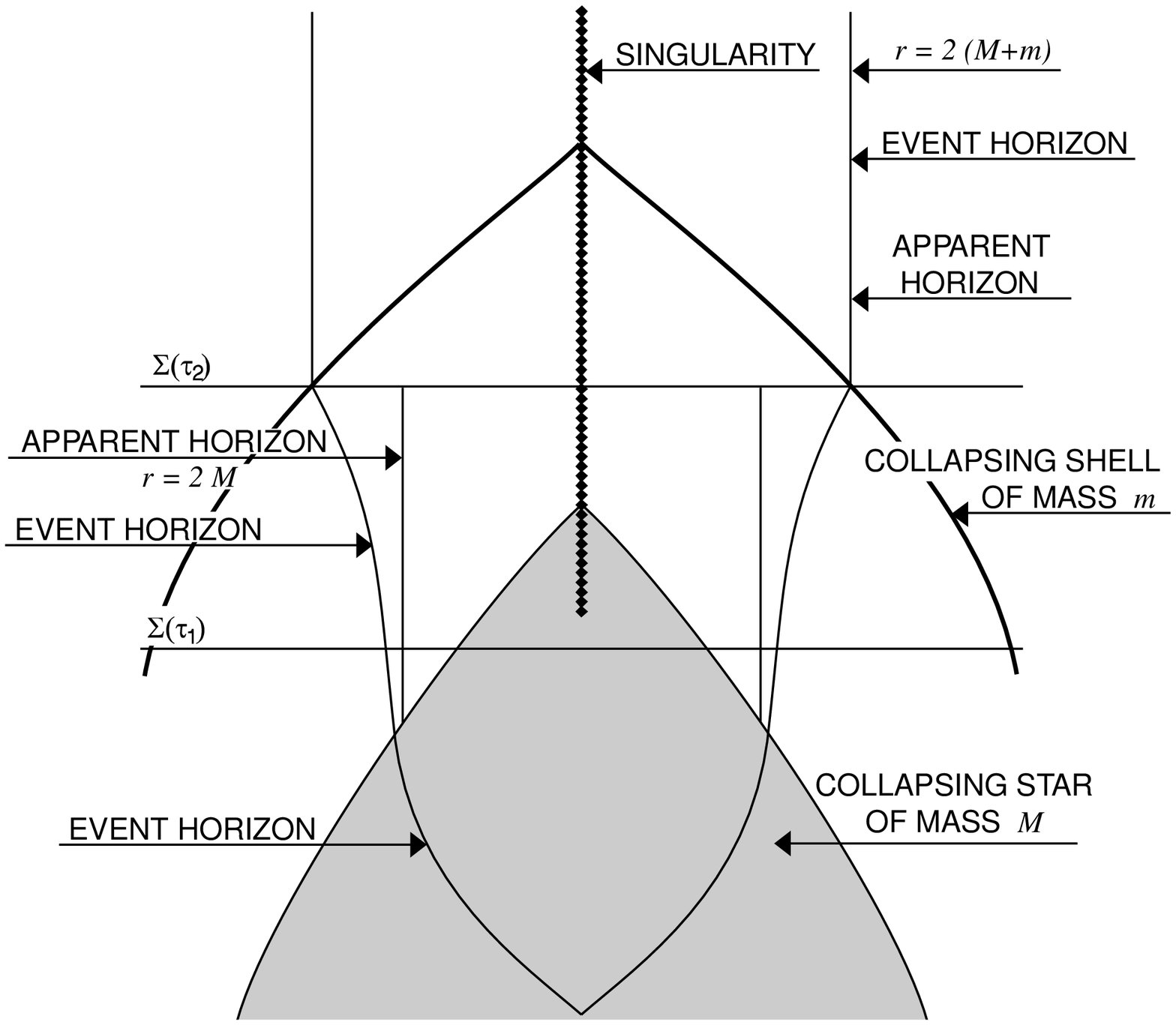}
\label{Figure 2}
\caption{The ``teleological'' behaviour of the event horizon during 
the gravitational collapse of a star, followed by the collapse of a 
shell. The event horizon moves outwards because it will be 
crossed by the shell. The apparent horizon moves outwards discontinuously
(adapted from \cite{Haw}).
}
\end{figure}

In general, the apparent horizon is
different from the intersection of the event horizon with
$\Sigma(\tau)$, as a nice simple example (based again on 
an exact solution) due to Hawking \cite{Haw} shows. 
Assume that after the spherical collapse of a star a
spherical thin shell of mass $m$ surrounding the star collapses
and eventually crashes at the singularity at $r=0$ (Fig. 2). In
the vacuum region inside the shell there is the Schwarzschild
metric (\ref{Equ4}) with mass $M$, and outside the shell with 
mass $M + m$. Hence the apparent horizon on $\Sigma(\tau_1)$ will be at $r=2M$
and will remain there until $\Sigma (\tau_2)$ when it
discontinuously jumps to $r=2(M+m)$. One can determine the
apparent horizon on a given hypersurface. In order to find the event
horizon one has to know the whole spacetime solution. The future
event horizon separates events which are visible 
from future infinity, from those which are not, 
and thus forms the boundary of a {\it black hole}.

From the above example of a shell collapsing onto a Schwarzschild
black hole we can also learn about the ``teleological'' nature of the
horizon: the motion of the horizon depends on what will happen to
the horizon in the future (whether a collapsing shell will cross
it or not). This {\it teleological behaviour of the horizon} has
later been discovered in a variety of astrophysically realistic
situations such as the behaviour of a horizon perturbed by a mass
orbiting a black hole (see \cite{TP} for enlightening discussions
of such effects).

By studying the Schwarzschild solution and spherical collapse it
became evident that one has to turn to {\it global methods} to
gain a full understanding of general relativity. The intuition
acquired from analyzing the Schwarzschild metric helped crucially
in defining and understanding such concepts as the trapped
surface, the event horizon, or the apparent horizon in general
situations without symmetry. Nowadays these concepts are
explained in several advanced textbooks and monographs (e.g.
\cite{MTW,Wa,HE,Jo,FN}).

Following from the example of spherical collapse one is led to
ask whether generic gravitational collapses lead to spacetime
singularities and whether these are always surrounded by an event
horizon. The Penrose-Hawking singularity theorems 
\cite{Wa,HE} show that singularities do arise under quite generic
circumstances (the occurrence of a closed trapped surface is most
significant for the appearance of a singularity).
The second question is the essence of the
cosmic censorship hypothesis. Various exact solutions have
played a role in attempts to ``prove'' or ``disprove'' this ``one
of the most important issues'' of classical relativity. We shall
meet it in several other places later on, in particular in Section 3.1.
There a more detailed formulation is given.

\subsection{The Schwarzschild-Kruskal spacetime}

In the remarks above we considered the Schwarzschild solution
outside a static (possibly oscillating, or expanding from $r>2M$)
star, and outside a star collapsing into a black hole.
It is not excluded that just these situations
will turn out to be physically relevant. Nevertheless, in
connection with the Schwarzschild metric it would be heretical not
to mention the enormous impact which its maximal vacuum
analytic extension into the Schwarzschild-Kruskal spacetime has had.
This is today described in detail in many places (see e.g. 
\cite{MTW,Wa,HE,FN}). We need two sets of the
Schwarzschild coordinates to cover the complete spacetime, and we
obtain two asymptotically flat spaces, i.e. the spacetime with
two (``right'' and ``left'') infinities. The metric in Kruskal
coordinates $U, V$, related to the Schwarzschild $r,t$ (in
the regions with $r>2M$) by
\begin{eqnarray}
U & = & \pm{\left(r/2M-1\right)}^{1/2} e^{r/4M} \cosh
\left(t/4M\right), \nonumber \\
V & = & \pm{\left(r/2M-1\right)}^{1/2} e^{r/4M} \sinh
\left(t/4M\right),
\end{eqnarray}
takes the form
\begin{equation}
\label{Equ6}
ds^2 = \frac{32M^3}{r} e^{-r/2M} \left(-dV^2 + dU^2\right) +
r^2\left(d\theta ^2
+ \sin ^2 \theta~d \varphi^2\right).
\end{equation}
The introduction of the Kruskal coordinates which remove the
singularity of the Schwarzschild metric (\ref{Equ2}) at the horizon
$r=2M$ and cover the complete spacetime manifold (every geodesic
either hits the singularity or can be continued to the infinite
values of its affine parameter), was the most influential example
which showed that one has to distinguish carefully between just a
coordinate singularity and the real, physical singularity. It
also helped us to realize that the definition of a singularity itself
is a subtle issue in which the concept of geodesic
completeness plays a significant role (see \cite{Cla} for a
recent analysis of spacetime singularities).

The character of the Schwarzschild-Kruskal spacetime is best seen
in the Penrose diagram given in Fig. 3, in which the spacetime
is compactified by a suitable conformal rescaling of the metric. Both
right and left infinities are represented, and the causal structure is
well illustrated because worldlines of radial light signals
(radial null geodesics) are 45-degree lines in the diagram. In
particular the black hole region {\it II} and a ``newly emerged'' (as a
consequence of the analytical continuation) {\it white hole}
region {\it IV} (with the white-hole singularity at $r=0$) are exhibited.
For more detailed analyses of the Penrose diagram of the
Schwarzschild-Kruskal spacetime the reader is referred to e.g.
\cite{MTW,Wa,HE,FN}. Here we wish to turn
in some detail to two very important concepts in black hole
theory which were first understood by the analytic extension of
the Schwarzschild solution, and which are not often treated in
standard textbooks. These are the concepts of the {\it bifurcate
horizon} and of the {\it horizon surface gravity}. 
J\"{u}rgen Ehlers played a somewhat indirect, but important 
and noble part in their introduction into literature.

\begin{figure}
\centering
\includegraphics[width=.7\textwidth]{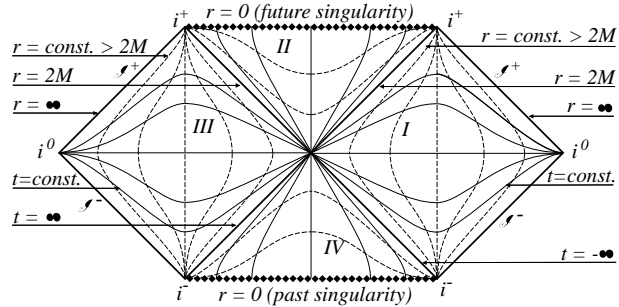}
\label{Figure 3}
\caption{The Penrose diagram of the compactified Schwarzschild-Kruskal spacetime.
Radial null geodesics are 45-degrees lines.
Timelike geodesics reach the future (or past) timelike infinities $i^+$ (or $i^-$),
null geodesics reach the future (or past) null infinities $\cal J^+$ (or $\cal J^-$) 
and spacelike geodesics lead to spatial infinities $i^0$.
(Notice that at $i^0$ the lines $t=\const$ are tangent to each other 
-- this is often not taken into account in the literature 
-- see e.g. \cite{HE,DIN}.)}
\end{figure}

These concepts were the main subject of the last work of Robert
Boyer who became one of the victims of a mass murder on August 1,
1966, in Austin, Texas. J\"{u}rgen Ehlers was authorized by Mrs.
Boyer to look through the scientific papers of her husband, and
together with John Stachel, prepared posthumously the paper
\cite{BO} from R. Boyer's
notes. Ehlers inserted his own discussions, generalized the
main theorem on bifurcate horizons, but the paper \cite{BO}
was published with R. Boyer as the only author.

In the Schwarzschild spacetime there exists the timelike Killing vector,
$\partial / \partial t$, which when analytically extended into all
Schwarzschild-Kruskal manifold, becomes null at the event horizon $r = 2M$,
and is spacelike in the regions
$II$ and $IV$ with $r < 2M$. In Kruskal coordinates it is given by
\begin{equation}
k^\alpha = \left(k^V = U/4M,\,\,\, k^U = V/4M,
\,\,\, k^\theta = 0, \,\,\, k^\varphi = 0\right).
\end{equation}
Hence it vanishes at all points with $U = V = 0,\; \theta \in [0,
\pi], \varphi \in [0, 2 \pi)$.
These points, forming a spacelike 2-sphere 
which we denote $B$ (in Schwarzschild coordinates given
by $r = 2M,\,\, t = \const$), are fixed points of the 1-dimensional group $G$ of
isometries generated by $k^\alpha$ (see Fig. 3). At the event horizon the
corresponding 1-dimensional orbits are null geodesics, with $k^\alpha$
being a tangent vector.
However, since $k^\alpha$ vanishes at $B$, these orbits are incomplete.

This (and similar observations for other black hole solutions) motivated
a general analysis of the bifurcate Killing horizons given in
\cite{BO}. There it
is proven for spacetimes admitting a general Killing vector
field $\xi ^\alpha$, which
generates a 1-dimensional group of isometries, that (i) a 1-dimensional
orbit is a complete geodesic if the gradient of the square $\xi ^2$
vanishes on the orbit,
(ii) if a geodesic orbit is incomplete, then it is null and
$(\xi ^2)_{,\alpha} \not= 0$. In addition, if $\xi ^\alpha = {dx^\alpha / dv }$
($v$ being the group parameter), the affine parameter along the geodesic is
$w = e^{\kappa v}$, where $\kappa = \const$ satisfies
\begin{equation}
\label{Equ8}
(-\xi ^2)^{,\alpha} = 2 \kappa \xi ^\alpha.
\end{equation}

In the Schwarzschild case, with 
$\xi^\alpha = k^\alpha = (\partial / \partial t)^\alpha$, and considering the part
$V = U$ of the horizon, we get $\kappa = 1/4M$. The relation $w = e^{\kappa v}$ is just
the familiar equation
$\tilde{V} = e^{v/4M}$, where $\tilde{V} = V + U$
is the Kruskal null coordinate and
$v$ is the Eddington-Finkelstein ingoing null coordinate used in
(\ref{Equ4}). (Notice that $\tilde{V}$ is indeed the affine parameter along the null geodesics
at the horizon $V = U$.) The quantity $\kappa$, first introduced in
\cite{BO}, has become
fundamental in modern black hole theory, and also in its generalizations in
string theory. It is the well-known {\it surface gravity} of the black hole horizon.

With $\kappa \not= 0$, the limit points corresponding to
$v \rightarrow - \infty, w = 0$
are fixed  points of $G$.
(Unless the spacetime is incomplete, there exists a continuation
of each null geodesic beyond these fixed points to $w < 0$.) One can show that
the fixed points form a spacelike 2-dimensional manifold $B$,
given by $U = V = 0$
in the Schwarzschild case; this ``bifurcation surface'' is a
totally geodesic submanifold. By the original definition
\cite{CA}, a Killing horizon
is a $G$ invariant null hypersurface $N$ on which $\xi ^2 = 0$.
(A recent definition
\cite{PC,HEU} specifies a Killing horizon to be any union of such hypersurfaces.) If
$\kappa \not= 0$, at each point of $B$ there is one null direction orthogonal to $B$
which is not tangent to $\bar{N} = N \cup B$. The null geodesics intersecting $B$ in
these directions form another null hypersurface, $\tilde{N}$, which is also a Killing
horizon. The union $N \cup \tilde{N}$
is called a {\it bifurcate Killing horizon} (Fig. 4).

\begin{figure}
\centering
\includegraphics[width=.52\textwidth]{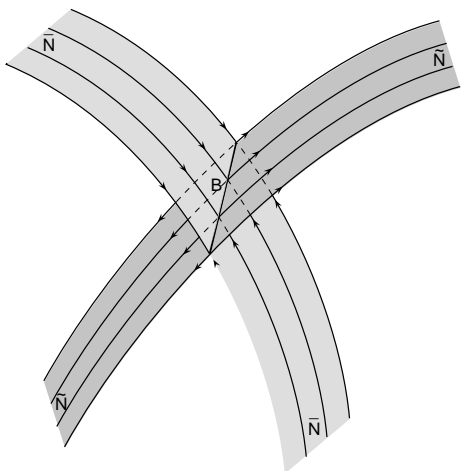}
\label{Figure 4}
\caption{
The bifurcate Killing horizon consisting of two null hypersurfaces
$\tilde N$ and $\bar N$ which intersect in the spacelike 
2-dimensional ``bifurcation surface'' $B$.
}
\end{figure}

Bifurcate Killing horizons exist also in flat and other curved spacetimes.
For example, in the boost-rotation symmetric spacetimes (Section
11), null
hypersurfaces $z = \pm t$ form the bifurcate Killing horizon corresponding to
the boost Killing vector; $B$, given by $z = t = 0$, is then not compact.
(As
in the Schwarzschild-Kruskal spacetime, a bifurcate Killing horizon locally
divides the spacetime into four wedges.) However, the first motivation for
analyzing Killing horizons came from the black hole solutions.

Both Killing horizons and surface gravity play an important role in
{\it black hole thermodynamics} and {\it quantum field theory on curved
backgrounds}
\cite{RW}, in particular in their two principal results: the {\it Hawking
effect} of particle
creation by black holes; and the {\it Unruh effect} showing that a thermal bath
of particles will be seen also by a uniformly accelerated observer in flat spacetime
when the quantum field is in its vacuum state with respect to inertial
observers. Recently, new results were obtained \cite{RAW} which support
the view that
a spacetime representing the final state of a black hole formed by collapse has
indeed a bifurcate Killing horizon, or the Killing horizon is degenerate ($\kappa = 0$).

\subsection{The Schwarzschild metric as a case 
            against Lorentz-covariant approaches}

There are many other issues on which the Schwarzschild solution has made an impact.
Some of astrophysical applications will be very 
briefly mentioned later on. As the last theoretical
point in this section I would like to discuss in some detail the causal structure
of the Schwarzschild spacetime including infinity. By analyzing this structure,
Penrose \cite{RPT} presented evidence against various Lorentz
(Poincar\'e)-covariant field theoretical approaches, which regard the physical metric tensor $g$
to be not much different from any other tensor in Minkowski spacetime with
flat metric $\eta$ (see e.g. \cite{WB,Ro}). I thought it appropriate to mention this point here, since J\"{u}rgen Ehlers, among others,
certainly does not share a field theoretical viewpoint.

The normal procedure of calculating the metric $g$ in these approaches is from
a power series expansion of Lorentz-covariant terms (in quantum theory this
corresponds to an infinite summation of Feynman diagrams). The derived field
propagation has to follow the true null cones of the curved metric $g$ instead
of those of $\eta$. However, as Penrose shows, in a satisfactory theory the
null cones defined by $g$ should not extend outside the null cones defined
by $\eta$, or ``the causality defined by $g$ should not violate the
background $\eta$-causality''. Following \cite{RPT}, let us write this condition as $g < \eta$. Now at first
sight we may believe that $g < \eta$ is satisfied in the Schwarzschild field since
its effect is to ``slow down'' the velocity of light (cf. ``signal retardation''
mentioned in 2.2). However, in the field-theoretical approaches one of the
main emphasis is in a consistent formulation of scattering theory. This
requires a good behaviour at infinity. But with the Schwarzschild metric, null
geodesics with respect to metric $g$ ``infinitely deviate'' from those with
respect to $\eta$: for example, the radial outgoing $g$ null geodesics
$\theta, \varphi = \const$, and $u = t - r -2M\log (r/2M - 1) = \const$ 
at $r \rightarrow \infty$ go ``indefinitely far'' into the
retarded time $t - r$ of $\eta$, and hence, do not correspond to
outgoing $\eta$-null geodesics $t-r = \const$. One can try to use
a different flat metric associated with the Schwarzschild metric $g$
which does not lead to pathological behaviour at infinity, but
then it turns out that $g<\eta$ is violated locally. In fact,
Penrose \cite{RPT} proves the theorem, showing that there is an
essential incompatibility between the causal structures in the
Schwarzschild and Minkowski spacetimes which appears either
asymptotically or locally.

This incompatibility is easily understood with the exact
Schwarzschild solution, but it is generic, since one is concerned
only with the behaviour of the space at large distances from a
positive-mass source, i.e. with the causal properties in the
neighbourhood of spacelike infinity $i^0$.

In the present post-Minkowskian approximation methods for the
generation of gravitational waves by relativistic sources, a
suitable (Bondi-type) coordinate system \cite{e}
is constructed at all orders in the far wave zone,
which in particular corrects for the logarithmic deviation of the
true light cones with respect to the coordinate flat light cones
(cf. contribution by L. Blanchet in this volume).

\subsection{The Schwarzschild metric and astrophysics}

In his introductory chapter ``General Relativity as a Tool for
Astrophysics'' for the Seminar in Bad Honnef in 1996 \cite{JRG},
J\"{u}rgen Ehlers remarks that ``The interest of black holes for
astrophysics is obvious... The challenge here is to find
observable features that are truly relativistic, related, for example, to
horizons, ergoregions... Indications exist, but -- as far as I am
aware -- no firm evidence.''

There are many excellent recent reviews on the astrophysical evidence
for black holes (see e.g. \cite{Re,Na,Ca}). It is true, 
that the evidence points towards the presence of dark massive
objects -- stellar-mass objects in binaries, and supermassive
objects in the centres of galaxies -- which are associated with
deep gravitational potential wells where Newtonian gravity cannot
be used, but it does not offer a clear diagnostic of general
relativity.

Many investigations of test particle orbits in the strong-gravity
regions $(r \leq 10M)$
have shown basic differences between the motion in the
Schwarzschild metric and the motion in the central field in
Newton's theory (e.g. \cite{MTW,FN,Chan}).
For example for $3M<r<6M$ unstable circular
particle orbits exist which are energetically unbound, and thus
perturbed particles may escape to infinity; at $r = 3M$ circular
photon orbits occur and there are no circular orbits for $r<3M$.
Particles are trapped by a Schwarzschild black hole if they reach
the region $r<3M$.

About ten years ago, the study of the {\it behaviour of particles and
gyroscopes in the Schwarzschild field} revived interest in
the ``classical'' problem of the definition of gravitational,
centrifugal, and other inertial ``forces'' acting on particles
and gyros moving on the Schwarzschild or on a more general
curved backgrounds, usually axisymmetric and stationary (see e.g.
\cite{Abr,Se}, and many references therein). One would
like to have a split of a covariantly defined quantity
(like an acceleration) into non-covariant parts,
the physical meaning of which would increase our intuition of
relativistic effects in
astrophysical problems. If, for example, we adopt the view that
the ``gravitational force'' is velocity-independent, then we
find that at the orbits outside the circular photon orbit $(r>3M)$,
the centrifugal force is as in classical physics, repulsive,
while it becomes attractive inside this orbit, being zero
exactly at the orbit.\footnote{Curiously enough, Feynman in his 1962-63 lectures on
gravitation \cite{Fey} writes that ``inside $r=2M$ [not
$3M!$]... the `centrifugal force' apparently acts as an
attraction rather than a repulsion''.}

Relativistic effects will, of course, play a role in many
astrophysical situations involving spherical accretion, the
structure of accretion disks around compact stars and black
holes, their optical appearance etc. They have become an
important part of the arsenal of astrophysicists, and they have entered
standard literature (see e.g. \cite{TeSh,Fr}). Though
this whole field of science lies beyond the scope of this article,
I would
like to mention three recent issues which provide us with hope
that we may perhaps soon meet the challenge
noted in J\"{u}rgen Ehlers' remarks made in Bad Honnef in 1996.

The first concerns our Galactic centre. Thanks to new
observations of stars in the near infrared band it was possible
to detect the transverse motions of stars (for which the radial
velocities are also observed) within 0.1 pc in our Galactic
centre. The stellar velocities up to $2000$ km/sec and their dependence
on the radial distance from the centre are consistent with a
black hole of mass $2.5\times10^6 M_{\odot}$. In the opinion of some
leading astrophysicists, our Galactic centre now provides ``the
most convincing case for a supermassive hole, with the single
exception of NGC 4258'' \cite{Re}. (In NGC 4258 a disk is
observed whose inner edge is orbiting at 1080 km/sec, implying a
black hole -- ``or something more exotic'' \cite{Re} -- with a
mass of $3.6 \times 10^7 M_{\odot}$.) Perhaps we shall be able to
observe relativistic effects on the proper motions of stars in our
Galactic centre in the not too distant future.

The second issue concerns the fundamental question of whether
observations can bring convincing proof of the existence of
black hole event horizons. Very recently some astrophysicists
\cite{Na} claimed that new observations, in particular of X-ray
binaries, imply such evidence. The idea is that thin disk
accretion cannot explain the spectra of some of X-ray binaries. One
has to use a different accretion model, a so called
advection-dominated accretion flow model (ADAF) in which most of the
gravitational energy released in the infalling gas is carried
(advected) with the flow as thermal energy, which falls on the central
object. (In thin disks most of this energy is radiated out from
the disk.) If the central compact object (for example a neutron star)
has a hard surface, the thermal energy stored in the flow is
re-radiated after the flow hits the surface. However, some of the
X-ray binaries show such low luminosities that a very large
fraction of the energy in the flow must be advected through 
an event horizon into a black hole \cite{Na}. Although Rees
\cite{Re}, for example, considers this evidence ``gratifyingly
consistent with the high-mass objects in binaries being black
holes'', he believes that it ``would still not convince an
intelligent sceptic, who could postulate a different theory of
strong-field gravity or else that the high-mass compact objects
were (for instance) self-gravitating clusters of weakly interacting
particles...''.

For a sceptical optimistic relativist, the most challenging
observational issue related to black holes probably is to find
astrophysical evidence for a Kerr metric. We shall come to this
point in Section 4.3.

The last (but certainly not the least) issue lies more in the future,
but eventually should turn out to be most promising. It is
connected with both the Numerical Relativity Great Challenge
Alliance and the ``great challenge'' of experimental relativity:
to calculate reliable gravitational wave-forms and to detect
them. When gravitational waves from stars captured by a
supermassive black hole, or from a newly forming supermassive
black hole, or, most importantly, from coalescing supermassive
holes will be detected and compared with the predictions of the
theory, we should learn significant facts about black holes
\cite{Re,KST}. Are these so general remarks entirely inappropriate
in the section on the Schwarzschild solution?

One of the most important roles of the Schwarzschild solution in
the development of mathematical relativity and especially of
relativistic astrophysics stems from its simplicity, in particular
from its
spherical symmetry. This has enabled us to develop the
mathematically beautiful theory of linear perturbations of the
Schwarzschild background and employ it in various astrophysically
realistic situations (see e.g. \cite{TP,FN,Chan},
and many references therein). Surprisingly enough, this theory
does not only give reliable results in such problems as the
calculation of waves emitted by pulsating neutron stars, or
waves radiated out from stars falling into a supermassive black
hole. Very recently we have learned that one can use perturbation theory of a
single Schwarzschild black hole as a ``close approximation'' to black
hole collisions. Towards the end of the collision of two black holes,
they will not in fact be two black holes, but will merge into a
highly distorted single black hole \cite{Pull}. When compared with
the numerical results on a head-on collision it has been found
that this approximation gives predictions for separations $\Delta$
as large as $\Delta/M \sim 7$.

%%%%%%%%%%%%%%%%%%%%%%%%%%%%%%%%%%%%%%%%%%%%%%%%%%%%%%%%%%%%%%%%%%%%%
%%%%%%%%\input reiss.tex
%%%%%%%%%%%%%%%%%%%%%%%%%%%%%%%%%%%%%%%%%%%%%%%%%%%%%%%%%%%%%%%%%%%%%
\section{The Reissner-Nordstr\"{o}m solution}

This spherically symmetric solution of the Einstein-Maxwell equations was
derived independently\footnote{In the literature one finds the solution to be
repeatedly connected only with the names of Reissner and Nordstr\"{o}m,
except for the ``exact-solutions-book'' \cite{KSH}: there in
four places the solution is called as everywhere else, but in one
place (p. 257) it is referred to as the ``Reissner-Weyl-solutions''. An
enlightening discussion on p. 209 in \cite{KSH} shows that the
solution belongs to a more general ``Weyl's electrovacuum class''
of electrostatic solutions discovered by Weyl (in 1917) which
follow from an Ansatz that there is a functional relationship
between the gravitational and electrostatic potentials. As will be
noticed also in the case of cylindrical waves in Section 9, if
``too many'' solutions are given in one paper, the name of the
author is not likely to survive in the name of an important subclass...}
by H. Reissner in 1916, H. Weyl in 1917,
and G. Nordstr\"{o}m in 1918. It represents a spacetime with no
matter sources except for a radial electric field, the energy of
which has to be included on the right-hand side of the Einstein
equations. Since Birkhoff's theorem, mentioned in connection with
the Schwarzschild solution in Section 2.2, can be generalized to
the electrovacuum case, the  Reissner-Nordstr\"{o}m solution is the
unique spherical electrovacuum solution. Similarly to the
Schwarzschild solution, it thus describes the exterior
gravitational and electromagnetic fields of an arbitrary --
static, oscillating, collapsing or expanding -- spherically
symmetric, charged body of mass $M$ and charge $Q$. The metric
reads
\begin{eqnarray}
\label{Equ9}
 ds^2 =  &-& \left( 1- \frac{2M}{r} + \frac{Q^2}{r^2}\right) dt^2 +
\left( 1-\frac{2M}{r} + \frac{Q^2}{r^2} \right)^{-1} dr^2
\nonumber \\
~&\lefteqn{~~~~~~~+~r^2 \left( d \theta ^2 + 
\sin ^2 \theta ~d \varphi ^2 \right),}&~
\end{eqnarray}
the electromagnetic field in these spherical coordinates is
described by the ``classical'' expressions for the time component of
the electromagnetic potential and  the (only non-zero) component
of the electromagnetic field tensor:
\begin{equation}
\label{Equ10}
A_t = -\frac{Q}{r}, \,\,\, F_{tr} = - F_{rt} = -\frac{Q}{r^2}.
\end{equation}
A number of authors have discussed spherically symmetric, static
charged dust configurations producing a Reissner-Nordstr\"{o}m
metric outside, some of them with a hope to construct a
``classical model'' of a charged elementary particle (see
\cite{KSH} for references). The main influence the metric has
exerted on the developments of general relativity, and more
recently in supersymmetric and superstring theories (see Section
3.2), is however in its analytically extended electrovacuum form when
it represents charged, spherical black holes.

\subsection{Reissner-Nordstr\"{o}m black holes and the
question of cosmic censorship}

The analytic extensions have qualitatively different character in
three cases, depending on the relationship between the mass $M$
and the charge $Q$. In the case $Q^2>M^2$ (corresponding, for
example, to
the field outside an electron), the complete electrovacuum
spacetime is covered by the coordinates $(t, r, \theta,
\varphi)$, $0<r< \infty$. There is a {\it naked singularity}
(visible from infinity) at $r=0$ in which the curvature
invariants diverge. If $Q^2<M^2$, the metric (\ref{Equ9})
describes a (generic) {\it Reissner-Nordstr\"{o}m black hole}; it
becomes singular at two radii:
\begin{equation}
\label{Equ11}
r=r_{\pm}= M\pm(M^2 - Q^2)^{\frac{1}{2}}.
\end{equation}
Similarly to the Schwarzschild case, these are only coordinate
singularities. Graves and Brill \cite{GB} discovered, however,
that the analytic extension and the causal structure of the
Reissner-Nordstr\"{o}m spacetime with $M^2>Q^2$ is fundamentally
different from that of the Schwarzschild spacetime. There are two
null hypersurfaces, at $r=r_+$ and $r=r_-$,  which are known as
the {\it outer (event) horizon} and the {\it inner horizon}; the
Killing vector $\partial/\partial t$ is null at the horizons,
timelike at $r>r_+$ and $r<r_-$, but spacelike at $r_-<r<r_+$.
The character of the extended manifold is best seen in the
Penrose diagram in Fig. 5, in which the spacetime is
compactified by a suitable conformal rescaling of the metric (see, e.g.
\cite{MTW,HE,DIN}). As in the compactified Kruskal-Schwarzschild diagram
in Fig. 3, the causal structure is well illustrated because
worldlines of radial light signals are 45-degree lines. There are
again two infinities illustrated - the right and left - in
regions {\it I} and {\it III}. However, the maximally extended
Reissner-Nordstr\"{o}m geometry consists of an {\it infinite
chain of asymptotic regions} connected by ``wormholes'' between the
real singularities (with divergent curvature invariants) at $r=0$.
In Fig. 5, the right and left (past null) infinities in regions
$I'$ and $III'$ are still seen - the others are obtained by
extending the diagram vertically in both directions.

\begin{figure}
\centering
\includegraphics[width=.75\textwidth]{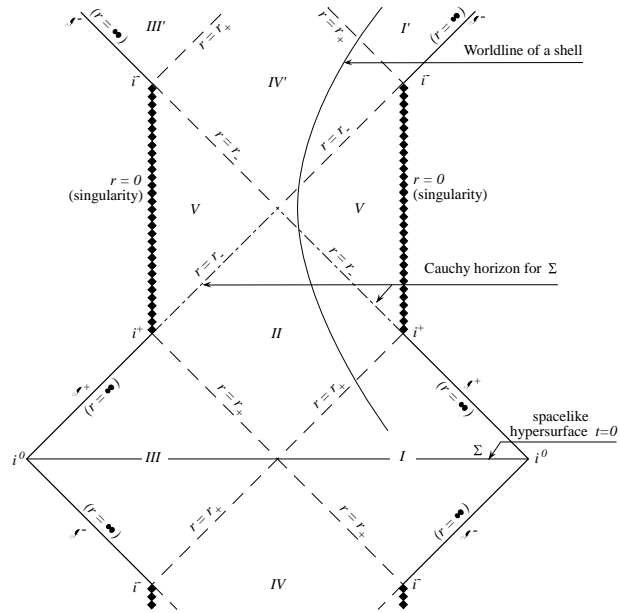}
\label{Figure 5}
\caption{
~~The compactified Reissner-Nordstr\"{o}m spacetime
representing a non-extreme black hole consists of
an indefinite chain of asymptotic regions (``universes'')
connected by ``wormholes'' between timelike singularities.
The worldline of a shell collapsing from ``universe'' $I$ and 
re-emerging in ``universe'' $I'$ is indicated. The inner horizon 
at $r=r_-$ is the Cauchy horizon for spacelike hypersurface $\Sigma$.
It is unstable and will thus very likely prevent such a process occuring.}
\end{figure}

An important lesson one has learned is that the character of the
singularity need not be spacelike as it is in the Schwarzschild
case, or with the big bang singularities in standard cosmological
models. Indeed, the {\it singularities} in the
Reissner-Nordstr\"{o}m geometry are {\it timelike}:  they do not
block the way to the future. By solving the geodesic equation one
can show that there are test particles which start in ``our
universe'' (region {\it I}), cross the outer horizon at $r=r_+$ and the
inner horizon at $r=r_-$, avoid the singularity and through a
``white hole'' (the outer horizon between regions $IV'$ and $I'$)
emerge into ``another universe'' $I'$ with its own asymptotically
flat region. Such a {\it gravitational bounce} can occur not only
with test particles. The studies of the gravitational collapse of
charged spherical shells (\cite{Bou} and references therein) and
of charged dust spheres (\cite{ZN} and references therein) have
shown that a bounce can take place also in fully dynamical 
cases.\footnote{An intuitive explanation \cite{ZN} of this bounce is
that as the sphere (the shell) contracts, the volume of the
exterior region increases, and hence also the total energy in
the electric field, which eventually exceeds the energy in the
sphere. However, the external plus internal energy does not
change during collapse (there are no waves), and so in the
neighborhood of a highly contracted charged object, the
gravitational field must have a repulsive character corresponding
to a negative mass-energy.}
The part of Fig. 5 which is ``left'' from the worldline of the
surface of the sphere or the shell is ``covered'' by the interior
of the sphere or flat space inside the shell.
As observed in \cite{Bou}, the outcome of the bounce of a shell
can be different, depending on the value of the shell's total mass,
charge and rest mass. The shell may crash into the ``right''
singularity or it may continue to expand and emerge in region $I'$.
If the rest mass of the shell is negative the collapse may even
lead to a naked singularity.

Now even if the shell collapses into a black hole, and after a
bounce, emerges in region $I'$, a {\it locally} naked singularity is
present: the timelike singularity at $r=0$, to the ``right'' from
the wordline of the shell. An observer travelling into the future
``between'' the shell and the singularity can be surprised by a
signal coming from the singularity (see Fig. 5). Penrose's {\it strong
cosmic censorship conjecture} (see e.g. \cite{ROG}) suggests that
this should not happen. In its physical formulation, as given by
Wald (see \cite{Wa} also for the precise formulation), it says
that all physically reasonable spacetimes are globally hyperbolic,
i.e. apart from a possible initial (big bang-type) singularity,
{\it no singularity is visible to any observer}. It was just the
example of the Reissner-Nordstr\"{o}m solution (and a similar
property of the Kerr solution) which inspired Penrose to
formulate the strong cosmic censorship conjecture, in addition to
its {\it weak version} which only requires that from generic
nonsingular initial data on a Cauchy hypersurface {\it no
spacetime singularity} develops which is {\it visible from
infinity}. As Penrose \cite{ROG} puts it, ``it seems to be
comparatively unimportant whether the observer himself can escape
to infinity''.

It is evident from Fig. 5 that the Reissner-Nordstr\"{o}m
spacetime is not globally hyperbolic, i.e. it does not possess a
Cauchy hypersurface $\Sigma$, the initial data on which (for a test
field, say) would determine the development of data in the entire
future. If data are given on the spacelike hypersurface $\Sigma$
``connecting'' left and right infinities of regions {\it III} and {\it I}, the
Cauchy development will predict what happens only in regions {\it III}
and {\it I} above $\Sigma$, and in region {\it II}, i.e. not beyond the null
hypersurfaces (inner horizons) $r=r_-$ between region {\it II} and
regions {\it V} in the figure. The inner horizons $r=r_-$ represent the
{\it Cauchy horizon} for a typical initial hypersurface like
$\Sigma$. As noticed above, what is happening at an event in
regions {\it V} is in general influenced not only by data on $\Sigma$
but also by what is happening at the (locally) naked
singularities (which cannot be predicted since the physics at a
singularity cannot be controlled).

Penrose was also the first who predicted that the {\it inner
(Cauchy) horizon is unstable} \cite{Pen}. If this is true, a null
singularity, or possibly even a spacelike singularity may arise
during a general collapse, so preventing a violation of the
strong cosmic censorship conjecture. The instability of the
Cauchy horizon can in fact be expected by using first the following
simple geometrical-optics argument.

Introduce the ingoing Eddington-Finkelstein null coordinate $v$
by
\begin{eqnarray}
\label{Equ12}
v = t + r^* = t + \int f(r) dr, \nonumber \\
f = 1 - 2M/r + Q^2/r^2,
\end{eqnarray}
which brings the metric (\ref{Equ9}) into the form as of
equation \hskip -0.3pt(\hskip -1pt \ref{Equ4}\hskip -0.3pt) in the 
Schwarzschild case. Consider a freely falling
observer who is approaching the inner horizon given by $r=r_-, v
= \infty$. Denoting the observer's constant specific
energy parameter (see e.g. \cite{MTW})
by $\tilde E = -{\vec {U \xi}}$, where the Killing
vector ${\vec \xi} = \partial/\partial v$, observer's four-velocity $\vec U = d/d\tau$
($\tau$ - observer's proper time), the geodesic equations
imply

\begin{equation}
\label{Equ13}
\dot r^2 + f = \tilde E^2, \,\,\,\, \dot v = f^{-1}[\tilde E -
(\tilde E^2 - f)^\frac{1}{2}],
\end{equation}
where a dot denotes $d/d\tau$. Between the horizons, $\vec \xi$
is spacelike and $\tilde E$ can be negative. Geodesic equations
(\ref{Equ13}) imply $dr/dv \cong \frac{1}{2} f$ for an observer
with $\tilde E<0$, approaching $r=r_-$ from region {\it II}. Expanding $f$
near $r=r_-$, $f \cong f'(r_-)(r-r_-) = -2 \kappa (r-r_-)$, where

\begin{equation}
\label{Equ14}
\kappa = (M^2 - Q^2)^{\frac{1}{2}}/r_-^2
\end{equation}
is the surface gravity of the inner horizon (as it follows from
definition (\ref{Equ8})), and integrating, we get $f$ near
$r_-$. From the second equation in (\ref{Equ13}) we then obtain the
``asymptotic formula''
\begin{equation}
\label{Equ15}
\dot v \simeq \const \mid \tilde E \mid e^{\kappa \tau} \cong
\frac{1}{\kappa \Delta \tau}, \,\,\, \nonumber \\
v \rightarrow \infty, \Delta \tau \rightarrow 0,
\end{equation}
where $\Delta \tau$ is the amount of proper time the observer
needs to reach the inner (Cauchy) horizon.

Imagine now two nearby events $A_{out}$ and $B_{out}$ in 
the outside world {\it I}, for example the emission of two 
photons from a given fixed $r>r_+$, which are connected by the ingoing null
geodesics $v =  \const$ and $v + dv = \const$ with events $A_{in}$ and
$B_{in}$ on the worldline of the observer approaching $r_-$. The
interval of proper time between $A_{out}$ and $B_{out}$ is $d
\tau_{out} \sim dv$, whereas (\ref{Equ15}) implies that the
interval of proper time between $A_{in}$ and $B_{in}$, as measured
by the observer approaching $r_-$, is $d\tau_{in} \sim e^{-\kappa
v}dv$. Therefore,
\begin{equation}
\label{Equ16}
\frac{d\tau_{in}}{d\tau_{out}} \sim e^{-\kappa v},
\end{equation}
so that as $v \rightarrow +\infty$ the events (clocks) in the
outside world are measured to proceed increasingly fast by the
inside observer approaching the inner horizon. In the limit, when
the observer crosses the Cauchy horizon, he sees the whole future
history (from some event as $A_{out}$) of the external universe
to ``proceed in one flash'': $d\tau_{in} \rightarrow 0$ with
$d\tau_{out} \rightarrow \infty$. An intuitive explanation is
given by the fact that the observers in region {\it I} need infinite
proper time to reach $v=\infty$, whereas inside observers only
finite proper time. The infalling radiation will thus be
unboundedly blue-shifted at the inner horizon which in general
will lead to a divergence of the energy density there. This {\it
infinite blueshift at the inner (Cauchy) horizon makes it
generally unstable to perturbations (``the blue-sheet
instability'')}.

There exists an extensive literature analyzing the exciting
questions of the {\it black hole interiors} (see for example
\cite{FN}, the introductory review \cite{Ori} in the proceedings
of the recent workshop devoted entirely to these issues, and
other contributions in the proceedings which give also many
further references). Some crucial questions are still the subject
of much debate. One of the following two approaches to the
problem is usually chosen: (i) a linear perturbation analysis
of the behaviour of fields at the Cauchy horizon, (ii) the
simplified nonlinear, spherically symmetric models of black hole
interiors.

In the first approach one considers the evolution of linear
perturbations, representing scalar, electromagnetic, or
gravitational fields, on the Reissner-Nordstr\"{o}m background.
Since there is a nonvanishing background electric field, the
{\it electromagnetic and gravitational} perturbations are {\it
coupled}.\footnote{This leads to various interesting phenomena. For
example, the scattering of incident electromagnetic and
gravitational waves by the Reissner-Nordstr\"{o}m black hole
allows for the partial conversion of electromagnetic waves into
gravitational waves and vice versa \cite{Chan}. When studying
stationary electromagnetic fields due to sources located outside
the Reissner-Nordstr\"{o}m black hole, one discovers that closed
magnetic field lines not linking any current source may exist,
since gravitational perturbations constitute, via the background
Maxwell tensor, an effective source \cite{BiDv}.}
It is a remarkable fact that ``wave equations'' for certain
gauge-invariant combinations of perturbations can be derived from
which all perturbations can eventually be constructed 
\cite{Chan,Mo,JiBi}. In the simplest case of the scalar field
$\Phi$ on the Reissner-Nordstr\"{o}m background, after resolving
the field into spherical harmonics and putting $\Phi= r\Psi$, the
wave equation has the form \cite{Bca}
\begin{equation}
\label{Equ17}
\Psi_{,tt} - \Psi_{,r^*r^*} + F_l(r^*) \Psi = 0,
\end{equation}
where the curvature-induced potential barrier is given by
\begin{equation}
\label{Equ18}
F_l(r^*) = \left( 1-\frac{2M}{r} + \frac{Q^2}{r^2}\right) \left[
\frac{2}{r^3} \left( M- \frac{Q^2}{r^2}\right) +
\frac{l(l+1)}{r^2} \right],
\end{equation}
where $r$ is considered to be a function of $r^*$ (cf. (\ref{Equ12})). 
In order to determine the evolution of the field below the outer
horizon in a real gravitational collapse, one first concentrates
on the evolution of the field outside a collapsing body
(star). A nonspherically symmetric scalar test field
(generated by a nonspherical distribution of ``scalar charge'' in
the star) serves as a prototype for (small) asymmetries in the
external gravitational and electromagnetic fields, which are generated by
asymmetries in matter and charge distributions inside the star.
Now when a slightly nonspherical star starts to collapse, the
perturbations become dynamical and propagate as waves. Their
evolution can be determined by solving the wave equation
(\ref{Equ17}). Because of the potential barrier (\ref{Equ18}) the
waves get backscattered and produce slowly decaying radiative
tails, as shown in the classical papers by Price 
\cite{Pri,Pri1}, and
generalized to the Reissner-Nordstr\"{o}m case in 
\cite{Bca,Bca1}. The tails decay in the vicinity of the outer event
horizon $r_+$ (i.e. between regions {\it I} and {\it II} in
Fig. 5) as $\Psi
\sim v^{-2(l+1)}$ for $l$-pole 
perturbations.\footnote{This result is true for a general charged
Reissner-Nordstr\"{o}m black hole with $Q^2<M^2$. In the
extremal case, $Q^2 = M^2$, the field decays only as $\Psi \sim
v^{-(l+2)}$ \cite{Bca}.}
The decaying tails provide the initial data for the ``internal
problem'' -- the behaviour of the field near the Cauchy horizon.
Calculations show (see \cite{Ori} and references therein)
that near the Cauchy
horizon the behaviour of the field remains qualitatively the
same: $\Psi(u, v\!\rightarrow\!\infty) \sim v^{-2(l+1)}$ $+$ \{slowly
varying function of $u$\}, where $u = 2r^*-v$ is constant along
outgoing radial null geodesics in region {\it II}. However, as a
consequence of the ``exponentially growing blueshift'', given by
formula (\ref{Equ15}), the {\it rate of change of the field
diverges as the observer approaches the Cauchy horizon}:
$d\Psi/d\tau = (\partial \Psi /\partial x^\alpha) U^\alpha \simeq
\Psi_{,v}\, \dot v \sim v^{-2l-3} e^{\kappa v}$. Therefore, the
measured energy density in the field would also diverge, causing
an instability of the Cauchy horizon, which would be expected to
create a curvature singularity. More detailed considerations
\cite{Ori} show that the singularity, at least for large $|u|$,
is null and weak (the metric is well-defined, only the
Riemann tensor is singular). Any definitive picture of the Cauchy
horizon instability can come however only from a fully
nonlinear analysis which takes into account the backreaction of
spacetime geometry to the growing perturbations.

The second approach to the study of the Cauchy horizon
instabilities employs a simplified, {\it spherically symmetric
model which treats the nonlinearities exactly} \cite{POI}. The
ingoing radiation is modelled by a stream of ingoing charged
null dust \cite{BOVA} which is infinitely blueshifted at the inner
horizon. There is, however, also an outgoing stream of charged
null dust considered to propagate into region {\it II} towards the inner
horizon. The outgoing flux may represent radiation coming from
the stellar surface below the outer horizon, as well as a portion
of the ingoing radiation which is backscattered in region {\it II},
and irradiates thus the inner horizon. A detailed analysis based on
exact spherically symmetric solutions revealed a remarkable
effect: an effective {\it internal} gravitational-mass parameter
of the hole unboundedly increases at the inner (Cauchy) horizon
(though the external mass of the hole remains finite). This
{\it ``mass inflation phenomenon''} causes the divergence of some
curvature scalars at the Cauchy horizon \cite{POI}. In reality,
the classical laws of general relativity will break down when the
curvature reaches Planckian values.

It is outside the scope of this review to discuss further the
fascinating issues of black hole interiors. They involve deep
questions of classical relativity, of quantum field theory on
curved background (as, for example, in discussions of
electromagnetic pair production and vacuum polarization effects
inside black holes), and they lead us eventually to quantum
gravity.  We refer again especially to \cite{FN} and \cite{Ori} for
more information. Let us only add three further remarks. We mentioned
above the work on the inner structure of Reissner-Nordstr\"{o}m
black holes because this is the most explored  (though not
closed) area. However, Kerr black holes (Section 4) possess also
inner horizons and there are many papers concerned with the
instabilities of the Kerr Cauchy horizons 
(see \cite{FN,Ori} for references). 
Secondly, at the beginning of 1990s, it
was shown that the inner horizons of the Reissner-Nordstr\"{o}m-de
Sitter and Kerr-de Sitter black holes are classically stable in
the case when the surface gravity at the inner horizons is
smaller than the surface gravity at the cosmological horizon
(\cite{Ori} and references therein, in particular, the review
\cite{Chamb}). Penrose \cite{PCH} even suggested that ``it may
well be that cosmic censorship requires a zero (or at least a
nonpositive) cosmological constant''. Very recently, however,
three experts in the field \cite{BMM} have claimed that outgoing
modes near to the black hole (outer) event horizon lead to
instability for all values of the parameters of
Reissner-Nordstr\"{o}m-de Sitter black holes. Let me borrow again
a statement from Penrose \cite{PCH}: ``My own feelings are left
somewhat uncertain by all these considerations''.

Finally, a new contribution \cite{VH} to the old problem of
testing the {\it weak} cosmic censorship by employing a Reissner-Nordstr\"{o}m
black hole indicates that one can overcharge a near extreme $(Q^2
\rightarrow M^2)$ black hole by throwing in a charged particle
appropriately. However, the backreaction effects remain to be
explored more thoroughly. The question of cosmic censorship thus
remains as interesting as ever.
\newpage

\subsection{On extreme black holes, ${\vec d}$-dimensional black holes,
string theory and ``all that''}

In the previous section we considered generic
Reissner-Nordstr\"om black holes with $M^2>Q^2$. They have outer and
inner horizons given by (\ref{Equ11}), with nonvanishing
surface gravities (cf. (\ref{Equ14}) for the inner horizon).
For $M^2 = Q^2$ the two horizons coincide at $r_+ = r_- = M$.
Defining the ingoing null coordinate $v$ as in (\ref{Equ12}), we
obtain the ingoing extension of the Reissner-Nordstr\"om metric
(\ref{Equ9}) in the form

\begin{equation}
\label{Equ19}
ds^2 = -\left( 1-\frac{M}{r} \right)^2 dr^2 + 2dv dr + r^2
(d\theta ^2 + \sin^2 \theta d\varphi ^2). \nonumber
\end{equation}
This is the metric of {\it extreme Reissner-Nordstr\"om black
holes}.
Frequently, these holes are called ``degenerate''. At the horizon
$r=M$, the Killing vector field $\vec k = \partial/\partial v$
obeys the equation $(k^\alpha k_\alpha)_{,\beta} = 0$, so that
regarding the general relation (\ref{Equ8}), the surface gravity
$\kappa = 0$, i.e. the {\it Killing horizon is degenerate}. Using
$(k^2)_{,\beta} = 0$ and the Killing equation, we easily deduce
that the horizon null generators with tangent $k^\alpha =
dx^\alpha/dv$ satisfy the geodesic equation with affine
parameter $v$. The generators have infinite affine length to the
past given by $v \rightarrow -\infty$ (in contrast to the
generators of a bifurcate Killing horizon -- cf. Section 2.4). This
part of the extreme Reissner-Nordstr\"om spacetime, given by
$r=M, v\rightarrow -\infty$, is called an {\it ``internal
infinity''}. That there is no ``wormhole'' joining two
asymptotically flat regions and containing a minimal surface
2-sphere like in the non-extreme case can also be seen from the
metric in the original Schwarzschild-type coordinates.
Considering an embedding diagram $t=\const, \theta = \pi /2$ in
flat Euclidean space one finds that an infinite ``tube'', or
an asymptotically cylindrical region on each $t = \const$ hypersurface
develops. The boundary of the cylindrical region is the internal
infinity. It is a compact 2-dimensional spacelike surface. The
hypersurfaces $t = \const$ do not intersect the horizon but only
approach such an intersection at the internal infinity. (See
\cite{CA} for the conformal diagram and a detailed discussion,
including analysis of the electrovacuum Robinson-Bertotti
universe as the asymptotic limit of the extreme
Reissner-Nordstr\"om geometry at the internal infinity.)

There has been much interest in the extreme Reissner-Nordstr\"om
black holes within standard Einstein-Maxwell theory. They
admit surprisingly simple solutions of the perturbation equations
\cite{BIK}. Some of them appear to be stable with respect to both
classical and quantum processes, and there are attempts to
interpret them as solitons \cite{Gib}. Also, they admit
supersymmetry \cite{PACH}.

The quotation marks in the title of this section play a double
role: the last two words are just ``quoting'' from the end of the
title of a general review on string theory and supersymmetry
prepared for the special March 1999 issue of the Reviews of
Modern Physics in honor of the centenary of the American Physical
Society by Schwarz and Seiberg \cite{SSE}, but they also should
``self-ironically'' indicate my ignorance in these issues. In
addition, unified theories of the type of string theory
appear to be somewhat outside the direct interest of J\"urgen
Ehlers, who has always emphasized the depth and economy of general
relativity because it is a ``background-independent'' theory:
string theories still suffer from the lack of a 
background-independent formulation. Nevertheless, they are
beautiful, consistent, and very challenging constructions,
representing one of the most active areas of theoretical physics.
Recently, string theory provided an explanation of the
Bekenstein-Hawking prediction of the entropy of extreme and
nearly extreme black holes. From the point of view of this review
we should emphasize that many of the techniques that have been
used to obtain exact solutions -- mostly exact black hole
solutions -- in generalized theories like string theory
were motivated by classical general relativity. There are also
results in classical general relativity which are finding
interesting generalizations to string theories, as we shall
see with one example below.

\vskip -0.2pt
Before making a few amateurish comments on new results concerning extreme
black holes in string theories, let us point out that in many
papers from the last 20 years, {\it black hole solutions}
were studied in spacetimes with the number of {\it
dimensions} either {\it lower} or {\it higher} than four. The
lower-dimensional cases are usually analyzed as ``toy models''
for understanding the complicated problems of quantum gravity. The
higher-dimensional models are motivated by efforts to
find a theory which unifies gravity with the other forces. The most
surprising and popular (2+1)-dimensional black hole is the BTZ
(Ba\~nados-Teitelboim-Zanelli) black hole in the Einstein theory
with a negative cosmological constant. Locally it is isometric to
anti de Sitter space but its topology is different. In
\cite{Carl} the properties of (2+1)-dimensional black holes are
reviewed. In (1+1)-dimensions one obtains black holes only if one
includes at least a simple dilaton scalar field; the motivation
for how to do this comes from string theory. In higher dimensions
one can find generalizations of all basic black hole solutions in
four dimensions \cite{MPe}. Interesting observations concerning
higher-dimensional black holes have been given a few years ago
\cite{GiHo}. Perhaps one does not need to quantize gravity in
order to remove the singularities of classical relativity. It may
well be true that some new classical physics intervenes below
Planckian energies. In \cite{GiHo} it is demonstrated that
certain singularities of the four-dimensional extreme dilaton
black holes can be resolved by passing to a higher-dimensional
theory of gravity in which usual spacetime is obtained only below
some compactification scale. A useful, brief pedagogical
introduction to black holes in unified theories is contained
in \cite{FN}.

\vskip -0.2pt
One of the most admirable recent results of string theory,
which undoubtedly converted some relativists and stimulated many
string theorists, has been the derivation of the {\it exact value
of the entropy of extreme and nearly extreme black holes}. I shall
just paraphrase a few statements from the March 1999 review for the
centenary of the American Physical Society by Horowitz and
Teukolsky \cite{HoTe}. There are very special states in 
string theory called BPS (Bogomol'ny-Prasad-Sommerfield) states
which saturate an equality $M\geq c |Q|$, with $M$ being
the mass, $Q$ the charge, and $c$ is a fixed constant. The mass of
these special states does not get any quantum corrections. The
strength of the interactions in string theory is determined by a
coupling constant $g$. One can count BPS states at large $Q$ and
small $g$. By increasing $g$ one increases gravity, 
and then all of these
states become black holes. (The BPS states are supersymmetric and
one can thus follow the states from weak to strong coupling.) But
they all become identical extreme Reissner-Nordstr\"om black 
holes, because there is only one black hole for given $M=|Q|$.
When one counts the number $N$ of BPS states in which an extreme
hole can exist, and compares this with the entropy $S_{bh} = \frac{1}{4} A$ of the
hole as obtained in black hole thermodynamics \cite{RW,WD}, 
where $A$ is the area of the event
horizon $(A=4 \pi M^2$ for the extreme Reissner-Nordstr\"om black
hole), one finds exactly the ``classical'' result: $S_{bh}=\log N!$.
The entropy of the classical black hole configuration is
given in terms of the number of quantum microstates associated
with that configuration, by the basic formula of statistical physics.
For more detailed recent reviews, see \cite{HORO,SKE}, and 
references therein. Remarkably, the results for
the black hole entropy have been obtained also
within the canonical quantization of gravity \cite{AA}.
A comprehensive review \cite{YO} of black holes and solitons in 
string theory appeared very recently.

Allow me to finish this ``all that'' section with a personal
remark. In 1980 L. Dvo\v{r}\'ak and I found that in the
Einstein-Maxwell theory, external magnetic flux lines are
expelled from the black hole horizon as the hole becomes an
extreme Reissner-Nordstr\"om black hole \cite{BiDv}. Hence,
extreme black holes exhibit some sort of {\it ``Meissner
effect''} known from superconductivity. Last year it was
demonstrated by Chamblin, Emparan and Gibbons \cite{CEG} that
this effect occurs also for black hole solutions in string
theory and Kaluza-Klein theory. Other extremal solitonic
objects in string theory (like {\it p}-branes) can also have
superconducting properties. Within the Einstein-Maxwell theory
this effect was first studied to linear order in magnetic field --
we analyzed Reissner-Nordstr\"om black holes in the presence of
magnetic fields induced by current loops. However, we also used
an exact solution due to Ernst \cite{ERN}, describing a charged
black hole in a background magnetic field, which asymptotically
goes over to a Melvin universe, and found the same effect
(see also \cite{KVO} for the case of the magnetized Kerr-Newman 
black hole).
In \cite{CEG} the techniques of finding exact solutions of
Einstein's field equations are employed within string theory
and Kaluza-Klein theory to demonstrate the ``Meissner effect'' in
these theories.
\newpage

%%%%%%%%%%%%%%%%%%%%%%%%%%%%%%%%%%%%%%%%%%%%%%%%%%%%%%%%%%%%%%%%%%%%%
%%%%%%%%%\input kerr.tex
%%%%%%%%%%%%%%%%%%%%%%%%%%%%%%%%%%%%%%%%%%%%%%%%%%%%%%%%%%%%%%%%%%%%%
\section{The Kerr metric}

The discovery of the Kerr metric in 1963 and the proof of its
unique role in the physics of black holes have made an immense
impact on the development of general relativity and
astrophysics. This can hardly be more eloquently demonstrated
than by an emotional text from Chandrasekhar \cite{SCH}: ``In my
entire scientific life, extending over forty-five years,
the most shattering experience has been
the realization that an exact solution of Einstein's equations of
general relativity, discovered by the New Zealand mathematician
Roy Kerr, provides the absolutely exact representation of untold
numbers of massive black holes that populate the Universe...''

In Boyer-Lindquist coordinates the Kerr metric \cite{KER}
looks as follows (see e.g. \cite{MTW,DIN}):

\begin{eqnarray}
\label{Equ20}
ds^2 &=&  - \left( 1- \frac{2Mr}{\Sigma}\right)dt^2 - 2\frac{2aMr \sin
^2 \theta}{\Sigma} dt~d \varphi + \nonumber \\
~&\lefteqn{
~+ ~\frac{\Sigma}{\Delta} dr^2 + \Sigma d \theta ^2  
+ ~\frac{{\cal{A}}}{\Sigma} \sin ^2 \theta \,\, d \varphi ^2,
}&~
\end{eqnarray}
where
\begin{eqnarray}
\label{Equ21}
~&\Sigma = r^2 + a^2 \cos ^2 \theta, \hspace{0,5cm} \Delta = r^2 -
2Mr + a^2,&~ \nonumber \\
~&{\cal{A}} = \Sigma (r^2 + a^2) + 2Mr a^2 \sin ^2 \theta,&~
\end{eqnarray}
where $M$ and $a$ are constants.

\subsection{Basic features}

The Boyer-Lindquist coordinates follow naturally from the
symmetries of the Kerr spacetime. The scalars $t$ and $\varphi$
can be fixed uniquely (up to additive constants) as parameters
varying along the integral curves of (unique) stationary and
axial Killing vector fields $\vec k$ and $\vec m$; and the scalars $r$ and
$\theta$ can be fixed (up to constant factors) as parameters
related as closely as possible to the (geometrically preferred)
principal null congruences, which in the Kerr spacetime exist
(see e.g. \cite{MTW,DIN}), and their projections on to
the two-dimensional spacelike submanifolds orthogonal to both
$\vec k$ and $\vec m$ (see \cite{STW} for details). The
Boyer-Lindquist coordinates represent the natural generalization
of Schwarzschild coordinates. With
$a=0$ the metric (\ref{Equ20}) reduces to the Schwarzschild
metric.

By examining the Kerr metric in the asymptotic region $r
\rightarrow \infty$, one finds that $M$ represents the mass
and $J=Ma$ the angular momentum pointing in the $z$-direction, so
that $a$ is the angular momentum per unit mass. One can arrive at
these results by considering, for example, the weak field and
slow motion limit, $M/r \ll 1$ and $a/r \ll 1$. The Kerr metric
(\ref{Equ20}) can then be written in the form

\begin{eqnarray}
\label{Equ22}
ds^2 = -\left( 1-\frac{2M}{r}\right) dt^2 + \left( 1+\frac{2M}{r}
\right)dr^2 \nonumber \\
+ r^2 \left(d\theta ^2 + \sin ^2\theta~d\varphi ^2
\right)-\frac{4aM}{r} \sin ^2 \theta~d\varphi~dt,
\end{eqnarray}
which is the weak field metric generated by a central body with
mass $M$ and angular momentum $J=Ma$. A general, rigorous way of
interpreting the parameters entering the Kerr metric starts from
the {\it definition of multipole moments} of asymptotically flat,
stationary vacuum spacetimes. This is given in physical space by
Thorne \cite{TRN}, using his ``asymptotically Cartesian and
mass centered'' coordinate systems, and by Hansen \cite{HN}, who,
generalizing the definition of Geroch for the static case, gives
the coordinate independent definition based on the conformal
completion of the 3-dimensional manifold of trajectories of a
timelike Killing vector $\vec k$. The exact Kerr solution has
served as a convenient ``test-bed'' for such 
definitions.\footnote{For the most complete, 
rigorous treatment of the
asymptotic structure of stationary spacetimes characterized
uniquely by multipole moments defined at spatial infinity, see
the work by Beig and Simon \cite{BeS}, the article by Beig and
Schmidt in this volume, and references therein.}
The mass monopole moment -- the mass -- is $M$, the mass dipole
moment vanishes in the ``mass-centered'' coordinates, the
quadrupole moment components are $\frac{1}{3}Ma^2$ and
$-\frac{2}{3}Ma^2$. The current dipole moment -- the angular
momentum -- is nonvanishing only along the axis of symmetry and
is equal to $J=Ma$, while the current quadrupole moment vanishes. All
other nonvanishing $l$-pole moments are proportional to $Ma^l$ 
\cite{TRN,HN}. Because these specific values of the multipole moments
depend on only two parameters, the Kerr solution clearly
cannot represent the gravitational field outside a general
rotating body. In Section 6.2 we indicate how the Kerr metric
with general values of $M$ and $a$ can be produced by special
disk sources. The fundamental significance of the Kerr spacetime,
however, lies in its role as the {\it only vacuum rotating black
hole solution}.

Many texts give excellent and thorough
discussions of properties of Kerr black holes from various
viewpoints \cite{MTW,Wa,HE,DIN,IS,TP,FN,CA,Chan,TeSh,FEC}.
The Kerr metric entered the new edition of ``Landau and Lifshitz''
\cite{LL}. A few years ago, a book devoted
entirely to the Kerr geometry appeared \cite{ONE}. Here we can
list only a few basic points.

\begin{figure}
\centering
\includegraphics[width=.7\textwidth]{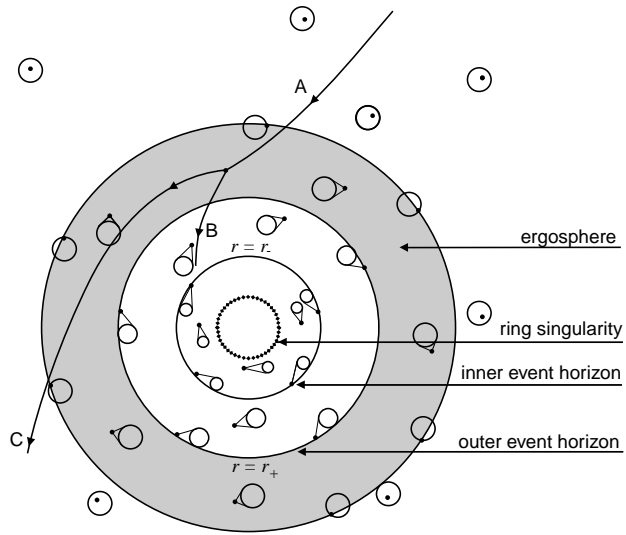}
\label{Figure 6}
\caption{A schematic picture of a Kerr black hole with two 
horizons, ergosphere, and the ring singularity; local light wavefronts are also indicated.
Particle $A$, entering the ergosphere from infinity, 
can split inside the ergosphere into particles $B$ and $C$ in 
such a manner that $C$ arrives at infinity with a higher energy
than particle $A$ came in.
}
\end{figure}

As with the Reissner-Nordstr\"{o}m spacetime, one can make the
maximal analytic extension of the Kerr geometry. This, in fact,
has much in common with the Reissner-Nordstr\"{o}m case. Loosely
speaking, the ``repulsive'' characters of both charge and
rotation have somewhat similar manifestations. When
$a^2<M^2$, the metric (\ref{Equ20}) has coordinate singularities
at $\Delta = 0$, i.e. at (cf. (\ref{Equ21}))
\begin{equation}
\label{Equ23}
r= r_{\pm} = M\pm (M^2 - a^2)^{\frac{1}{2}}.
\end{equation}
The intrinsic three-dimensional geometry at $r=r_{\pm}$ reveals
that these are null hypersurfaces -- the (outer) event horizon and
the inner horizon (Fig. 6). As with the Reissner-Nordstr\"{o}m metric, the
inner horizon -- the Cauchy hypersurface -- is unstable (see a
more detailed discussion in Section 3.2 and references there). And
as in
the Penrose diagram in Fig. 5, one finds infinitely many
asymptotically flat regions in the analogous Penrose diagram for
the Kerr black hole spacetime.

A crucial difference between the Reissner-Nordstr\"{o}m and Kerr
geometry is the existence of the {\it ergosphere} (or, more
precisely, ergoregion) in the Kerr case. This is caused by
the {\it dragging of inertial frames} due to a nonvanishing angular
momentum. The timelike Killing vector $\vec k$, given in the
Boyer-Lindquist coordinates by $\partial/\partial t$, becomes
null ``sooner'', at $r = r_0$, than at the event horizon,
$r_0>r_+$ at $\theta \neq 0, \pi$, as a consequence of
this dragging:
\begin{eqnarray}
\label{Equ24}
k^\alpha k_\alpha = -g_{tt} = 1-2Mr/\Sigma = 0, \nonumber \\
r = r_0 = M+(M^2-a^2\cos^2 \theta)^{\frac{1}{2}}.
\end{eqnarray}
This is the location of the {\it ergosurface}, the ergoregion being
between this surface and the (outer) horizon. In the ergosphere,
schematically illustrated in Fig. 6, the ``rotating geometry''
drags the particles and light (with wavefronts indicated in the
figure) so strongly that all physical observers must corotate
with the hole, and so rotate with respect to distant observers --
``fixed stars'' -- at rest in the Boyer-Lindquist 
coordinates.\footnote{It is instructive to analyze 
the somewhat ``inverse
problem'' of gravitational collapse of a slowly rotating dust
shell which produces the Kerr metric, linearized in $a/M$,
outside (cf. Eq.(\ref{Equ22})), and has flat space inside. Fixed
distant stars seen from the centre of such shell appear to rotate
due to the dragging of inertial frames, as was discussed in
detail recently \cite{KLB}.}
Static observers, whose worldlines 
$(r, \theta, \varphi) = \const$ would have $\vec k$ 
as tangent vectors, cannot exist since
$\vec k$ is spacelike in the ergosphere. Indeed, a non-spacelike
worldline with $r,\theta$ fixed must satisfy
the condition
\begin{equation}
\label{Equ25}
g_{tt} dt^2 + g_{\varphi \varphi} d\varphi ^2 + 2 g_{\varphi t}
d\varphi dt \leq 0,
\end{equation}
in which $g_{tt} = -k^\alpha k_\alpha,\,\,\, g_{\varphi \varphi} =
m^\alpha m_\alpha,\,\,\, g_{\varphi t}=k^\alpha m_\alpha$ are
invariants. In the ergosphere, the metric
(\ref{Equ20}), (\ref{Equ21}) yields $g_{tt}>0$, $g_{\varphi
\varphi}>0, g_{\varphi t}<0$, so that $d\varphi/dt>0$ -- an
observer moving along a non-spacelike worldline must corotate with
the hole. The effect of dragging on the forms of photon escape
cones in a general Kerr field (without restriction $a^2<M^2$) has
been numerically studied and carefully illustrated in a number of
figures only recently \cite{OLS}.

In order to ``compensate'' the dragging, the congruence of
{\it``locally nonrotating frames''} (LNRFs), or {\it
``zero-angular momentum observers''} (ZAMOS), has been
introduced. These frames have also commonly been used outside
relativistic, rapidly rotating stars constructed numerically, but
the Kerr metric played an inspiring role (as, after all, 
in several other issues, such as in understanding the ergoregions,
etc.). The four-velocity of these (not freely falling!)
observers, given in Boyer-Lindquist coordinates by
\begin{equation}
\label{Equ26}
e^\alpha_{(t)}= \left[   ({\cal{A}}/\Sigma \Delta)^{\frac{1}{2}},
0, 0, 2aMr/({\cal{A}}\Sigma \Delta)^{\frac{1}{2}} \right],
\end{equation}
is orthogonal to the hypersurfaces $t = \const$. The particles
falling from rest at infinity with zero total angular momentum
fall exactly in the radial direction in the locally
nonrotating frames with an orthogonal triad tied to the $r, \theta,
\varphi$ coordinate directions (see \cite{BST} for the study of
the shell of such particles falling on to a Kerr black hole).

Now going down from the ergosphere to the outer horizon, we find
that both Killing vectors $\vec k$ and $\vec m$ are tangent to
the horizon, and are spacelike there (with $\vec k$ ``rotating'' with
respect to infinity). The null geodesic generators of the horizon
are tangent to the null vectors $\vec l = \vec k + \Omega \vec
m$, where $\Omega = a/2Mr_+ = \const$ is called the {\it angular
velocity of the hole}. $\Omega$ is constant over the horizon so
that the {\it horizon rotates rigidly}. Since $\vec l$ is a
Killing vector, the horizon is a Killing horizon (cf. Section
2.4).

Another notable difference from the Reissner-Nordstr\"{o}m metric
is the character of the singularity at $\Sigma = 0$, i.e.
at $r=0, \,\, \theta = \pi/2$. It is timelike, as in the
Reissner-Nordstr\"{o}m case, but it is a {\it ring} singularity
(see Fig. 6).
In the maximal analytic extension of the Kerr metric one can go
through the ring to negative values of the coordinate $r$ and
discover {\it closed timelike lines} since $g_{\varphi
\varphi}<0$ there.
If the Kerr parameters are such that $a^2>M^2$, the Kerr metric
does not represent a black hole. It describes the gravitational
field with a {\it naked ring singularity}. The Kerr ring
singularity has a repulsive character near the rotation axis. It
gives particles outward accelerations and collimates them along
the rotation axis \cite{BISH}, which might be relevant in the
context of the formation and precollimation of cosmic jets.
However, the cosmic censorship conjecture is a very plausible,
though difficult ``to prove'' hypothesis, and Kerr naked
singularities are unlikely to form in nature. However, the Kerr
geometry with $a^2>M^2$, with a region containing the ring
singularity ``cut out'', can be produced by thin disks; 
though if they should be composed of physical matter, 
they cannot be very relativistic 
(see Section 6.2 and references therein).

If $a^2 = M^2$, the Kerr solution represents an {\it extreme
Kerr black hole}, as is the analogous Reissner-Nordstr\"{o}m
case with $Q^2=M^2$. The inner and outer horizons then coincide at
$r=M$. The horizon is degenerate with infinite affine length.
Almost extreme Kerr black holes probably play the most important
role in astrophysics (see below). 
In realistic astrophysical situations accreting matter will very
likely have a sufficient amount of angular momentum 
to turn a Kerr hole to an almost extreme state.

There exists a charged, electrovacuum generalization of the Kerr
family found by Newman et al. \cite{KN}. The {\it Kerr-Newman
metric} in the Boyer-Lindquist coordinates can be obtained
from the Kerr metric (\ref{Equ20}) if all of the terms $2Mr$
explicitly appearing in (\ref{Equ20}), (\ref{Equ21}) are replaced
by $2Mr-Q^2$, with $Q$ being the charge. The metric describes
{\it charged, rotating black holes} if $M^2>a^2+Q^2$,
with two horizons located at $r_\pm = M\pm(M^2 - a^2 -
Q^2)^{\frac{1}{2}}$. These become extreme when $M^2 = a^2 +
Q^2$, and with $M^2<a^2+Q^2$ one obtains naked (ring) singularities.
The analytic extension, the presence of ergoregions and the
structure of the singularity is similar to the Kerr case.

In addition to the gravitational field, there exists a stationary
electromagnetic field which is completely determined by the
charge $Q$ and rotation parameter $a$. The vector potential of
this field is given by the 1-form

\begin{equation}
\label{Equ27}
A_\alpha dx^\alpha = -(Qr/\Sigma)(dt-a\sin^2\theta~d\varphi),
\end{equation}
so that if $a\not= 0$ the electric field is supplemented by a
magnetic field. At large distances $(r \rightarrow \infty)$ the
field corresponds to a {\it monopole electric field} with charge
$Q$ and a {\it dipole magnetic field} with magnetic moment
$\mu=Qa$. Since the gyromagnetic ratio of a charged system with
angular momentum $J$ is defined by $\gamma = \mu/J$, one finds
the charged-rotating-black hole {\it gyromagnetic ratio} to
satisfy the relation $\gamma=Q/M$, i.e. it is twice as large as
that of classical matter, and the same as that of an electron. By
examining a black hole with a loop of rotating charged matter
around it, the radius of the loop changing from large values
to the size of the horizon, it is possible to gain some
understanding of this result \cite{GAT}.

\subsection{The physics and astrophysics around rotating black holes}

In the introduction to their new 770 page monograph on
black hole physics, Frolov and Novikov 
\cite{FN} write: ``... there are
a lot of questions connected with black hole physics and its
applications. It is now virtually impossible to write a book
where all these problems and questions are discussed in detail.
Every month new issues of Physical Review D, Astrophysical
Journal, and other physical and astrophysical journals add scores
of new publications on the subject of black holes...''  Although
Frolov and Novikov have also black hole-like solutions in
superstring and other theories on their minds, we would not
probably be much in error, in particular in the context of
astrophysics, if we would claim the same just about Kerr black
holes. Hence, first of all, we must refer to the same literature
as in the previous section 4.1. A few more references will be
given below.

A remarkable fact which stands at the roots of these developments
is that the wave equation is separable, and the geodesic equations
are integrable in the Kerr geometry. Carter \cite{CA}, who
explicitly demonstrated the separability of the Hamilton-Jacobi
equation governing the geodesic motion, has emphasized that one
can in fact {\it derive} the Kerr metric as the simplest nonstatic
generalization
of the Schwarzschild solution, by requiring the separability of
the covariant Klein-Gordon wave equation.\footnote{Although 
this is still not a ``constructive, analytic
derivation of the Kerr metric which would fit its physical
meaning'', as required by Landau and Lifshitz \cite{LL}, it is
certainly more intuitive than the original derivation by Kerr. On
the other hand, despite various hints like the existence of the
Killing tensor field (in addition to the Killing vectors) in the
Kerr geometry (see e.g. \cite{STW}), it does not seem to be clear
{\it why} the Kerr geometry makes it possible to separate these
equations.}

A thorough and comprehensive analysis of the behaviour of freely
falling particles in the Kerr field would produce material for a
book. We refer to e.g. \cite{MTW,FN,Chan,STW,FEC,BISH,BAR} for fairly
detailed accounts and a number of further references. From the
point of view of astrophysical applications the following items
appear to be most essential: in contrast to the Schwarzschild
case, where the stable circular orbits exist only up to
$r=3r_+=6M$, in the field of rotating black holes, the stable
direct (i.e. with a positive angular momentum) circular orbits in
the equatorial plane can reach regions of ``deeper potential
well''. With an extreme Kerr black hole the last stable direct
circular orbit occurs at $r=r_+=M$. (See \cite{BAR} for a clear
discussion of the positions of the innermost stable, innermost
bound, and photon orbits as the hole becomes extreme and a long
cylindrical throat at the horizon develops.) A {\it
``spin-orbit-coupling''} effect {\it increases the binding energy
of the direct orbits} and decreases the binding energy of the
retrograde (with a negative angular momentum) orbits relative to
the Schwarzschild values. The binding energy of the last stable
direct circular orbit is $\Delta E = 0.0572 \mu$
($\mu$ is the particle's proper mass) in the Schwarzschild case, whereas
$\Delta E = 0.4235 \mu$ for an extreme Kerr hole. A particle
slowly spiralling inward due the emission of gravitational waves
would radiate the total energy equal to this binding energy;
hence much more -- 42\% of its rest energy -- in the Kerr case. The second
significant effect is the dragging of the particles moving on
orbits outside of the equatorial plane. 
The {\it dragging}\footnote{Relativists 
often consider the effects produced by
moving mass currents as ``the dragging of inertial frames'', but
the concept of the gravomagnetic field, or {\it gravomagnetism}
has some advantages, as has been stressed recently \cite{Rin}.
The gravomagnetic viewpoint, however, has also been used in many
works in the past -- see, e.g. \cite{CW,TP}, and in
particular \cite{JT} and references therein.}
will make the orbit of a star around a supermassive black hole to
precess with angular velocity $\sim 2J/r^3$. The star may go
through a disk around the hole, subsequently crossing it at 
different places \cite{KaVo}. One can also show that as a result
of the joint action of the gravomagnetic effect and the viscous
forces in an accretion disk, the disk tends to be oriented in
the equatorial plane of the central rotating black hole (the
``Bardeen-Petterson effect'').

The above examples demonstrate specific effects in the Kerr
background which very likely play a significant role in
astrophysics (see also Section 4.3 below). The best known
process in the field of a rotating black hole is probably
astrophysically unimportant, but is of principal significance
in the black hole physics. This is the {\it Penrose process for
extracting energy from rotating black holes}. It is illustrated
schematically in Fig. 6: particle $A$ comes from infinity into the
ergosphere, splits into two particles, $B$ and $C$. Whereas $C$ is
ejected back to infinity, $B$ falls inside the black
hole. The process can be arranged in such a way that particle $C$
comes back to infinity with higher energy than with which
particle $A$ was coming in. The gain in the energy is caused by
the decrease of rotational energy of the hole. Such process is
possible because the Killing vector $\vec k$ becomes spacelike in
the ergosphere, so that the (conserved) energy of particle $B$
``as measured at infinity'' (see e.g. \cite{MTW}), $E_B =
-k_\alpha p^\alpha _B$, can be negative. Unfortunately, the
``explosion'' of particle $B$ requires such a big internal energy that
the process is not realistic astrophysically.

More general considerations of the interaction of black holes
with matter outside have led to the formulation of the four laws of
{\it black hole thermodynamics} \cite{Wa,FN,RW,WD}. 
These issues, in particular after the discovery of the
{\it Hawking effect} that black holes emit particles thermally
with temperature $T=\kappa \hbar/2\pi k c$ ($\kappa$-surface
gravity, $k$-Boltzmann's constant), have been an inspiration
in various areas of theoretical physics, going from general
relativity and statistical physics, to quantum gravity and string
theory (see \cite{WD,HORO} and some remarks and references in Section
3.2). The Kerr solution played indeed the most crucial role in
these developments. I recall how during my visits to Moscow in
the middle of the 1970s Zel'dovich and his colleagues were somewhat
regretfully admitting that they were on the edge of
discovering the Hawking effect. They realized that an analogue of
the Penrose process occurs with the waves (in so called {\it
superradiant scattering}) which get amplified if their energy per
unit angular momentum is smaller than the angular velocity
$\Omega$ of the horizon. Zel'dovich then suggested that there
should be spontaneous emission of particles in the corresponding
modes but did not study quantum fields on a nonrotating
background (cf. an account of these developments,
including the visit of Hawking to Moscow in 1973, by Israel
\cite{IS}).

Returning back to Earth or, rather, up to heavens, it is not so
well known that an astrophysically more realistic example of the
Penrose process exists: this is the {\it Blandford-Znajek
mechanism} -- see \cite{TP,Re,Bl} -- in which a
magnetic field threading a rotating hole (the field being
maintained by external currents in an accretion disk, for
example) can extract the hole's rotational energy and convert it into
a Poynting flux and electron-positron pairs. A Kerr black hole with
angular momentum parallel to an external magnetic field acts (by
``unipolar induction'') like a rotating conductor in an
external field. There will be an induced electric field and a
potential difference between the pole and the equator. If these
are connected, an electric current will flow and power will be
dissipated. In fact, this appears to be until now the most
plausible process to explain gigantic relativistic jets emanating
from the centres of some of the most active galaxies. The $BZ$-mechanism
has its problems: extremely rotating black holes expel magnetic
flux \cite{TP,BiJ} -- there probably exists a value of the
angular momentum $J_0<J_{max}$ for which the power extracted will
be greatest. It is not clear whether the process
can be efficiently maintained \cite{Puns}; and perhaps more
importantly, new astrophysical estimates of seed magnetic fields
seem to be too low to make the mechanism efficient \cite{MAB}.
The $BZ$-mechanism will probably attract more attention in the
coming years, in particular in view of the recent discovery of
two ``microquasars'' in our own Galaxy, which generate double radio
structures and apparent superluminal jets similar to 
extragalactic strong radio sources \cite{Mir}.

A remarkable achievement of pure mathematical physics,
with a great impact on astrophysics, has not only been the
discovery of the Kerr solution itself but also the development of
{\it the theory of Kerr metric perturbations} 
\cite{TP,FN,Chan,FuMa}. By employing the
Newman-Penrose null tetrad formalism, invented and extensively
used in mathematical relativity, in particular in gravitational
radiation theory, it has been possible to separate completely all
perturbation equations for non-zero spin fields. In particular, a
{\it single} ``master equation'' -- called the {\it Teukolsky
equation} -- governs scalar, electromagnetic and gravitational
perturbations of a Kerr black hole.\footnote{In the 
case of a Kerr-Newman black hole, the
electromagnetic and gravitational perturbations necessarily
couple. Until now, in contrast to the spherical
Reissner-Nordstr\"{o}m case, a way of how to decouple them has not
been discovered.}
If no sources are present on the right-hand side, the equation
looks as follows:
\begin{eqnarray}
\label{Equ28}
~&{\displaystyle
\left[ \frac{(r^2+a^2)^2}{\Delta} - a^2\sin ^2\theta \right]
\frac{\partial ^2\psi}{\partial t^2} + \frac{4Mar}{\Delta}
\frac{\partial ^2\psi}{\partial t\partial \phi} +
\left[ \frac{a^2}{\Delta} - \frac{1}{\sin ^2\theta} \right]
\frac{\partial ^2\psi}{\partial \phi ^2} \nonumber }&~\\
~&{\displaystyle
- \Delta ^{-s} \frac{\partial}{\partial r}\!\! \left( \Delta ^{s+1}
\frac{\partial \psi}{\partial r} \right) \!-\! \frac{1}{\sin \theta}
\frac{\partial}{\partial \theta} \!\left( \sin \theta \frac{\partial
\psi}{\partial \theta} \right) \nonumber 
\!-\! 2s\! \left[ \frac{a(r-M)}{\Delta}
\!+\! \frac{i\cos \theta}{\sin ^2 \theta} \right] \frac{\partial
\psi}{\partial \phi} \nonumber }&~\\
~&{\displaystyle
- 2s \left[ \frac{M(r^2-a^2)}{\Delta} - r - ia \cos \theta
\right] \frac{\partial \psi}{\partial t} + (s^2 \cot ^2 \theta -
s) \psi  =  0.} & ~
\end{eqnarray}
The coordinates are the Boyer-Lindquist coordinates used in
(\ref{Equ20}), $\Delta$ is defined in (\ref{Equ21}), and $s$ is
the spin weight of the perturbing field; $s=0, \pm 1, \pm 2$. The
variables in the Teukolsky equation can be separated by
decomposing $\psi$ according to
\begin{equation}
\label{Equ29}
{{}_s} \psi_{lm} = (1/\sqrt{2\pi}) {{}_s} R_{lm} (r, \omega)
{{}_s} S_{lm}
(\theta) e^{im\varphi} e^{-i\omega t},
\end{equation}
where ${{}_s} S_{lm}$ are so called spin weighted spheroidal harmonics.
By solving the radial Teukolsky equation for ${{}_s} R_{lm}$ with
appropriate boundary conditions one can find answers to a number
of (astro)physical problems of interest like the structure of
stationary electromagnetic or gravitational fields due to
external sources around a Kerr black hole (e.g. \cite{TP,BDV}),
 the emission of gravitational waves from particles
plunging into the hole (e.g. \cite{FN,KST,SNA}),
or the scattering of the waves from a rotating black hole (e.g.
\cite{FN,FuMa} and references therein). At present, the
Teukolsky equation is being used to study the formation of a
rotating black hole from a head-on collision of two holes of 
equal mass and spin, initially with small separation,
to find the wave forms of gravitational radiation 
produced in this process \cite{KrP}.
The first studies of second-order perturbations of a Kerr
black hole are also appearing \cite{CLO}.

To find all gravitational (metric) perturbations by solving the
complete system of equations in the Newman-Penrose formalism is
in general a for\-mi\-dable task. As Chandrasekhar's ``last
observation'' at the end of his chapter on gravitational
perturbations of the Kerr black hole reads \cite{Chan}:
``The treatment of the perturbations of the Kerr spacetime in
this chapter has been prolixious in its complexity. Perhaps, at a
later time, the complexity will be unravelled by deeper insights.
But mean time, the analysis has led us into a realm of the
rococo: splendorous, joyful, and immensely ornate.''

\subsection{Astrophysical evidence for a Kerr metric}

 Very recently new observations seem to have opened up real
possibilities of testing gravity in the strong-field regime. In
particular, it appears feasible to distinguish the Kerr metric
from the Schwarzschild, i.e. to measure $a/M$. Our following remarks
on these developments are based on the review by M. Rees \cite{Re},
and in particular, on the very recent survey by A. Fabian \cite{Fa},
an authority on diagnosing relativistic rotation from the
character of the emission lines of accretion disks around black
holes.

The interest here is not in optical lines since the optical band
comes from a volume much larger than the hole. However, the
X-rays are produced in the innermost parts of an accretion flow,
and should thus display substantial gravitational redshifts as
well as Doppler shifts. This only became possible to observe quite recently, 
when the ASCA X-ray satellite started to operate, 
and the energy resolution and sensitivity became 
sufficient to analyze line shapes.

Typically the profile of a line emitted by a disk from gas
orbiting around a compact object has a double-horned shape. The
disk can be imagined to be composed of thin annuli of orbiting
matter --  the total line is then the sum of contributions from
each annulus. If the disk is not perpendicular to our line of
sight its approaching sides will -- due to classical Doppler
shifts -- produce blue peaks, receding sides red peaks. The
broadest parts of the total line come from the innermost annuli
because the motion there is fastest. In addition, there are
relativistic effects: they imply that the emission is beamed in the
direction of motion, transverse Doppler shifted, and gravitationally
redshifted. As a result, the total line is broad and skewed in a
characteristic manner. Such lines are seen in the X-ray spectra
of most Seyfert 1 galaxies. In the Seyfert galaxy MCG-6-30-15 the
fluorescent iron line was observed to be (red)shifted further to
lower energies.\footnote{In Seyfert 1 galaxies 
hard flares occur which irradiate the accretion disk, and 
produce a reflection component of continuum peaking 
at $\sim 30$ keV and the fluorescent iron line at about $6.5$ keV.}
\begin{figure}
\centering
\includegraphics[width=.6\textwidth,angle=270]{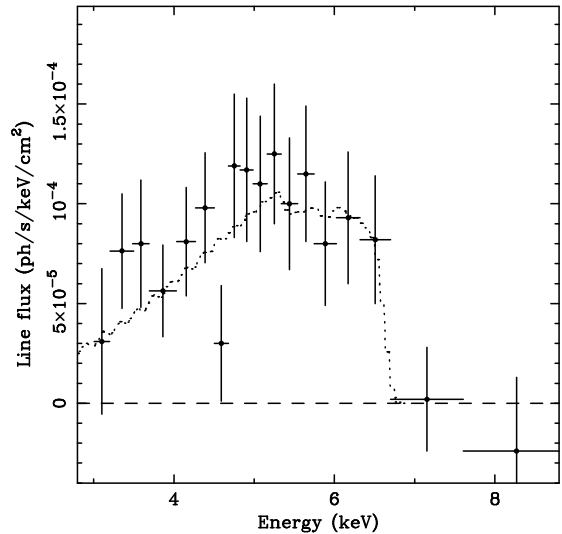}
\label{Figure 7}
\caption{The broad iron line from MCG-6-30-15. The best-fitting, maximally
spinning Kerr black hole model is shown (from Iwasawa, 
K. et al. (1996), Mon. Not. Roy. Astron. Soc. {\bf 282}, 1038).}
\end{figure}

This suggests that the emission took place below
$3R_s = 6M$, i.e. below the innermost stable orbit for a
Schwarzschild black hole. In 1996, the line shape was well
fitted by the assumption that the line is produced in a close
orbit around maximally rotating (extreme) Kerr black hole. In
1997, the parameter $a/M$ was quantified as exceeding 0.95.
Hence, it has been ``tentatively concluded that the line was the
first spectroscopic evidence for a Kerr hole'' \cite{Fa}.

There are alternative models for a broad skew iron line, including
Comptonization by cold electrons, or the emission from
irradiated matter falling from the inner edge of the disk
around a nonrotating Schwarzschild black hole. It
appears, however, that the data speak against these possibilities
\cite{Fa}. In any case, with future X-ray detectors, which will
yield count rates orders of magnitude higher than ASCA, the
line shapes should reveal in a much greater detail specific
features of the Kerr metric.

In addition, other possibilities to determine $a/M$ exist. These
include:\\
(i) Observations of stars in relativistic orbits going
through a disk around a supermassive rotating black hole
\cite{Re,KaVo}.\\
(ii) Characteristic frequencies of the
vibrational modes in disks or tori around rotating
black holes \cite{Re,Ips}.\\
(iii) The precession of a disk which is tilted with respect to
the hole's spin axis. This precession arises because of frame
dragging and produces a periodic modulation of the X-ray
luminosity.\\
(iv) Astrophysically most important would be a discovery showing
that the properties of cosmic jets depend on the value of $a/M$. 
This could indicate
that the Blandford-Znajek mechanism (see section 4.2) is really
going on. Its likelihood would increase if jets were found with
Lorentz factors $\gamma$ significantly exceeding 10 (see \cite{Re}
for more details).\\
(v) Last but not least, future observations of gravitational waves from black hole
collisions \cite{KST} offer great hopes of a clean observation of a
black hole geometry, without astrophysical complications.

It is hard to point out any other exact solution of Einstein's
field equations (or of any kind of field equations?)
discovered in the second half of the 20th century which has
had so many impacts on so many diverse areas of physics, astrophysics,
astronomy, and even space science as has had the Kerr metric.

%%%%%%%%%%%%%%%%%%%%%%%%%%%%%%%%%%%%%%%%%%%%%%%%%%%%%%%%%%%%%%%%%%%%%
%%%%%%\input unique.tex
%%%%%%%%%%%%%%%%%%%%%%%%%%%%%%%%%%%%%%%%%%%%%%%%%%%%%%%%%%%%%%%%%%%%%
\section{Black hole uniqueness and multi-black hole solutions}

Since black holes can be formed from the collapse of various matter
configurations, it is natural to expect that there will be many
solutions of Einstein's equations describing black holes. It is
expected that the asymptotic final state of a collapse can be
represented by a {\it stationary} spacetime, i.e. one which
admits a 1-dimensional group of isometries whose orbits are
timelike near infinity. Strong arguments show \cite{HE} that the
event horizon of a stationary black hole must be a Killing
horizon. One of the most remarkable and surprising results of 
black hole theory are the sequence of theorems showing
rigorously that the only stationary  solution of the Einstein
electrovacuum equations that is asymptotically flat and has a
regular event horizon is the Kerr-Newman solution. There is a
number of papers on this issue -- recent detailed reviews are
given in \cite{PC,HEU}. The intuition gained from exact 
black hole solutions in proving the theorems has been essential.

Roughly speaking, the uniqueness proof consists of the following
three parts. First, one demonstrates the ``rigidity theorem'',
which claims that nondegenerate $(\kappa \not= 0)$ stationary
electrovacuum analytic black holes are either static or axially symmetric.
One then establishes that the Reissner-Nordstr\"{o}m
nondegenerate electrovacuum black holes are all static (nonrotating)
nondegenerate black holes in electrovacuum. Finally, one
separately proves that the nondegenerate Kerr-Newman black holes
represent all nondegenerate axially symmetric stationary
electrovacuum black holes.

Although such results were proved more than 10 years ago,
recently there has been new progress in the understanding of
the global structure of stationary black holes. Again, exact
solutions have been inspiring:  by gluing together two copies
of the Kerr spacetime in a certain way, Chru\'{s}ciel \cite{PC}
constructed a black hole spacetime which is stationary but not
axisymmetric, demonstrating thus that the standard formulation and
proof of the rigidity theorem \cite{HE} is not correct. (The
reason being essentially that when one extends the isometries
from a neighbourhood of the horizon by analytic continuation one
has no guarantee that the maximal analytic extension is unique.)
Chru\'{s}ciel proved ``a corrected version of the black hole
rigidity theorem''; in the connected case one can prove a
uniqueness theorem for static electrovacuum black holes with {\it
degenerate} horizons. The uniqueness theorem for static
degenerate black holes which demonstrates that the extreme
Reissner-Nordstr\"{o}m black hole is the only case, is of
importance also in string theory. The most
unsatisfactory feature of the rigidity theorem 
is the assumption of analyticity of the metric in a
neighborhood of the event horizon. 
In this context, Chru\'{s}ciel \cite{PC} mentions
the case of Robinson-Trautman exact analytic metrics, which can be
smoothly but not analytically extended through an event horizon
\cite{BiPo}. We shall discuss this issue in somewhat greater
detail in Section 10.

The black hole uniqueness theorems indicated above are
concerned with only single black holes. 
(Corresponding spacetimes contain an asymptotically flat 
hypersurface $\Sigma$ with compact interior and
compact connected boundary $\partial \Sigma$ which is located
on the event horizon.) Consequently a question naturally arises 
as to whether one can generalize the theorems 
to some multi-black hole solutions.
In classical physics a solution exists in which a system of $n$
arbitrarily located charged mass points with charges $q_i$ and
masses $m_i$, such that $|q_i| = \sqrt{G} m_i$, is in static
equilibrium. In relativity the metric

\begin{equation}
\label{Equ30}
ds^2 = -V^{-2} dt^2 + V^2 \left(dx^2 + dy^2 + dz^2 \right),
\end{equation}
with time-independent $V$ satisfying Laplace's equation

\begin{equation}
\nabla^2V = \frac{\partial^2V}{\partial x^2} +
\frac{\partial^2V}{\partial y^2} + \frac{\partial^2V}{\partial
z^2} = 0\;,
\end{equation}
is a solution of the Einstein-Maxwell equations with the electric
field %given by

\begin{equation}
E = \nabla V^{-1},
\end{equation}
where $\nabla = (\partial/\partial x, \partial/\partial y,
\partial/\partial z)$. (In standard units $E = \sqrt{G} \nabla
V^{-1}$.) The simplest solution of this form is the
Majumdar-Papapetrou metric, corresponding to a linear combination
of $n$ ``monopole sources'' with masses $m_i>0$ and charges $q_i =
m_i$, located at arbitrary points $\vec x_i$:

\begin{equation}
V = 1 + \sum_{i=1}^{n} \frac{m_i}{\mid \vec x - \vec x_i \mid}.
\end{equation}
Hartle and Hawking \cite{HH} have shown that every such spacetime
can be analytically extended to a spacetime representing $n$
degenerate charged black holes in static equilibrium. The points
$\vec x = \vec x_i$ are actually event horizons of area $4 \pi
m^2_i$. For the case of one black hole, the metric (\ref{Equ30}) is just
the extreme Reissner-Nordstr\"{o}m black hole in isotropic
coordinates.

A uniqueness theorem for the Majumdar-Papapetrou metrics is not
available, although some partial answers are known (see
\cite{HS,CI} for more details). It is believed that
these are the only asymptotically flat, regular multi-black hole
solutions of the Einstein-Maxwell equations. In fact, such a
result would exclude an interesting possibility that {\it a repulsive
gravitational spin-spin interaction} between two (or more)
rotating, possibly charged, black holes can overcome their
gravitational attraction and thus that there exists in Einstein's
theory of gravitation -- in contrast to Newton's theory -- a
stationary solution of the two-body problem.

Among new solutions discovered by modern generating techniques,
there are the solutions of Kramer and Neugebauer which represent 
a nonlinear superposition of Kerr black holes (see \cite{KrNe} for
a review). These solutions have been the subject of a number of
investigations which have shown that spin-spin repulsion is not
strong enough to overcome attraction. In particular, two
symmetrically arranged equal black holes cannot be in stationary
equilibrium. The situation might change if one considers two
Kerr-Newman black holes \cite{BiHo}. Here one has four forces to
reckon with: gravitational and electromagnetic Coulomb-type
interactions, and gravitational and electromagnetic spin-spin
interactions. One can then satisfy the conditions which render
the system of two Kerr-Newman black hole free of singularities on
the axis, and make the total mass of the system positive. However,
there persists a singularity in the plane of symmetry away from
the axis \cite{BiHo}. In view of this result we have conjectured that
even with electromagnetic forces included one cannot achieve
balance for two black holes, except for the exceptional case of
two nonrotating extreme (degenerate) Reissner-Nordstr\"{o}m
black holes. Recently, some new rigorous results concerning
the (non)existence of multi-black hole stationary, axisymmetric
electrovacuum spacetimes have been obtained
\cite{Wein} (see also \cite{PC}), but the
``decisive theorem'' is still missing.

In connection with the problem of the balance of gravity by a
gravitational spin-spin interaction, we should mention that there
exists the solution of Dietz and Hoenselaers \cite{DiHo} in which
balance of the two rotating ``particles'' is achieved.
However, the ``sources'' are complicated naked singularities
which become Curzon-Chazy ``particles'' (see Section 6.1) if the
rotation goes to zero, and it is far from clear whether
appropriate physical interior solutions can be constructed.

In 1993, Kastor and Traschen \cite{KaTr} found an interesting family 
of solutions to the Einstein-Maxwell equations
with a non-zero cosmological constant $\Lambda$.
They describe an arbitrary number of charged black holes in a 
``background'' de Sitter universe. In the limit of $\Lambda = 0$
these solutions become Majumdar-Papapetrou static metrics.
In contrast to these metrics, the {\it cosmological multi-black hole
solutions with $\Lambda > 0$ are dynamical}. Remarkably,
one can construct solutions which describe coalescing black holes.
In some cases cosmic censorship is violated -- a naked singularity
is formed as a result of the collision \cite{BHK}.
Although these solutions do not have smooth horizons,
the singularities are mild, and geodesics can be extended through them.
The metric is always at least $C^2$. Since the solutions are dynamical,
one may interpret the non-smoothness of the horizons
as a consequence of gravitational and electromagnetic radiation.
In this sense, the situation is analogous to the case of 
the Robinson-Trautman spacetimes discussed in Section 10.
In five or more dimensions, however, one can construct {\it static}
multi-black hole solutions with $\Lambda = 0$, which 
do not have smooth horizons \cite{WLC}.
The solutions of Kastor and Traschen also inspired a new and careful
analysis \cite{BHA} of the global structure of the Reissner-Nordstr\"{o}m-de Sitter 
spacetimes characterized by mass, charge, and cosmological constant.
The structure is considerably richer than that with $\Lambda = 0$.
Most recently, the hoop conjecture 
(giving the criterion as to whether a black hole forms from a collapsing system)
was discussed \cite{INSH} by analyzing the solution of Kastor and Traschen.

%%%%%%%%%%%%%%%%%%%%%%%%%%%%%%%%%%%%%%%%%%%%%%%%%%%%%%%%%%%%%%%%%%%%%%%%
\section{On stationary axisymmetric fields and relativistic disks}
%%%%%%%%%%%%%%%%%%%%%%%%%%%%%%%%%%%%%%%%%%%%%%%%%%%%%%%%%%%%%%%%%%%%%%%%
\subsection{Static Weyl metrics}
The static axisymmetric vacuum metrics in Weyl's canonical
coordinates $\rho \in [0, \infty), z, t \in \RRe, \varphi \in
[0, 2\pi)$ have the form

\begin{equation}
ds^2 = e^{-2U} \left[ e^{2k} \left( d\rho ^2 + dz ^2 \right) + \rho
^2 d \varphi ^2 \right] - e^{2U} dt^2.
\end{equation}
The function $U(\rho,z)$ satisfies flat-space Laplace's equation
\begin{equation}
\frac{\partial ^2U}{\partial \rho ^2} + \frac{1}{\rho}
\frac{\partial U}{\partial \rho} + \frac{\partial ^2 U}{\partial
z^2} = 0.
\end{equation}
The function $k(\rho, z)$ is determined from $U$ by quadrature up to an additive
constant. The axis $\rho = 0$ is free of conical singularities
at places where $\lim_{\rho \rightarrow 0}k = 0$.

The mathematically simplest example is the Curzon-Chazy solution
in which $U = -m/\sqrt{\rho ^2 + z^2}$ is the Newtonian potential
of a spherical point particle. The spacetime, however, is not
spherically symmetric. In fact, one of the lessons 
which one has learned from
this solution is the {\it directional character of the
singularity} at
$\rho^2 + z^2 = 0$. For example, the limit of the invariant
$R_{\alpha \beta \gamma \delta} R^{\alpha \beta \gamma \delta}$
depends on the direction of approach to the singularity. The
singularity has a character of a ring through which some timelike
geodesics may pass to a Minkowski region \cite{SS}.

\vskip -0.15pt
Various studies of the Weyl metrics indicated explicitly how
important it is always to check whether a result is not
just a consequence of the choice of coordinates. There is the
subclass of Weyl metrics generated by the Newtonian potential of
a constant density line mass (``rod'') with total mass $M$ and
(coordinate) length $l$, which is located along the $z$-axis with
the middle point at the origin. These are Darmois-Zipoy-Vorhees
metrics, called also the $\gamma$-metrics \cite{Bo}. The Schwarzschild
solution (a spherically symmetric metric!) is a special case in
this subclass: it is given by the potential of the rod with
$l=2M$. Clearly, in general there is no correspondence between
the geometry of the physical source and the geometry of the
Newtonian ``source'' from the potential of which a Weyl metric is
generated.

\vskip -0.15pt
A survey of the best known Weyl metrics, including some specific
solutions describing fields due to circular disks is contained in
\cite{Bo}.
More recently, Bi\v c\'ak, Lynden-Bell and Katz \cite{BLK} have
shown that most vacuum static Weyl solutions, including the
Curzon and the Darmois-Vorhees-Zipoy solutions, can arise as the
metrics of
counterrotating relativistic disks (see \cite{BLK} also for
other references on
relativistic disks). The simple idea which inspired their work
is commonly used in Newtonian {\it galactic dynamics} \cite{BT}:
imagine a point mass placed at a distance $b$
below the centre $\rho = 0$ of a plane $z=0$. This gives a solution
of Laplace's
equation above the plane. Then consider the potential obtained by
reflecting this $z
\ge 0$ potential in $z = 0$ so that a symmetrical solution both above
and below the
plane is obtained. It is continuous but has a discontinuous normal
derivative on $z
= 0$, the jump in which gives a positive surface density on the plane.
In galactic
dynamics one considers general line distributions of mass along
the negative
$z$-axis and, employing the device described above, one finds the
potential-density pairs for general axially symmetric disks.
In \cite{BLP}, an infinite number of new static solutions of
Einstein's equations were found starting from realistic potentials
used to describe flat
galaxies, as given recently by Evans and de Zeeuw \cite{EZ}.

\vskip -0.15pt
Although these disks are Newtonian at large distances, in their
central regions interesting relativistic features arise, such as
velocities close to the velocity of light, and large redshifts. In
a more mathematical context, some particular cases are so far the
only explicit examples of spacetimes with a ``polyhomogeneous''
null infinity (cf. \cite{Chr} and Section 9), and spacetimes
with a meaningful, but infinite ADM mass \cite{BLP}. 
New Weyl vacuum solutions
generated by Newtonian potentials of flat galaxies correspond to
both finite and semi-infinite rods, with the line mass densities
decreasing according to general power laws. It is an open
question what kinds of singularities rods with different 
density profiles represent.

Very recently, new interesting examples of the static solutions describing
self-gravitating disks or rings, and disks or rings around static black holes have been
constructed \cite{SZZ,LEL,GOL} and the effects of the fields on freely moving
test particles studied \cite{SZZ}. Exact disks with electric 
currents and magnetic fields have also been considered \cite{PLE}. 

Employing the Weyl formalism, one can describe nonrotating 
{\it black holes strongly distorted} by the surrounding matter.
The influence of the matter
can be so strong that it may even cause the horizon topology to
be changed from spherical to toroidal (see \cite{TP} and
references therein).

Finally, we have to mention two solutions in the Weyl class,
which were found soon after the birth of general relativity, and
have not lost their influence even today. The first, discovered by
Bach and Weyl, is assigned by Bonnor \cite{Bo} as ``probably the most
perspicacious of all exact solutions in GR''. It refers to two
Curzon-Chazy ``monopoles'' on the axis of symmetry. One finds 
that the metric function $k$ has the property that $\lim_{\rho
\rightarrow 0} k \not= 0$, so that there is a stress described by
a conical singularity between the particles, which holds particles
apart. A similar solution can be constructed for the
Schwarzschild ``particles'' (black holes) held apart by a stress.
These cases can serve as one of the simplest demonstrations of
the difference between the Einstein theory and field theories like the
Maxwell theory: it is only in general relativity in which 
field equations involve also {\it equations of motion}.

The second ``old'' solution which has played a very significant
role is the metric discovered by Levi-Civita. 
It belongs to the class of degenerate (type {\it D}) static vacuum 
solutions which form a subclass of the Weyl solutions. 
In the invariant classification of the
degenerate solutions by Ehlers and Kundt \cite{EK}, this solution
is contained in the last, third subclass. That is why Ehlers and
Kundt called it the {\it C}-metric, and it is so well known today. We
shall discuss the {\it C}-metric later (Section 11) in greater
detail since, as it has been learned in the 1970s, it is actually a
radiative solution representing uniformly accelerated black
holes. What Levi-Civita found and Ehlers and Kundt analyzed is
only a portion of spacetime in which the boost Killing vector is
timelike, and the coordinates can thus be found there (analogous to the
coordinates in a uniformly accelerated frame in special
relativity) in which the metric is time-independent.

\subsection{Relativistic disks as sources of the Kerr metric and
other stationary spacetimes}
Thanks to the black hole uniqueness theorems (Section 5), the Kerr metric represents the unique solution
describing all
rotating vacuum black holes. Nevertheless, although the cosmic censorship
conjecture, on which the physical relevance of the Kerr metric rests,
is a very plausible hypothesis, it remains, as was noted in several places above,
one of the central unresolved issues in relativity. 
It would thus support the significance of the Kerr metric if a physical
source were found which produces the Kerr field. The situation would then
resemble the case of the spherically symmetric Schwarzschild metric 
which can represent both a black hole and the external field due to matter.

This has been realized by many workers. The review on the
``Sources for
the Kerr Metric'' \cite{Kr} written in 1978, contains 71 references, and
concludes
with: ``Destructive statements denying the existence of a material
source for the
Kerr metric should be rejected until (if ever) they are reasonably
justified.''
The work from 1991 gives ``a toroidal source'',
consisting of ``a toroidal shell \dots, a disk \dots and an annulus
of matter
interior to the torus'' \cite{Mc}. The masses of the disk and annulus
are negative. To
summarize in Hermann Bondi's way, the sources suggested for the Kerr metric
have not been the easiest materials to buy in the shops \dots

The situation is somewhat different in the special case of the {\it extreme}
Kerr metric,
where there is a definite relationship between mass and angular momentum.
The numerical study \cite{BW} of uniformly rotating disks indicated how
the extreme Kerr
geometry forms around disks in the ``ultrarelativistic'' limit. These numerical results
have been supported by important analytical work
(see Section 6.3). However, in the case
of a general Kerr metric physical sources had not been found before 1993.

A method similar to that of constructing disk sources of
static Weyl spacetimes (described in Section 5.1) has been
shown to work also for
axisymmetric, reflection symmetric, and {\it stationary} spacetimes
\cite{BiLe,PL}. It is
important to realize that although now no metric function solves
Laplace's equation
as in the static case, we may view the procedure described
in Section 5.1 as the {\it
identification} of the surface $z = b$ with the surface $z = -b$. The
field then remains
continuous, but the jump of its normal derivatives induces a matter
distribution in
the disk which arises due to the identification of the surfaces.
What remains to be seen, is whether
the material can be ``bought in the shops''.
This idea can be employed for all known asymptotically flat
stationary vacuum
spacetimes, for example for the Tomimatsu-Sato solutions, for the
``rotating'' Curzon
solution, or for other metrics (cf. \cite{KSH} for references).

Any stationary axisymmetric vacuum metric can be written in
canonical coordinates
$(t,\varphi,\rho,z)$ in the form \cite{KSH}
\begin{equation}
\label{Equ36}
ds^2 = e^{-2U} \left[ e^{2k} \, (d\rho ^2+dz^2) + \rho ^2 d\varphi^2)
\right] -
e^{2U} (dt +
Ad\varphi)^2 \ ,
\end{equation}
where $U$, $k$, and $A$ are functions of $\rho$, $z$.
For the Kerr solution (mass $M$, specific angular momentum $a \ge 0$),
the functions
 $U, k, A$ are ratios of polynomials when expressed in spheroidal
coordinates \cite{KSH}.

Now, identify the ``planes'' $z = b = \const > 0$ and $z = -b$
(this identification leads to disks
with zero radial pressure). With the Kerr
geometry the matching is more complicated than in the static
cases, and therefore,
one has to turn to Israel's covariant formalism (see \cite{Bar} for its
recent exposition).
Using this formalism one is able to link the surface
stress-energy tensor of the disk
arising from this identification, to the jump of normal
extrinsic curvature
across the timelike hypersurface given by $z = b$ (with the jump
being determined by the discontinuities in the normal
derivatives in functions $U, k, A$).

The procedure leads to physically plausible disks made of two
streams of collisionless particles, that  circulate in opposite
directions with differential velocities \cite{BiLe,PL}.
Although extending to infinity, the disks have finite mass and
exhibit interesting relativistic properties such as high velocities,
large redshifts, and dragging effects, including ergoregions.
Physical disk sources of Kerr spacetimes with $a^2>M^2$ can be
constructed (though these are ``less relativistic''). And the
procedure works also for electrovacuum stationary spacetimes. The
disks with electric current producing Kerr-Newman spacetimes are
described in \cite{LeB}, where the conditions for 
the existence of (electro)\-geodesic streams are also discussed. 

The power and beauty of the Einstein field equations is again
illustrated: the character of exact vacuum fields determines
fully the physical characteristics of their sources. In a more
sophisticated way, this is seen in the problem of
relativistic rigidly (uniformly) rotating disks of dust.

\subsection{Uniformly rotating disks}

The structure of an infinitesimally thin, finite relativistic
disk of dust particles which rotate uniformly around a common
centre was first explored by J. Bardeen and R. Wagoner 25 years
ago \cite{BW}. By developing an efficient expansion technique in
the quantity $\delta = z_c/(1+z_c)$, $z_c$ denoting the central
redshift, they obtained numerically a fairly complete picture of
the behaviour of the disk, even in the ultrarelativistic regime
$(\delta \rightarrow 1)$. In their first letter from 1969 they
noted that ``there may be some hope of finding an analytic
solution''. Today such a hope has been substantiated, thanks to the work
of G. Neugebauer and R. Meinel (see \cite{GN1,GN2} and
references therein). The solution had, in fact, to wait until the
``soliton-type-solution generating techniques'' for nonlinear
partial differential equations had been brought over from
applied mathematics and other branches of physics to general
relativity, starting from the end of the 1970s.

These techniques have been mainly applied only in the vacuum
cases so far, but this is precisely what is in this case
needed: the structure of the thin disk enters the field equations
only through the boundary conditions at $z = 0$, $0 \leq \rho \leq
a$ ($a$ is the radius of the disk). The specific procedures which
enabled Neugebauer and Meinel to tackle the problem are
sophisticated and lengthy. Nevertheless, we wish to mention them
telegraphically at least, since they represent the first example
of solving {\it the boundary value problem} for a {\it rotating}
object in Einstein's theory by analytic methods.

In the stationary axisymmetric case, Einstein's vacuum field
equations for the metric (\ref{Equ18}) imply the 
well-known Ernst equation (see e.g.
\cite{KSH}) --  a nonlinear partial differential equation for a
complex function $f$ of $\rho$ and $z$:

\begin{equation}
\label{Equ37}
({\rm Re} f) \left[ f_{,\rho \rho} + f_{,zz} + \frac{1}{\rho} f_{,\rho}
\right] = f^2_{,\rho} + f^2_{,z},
\end{equation}
where the Ernst potential
\begin{equation}
f(\rho,z) = e^{2U} + ib,
\end{equation}
with $U(\rho, z)$ being the function entering the metric
(\ref{Equ36}),
function $b(\rho,z)$ is a ``potential'' for $A(\rho, z)$ in
(\ref{Equ37}),

\begin{equation}
A_{,\rho} = \rho \, e^{-4U} b_{,z}, \hspace{0.5cm} A_{,z} = -\rho
\,
e^{-4U} b_{,\rho},
\end{equation}
and the last function $k(\rho, z)$ in (\ref{Equ37}) can be determined
from $U$ and $b$ by quadratures.

{\it The Ernst equation} can be regarded {\it as the integrability
condition}
of a system of {\it linear} equations for a complex matrix
$\Phi$, which is a function of $\rho + iz, \rho - iz$, and of a
(new) complex parameter $\lambda$. Knowing $\Phi$, one can
determine $f$ from $\Phi$ at $\lambda = 1$. Now the problem of 
solving the linear system can be reformulated as the so called
{\it Riemann-Hilbert problem} in complex function theory. (This,
very roughly, means the following: let $K$ be a closed curve in
the complex plane and $F(K)$ a matrix function given on $K$; find
a matrix function $\Phi_{in}$ which is analytic inside $L$, and
$\Phi_{out}$ analytic outside $K$ such that $\Phi_{in} \Phi_{out}
= F$ on $K$.) The Riemann-Hilbert problem can be formulated as an
integral equation. The hardest problem with which Neugebauer and
Meinel were faced was in connecting the specific physical
boundary values of $f$ on the disk with the functions entering the
Riemann-Hilbert problem (with contour $K$ being determined by the
position of the disk in the $\rho,z$ plane), and 
with the corresponding integral equation.
The fact that they succeeded and found the solution of their integral
equation is a remarkable achievement in mathematical physics. The
gravitational field and various physical characteristics of the
disk (e.g. the surface density) are given up to quadratures in
terms of ultraelliptic functions \cite{GN1}, which can be
numerically evaluated without difficulties. This result, however,
may appear as a ``lucky case'': it does not imply that one will
be able to tackle similarly more complicated situations as, for
example, thin disks with pressure, with non-uniform rotation,
or $3$-dimensional rotating bodies such as neutron stars.

\begin{figure}
\centering
\includegraphics[width=.740\textwidth]{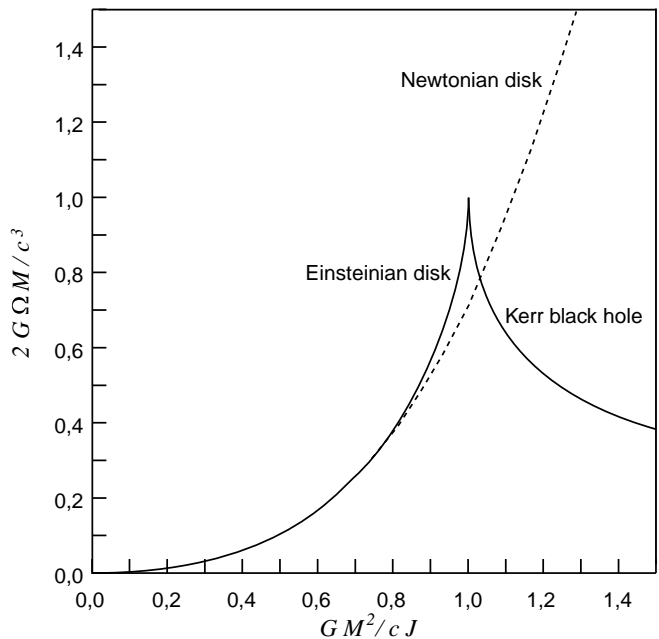}
\label{Figure 8}
\caption{The general relativistic (``Einsteinian'') thin disk of 
rigidly rotating dust constructed by Neugebauer and Meinel, 
compared with the analogous disk in Newtonian theory.
If the angular momentum is too low, the disk forms a rotating 
(Kerr) black hole. (From \cite{GN2}.)}
\end{figure}

Many physical characteristics of uniformly rotating relativistic
disks  such as their surprisingly high binding energies, the high
redshifts of photons emitted from the disks, or the dragging of
inertial frames in the vicinity of the disks, were already 
obtained with remarkable accuracy in \cite{BW}, as the exact solution
now verifies. Here we only wish to demonstrate the fundamental
difference between the Newtonian and relativistic case, as it is
illustrated in Fig. 8. 
The rigidly rotating disk of dust 
of Neugebauer and Meinel represents the relativistic
analogue of a classical Maclaurin disk. For the Maclaurin disk,
it is easy to show that the (dimensionless) quantities $y=2G\Omega
M/c^3$ and $ x = GM^2/cJ$ ($M$ and $J$ are the total mass and angular
momentum respectively, and $\Omega$ is the angular velocity) 
are related by $y=(9\pi^2/125)x^3$. 
For a fixed $M$ the angular velocity $\Omega \sim y$
can be increased arbitrarily, with $J$ being correspondingly
decreased. For relativistic disks, however, there is an upper 
bound on $\Omega$ given by $\Omega_{max} = c^3/2GM$,
whereas $J$ is restricted by the lower bound $J_{min} = GM^2/c$.
With an angular momentum too
low, a {\it rigidly} rotating disk cannot exist. If we
``prepare'' such a disk, it immediately begins to collapse and
forms -- assuming the cosmic censorship -- a rotating Kerr black hole
with $x=GM^2/cJ>1$. (Notice that the assumption of rigid
rotation is here crucial: the differentially rotating disks
considered in the preceding section can have an arbitrary value of
$x$.) Since one can define the angular velocity $\Omega(M,J)$ of
the horizon of a Kerr hole, one may consider $y(x)$ for
black hole states with $x>1$ (cf. Fig 8). The
rigidly rotating disk states and the black hole states just
``meet'' at $y=x=1$. In the ultrarelativistic limit the
gravitational field outside the disk starts to be unaffected by
the detailed structure of the disk -- it approaches the field of
an extremely rotating Kerr black hole with $x = 1$. Such a result
had already been obtained by Bardeen and Wagoner. However, it is only
now, with the exact solution available, that it can be
investigated with full rigor. It gives indirect evidence
that Kerr black holes are really formed in the gravitational
collapse of rotating bodies. 

As noticed also by Bardeen and Wagoner, in the ultrarelativistic limit 
the disk itself ``becomes buried in the horizon of the extreme Kerr
metric, surrounded by its own infinite, non asymptotically flat universe''
(see \cite{Me} for a recent detailed analysis of the ultrarelativistic limit). Similar phenomena arise also 
in the case of some spherical solutions of Einstein-Yang-Mills-Higgs
equations (cf. \cite{COR} and Section 13).
%%%%%%%%%%%%%%%%%%%%%%%%%%%%%%%%%%%%%%%%%%%%%%%%%%%%%%%%%%%%%%%%%%%%%
%%%%%%%%%\input taubnut.tex
%%%%%%%%%%%%%%%%%%%%%%%%%%%%%%%%%%%%%%%%%%%%%%%%%%%%%%%%%%%%%%%%%%%%%
\section{Taub-NUT space}
The name of this solution of vacuum Einstein's equations fits
both to the names of its discoverers (Taub-Newman-Unti-Tamburino)
and to its curious properties. Owing to these
properties (which induced Misner \cite{Mi} to consider the solution
``as a counterexample to almost anything''), this spacetime has
played a significant role in exhibiting the type of
effects that can arise in strong gravitational fields.

Taub \cite{Tau} discovered an empty universe with four global Killing
vectors almost half a century ago, during his pioneering study of
metrics with several symmetries. By continuing the Taub universe
through its horizon one arrives in NUT space. NUT space, however,
was only discovered in 1963 by a different method \cite{NTU}. 
In fact, it could
have been obtained earlier by applying the transformation given in
J\"{u}rgen Ehlers' dissertation \cite{EHD} and his talk at the GR2
conference in Royaumont in 1959 \cite{EHRO}. This transformation gives
the recipe for obtaining stationary solutions from static ones.
How the NUT space can be obtained by applying this transformation
to the Schwarzschild metric was demonstrated explicitly by Ehlers
at GR4 in London in 1965 \cite{EHLO}.

\subsection{A new way to the NUT metric}

Here we shall briefly mention a simple, physically appealing new
derivation of the NUT metric given recently by Lynden-Bell and
Nouri-Zonoz (LBNZ) \cite{LBNZ}. Their work also shows how even
uncomplicated solutions may still be of interest in unexpected
contexts. LBNZ's inspiration to study the NUT space has in fact
come from Newton's Principia! In one of his scholia Newton
discusses motion under the standard central force plus a force
which is normal to the surface swept out by the radius vector to
the body which is describing the non-coplanar path. A simple
interesting case is the motion of mass $m_0$ satisfying the equation

\begin{equation}
\label{Equ40}
m_0~ {d^2 \vec r \over dt^2} = -V^{'}\!(r)~{\bf \hat r} + \frac{m_0}{c}~ \vec v
\times \vec B_g,
\end{equation}
where ${\bf \hat r} = \vec r/r$,
\begin{equation}
\vec B_g = - Q~{\bf \hat r}/r^2, \\\\\\\\\ Q = \tilde Q~ c/m_0,
\end{equation}
$Q$ and $\tilde Q$ are constants, and $c$ is the velocity of light. 
Here we write $c$ explicitly though $c$ = 1, 
to make the analogy with magnetism. Indeed, $\vec B_g$
is the field of a ``gravomagnetic'' monopole of strength $Q$. The
classical orbits of particles lie on cones which, if the monopole
is absent, flatten into a plane \cite{LBNZ}.

It was known that NUT space corresponds to the mass with a
gravomagnetic monopole, but this was never used in such a
physical way as by LBNZ for its derivation. The main point is to
start from the well-known split of the stationary metrics as
described in Landau and Lifshitz \cite{LL} (see \cite{Ger} for a
covariant approach, and the contribution of Beig and Schmidt in the present volume)
\begin{equation}
\label{Equ42}
ds^2 = - e^{-2\nu}{(dt - A_idx^i)}^2 + \gamma_{ij} dx^i dx^j,
\end{equation}
where $\nu, A_i, \gamma_{ij}$ are independent of $t$. This form
is unique up to the choice of time zero: $t^{'} =  t + \chi (x^i)$
implies again the metric in the form (\ref{Equ42}) in $(t^{'},
x^i$), with
the ``vector potential'' undergoing a gauge transformation $A'_i =
A_i + \nabla_i \chi $. Writing down the equation of motion of a
test particle in metric (\ref{Equ42}), in analogy with the equation of
motion of a charged particle in an electromagnetic field, one is
naturally led to define the ``gravoelectric'' and
``gravomagnetic'' fields by
\begin{equation}
\label{Equ43}
\vec E_g = \nabla \nu, \,\,\,   \vec B_g = \nabla \times \vec A,
\end{equation}
where ``$\nabla \times$'' is with respect to $\gamma_{ij}$. Following the problem of
$\S 95$ in \cite{LL} one then rewrites all Einstein's equations in
terms of the fields (\ref{Equ43}), the metric $\gamma_{ij}$, and their
derivatives. To find the vacuum {\it spherically symmetric} spatial $\gamma$-metric
one takes $\gamma_{ij} dx^i dx^j = e^{2\lambda} dr^2 +
r^2(d\theta^2 + \sin^2 \theta~d\varphi^2)$, and one assumes $\nu = \nu (r)$,
and $B^r_g = - Q e^{- \lambda}/ r^2$. The Einstein equations then
imply the spacetime metric, which is {\it not} spherically symmetric, in the form
\begin{eqnarray}
\label{Equ44}
ds^2 &=& -e^{-2 \nu}{
\left({\begin{array}{c}\mbox{}\\\mbox{}\end{array} }\! dt -2q \left(1 + \cos \theta \right) d\varphi \right)}^2 +
{\left(1 - q^2/r^2\right)}^{-1} e^{2\nu} dr^2  \nonumber \\
~&\lefteqn{~~~~~~~~~~~~+~r^2 \left(d\theta^2 + \sin ^2 \theta~d \varphi^2\right),}&~
\end{eqnarray}
where $q =Q/2 = \const$,
\begin{equation}
e^{-2 \nu} = 1 - 2 r^{-2} \left(q^2 + m {\sqrt{r^2 - q^2}}
\right),
\end{equation}
and the vector potential $A_{\varphi} = 2q (1 + \cos \theta)$
satisfies (\ref{Equ43}). (The factor $(1-q^2/r^2)$ should be raised to the
power $-1$ in equation (3.22) in \cite{LBNZ}, as it is clear from (3.20).) Equation (\ref{Equ44}) 
is the NUT metric, with
$r$ being the curvature coordinate of spheres $r = \const$. With
$q = 0$ the metric (\ref{Equ44}) becomes the Schwarzschild metric in the
standard Schwarzschild coordinates.

More commonly the metric (\ref{Equ44}) is written in the form
\begin{equation}
\label{Equ46}
ds^2 = - V {\left(d \tilde t + 4q \sin^2 \frac {\theta}{2}~d
\varphi\right)}^2 + V^{-1}d \tilde r^2 + \left(\tilde r^2 +
q^2\right) \left(d \theta^2
+ \sin^2 \theta~d \varphi^2\right),
\end{equation}
\begin{equation}
\label{Equ47}
%V = 1 - 2 \left(m \tilde r + q^2\right)/\left(\tilde r^2 + q^2\right),
V = 1 - 2 ~{m \tilde r + q^2 \over \tilde r^2 + q^2},
\end{equation}
which can be obtained from (\ref{Equ44}) by putting
\begin{equation}
\label{Equ48}
\tilde r = \sqrt {r^2 - q^2}, \hspace{0.5cm}
\tilde t = t - 4q \varphi.
\end{equation}

Recently, Ehlers \cite{ETN} considered a Newtonian limit of NUT space
within his frame theory which encompasses general relativity and
Newton-Cartan theory (a slight generalization of Newton's
theory). The main purpose of Ehlers' frame theory is to define
rigorously what is meant by the statement that a one parameter family of
relativistic spacetime models converges to a Newton-Cartan model
or, in particular, to a strictly Newtonian model.

The strictly Newtonian limit occurs when the Coriolis angular
velocity field $\vec \omega$, related to the connection
coefficients $\Gamma^i_{tj}$ in the Newton-Cartan theory, depends
on time only. NUT spacetimes approach a truly Newton-Cartan limit
with spatially {\it non}-constant radial Coriolis field
$\omega ^{\tilde r} = -q/\tilde r^2$, which in this limit coincides
with the Newtonian gravomagnetic field. As in the analogous
classical problem with the equation of motion (\ref{Equ40}), the geodesics in NUT
space lie on cones. This result has been used to study 
gravitational lensing by gravomagnetic monopoles \cite{LBNZ}: they
twist the rays that pass them in a characteristic manner,
different from that due to rotating objects.

The metrics (\ref{Equ44}) and (\ref{Equ46}) appear to 
have a preferred axis of fixed points of symmetry. This
is a false impression since we can switch the axis into any direction by a
gauge transformation. For example, the metric (\ref{Equ44}) has a conical
singularity at $\theta = 0$ but is regular at $\theta = \pi$,
whereas the metric (\ref{Equ46}) has a conical singularity at
$\theta = \pi$ but is
regular at $\theta = 0$. The metrics are connected by the simple
gauge transformation, i.e. $t \rightarrow \tilde t = t - 4q
\varphi$. 
A mass endowed with a gravomagnetic monopole appears as a spherically
symmteric object but the spacetime is {\it not} spherically
symmetric according to the definition given in Section 2.1.
Nevertheless, there exist equivalent coordinate
systems in which the axis can be made to point in any direction --
just as the axis of the vector potential of a magnetic monopole
can be chosen arbitrarily. For further references on interpreting
the NUT metric as a gravomagnetic monopole, see \cite{LBNZ} and the
review by Bonnor \cite{Bo}.

\subsection{Taub-NUT pathologies and applications}

By introducing two coordinate patches, namely the coordinates of
metric (\ref{Equ44}) to
cover the south pole $(\theta = \pi)$ and those of (\ref{Equ46}) the north
pole $(\theta = 0)$, the rotation axis can be made regular.
However since $\varphi$ is identified 
with period $2\pi$, equation (\ref{Equ48})
implies that $t$ and $\tilde t$ have to be identified with the period
$8\pi q$. Then observers with $({\tilde r}, \theta, \varphi) = \const$
follow {\it closed timelike lines} if $V$ in (\ref{Equ47}) is positive, i.e. if
${\tilde r} > {\tilde r}_0 = m + (m^2 + q^2)^{\frac{1}{2}}$. 
The hypersurface ${\tilde r}={\tilde r}_0$ is the null hypersurface --
horizon, below which lines $(t, \theta, \varphi) = \const$
become spacelike. Because of the periodic identification of $t$
and $\tilde t$, the hypersurfaces of constant ${\tilde r}$ change the
topology from $S^2 \times R^1$ to $S^3$, on which 
$t/2q,~\theta,~\varphi$ become Euler angle coordinates.

\vskip -0.1pt
The region with $V<0$ is the Taub universe: it has homogeneous
but non-isotropic space sections ${\tilde r} = \const$. The coordinate ${\tilde r}$,
allowed to run from $-\infty$ to $+\infty$, is a timelike
coordinate, and is naturally denoted by $t$ in the Taub region.

\vskip -0.1pt
In addition to the closed timelike lines in the NUT region there
are further intriguing pathologies exhibited by the Taub-NUT
solutions. Here we just list some of them and refer to the relevant
literature \cite{HE,Pen,Mi}. The Taub region is globally hyperbolic:
its entire future and past history can be determined from
conditions given on a spacelike Cauchy hypersurface. However,
this is not the case with the whole Taub-NUT spacetime. As in the
Reissner-Nordstr\"{o}m spacetimes (Section 3), there are
Cauchy horizons $H^{\pm}(\Sigma)$ of a particular spacelike
section $\Sigma$ of maximal proper volume lying between the globally
hyperbolic Taub regions and the causality violating NUT regions.
$H^{\pm}(\Sigma)$ are smooth, compact null hypersurfaces
diffeomorphic to $S^3$ -- the generators of such null surfaces are
{\it closed null geodesics}. The Taub region is limited between
$t_- \leq t \leq t_+$, where $t_{\pm}$ are roots of $V$ in equation
(\ref{Equ47})
(with the interchange $t\leftrightarrow {\tilde r}$). This region is
compact but there are timelike and null geodesics which remain
within it and are not complete. (See \cite{Pen} for a nice picture of
these geodesics spiralling around and approaching $H_+(\Sigma)$
asymptotically.) This pathological behaviour of ``the incomplete
geodesics imprisoned in a compact neighbourhood of the horizon''
was inspirational in the definition of singularities \cite{Cla} -- one meets
here the example in which the geodesic incompleteness is not
necessarily connected with strong gravitational fields. 
It can be shown, however, that after an addition of even the slightest
amount of matter this pathological behaviour will not take place
-- true singularities arise.

\vskip -0.1pt
This enables one to consider the time between $t_-$ and $t_+$ as
the lifetime of the Taub universe. Wheeler \cite{JAW} constructed a
specific case of the Taub universe which will live as long as a
typical Friedmann closed dust model ($\sim 10^{10}$ years) but
will have a volume at maximum expansion smaller by a factor of $5
\times 10^{10}$. This example thus appears to be a difficulty for the
anthropic principle.

\vskip -0.1pt
Taub space seems also to be the only known example giving the
possibility of making inequivalent NUT-like extensions which lead
to a non-Hausdorff spacetime manifold \cite{HE,Ger,Haj}.

The Taub-NUT solution plays an important role in
cosmology and quantum gravity. Here we wish to
note yet two other recent applications of this space. About ten years
ago, interest was revived in closed timelike lines, time
machines, and wormholes. One of the leaders in this activity, Kip
Thorne, explains in \cite{KT} in a pedagogical way the main recent results
on closed timelike curves and wormholes by using ``Misner space'' --
Minkowski spacetime with identification under a boost, which
Misner introduced as a simplified version of Taub-NUT space.

The second application of the Taub-NUT space is still more 
remarkable -- it plays an important role outside general
relativity. The asymptotic motion of monopoles in 
(super-)Yang-Mills theories corresponds to the geodesic motion in {\it
Euclidean} Taub-NUT space \cite{GiMa}. Euclidean Taub-NUT spaces have
been discussed in many further works on monopoles in gauge
theories. One of the latest of these works \cite{KrB}, on the exact $T$-duality (which
relates string theories compactified on large and small tori)
between Taub-NUT spaces and so called ``calorons'' (instantons at
finite temperature defined on $R^3 \times S^1)$, gives also
references to previous contributions.

%%%%%%%%%%%%%%%%%%%%%%%%%%%%%%%%%%%%%%%%%%%%%%%%%%%%%%%%%%%%%%%%%%%%%
%%%%%%%%\input pwaves.tex
%%%%%%%%%%%%%%%%%%%%%%%%%%%%%%%%%%%%%%%%%%%%%%%%%%%%%%%%%%%%%%%%%%%%%
\section{Plane waves and their collisions}

\subsection{Plane-fronted waves}
The history of gravitational plane waves had began already by 1923
with the paper on
spaces conformal to flat space by Brinkmann. Interest in these waves
was revived in 1937 by Rosen, and in the late 1950s by
Bondi, Pirani and Robinson, Holy, and Peres 
(see \cite{EK,KSH} for references). A comprehensive geometrical approach
to these spacetimes soon followed in the classical treatise by
Jordan, Ehlers and Kundt \cite{JEK}, and in the subsequent well-known
chapter by Ehlers and Kundt \cite{EK}. As an application of
various newly developed methods to analyze gravitational
radiation, and as a simple background to test various physical
theories, plane waves have proved to be a useful and stimulating
arena which offers interesting contests even today, as we shall
indicate by a few examples in Section 8.2.

Consider a congruence of null geodesics (rays) $x^\alpha (v)$
such that $dx^\alpha / dv = k^\alpha, k_\alpha k^\alpha = 0,
k_{\alpha; \beta} k^\beta = 0$, $v$ being an affine parameter. In
general a geodesic congruence is characterized by its expansion
$\theta$, shear $|\sigma|$ and twist $\omega$ given by
(see e.g. \cite{Wa})
\begin{eqnarray}~&\displaystyle
\label{Equ49}
\theta = \frac{1}{2} k^\alpha_{;\alpha}, \hspace{0.5cm} |\sigma|
= \sqrt{\frac{1}{2} k_{(\alpha ; \beta )} k^{\alpha ; \beta}
- \theta ^2},&~\\
~&\displaystyle
\label{Equ50}
\omega = \sqrt{\frac{1}{2} k_{[\alpha ; \beta]} k^{\alpha ;
\beta}}.&~
\end{eqnarray}

According to the definition given by Ehlers and Kundt \cite{EK} a
vacuum spacetime is a {\it ``plane-fronted gravitational wave''} if it
contains a shearfree $|\sigma| = 0$ geodesic null
congruence, and if it admits ``plane wave surfaces'' (spacelike
2-surfaces orthogonal to $k^\alpha$). This definition is inspired by
plane electromagnetic waves in Maxwell's theory. Electromagnetic
plane waves are {\it null fields} (``pure radiation fields''):
there exists a null vector $k^\alpha$, tangent to the rays,
which is transverse to the electromagnetic field $F_{\alpha \beta}$,
i.e. $F_{\alpha\beta} k^\beta = 0$, $F^*_{\alpha \beta} k^\beta = 0$,
and the quadratic invariants of which vanish,
$F_{\alpha \beta} F^{\alpha \beta} = 0 = F_{\alpha \beta} F^{* \alpha \beta}$,
where $F^*_{\alpha \beta}$ is dual to $F_{\alpha \beta}$. Analogously, Petrov type $N$
gravitational fields (see \cite{KSH}) are null fields with rays
tangent to $k^\alpha$ (the ``quadruple Debever-Penrose null vector''),
and with the Riemann tensor satisfying 
$R_{\alpha \beta \gamma \delta} k^\delta = 0$, 
$R_{\alpha \beta \gamma \delta} R^{\alpha\beta \gamma \delta} = 0$,
and  $R^{\alpha \beta \gamma \delta} R^*_{\alpha \beta \gamma \delta} = 0$.\footnote{This 
algebraic (local) analogy between null fields
exists also between electromagnetic and gravitational shocks
(possible discontinuities across null hypersurfaces), and in the
asymptotic behaviour of fields at large distances from sources
(the ``peeling property'' -- see e.g. \cite{Wa,Pen}).}
Then the Bianchi identities and the Kundt-Thompson theorem for
type $N$ solutions in vacuum spacetimes (also more generally, 
under the presence of a nonvanishing cosmological constant) imply
that the shear of $k^\alpha$ must necessarily vanish (see
\cite{KSH,BiPy}). Because of the existence of plane
wave surfaces, the expansion (\ref{Equ49}) and twist (\ref{Equ50}) must vanish as well,
$\theta = \omega = 0$. In this way we arrive at the Kundt class
of nonexpanding, shearfree and twistfree gravitational waves
\cite{KSH}. The best known subclass of these waves are
{\it ``plane-fronted gravitational waves with parallel rays''
(pp-waves)} which are defined by the condition that the null
vector $k^\alpha$ is covariantly constant, $k_{\alpha ; \beta} =
0$. Thus, automatically $k^\alpha$ is the Killing vector,
and $\theta = |\sigma| = \omega = 0$.

Ehlers and Kundt \cite{EK} give several equivalent
characterizations of the pp-waves and show, following their
previous work \cite{JEK}, that in suitable null coordinates with
a null coordinate $u$ such that $k_\alpha = u_{,\alpha}$ and
$k^\alpha = \left(\partial/\partial v\right)^\alpha$, the metric has the form
\begin{equation}
\label{Equ51}
ds^2 = 2d \zeta d \bar \zeta - 2 du dv - 2H(u, \zeta, \bar
\zeta) du^2,
\end{equation}
where $H$ is a real function dependent on $u$, and on the
complex coordinate $\zeta$ which spans the wave 2-surfaces
$u = \const$, $v = \const$. These 2-surfaces with Euclidean geometry are thus
contained in the wave hypersurfaces $u = \const$ and cut the rays
given by $(u, \zeta) = \const, v$ changing. The vacuum field
equations imply 2-dimensional Laplace's equation
\begin{equation}
\label{Equ52}
H_{, \zeta \bar \zeta} = 0,
\end{equation}
so that we can write
\begin{equation}
\label{Equ53}
2H = f(u, \zeta) + \bar f (u, \bar \zeta),
\end{equation}
where $f(u, \zeta)$ is an arbitrary function of $u$, analytic in
$\zeta$. To characterize the curvature in the waves and their
effect on test particles it is convenient to introduce the null
complex tetrad, such that at each spacetime
point, together with the preferred null vector $k^\alpha$, we
have a null vector $l^\alpha, l^\alpha k_\alpha = -1$, and complex
spacelike vector $m^\alpha$ satisfying $m_\alpha \bar m^\alpha =
1, m_\alpha k^\alpha = m_\alpha l^\alpha = 0.$ For the metric
(\ref{Equ51}) the only nonvanishing projection of the Weyl 
(in the vacuum case, the Riemann) tensor onto this 
tetrad is the (Newman-Penrose) scalar 
\begin{equation}\label{Equ54}
\Psi_4 = C_{\alpha \beta \gamma \delta} l^\alpha \bar m^\beta
l^\gamma m^\delta = H_{, \bar \zeta \bar \zeta}~,
\end{equation}
which denotes a {\it transverse} component of the wave
propagating in the $k^\alpha$ direction. As shown by Ehlers and
Kundt \cite{EK} (see also e.g. \cite{KSH}), though in a somewhat
different notation, we can use again an analogy with the
electromagnetic field -- described for an analogous plane wave by
the transverse component $\phi_2 = F_{\alpha \beta} \bar m^\alpha
l^\beta $ -- and write $\Psi_4 = A\,e^{i \Theta}$, where real $A>0$ is
considered as the {\it amplitude} of the wave, and at each
spacetime point associate $\Theta$ with the plane of
polarization. Vacuum pp-waves with $\Theta = \const$ are
called {\it linearly polarized}.

Consider a free test particle (observer) with 4-velocity ${\bf u}$
and a neighbouring free test particle displaced by a
``connecting'' vector $Z^\alpha(\tau)$. Introducing then the
physical frame ${\bf e}_{(i)}$ which is connected with the observer such
that ${\bf e}_{(0)} = {\bf u}$ and ${\bf e}_{(i)}$ are connected
with the null tetrad vectors by
%\begin{eqnarray}
%&&{\bf m}    ={\textstyle{1\over\sqrt 2}}\left({\bf e}_{(1)}+i {\bf %e}_{(2)}\right)
%                        ,\
%{\bf \bar{ m}}={\textstyle{1\over\sqrt 2}}\left({\bf e}_{(1)}-i {\bf %e}_{(2)}\right)
%                        \ ,\label{Equ55} \nonumber \\
%&&{\bf l} ={\textstyle{1\over\sqrt 2}}\left({\bf u}-{\bf e}_{(3)}\right)
%                        ,\
%~~~~~{\bf k}  ={\textstyle{1\over\sqrt 2}}\left({\bf u}+{\bf e}_{(3)}\right),
%\end{eqnarray}
\begin{equation}
\label{Equ55} 
\begin{array}{ccccccc}
{\bf m}&=&{\textstyle{1\over\sqrt 2}}\left({\bf e}_{(1)}+i {\bf e}_{(2)}\right),
&~~~&{\bf \bar{ m}}&=&{\textstyle{1\over\sqrt 2}}\left({\bf e}_{(1)}-i {\bf e}_{(2)}\right), \\
{\bf l}&=&{\textstyle{1\over\sqrt 2}}\left({\bf u}-{\bf e}_{(3)}\right),
&~~~&{\bf k}&=&{\textstyle{1\over\sqrt 2}}\left({\bf u}+{\bf e}_{(3)}\right),
\end{array}
\end{equation}
we find that the equation of geodesic deviation in spacetime with
only $\Psi_4 \not= 0$ implies (see \cite{BiPy})
\begin{equation}
\label{Equ56}
\ddot Z^{(1)} = -A_+ Z^{(1)} + A_{\times}Z^{(2)}, \,\,
\ddot Z^{(2)} = A_+Z^{(2)} + A_{\times}Z^{(1)}, \,\,
\ddot Z^{(3)} = 0,
\end{equation}
where $A_+ = \frac{1}{2}$ Re $\Psi_4, A_{\times} = \frac{1}{2}$ Im
$\Psi_4$
are amplitudes of ``+'' and ``$\times$'' polarization modes, and $Z^{(i)}$
are the frame components of the connecting vector ${\bf Z}$.
Since the frame vector ${\bf e}_{(3)}$ is chosen in the longitudinal
direction (the direction of the rays), equation (\ref{Equ56}) clearly exhibits the
transverse character of the wave. If particles, initially at
rest, lie in the $({\bf e}_{(1)}, {\bf e}_{(2)})$ plane, there is
no motion in the longitudinal direction of ${\bf e}_{(3)}$. The
ring of particles is deformed into an ellipse, the axes of
different polarizations are shifted one with respect to the other
by $\frac{\pi}{4}$ (such behaviour is typical for linearized
gravitational waves -- cf. e.g. \cite{MTW}). Making a rotation in
the transverse plane by an angle $\vartheta$,
\begin{equation}
{\bf e'}_{(1)}= \cos \vartheta\, {\bf e}_{(1)}+\sin \vartheta\, {\bf e}_{(2)}
  \ ,\quad
{\bf e'}_{(2)}= -\sin \vartheta\, {\bf e}_{(1)}+\cos \vartheta\, {\bf e}_{(2)}~,
 \label{Equ57}
\end{equation}
and taking $\vartheta = \vartheta_+(\tau) = -\frac{1}{2}$ Arg $\Psi _4
= -\frac{1}{2} \Theta$, then
$A'_+ = \frac{1}{2} |\Psi|$, $A'_{\times} = 0$ --
the wave is purely ``+'' polarized. If $\Theta
= \const$, the rotation angle is independent of time -- the wave
is rightly considered as linearly polarized.

Hence, with the discovery of pp-waves, the understanding of the
properties of gravitational radiation has become deeper and
closer to physics. In addition, the pp-waves can easily be
``linearized'' by taking the function $H$ in the metric (\ref{Equ51}) 
to be so small that the spacetime can be considered as a perturbation of
Minkowski space within the linearized theory. Such an ``easy
way'' from the linear to fully nonlinear spacetimes is of course
paid by their simplicity.

In general, in fact, the pp-waves have only the single isometry
generated by the Killing vector $k^\alpha = \left(\partial/\partial v\right)^\alpha$.
However, a much larger {\it group of symmetries} may exist for
various particular choices of the function $H(u, \zeta, \bar \zeta)$.
Jordan, Ehlers and Kundt \cite{JEK} (see also 
\cite{EK,KSH}) gave a complete classification of the pp-waves in
terms of their symmetries and corresponding special forms of $H$.
For example, in the best known case of plane waves to which we
shall turn in greater detail below, $\Psi_4$ is independent of
$\zeta$, so that after removing linear terms in $\zeta$ by a
coordinate transformation, we have
\begin{equation}
\label{Equ58}
H(u, \zeta, \bar \zeta) = A (u) \zeta^2 + \bar A (u) \bar \zeta^2,
\end{equation}
with $A(u)$ being an arbitrary function of $u$. 
This spacetime admits five Killing vectors.

Recently, Aichelburg and Balasin \cite{AiB,AiB2}
generalized the classification given in \cite{JEK} by admitting
distribution-valued profile functions and allowing for
non-vacuum spacetimes
with metric (\ref{Equ51}), but with $H$ which in general does not satisfy
(\ref{Equ52}). They have shown that with $H$ in the form of
delta-like pulses,
\begin{equation}
\label{Equ59}
H(u, \zeta, \bar \zeta) = f(\zeta, \bar \zeta) \delta(u),
\end{equation}
new symmetry classes arise even in the vacuum case.

The main motivation to consider impulsive pp-waves stems from
the metrics describing a black hole or a ``particle'' boosted to
the speed of light. The simplest metric of this type, given by
Aichelburg and Sexl \cite{AS}, is a Schwarzschild black
hole with mass $m$ boosted in such a way that $\mu = m/\sqrt {1-w^2}$ is
held constant as $w \rightarrow 1$. It reads
\begin{equation}
ds^2 = 2d \zeta d \bar \zeta - 2 du dv - 4 \mu \log (\zeta \bar
\zeta) \delta (u) du^2,
\end{equation}
with $H$ clearly in the form (\ref{Equ59}). This is not a vacuum metric:
the energy-momentum tensor $T_{\alpha \beta} = \mu \delta(u)
\delta (\zeta)k_\alpha k_\beta$ indicates that there is a
``point-like particle'' moving with the speed of light along $u=0$.
The Aichelburg-Sexl metric and its more recent generalizations
have found interesting applications even outside of general
relativity. Some of them will be briefly mentioned in Section
8.2.

Let us now turn to the simplest class of pp-waves, which
comprises of the best known and illuminating examples of exact
gravitational waves. These are the {\it plane waves}. They are
defined as homogeneous pp-waves in the sense that the curvature
component $\Psi_4$ (see (\ref{Equ54})) is constant along the wave surfaces
so that function $H$ is in the form (\ref{Equ58}). One can write $H$
as in
(\ref{Equ53}) where
\begin{equation}
\label{Equ61}
f(u, \zeta) = \frac{1}{2} {\cal A}(u) e^{i\Theta (u)} \zeta^2,
\end{equation}
with linear terms being removed by a coordinate transformation.
Just as a plane electromagnetic wave, a plane gravitational wave
is thus completely represented by its amplitude ${\cal A}(u)$ and
polarization angle $\Theta (u)$ as functions of the phase $u$.

The plane waves, including their generalization into the
Einstein-Maxwell theory (an additional term $B(u) \zeta \bar
\zeta$ then appears in $H$, both $\Psi_4$ and the electromagnetic
quantity $\Phi_2$ being independent of $\zeta$), were already 
studied in 1926 (see \cite{KSH}). A real understanding however
came only in the late 1950s. Ehlers and Kundt \cite{EK} give various
characterizations of this class. For example, they prove that a
non-flat vacuum field is a pp-wave if and only if the curvature
tensor is complex recurrent, i.e. if $P_{\alpha \beta \gamma
\delta, \mu} = P_{\alpha \beta \gamma \delta} q_{\mu}$, where
$P_{\alpha \beta \gamma \delta} = R_{\alpha \beta \gamma \delta}
+ i\, {}^{{}^*}\!\!R_{\alpha \beta \gamma \delta}$; and it is a plane wave
if and only if the recurrence vector $q_\mu$ is collinear with a
real null vector. They also state a nice theorem showing that the
plane wave spacetimes defined by the metric (\ref{Equ51}), $H$ and $f$ given
by (\ref{Equ53}), (\ref{Equ69}), $\zeta = x + iy$, 
and with coordinate ranges $-\infty < x,y,u,v < \infty$,
are geodesically complete if functions ${\cal A}(u)$
and $\Theta (u)$ are $C^1$-functions. Quoting directly from
\cite{EK}, {\it ``there exist ... complete solutions free of sources
(singularities), proving to think of a graviton field independent
of any matter by which it be generated. This corresponds to the
existence of source-free photon fields in electrodynamics''}.
Ehlers and Kundt \cite{EK} also state an open problem which, as
far as I am aware, has not yet been solved: to prove that plane waves
are the only geodesically complete pp-waves.

The most telling examples of plane waves are {\it sandwich
waves}. The amplitude ${\cal A}(u)$ in (\ref{Equ61}) need not be smooth: 
either it can only be continuous and nonvanishing on a finite interval of $u$
(sandwich), or a step function (shock), or a delta function
(impulse). A physical interpretation of such waves is better
achieved in other coordinate systems, in which the metric
``before'' and ``after'' the wave is not Minkowskian but has a
higher degree of smoothness. For linearly polarized waves 
($\Theta$ equal to zero), a convenient coordinate
system can be introduced by setting (see e.g. \cite{Prs}) 
$\zeta = (1/\sqrt{2}) (px + iqy)$, $v = (1/2)(t + z + pp'x^2 +
qq'y^2)$, $u = t - z$, where $'\!= d/du$, and functions 
$p = p(u)$ and $q = q(u)$ solve equations 
$p'' + {\cal A}(u) p = 0$ and $q'' - {\cal A}(u) q = 0$.
In these coordinates the metric turns out to be
\begin{equation}
\label{Equ62}
ds^2 = -dt^2 + p^2dx^2 + q^2dy^2 + dz^2.
\end{equation}
In double-null coordinates $\tilde u$, $\tilde v$, with $\tilde u =
u = t - z$, $\tilde v = t + z$, and with a general polarization, the
metric can be cast into the form (see e.g. \cite{BoGM,Gr})
\begin{equation}
\label{Equ63}
ds^2 = - d \tilde u d \tilde v + e^{-U} (e^V \cosh W dx^2 \!+\!
e^{-V} \cosh W dy^2 \!-\! 2 \sinh Wdx dy),
\end{equation}
where $U, V, W$ depend on $\tilde u$ only. This so called Rosen
form was used in the classical paper on exact plane waves by
Bondi, Pirani and Robinson \cite{ae}.

A simple, textbook example \cite{Ri} of a sandwich wave is the
wave with a ``square profile'': ${\cal A}(u)=0$ for $u<0$ and
$u>a^2, {\cal A}(u)=a^{-2} = \const$ for $0\leq u \leq a^2 $. The
functions $p$ and $q$ which enter (\ref{Equ62}) are then $p=q=1$ at $u
\leq 0, p=\cos (u/a), q=\cosh (u/a)$ at $0 \leq u \leq a^2$, and $p=-(u/a)
\sin a+\const, q=(u/a) \sinh a+\const$ at $a^2 \leq u$. This example
can be used to demonstrate explicitly various typical features of
plane sandwich gravitational waves within the exact theory: (i)
the wave fronts travel with the speed of light; (ii) the
discontinuities of the second derivatives of the metric tensor
are permitted along a null hypersurface, but must have a special
structure;  (iii) the waves have a transverse
character and produce relative accelerations in test
particles; (iv) the waves focus astigmatically initially
parallel null congruences (rays) that are pointing in other
directions than the waves themselves; (v) as a
consequence of the focusing, Rosen-type line elements contain
coordinate singularities on a hypersurface behind the waves,
and in general caustics will develop there \cite{Ri}.

The focusing effects imply a remarkable property of plane wave
spacetimes: no spacelike global hypersurface exists on which initial
data can be specified, i.e. {\it plane wave spacetimes contain
no global Cauchy hypersurface}. This can be understood from
Fig. 9. Considering a point $Q$ in flat space in front of the
wave, Penrose \cite{RPE} has shown that its future null cone is
distorted as it passes through the wave in such a manner that it
is refocused to either a point $R$ or a line passing through $R$
parallel to the wave front. Any possible Cauchy hypersurface
going through $Q$ must lie below the future null cone through
$Q$, i.e. below the past null cone of $R$. Hence, it cannot
extend as a spacelike hypersurface to spatial infinity.

\begin{figure}
\centering
\includegraphics[width=.7\textwidth]{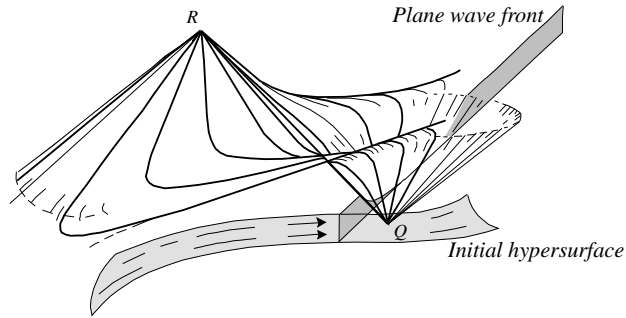}
\label{Figure 9}
\caption{The future null cone of the event $Q$ is distorted 
as it passes through the plane wave, and refocused at 
the event $R$ in such a manner that no Cauchy initial 
hypersurface going through $Q$ exists. (From \cite{RPE}.)
}
\end{figure}

\subsection{Plane-fronted waves: new developments and applications}

The interest in impulsive waves generated by boosting a
``particle'' at rest to the velocity of light by means of an
appropriate limiting procedure persists up to the present. The
ultrarelativistic limits of Kerr and Kerr-Newman black holes were
obtained in \cite{LoSa,FePe,BaNa}, and recently, boosted
static multipole (Weyl) particles were studied \cite{PoGr1}.
Impulsive gravitational waves were also generated by boosting 
the Schwarzschild-de Sitter and  Schwarzschild-anti de Sitter
metrics to the ultrarelativistic limit \cite{HoTa,PoGr2}.

These types of spacetimes, especially the simple Aichelburg-Sexl
metrics, have been employed in current problems of the
generation of gravitational radiation from 
axisymmetric black hole collisions and black hole encounters. The recent
monograph by d'Eath \cite{PE} gives a comprehensive survey,
including the author's new results. There is good reason to believe
that spacetime metrics produced in high speed collisions will be
simpler than those corresponding to (more realistic) situations
in which black holes start to collide with low relative
velocities. The spacetimes corresponding to the collisions at
exactly the speed of light is an interesting limit which can be
treated most easily. Aichelburg-Sexl metrics are used to describe
limiting ``incoming states'' of two black holes, moving one
against the other with the speed of light. An approximation
method has been developed in which a large Lorentz boost is
applied so that one has a weak shock propagating on a strong
shock. One finds an estimate of 16.8 \% for the efficiency of
gravitational wave generation in a head-on speed-of-light
collision \cite{PE}.

Great interest has been stimulated by 't Hooft's \cite{Hoo} work
on the quantum scattering of two pointlike particles at
centre-of-mass energies higher or equal to the Planck energy.
This quantum process has been shown to have close connection with
classical black hole collisions at the speed of light 
(see \cite{PE,Fab} and references therein).

Recently, the Colombeau algebra of generalized functions, which
enables one to deal with singular products of distributions, has
been brought to general relativity and used in the description
of impulsive pp-waves in various coordinate systems
\cite{Kus1}, and also for a rigorous solution of the geodesic and
geodesic deviation equations for impulsive waves \cite{KuS}.
The investigation of the equations of geodesics in
non-homogeneous pp-waves (with $f \sim \zeta ^3$) has shown
that the motion of test particles in these spacetimes is
described by the H\'enon-Heiles Hamiltonian which implies that the
motion is chaotic \cite{PoVe}.

Plane-fronted waves have been used as simple metrics in various
other contexts, for example, in quantum field theory on a given
background (see \cite{Per} for recent work), and in
string theory \cite{AKl}. As emphasized very recently by Gibbons
\cite{Gi}, since for pp-waves and type {\it N} Kundt's class (see the
beginning of Section 8.1) all possible invariants formed from the
Weyl tensor and its covariant derivatives vanish \cite{BiPr},
these metrics suffer no quantum corrections to all loop orders.
Thus they may offer insights into the behaviour of a full
quantum theory. The invariants vanish also in type {\it III}
spacetimes with nonexpanding and nontwisting rays \cite{PRD}.

\subsection{Colliding plane waves}

As with a number of other issues in gravitational (radiation)
theory, the pioneering ideas on colliding plane gravitational
waves are connected with Roger Penrose. It does not seem to be
generally recognized that the basic idea appeared six years
before the well-known paper by Khan and Penrose \cite{KhPe} in
which the metric describing the general spacetime representing
a collision of two parallel-polarized impulsive gravitational
waves was obtained. Having demonstrated the surprising fact
that general relativistic plane wave spacetimes admit no Cauchy
hypersurface due to the focusing effect the waves exert on null
cones, Penrose \cite{RPE} (in footnote 12) remarks: ``This fact
has relevance to the question of two colliding weak plane
sandwich waves. Each wave warps the other until singularities in
the wave fronts ultimately appear. This, in fact, causes the
spacetime to acquire genuine physical singularities in this
case. The warping also produces a scattering of each wave after
collision so that they cease to be sandwich waves when they
separate (and they are no longer plane -- although they have a
two-parameter symmetry group).''

The first detailed study of colliding plane waves, independently
of Khan and Penrose, was also undertaken by Szekeres (see
\cite{Se1,Se2}). He formulated the problem as a
characteristic initial value problem for a system of hyperbolic
equations in two variables (null coordinates) $u, v$ with data
specified on the pair of null hypersurfaces $u=0, v=0$
intersecting in a spacelike 2-surface (Fig. 10). In the particular
case of spacetimes representing plane waves propagating before
the collision in a flat background, Szekeres has shown that
coordinates (of the ``Rosen type'', as known from the case of one
wave -- see Eq. (\ref{Equ63})) exist in which the metric reads
\begin{eqnarray}
\label{Equ64}
ds^2 = &-& e^{-M} du\;dv + \nonumber\\
&+& e^{-U} \left[\, e^V \cosh W dx^2 
  + e^{-V} \cosh W dy^2 - 2 \sinh W dx\;dy \right],
\end{eqnarray}
where $M$, $U$, $V$ and $W$ are functions of $u$ and $v$. Coordinates
$x$ and $y$ are aligned along the two commuting Killing vectors
$\partial/\partial x$ and $\partial/\partial y$, which are
assumed to exist in the whole spacetime representing the colliding
waves (cf. the note by Penrose above). In almost all recent work
on colliding waves, region {\it IV} in Fig. 10, where $u<0$, $v<0$, is
assumed to be flat. The null lines $u=0, v<0$ and $v = 0, u<0$
are wavefronts, and in regions {\it II} $(u<0, v>0)$ and {\it III} $(u>0, v<0)$
one has the standard plane wave metric corresponding to two
approaching plane waves from opposite directions. In region {\it II},
functions $M,~U,~V,~W$ depend on $v$ only, and in region {\it III} only on
$u$. The waves collide at the 2-surface $u = v = 0$, 
in region {\it I} they interact. The spacetime here can be
determined by the initial data posed on the $v \geq 0$ portion of
the hypersurface $u=0$ (which in Fig. 10 are ``supplied'' by the
wave propagating to the right) and by the data on the $u \geq 0$
portion of the hypersurface $v=0$ (given by the wave propagating
to the left). Unfortunately, the integration of such an initial
value problem does not seem to be possible for general incoming
wave forms and polarizations. If, however, the approaching waves
have constant and aligned (parallel) polarizations, one may set
the function $W=0$ globally. The solution of the initial value
problem then reduces to a one dimensional integral for the function
$V$, and two quadratures for the function $M$. (The function $\exp(-U)$
must have the form $f(u) + g(v)$ as a consequence of the field
equations everywhere, and it can be determined easily from the initial
data.) Despite these simplifications it is very difficult to
obtain exact solutions in closed analytic form. Szekeres
\cite{Se2} found a solution (as he puts it ``more or less by
trial and error'') which, as special cases, includes the solution
given by himself earlier \cite{Se1} and the solution obtained
independently and simultaneously by Khan and Penrose \cite{KhPe}.
Although Szekeres' formulation of a general solution for the
problem of colliding parallel-polarized waves is difficult to use
for constructing other specific explicit examples, it has been
employed in a general analysis of the structure of the
singularities produced by the collision \cite{Yu1}, which will be
discussed in the following.

\begin{figure}
\centering
\includegraphics[width=.7\textwidth]{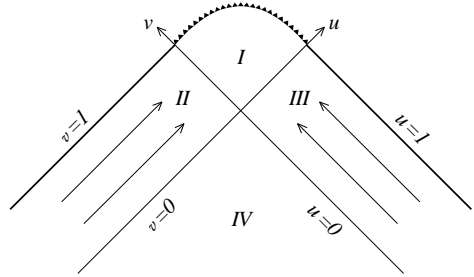}
\label{Figure 10}
\caption{
The spacetime diagram indicating the collision of two plane-fronted
gravitational waves which come from regions {\it II} and {\it III},
collide in region $I$, and produce a spacelike singularity.
Region {\it IV} is flat.}
\end{figure}

It has also inspired an important, difficult piece of
mathematical physics which was developed at the beginning of
the 1990s in the series of papers by Hauser and Ernst \cite{HET}.
Their new method of analyzing the initial value problem can be
used also for the case when the polarization of the approaching
waves is not aligned. They formulated the initial value problem in terms
of the equivalent matrix Riemann-Hilbert
problem in the complex plane. Their techniques are related to
those used by Neugebauer and Meinel to analyze and construct the
rotating disk solution as a boundary value problem (Section 6.3).
No analogous solution for colliding waves in the noncollinear case is
available at present, but investigations in this direction are
still in progress. Most recently, Hauser and Ernst prepared an 
extensive treatise \cite{HET1} in which they give a general description
and detailed mathematical proofs of their study of the solutions of the hyperbolic 
Ernst equation.

The approach of Khan and Penrose for obtaining exact solutions
describing colliding plane waves starts in the region {\it I} where the
waves interact: (i) find a solution with two commuting spacelike
Killing vectors $\partial/\partial x$ and $\partial/\partial y$,
transform to null coordinates, and look back in time whether this
solution can be extended across the null hypersurface $u=0, v=0$ so
that it describes a plane wave propagating in the $u$-direction
in region {\it II} and another plane wave propagating in the
$v$-direction in region {\it III}$\;$; (ii) satisfy boundary conditions not
only across boundaries between regions {\it I} and {\it II},
and regions {\it I} and {\it III}, but also across the boundaries between 
{\it II} and {\it IV}, and {\it III} and {\it IV} in such
a manner that {\it IV} is flat. The original prescription of Khan and
Penrose for extending the solution from region {\it I} to regions {\it II} and
{\it III} consists in the substitutions $u H(u)$ and $vH(v)$ in place
of $u$ and $v$ everywhere in the metric coefficients; here 
$H(u) = 1$ for $u \geq 0$, $H = 0$ for $u<0$ is the usual Heaviside
function. We then get the metric as a function of $v$ (respectively
$u$) in region {\it II} (respectively {\it III}) corresponding to the wave
propagating to the right (respectively to the left) in Fig. 10. Finally, it
remains to investigate carefully the structure of discontinuities
and possible singularities on the null boundaries between these
regions. In the original Khan and Penrose solutions the Riemann
tensor has a $\delta$-function character on the boundaries
between {\it II} and {\it IV}, and {\it III} and {\it IV}; 
but inside regions {\it II} and {\it III}
themselves the spacetime is flat (the collision of impulsive
plane waves). In the solution obtained by Szekeres \cite{Se2},
regions {\it II} and {\it III} are not flat, and the Riemann tensor at the
boundaries between {\it II} (respectively {\it III}$\;$) and {\it IV} is just discontinuous
(the collision of shock waves).

\vskip -0.1pt
Nutku and Halil \cite{NH} constructed an exact solution
describing the collision of two impulsive plane waves with
non-aligned polarizations. In the limit of collinear
polarizations their solution reduces to the solution of Khan and
Penrose. All of these solutions reveal that the spacelike
singularity always develops in region {\it I} (given by $u^2 + v^2
= 1$ in Fig. 10) -- in agreement with the original suggestion of
Penrose. Moreover, the singularity ``propagates backward'' and so
called {\it fold singularities}, analyzed in detail in 1984 by
Matzner and Tipler \cite{MT}, appear also at $v = 1$ and $u = 1$
in regions {\it II} and {\it III}. This new type of singularity provides
evidence of how even relatively recent studies of explicit
exact solutions may reveal unexpected global features of
relativistic spacetimes.

\vskip -0.1pt
The remarkable growth of interest in colliding plane waves owes
much to the systematic (and symptomatic) effort of S.
Chandrasekhar who, since 1984, together with V. Ferrari, and 
with B. Xanthopoulos, published a number of papers on
colliding plane vacuum gravitational waves 
\cite{ChF,ChX2}, and on gravitational waves coupled with electromagnetic waves,
with null dust, and with perfect fluid (see \cite{Gr} for
references). The basic strategy of their approach follows that of
Khan and Penrose: first a careful analysis of the possible
solution is done in the interaction region {\it I}, and then one works
backward in time, extending the solutions to regions {\it II}, {\it III}
and {\it IV}.

\vskip -0.1pt
The main new input consists in carrying over the techniques known
from stationary, axisymmetric spacetimes with one timelike and
one spacelike Killing vector to the case of two spacelike Killing
vectors, $\partial/\partial x, \partial/\partial y$, and exploring
new features.

\vskip -0.1pt
Taking a simple linear solution  of the Ernst equation,
$E=P\eta + iQ\mu$, where $P$ and $Q$ are real constants which satisfy
$P^2 + Q^2 = 1$, and $\eta$, $\mu$ are suitable time and space coordinates, 
Chandrasekhar and Ferrari \cite{ChF} show that one arrives at the Nutku-Halil
solution. In particular, if $Q = 0$, the Khan-Penrose solution
emerges. Since by starting from the same simplest form of the
Ernst function in the axisymmetric stationary case one arrives at
the Kerr solution (or at the Schwarzschild solution for the real
Ernst function), we may conclude that in region {\it I} the solutions
of Khan and Penrose and of Nutku and Halil are, for spacetimes
with two spacelike Killing vectors, the analogues of the
Schwarzschild and Kerr solutions. This mathematical analogy can
be generalized to colliding electromagnetic and gravitational
waves within the Einstein-Maxwell theory -- Chandrasekhar and
Xanthopoulos \cite{ChX1} found the analogue of the charged
Kerr-Newman solution. Such a generalization is also of interest
from a conceptual viewpoint: the $\delta$-function singularity
in the Weyl tensor of an impulsive gravitational wave might imply
a similar singularity in the Maxwell stress tensor, which would
seem to suggest that the field itself would contain
``square roots of the $\delta$-function''.

\vskip -0.1pt
In the most important paper \cite{ChX2} of the series,
Chandrasekhar and Xanthopoulos, starting from the simplest
linear solution for the Ernst conjugate function $E^+ = P\eta +
iQ\mu, P^2 + Q^2 = 1$, obtained a new exact solution for
colliding plane impulsive gravitational waves accompanied by
shock waves. This solution results in the development of a
nonsingular Killing-Cauchy horizon instead of a spacelike
curvature singularity. The metric can be analytically extended
across this horizon to produce a maximal spacetime which
contains timelike singularities. (The spacelike singularity in
region {\it I} in Fig. 10 is changed into the horizon, to the future of
which timelike singularities occur.) In the region of 
interaction of the colliding waves, the spacetime is isometric to
the spacetime in a region interior to the ergosphere.

\vskip -0.1pt
Many new interesting solutions were discovered by using the Khan
and Penrose approach. In addition, inverse scattering (soliton)
methods and other tools from the solution generation techniques
were applied. They are reviewed in detail in \cite{BoGM,Gr,r}.

\vskip -0.1pt
Although very attractive mathematical methods are contained in
these works, one feels that physical interpretation has 
receded into the background -- as seemed to be the
case when the new solution generating techniques were exploited
in all possible directions for stationary axisymmetric
spacetimes. It is therefore encouraging that a more physical
and original approach to the problem has been initiated by
Yurtsever. In a couple of papers he discusses Killing-Cauchy
horizons \cite{Y1} and the structure of the singularities
produced by colliding plane waves \cite{Yu1}. Similar to the
Cauchy horizons in black hole physics, one finds that the
Killing-Cauchy horizons are unstable. We thus expect that the
horizon will be converted to a spacelike singularity. By using the
approach of Szekeres described at the beginning of this section, it
is possible to relate the asymptotic form of the metric near the
singularity -- which approaches an inhomogeneous Kasner solution
(see Section 12.1) -- to the initial data given along the wavefronts
of the incoming waves. For specific choices of initial data the
singularity degenerates into a coordinate singularity and a
Killing-Cauchy horizon arises. However, Yurtsever's analysis
\cite{Yu1} shows that such horizons are unstable (within {\it
full nonlinear} theory) against small but generic perturbations
of the initial data. These results are stronger than those on the
instability of the inner horizons of the Reissner-Nordstr\"{o}m
or Kerr black holes. {In particular, Yurtsever constructs
an interesting (though unstable) solution which, when
analytically extended across its Killing-Cauchy horizon,
represents a Schwarzschild black hole created out of the
collision between two plane sandwich waves propagating in a
cylindrical universe \cite{Yu1}.}

Yurtsever also introduced ``almost plane wave spacetimes'' and
analyzed collisions of almost plane waves \cite{Y2}. These waves
have a finite but very large transverse sizes. Some general
results can be proved (for example, that almost plane waves
cannot have a sandwich character, but always leave tails behind),
and an order-of-magnitude analysis can be used in the discussion
of the outcome of the collision of two almost plane waves;
i.e. whether they will focus to a finite minimum size and then
disperse, or whether a black hole will be created. Although in
the case of almost plane waves one can hardly hope to find
an exact spacetime in an explicit form,
this is a field which was inspired by exact explicit
solutions, and may play a significant role in other parts of general
relativity.

%%%%%%%%%%%%%%%%%%%%%%%%%%%%%%%%%%%%%%%%%%%%%%%%%%%%%%%%%%%%%%%%%%%%%%%%
\section{Cylindrical waves}
%%%%%%%%%%%%%%%%%%%%%%%%%%%%%%%%%%%%%%%%%%%%%%%%%%%%%%%%%%%%%%%%%%%%%%%%
In 1913, before the final formulation of general relativity,
Einstein remarked in a discussion with Max Born that, in the
weak-field limit, gravitational waves exist and propagate with the
velocity of light (Poincar\'{e} pioneered the idea
of gravitational waves propagating with the velocity of light
in 1905 -- see \cite{Pai}). Yet, in 1936 Einstein wrote to Born
\cite{CHR}: ``... gravitational waves do not exist, though they
had been assumed a certainty to the first approximation. This
shows the nonlinear general relativistic field equations can
tell us more, or, rather, limit us more than we have believed up
to now. If only it were not so damnably difficult to find
rigorous solutions''. However, after finding a mistake in his
argumentation (with the help of H. Robertson) and discovering 
with Nathan Rosen cylindrical gravitational waves \cite{ER} as the first
exact radiative solutions to his vacuum field equations,
Einstein changed his mind. In fact, cylindrical waves were
found more than 10 years before Einstein and Rosen by Guido
Beck in Vienna \cite{Be}. Beck was mainly interested in
time-independent axisymmetric Weyl fields, but he realized that
through a complex transformation of coordinates $(z
\rightarrow it, t \rightarrow iz)$ one obtains cylindrically
symmetric time-dependent fields which represent cylindrical
gravitational waves, and wrote down equations (\ref{Equ71}) and
(\ref{Equ72}) below. The work
of Einstein and Rosen is devoted explicitly to gravitational
waves. It investigates conditions for the existence of
standing and progressive waves, and even notices  
that the waves carry away energy from the mass located at
the axis of symmetry. We shall thus not modify the tradition and
will call this type of waves Einstein-Rosen waves (which some
readers may wish to shorten to {\small EROS}-waves).

This type of waves, symmetric with respect to the transformation
$z \rightarrow -z$ ({\it z} -- the axis of symmetry), contains one
degree of freedom of the radiation field and corresponds to a fixed
state of polarization. The metric can be written in the form
\begin{equation}
\label{Equ65}
ds^2 = e^{2(\gamma - \psi)}(-dt^2+d\rho ^2)+e^{2\psi} dz^2+\rho ^2
e^{-2\psi} d\varphi^2,
\end{equation}
where $\rho$ and $t$ are invariants (``Weyl-type canonical
coordinates''), and $\psi = \psi(t, \rho)$, 
$\gamma = \gamma(t,\rho)$. The Killing vectors $\partial/\partial\varphi$ and
$\partial/\partial z$ are both spacelike and hypersurface
orthogonal.

The metric containing a second degree of freedom was
discovered by J\"{u}rgen Ehlers (working in the group of Pascual Jordan),
who used a trick similar to Beck's on the generalized
(stationary) Weyl metrics, and independently by Kompaneets (see
the discussion in \cite{St}). In the literature 
(e.g. \cite{RI,Pi}) one refers to the Jordan-Ehlers-Kompaneets form of the
metric:
\begin{equation}
\label{Equ66}
ds^2 = e^{2(\gamma - \psi)} \left( -dt^2 + d \rho^2 \right) +
e^{2 \psi} \left(dz + \omega d\varphi \right)^2 + \rho^2
e^{-2\psi}d\varphi^2.
\end{equation}
Here, the additional function $\omega(t, \rho)$ represents the
second polarization.

Despite the fact that cylindrically symmetric waves cannot
describe exactly the radiation from bounded sources, both the
Einstein-Rosen waves and their generalization (\ref{Equ66}) have played
an important role in clarifying a number of complicated issues, 
such as the energy loss due to gravitational waves \cite{KiT},
the interaction of waves with cosmic strings \cite{GVe,X},
the asymptotic structure of radiative spacetimes \cite{St}, the dispersion of waves
\cite{ChV}, testing the quasilocal mass-energy \cite{To}, testing
codes in numerical relativity \cite{RI}, investigation of the
cosmic censorship  \cite{BER}, and quantum gravity in a
simplified but field theoretically interesting context of
midisuperspaces \cite{Ku,AP,KOS}.

In the following we shall discuss in some detail the asymptotic
structure and midisuperspace quantization since in these two
issues cylindrical waves have played the pioneering role.
Some other applications of cylindrical waves will be briefly mentioned
at the end of the section.

\subsection{Cylindrical waves and the asymptotic structure\\
of 3-dimensional general relativity}

In recent work with Ashtekar and Schmidt \cite{ABS1,ABS2},
which started thanks to the hospitality of
J\"{u}rgen Ehlers' group, we considered gravitational waves with a
space-translation Killing field (``generalized Einstein-Rosen
waves''). In (2+1)-dimensional framework the
Einstein-Rosen subclass forms a simple instructive example of
{\it explicitly given spacetimes which
admit a smooth global null (and timelike) infinity even for
strong initial data}. Because of the symmetry, the 4-dimensional
Einstein vacuum equations are equivalent to the 3-dimensional
Einstein equations with certain matter sources. This result has
roots
in the classical paper by Jordan, Ehlers and Kundt \cite{JEK}
which includes ``reduction formulas'' for the calculation of the
Riemann tensor of spaces which admit an Abelian isometry group.

Vacuum spacetimes which admit a spacelike, hypersurface
orthogonal Killing vector $\partial/\partial z$ can be
described conveniently in coordinates adapted to the symmetry:
\begin{equation}
\label{Equ67}
ds^2=V^2(x)dz^2+\bar{g}_{ab}(x)dx^adx^b, \,\,\,a,b,\ldots=0,1,2,
\end{equation}
where $x\equiv x^a$ and $\bar{g}_{ab}$ is a metric with Lorentz
signature. The field equations can be simplified if one uses
a metric in the 3-space
which is rescaled by the norm of the Killing vector, and writes
the norm of the Killing vector as an exponential. Then (\ref{Equ67}) becomes
\begin{equation}
ds^2 = e^{2\psi(x)}dz^2 + e^{-2\psi(x)}g_{ab}(x)dx^adx^b,
\end{equation}
and the field equations,
\begin{equation}
\label{Equ69}
R_{ab} - 2\nabla_a \psi\nabla_b \psi = 0, \hspace{0.5cm}
g^{ab}\nabla_a \nabla_b \psi = 0,
\end{equation}
where $\nabla$ denotes the derivative with respect to the metric
$g_{ab}$,
can be reinterpreted as Einstein's equations in
3 dimensions with a scalar field  $\Phi = \sqrt{2} \psi$ as
source. Thus, 4-dimensional vacuum gravity is equivalent to 3-dimensional
gravity coupled to a scalar field. In 3 dimensions, there is no
gravitational radiation. Hence, the local degrees of freedom are
all contained in the scalar field. One therefore expects that 
Cauchy data for the scalar field will suffice to determine the
solution. For data which fall off appropriately, we thus expect
the 3-dimensional Lorentzian geometry to be asymptotically flat
in the sense of Penrose \cite{Pen,PEN}, i.e. that there
should exist a 2-dimensional boundary representing null infinity.

In general cases, this is analyzed in \cite{ABS1}. Here we shall restrict
ourselves to the Einstein-Rosen waves by assuming that
there is a further spacelike, hypersurface orthogonal Killing
vector $\partial /\partial\varphi$ which commutes with
$\partial/\partial z$. Then, as is well known, the equations
simplify drastically. The 3-metric is given by
\begin{equation}
\label{Equ70}
d\sigma^2= g_{ab}dx^adx^b = e^{2\gamma}(-dt^2+d\rho^2)+
\rho^2 d\varphi^2 ,
\end{equation}
the field equations (\ref{Equ69}) become
\begin{equation}
\label{Equ71}
\gamma' = \rho (\dot{\psi}^2+\psi'^2), \hspace{0.5cm}
\dot \gamma = 2\rho \dot \psi \psi',
\end{equation}
and
\begin{equation}
\label{Equ72}
- \ddot\psi + \psi'' + \rho^{-1}\psi' = 0,
\end{equation}
where the dot and the prime denote derivatives with respect to $t$
and $\rho$ respectively. The last equation is the wave equation
for the non-flat 3-metric (\ref{Equ70}) as well as for the flat metric
obtained by setting $\gamma = 0$.

Thus, we can first solve the axisymmetric wave equation
(\ref{Equ72})
for $\psi$ on Minkowski space and then solve (\ref{Equ71}) for
$\gamma$ -- the only unknown metric coefficient -- by quadratures.
The ``method of descent'' from the Kirchhoff formula in 4
dimensions
gives the representation of the solution of the wave
equation in 3 dimensions in terms of Cauchy data $\Psi_0 =
\psi(t=0,x,y), \Psi_1 = \psi_{,t} (t=0,x,y)$ (see \cite{ABS1}).
We assume that the Cauchy data are axially symmetric 
and of compact support.

Let us look at the behaviour of the solution at future null
infinity $\cal J^+$. Let $\rho, \varphi$ be polar coordinates in the
plane, and introduce the retarded time coordinate $u = t - \rho$
to explore the fall-off along constant $u$ null
hypersurfaces. For large $\rho$, the function $\psi$ at $u = \const$ admits a
power series expansion in $\rho^{-1}$:
\begin{equation}
\label{Equ73}
\psi (u, \rho) = \frac {f_0(u)}{\sqrt{\rho}} + \frac
{1}{\sqrt{\rho}} \sum_{k=1}^{\infty} \frac{f_k(u)}{\rho^k}  .
\end{equation}
The coefficients in this expansion are determined by integrals
over the Cauchy data. At $u\gg \rho_0$, $\rho_0$ being the radius
of the disk in the initial Cauchy surface in which the data
are non-zero, we obtain
\begin{equation}
\label{Equ74}
f_0(u) = \frac {k_0}{u^{\frac {3}{2}}} + \frac {k_1}{u^ \frac
{1}{2}} + \ldots,
\end{equation}
where $k_0, k_1$ are constants which are determined by the data.
If the solution happens to be time-symmetric, so that $\Psi_1$
vanishes, we find $f_0 \sim u^{-\frac {3}{2}}$ for large $u$.
Similarly, we can also study the behaviour of the solution
near the timelike infinity $i^+$ of 3-dimensional Minkowski space
by setting $t = U +\kappa\rho$, $\kappa > 1$, and investigating
$\psi$ for $\rho\rightarrow\infty$ with $U$ and $\kappa$ fixed.
We refer to \cite{ABS1} for details.

In Bondi-type coordinates ($u = t - \rho, \rho, \varphi)$,
equation (\ref{Equ70}) yields
\begin{equation}
\label{Equ75}
d\sigma^2 = e^{2\gamma} (-du^2 - 2dud\rho) + \rho^2 d\varphi^2.
\end{equation}
The Einstein equations take the form
\begin{equation}
\label{Equ76}
\gamma_{,u} = 2\rho\;{\psi_{,u}}(\psi_{,\rho} - \psi _{,u}),
\hspace{0.5cm}
\gamma_{,\rho} = \rho\;\psi ^2_{,\rho},
\end{equation}
and the wave equation on $\psi$ becomes
\begin{equation}
\label{Equ77}
-2\psi_{,u\rho} + \psi_{,\rho \rho} + \rho ^{-1} (\psi_{,\rho} -
\psi_{,u}) = 0.
\end{equation}
The asymptotic form of $\psi (t,\rho)$ is given by the expansion
(\ref{Equ73}). Since we can differentiate (\ref{Equ73}) term by term, the field
equations (\ref{Equ76}) and (\ref{Equ77}) imply
\begin{equation}
\label{Equ78}
\gamma_{,u} = -2 \lbrack \dot f_0(u)\rbrack^2 + \sum
_{k=1}^{\infty} \frac {g_k(u)}{\rho^k},
\end{equation}
\begin{equation}
\label{Equ79}
\gamma_{,\rho} =
%\frac {1}{4} \lbrack \dot f_0(u)\rbrack^2 \frac{1}{\rho^2}+
\sum
_{k=0}^{\infty} \frac {h_k(u)}{\rho^{k+2}},
\end{equation}
where the functions $g_k, h_k$ are products of the functions
$f_0, f_k, \dot f_0, \dot f_k$. Integrating (\ref{Equ79}) and
fixing the arbitrary function of $u$ in the result using
(\ref{Equ78}), we obtain
\begin{equation}
\label{Equ80}
\gamma = \gamma_0 - 2 \int_{-\infty} ^{u}\left[\dot f_0(u)\right]^2
du - \sum_{k=1}^{\infty} \frac{h_k(u)}{(k+1)\rho^{k+1}} .
\end{equation}
Thus, $\gamma$ also admits an expansion in $\rho^{-1}$, where the
coefficients depend smoothly on $u$. It is now straightforward to
show that the spacetime admits a smooth future null infinity,
$\cal J^+$. Setting $\tilde \rho= \rho ^{-1}, \tilde u = u, \tilde
\varphi= \varphi$ and rescaling $g_{ab}$ by a conformal factor $\Omega
= \tilde \rho$, we obtain
\begin{equation}
\label{Equ81}
d\tilde \sigma^2 = \Omega^2 d\sigma^2 = e^{2\tilde \gamma}
(-\tilde \rho^2 d \tilde u^2 + 2d\tilde u d \tilde \rho) + d
\tilde \varphi^2,
\end{equation}
where $\tilde \gamma (\tilde u, \tilde \rho) = \gamma(u,\tilde
\rho^{-1})$. Because of (\ref{Equ80}), $\tilde \gamma$ has a smooth
extension through $\tilde \rho = 0$. Therefore, $\tilde g _{ab}$
is smooth across the surface $\tilde \rho = 0$. This surface is
the future null infinity, $\cal J^+$.
Hence, the (2+1)-dimensional curved spacetime
has a smooth (2-dimensional) null infinity.  Penrose's picture
works for arbitrarily strong initial data $\Psi_0$, $\Psi_1$.

Using (\ref{Equ81}), we find that at $\cal J^+$ we have:
\begin{equation}
\label{Equ82}
\gamma (u, \infty) = \gamma_0 - 2 \int_{{-\infty}}^{u} \dot f_0^2
du.
\end{equation}
Since one can make sure that $\gamma = 0$ at $i^+$ \cite{ABS2},
one finds the simple result that
\begin{equation}
\gamma_0 = 2 \int _{{-\infty}} ^{{+\infty}}\dot f _0^2 du.
\end{equation}
At spatial infinity $(t = \const$, $\rho
\rightarrow \infty)$, the metric is given by
\begin{equation}
d\sigma^2 = e^{2\gamma_{0}} (-dt^2 + d\rho^2) + \rho^2
d\varphi^2.
\end{equation}
For a non-zero data, constant $\gamma_0$ is positive, whence the
metric has a ``conical singularity'' at spatial infinity. This
conical singularity, present at spatial infinity, is ``radiated
out'' according to equation (\ref{Equ82}). The future timelike infinity,
$i^+$, is smooth. In (2+1)-dimensions, modulo some 
subtleties \cite{ABS1}, equation (\ref{Equ82})
plays the role of the {\it Bondi mass loss formula} in
(3+1)-dimensions, relating the decrease of the total (Bondi)
mass-energy at null infinity to the flux of gravitational
radiation. We can thus conclude that {\it cylindrical waves in
(2+1)-dimensions give an explicit model of the 
Bondi-Penrose radiation theory which admits smooth null and timelike 
infinity for arbitrarily strong initial data}.
There is no other such model available.
The general results on the existence of $\cal J$ in 4 dimensions,
mentioned at the end of Section 1.3, assume weak data.

It is of interest to investigate cylindrical waves also in
a (3+1)-dimensional context. The asymptotic behaviour of these
waves was discussed by Stachel \cite{St} many years ago. However, his
work deals solely with asymptotic directions, which are
perpendicular to the axis of symmetry, i.e. to the
$\partial/\partial z$ -- Killing vector. Detailed calculations show
that, in contrast to the perpendicular directions, where null
infinity in the (3+1)-dimensional framework does not exist, it
{\it does} exist in other directions for data of compact
support. If the data are not time-symmetric, the fall-off is so
slow that (local) null infinity has a {\it polyhomogeneous} (logarithmic)
character \cite{Chr} -- see \cite{ABS2} for details.

We have concentrated on the simplest case of Einstein-Rosen
waves. They served as a prototype for developing a general
framework to analyze the asymptotic structure of spacetime at
null infinity in three spacetime dimensions. This structure has a
number of quite surprising features which do not arise in the
Bondi-Penrose description in four dimensions \cite{ABS1}. One of
the motivations for developing such a framework is to provide a
natural point of departure for constructing the stage for
asymptotic quantization and the S-matrix theory of an interesting
midisuperspace in quantum gravity.

\subsection{Cylindrical waves and quantum gravity}

As the editors of the Proceedings of the 117th WE Heraeus
Seminar on canonical gravity in 1993 \cite{CG}, J\"{u}rgen Ehlers
and Helmut Friedrich start their Introduction realistically:
``When asking a worker in the field about the progress in quantum
general relativity in the last decade, one shouldn't be surprised
to hear: `We understand the problems better'. If it referred to a
lesser task, such an answer would sound ironic. But the search
for quantum gravity... has been going on now for more than half a
century and in spite of a number of ingenious proposals, a
satisfactory theory is still lacking...'' Although I am following
the subject from afar, I believe that one would not be
too wrong if one repeated the same words in 1999. However, 
apart from general theoretical developments, many interesting
quantum gravity models have been studied,
and exact solutions have played a basic role in them. In particular, in
the investigations of (spherical) gravitational collapse and in
quantum cosmology based typically on homogeneous cosmological
models (cf. Section 12.1), one starts from simple classical solutions -- see e.g.
\cite{Ry,MC,HL} for reviews  and \cite{HLL}
for a bibliography up to 1990.
A common feature of such models is the reduction of infinitely
many degrees of freedom of the gravitational field to a
{\it finite} number. In quantum field theory (such as
quantum electrodynamics) a typical object to be quantized 
is a wave with an infinite number of degrees of freedom. The first
radiative solutions of the gravitational field equations which were
subject to quantization were the Einstein-Rosen waves. Kucha\v{r} \cite{Ku} applied the
methods of canonical quantization of gravity to these waves,
using the methods employed earlier in the minisuperspace models,
i.e. restricting himself only to geometries (fields) preserving
the symmetries.

The Einstein-Rosen cylindrical waves have an {\it infinite}
number $\infty^1$ of degrees of freedom contained in one polarization, one
degree of freedom for each cylindrical surface drawn around the
axis of symmetry. Moreover, the slicing of spacetime by spacelike
(cylindrically symmetric) hypersurfaces is not fixed completely 
by the symmetry -- an arbitrary cylindrically symmetric deformation of a given slice
leads again to an allowed slice. Such a deformation represents an
$\infty^{1}$ ``fingered time''. Hence, the resulting space of
3-geometries on cylindrically symmetric slices 
is infinitely richer than the minisuperspaces of quantum
cosmology. The exact Einstein-Rosen waves thus inspired the first
construction of what Kucha\v{r} \cite{Ku} called the ``{\it midisuperspace}''.

Let us briefly look at the main steps in Kucha\v{r}'s 
procedure.\footnote{For the basic concepts 
and ideas of canonical gravity, we
refer to e.g. \cite{MTW,Wa} and especially
to Kucha\v{r}'s review \cite{KK}, where the canonical
quantization of cylindrical waves is also analyzed.}
The symmetry of the problem implies that the spatial metric 
has the form
\begin{equation}
g_{11} = e^{\gamma - \Phi}, \hspace{0.5cm} g_{22} = R^2
e^{-\Phi}, \hspace{0.5cm} g_{33} = e^{\Phi},
\end{equation}
where $\gamma, \Phi$, and $R$ are functions of a single cylindrical coordinate
$x^1 = r$~$(x^2 = \varphi$, $x^3 = z)$. 
Similarly the lapse function $N=N(r)$ depends only on $r$,
and the shift vector has the only nonvanishing radial component
$N^1 = N^1(r), N^2 = N^3 = 0$. We have adopted here Kucha\v{r}'s
notation. When we put $R = r = \rho$,~$\Phi = 2\psi$,~$\gamma
\rightarrow 2\gamma$, $N = e^{\gamma-\Phi}$, and $N^1 = 0$, we
recover the standard Einstein-Rosen line element (\ref{Equ65}); however,
in general the radial and time coordinates $t$ and $r$ differ from the
canonical Einstein-Rosen radial and time coordinates in which the
metric has the standard form (\ref{Equ65}).
The symmetry implies that the canonical momentum $\pi^{ik}$ is
diagonal and expressible by three functions $\pi_\gamma, \pi_R,
\pi_\Phi$ of $r$; for example, $\pi^{11} = \pi_\gamma e^{\Phi-\gamma}$,
 and similarly for the other components. 
After the reduction to cylindrical symmetry, the
action functional assumes the canonical form
\begin{equation}
\label{Equ86}
S=2\pi \int_{-\infty}^{\infty} dt \int_{0}^{\infty} dr
(\pi_\gamma \dot{\gamma} + \pi_R \dot{R} + \pi_\Phi \dot{\Phi} -
N {\cal{H}} - N^1 {\cal{H}}_1)~,
\end{equation}
in which $\gamma, R, \Phi$ are the canonical coordinates and
$\pi_\gamma, \pi_R, \pi_\Phi$ the conjugate momenta (the integration
over $z$ has been limited by $z=z_0$ and $z=z_0+1$). The
superhamiltonian ${\cal H}$ and supermomentum ${\cal H}_1$ are
rather complicated functions of the canonical variables:
\begin{eqnarray}
{\cal {H}}_{~} &=&  e^{\frac{1}{2}(\Phi-\gamma)} \left( -\pi_\gamma \pi_R +
{\textstyle\frac{1}{2}}R^{-1} \pi_\Phi^2 + 2R'' - \gamma 'R' +
{\textstyle\frac{1}{2}}R\Phi'^2 \right), \\
{\cal {H}}_1 &=&  -2\pi_\gamma' + \gamma '\pi_\gamma + R' \pi_R +
\Phi' \pi_\Phi.
\end{eqnarray}

The most important step now is the replacement of the old canonical 
variables $\gamma,\pi_\gamma, R, \pi_R$ 
by a new canonical set $T, \Pi_T, R, \Pi_R$ 
through a suitable canonical transformation. We
shall write here only one of its components (see \cite{Ku,KK}
for the complete transformation):
\begin{equation}
\label{Equ89}
T(r) = T (\infty) + \int_{\infty}^{r} \left[ -\pi_\gamma (r)
\right] dr.
\end{equation}
By integrating the Hamilton equations following from the
action (\ref{Equ86}), rewritten in the new canonical coordinates, one finds
that $T$ and $R$ are the Einstein-Rosen privileged time and radial
coordinates, i.e. those appearing in the canonical form (\ref{Equ65}) of
the Einstein-Rosen metric (with $T = t, R = \rho$). According to
(\ref{Equ89}), the Einstein-Rosen time can be reconstructed, in a non-local
way, from the momentum $\pi_{\gamma}$, which characterizes the extrinsic
curvature of a given hypersurface. In this way, the concept of the 
{\it ``extrinsic time representation''} entered canonical gravity 
with cylindrical gravitational waves.

In terms of the new canonical variables, the superhamiltonian and
supermomentum become
\begin{eqnarray}
{\cal{H}} &=& e^{\frac{1}{2}(\Phi - \gamma)} \left( R'\Pi_T + T'
\Pi_R + {\textstyle\frac{1}{2}} \left( R^{-1} \pi_{\Phi}^2 + R \Phi '^2 \right)
\right),\\
{\cal{H}}_1 &=& T' \Pi_T + R' \Pi_R + \Phi ' \pi_\Phi .
\end{eqnarray}
Since ${\cal {H}}$ and ${\cal {H}}_1$ are linear in $\Pi_T$ and
$\Pi_R$, the classical constraints ${\cal {H}} = 0, {\cal {H}}_1
= 0$ can immediately be resolved with respect to these momenta,
conjugate to the ``embedding'' canonical variables $T(r)$ and
$R(r)$:
\begin{equation}
-\Pi_T = \left(R'^2 - T'^2 \right)^{-1} \left[ {\textstyle\frac{1}{2}} (R^{-1}
\pi_\Phi^2 + R \Phi'^2) R' - \Phi' \pi_\Phi T' \right] = 0,
\end{equation}
and similarly for $\Pi_R$. It is easy to see \cite{Ku,KK}
that the constraints have the same form as the constraints for a
massless scalar field $\Phi$ propagating on a flat background 
foliated by arbitrary spacelike hypersurfaces $T = T(r),
R=R(r)$. The canonical variables $\Phi, \pi_\Phi$ represent the
true degrees of freedom, and the remaining canonical variables play
the role of spacelike embeddings of a Cauchy hypersurface into
spacetime.

After turning the canonical momenta $\Pi_T, \Pi_R, \pi_\Phi$,
into variational derivatives, e.g. $\Pi_T = -i
\delta/\delta T(r)$, one can impose the classical constraints ${\cal{H}}
= 0, {\cal {H}}_1 = 0$ as restrictions on the state functional $\Psi (T, R, \Phi)$:
${\cal {H}} \Psi = 0$, ${\cal {H}}_1 \Psi = 0$.
In particular, the Wheeler-DeWitt
equation ${\cal {H}} \Psi = 0$ in the extrinsic time representation 
assumes the form of a many-fingered time counterpart of an
ordinary Schr\"{o}dinger equation. This reduces to the ordinary
Schr\"{o}dinger equation for a single massless scalar field in
Minkowski space if we adopt the standard foliation $T = \const$
(see \cite{Ku,KK} for details).

The described procedure, first realized in the case of the
Einstein-Rosen waves, has opened a new route in canonical and
quantum gravity. In contrast to the Arnowitt-Deser-Misner
approach, in which the gravitational dynamics is
described relative to a fixed foliation of spacetime, in this 
new approach (called ``bubble time'' dynamics of the gravitational 
field or the ``internal time formalism'' \cite{KU}) one tries to extract 
the many-fingered time (i.e. embeddings of Cauchy hypersurfaces) from
the gravitational phase space, but does not fix the foliation
in the ``target manifold'' by coordinate conditions.
However, the definition of the target manifold by a gauge
(coordinate) condition is needed.

This new approach has been so far successfully applied to a few
other models (based on exact solutions) with infinite degrees of
freedom, for example, plane gravitational waves,
bosonic string, and as late as 1994, to spherically symmetric vacuum
gravitational fields \cite{KVK}.
The internal time formalism for spacetimes 
with two Killing vectors was developed in \cite{RTo}
(therein references to previous works can also be found).
Recently, canonical
transformation techniques have been
applied to Hamiltonian spacetime dynamics with a thin
spherical null-dust shell \cite{LW}. One would like to construct
a midisuperspace model of spherical gravitational collapse, or
more specifically, a model for Hawking radiation with
backreaction. The extensive past work on Hamiltonian
approaches to spherically symmetric geometries (see \cite{LW} for
more than 40 references in this context) have not yet led to
convincing insights. The very basic question of existence of 
the ``internal time'' formalism in a general situation has
been most recently addressed by H\'{a}j\'{\i}\v{c}ek \cite{HP};
the existence has been proven, and shown to be related to the choice of gauge.

\subsection{Cylindrical waves: a miscellany}

Chandrasekhar \cite{CHR} constructed a formalism for cylindrical
waves with two polarizations (cf. the metric (\ref{Equ66})), similar to
that used for the discussion of the collision of plane-fronted
waves (Section 8.3). He obtained the ``cylindrical'' Ernst equation
and corroborated (following the suggestion of O. Reula) the
physical meaning of Thorne's {\it C}-{\it energy} \cite{KiT} -- the
expression for energy suggested for cylindrical fields -- by
defining a Hamiltonian density corresponding to the Lagrangian
density from which the Ernst equation can be derived. A brief
summary of older work on the mass loss of a cylindrical source
radiating out cylindrical waves and its relation to the {\it
C}-energy
is given in \cite{BoGM}. It should be pointed out, however, that
although {\it C}-energy is a useful quantity, it was constructed by
exploiting the local field equations, without direct reference to
asymptotics. The physical energy (per unit $z$ length) at both
spatial and null infinity, which is the generator of the time
translation, is in fact a non-polynomial function of the {\it C}-energy.
In the weak field limit the two agree, but in strong fields they
are quite different \cite{ABS1}.

In \cite{ChV}, an exact solution was constructed with
which one can study the {\it dispersion} of waves: a cylindrical
wave packet, which though initially impulsive, after reflection
at the axis disperses, and develops shock wave fronts when the
original wave meets the waves that are still ingoing. Cylindrical
waves have been also analyzed in the context of {\it phase shifts}
occurring in gravitational soliton interactions  (see \cite{GM}
and references therein).

An exact explicit solution for cylindrical waves with two degrees
of polarization has been obtained \cite{Pi} from the Kerr
solution after transforming the metric into ``cylindrical''
coordinates and using the substitution $t \rightarrow i \tilde z, z
\rightarrow i \tilde t, a \rightarrow i \tilde a$. Both this
solution and the well-known Weber-Wheeler-Bonnor pulse
\cite{BoGM} have been employed as {\it test beds in numerical
relativity} \cite{PiR}, in particular in the approach which
combines a Cauchy code for determining the dynamics of the
central source with a characteristic code for determining the
behaviour of radiation \cite{RI}.

In a number of works cylindrical waves have been considered in
interaction with {\it cosmic strings} \cite{GVe,X}. The
strings are usually modelled as infinitely thin conical
singularities. Recently Colombeau's theory of generalized
functions was used to calculate the distributional curvature
at the axis for a time-dependent cosmic string \cite{Wil}.

A somewhat surprising result concerning cosmic strings and
radiation theory should also be noted: although an infinite,
static cylindrically symmetric string does not, of course,
radiate, it generates a nonvanishing (though ``non-radiative'')
contribution to the Bondi news function \cite{BiS,BPa}.
Recently, the asymptotics at null infinity of cylindrical 
waves with both polarizations (and, in general, an infinite cosmic string along the axis)
has been analyzed in the context of axisymmetric
electrovacuum spacetimes with a translational Killing 
vector at null infinity \cite{BAP}.

Finally, the cylindrically symmetric electrovacuum spacetimes
with both polarizations, satisfying certain completeness and
asymptotic flatness conditions in spacelike directions have been
shown rigorously to imply that strong cosmic censorship
holds \cite{BER}. This means that for generic (smooth) initial
data the maximal globally hyperbolic development of the data is
inextendible (no Cauchy horizons as for example, those discussed
in Section 3.1 for the Reissner-Nordstr\"{o}m spacetime arise).
This global existence result is non-trivial since with two
polarizations and electromagnetic field present, all field
equations are nonlinear.

%%%%%%%%%%%%%%%%%%%%%%%%%%%%%%%%%%%%%%%%%%%%%%%%%%%%%%%%%%%%%%%%%%%%%%%%
\section{On the Robinson-Trautman solutions}
%%%%%%%%%%%%%%%%%%%%%%%%%%%%%%%%%%%%%%%%%%%%%%%%%%%%%%%%%%%%%%%%%%%%%%%%
Robinson-Trautman metrics are the general radiative
vacuum solutions which
admit a geodesic, shearfree and twistfree null congruence of
diverging rays. In the standard coordinates the metric has the
form \cite{aj}
\begin{equation}
ds^2 = 2r^2P^{-2}d\zeta d\bar{\zeta} - 2du\;dr -\lbrack\Delta \ln P
- 2r (\ln P)_{,u} - 2 mr^{-1} \rbrack\;du^2,
\end{equation}
where $\zeta$ is a complex spatial (stereographic) coordinate
(essentially $\theta$ and $\varphi$),
$r$ is the affine parameter along the rays, $u$ is a retarded time,
$m$ is a function of $u$ (which can be in some cases interpreted as
the mass of the system), $\Delta = 2P^2
(\partial^2/\partial \zeta \partial \bar {\zeta})$, and 
$P = P (u, \zeta, \bar {\zeta})$ satisfies the fourth-order
Robinson-Trautman equation
\begin{equation}
\label{Equ94}
\Delta \Delta (\ln P) + 12\; m\; (\ln P)_{,u} - 4 m_{,u} = 0.
\end{equation}

The best candidates for describing radiation from isolated
sources are the Robinson-Trautman metrics of type {\it II} with the
2-surfaces $S^2$ given by $u, r = \const$ and having spherical
topology. The Gaussian curvature of $S^2$ can be expressed as $K
= \Delta \ln P$. If $K = \const$, we obtain the Schwarzschild
solution with mass equal to $K^{-\frac {3}{2}}$.

These spacetimes have attracted increased attention in the
last decade -- most recently in the work by Chru\'{s}ciel, and
Chru\'{s}ciel and Singleton \cite{ak}. In these studies the
Robinson-Trautman spacetimes have been shown to exist globally for all
positive ``times'', and to converge asymptotically to a
Schwarzschild metric. Interestingly, the extension of these
spacetimes across the ``Schwarz\-schild-like'' event horizon can only
be made with a finite degree of smoothness. All these rigorous
studies are based on the derivation and analysis of an asymptotic
expansion describing the long-time behaviour of the solutions of
the nonlinear parabolic equation (\ref{Equ94}).
\begin{figure}
\centering
\includegraphics[width=.7\textwidth]{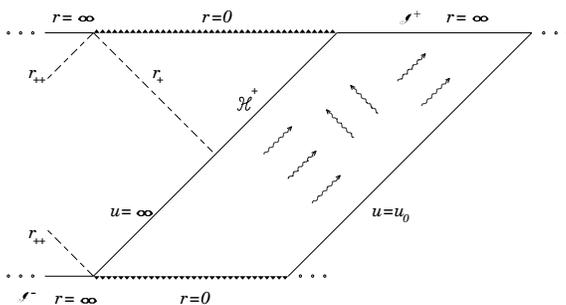}
\label{Figure 11}
\caption{The evolution of the cosmological Robinson-Trautman solutions with a 
positive cosmological constant. A black hole with the horizon 
$\cal H^+$ is formed; at future infinity $\cal J^+$ the spacetime approaches 
a de Sitter spacetime exponentially fast, in accordance with the cosmic 
no-hair conjecture.}
\end{figure}

In our recent work \cite{BiPo,al} we studied Robinson-Trautman radiative
spacetimes with a positive cosmological constant $\Lambda$. The
results proving the global existence and convergence of the
solutions of the Robinson-Trautman equation (\ref{Equ94}) can be taken over
from the previous studies since $\Lambda$ does not explicitly
enter this equation. We have shown that,
starting with arbitrary, smooth initial data at $u=u_0$ (see Fig.
11), these cosmological
Robinson-Trautman solutions converge exponentially fast
to a Schwarzschild-de Sitter solution at large retarded times
($u\to \infty$).
The interior of a
Schwarzschild-de Sitter black hole can be joined to an ``external''
cosmological Robinson-Trautman spacetime across the horizon
$\cal H^+$ with
a higher degree of smoothness than in the corresponding case with
$\Lambda = 0$. In particular, in the extreme case with
$9 \Lambda m^2 = 1$, in which the black hole and cosmological horizons coincide,
the Robinson-Trautman spacetimes can be extended {\it smoothly}
through $\cal H^+$ to the extreme Schwarzschild-de Sitter spacetime
with the same values of $\Lambda$ and $m$. However, such an extension 
is {\it not analytic} (and not unique). 

We have also demonstrated that the cosmological
Robinson-Trautman solutions represent explicit models exhibiting
the cosmic no-hair conjecture: a geodesic observer outside of the
black hole horizon will see, that inside his past light cone, these
spacetimes approach the de Sitter spacetime exponentially fast as
he approaches the future (spacelike) infinity ${\cal J}^+$.
For a freely
falling observer the observable universe thus becomes quite bald. This
is what the cosmic no-hair conjecture claims. As far as we are aware,
these models
represent the only exact analytic demonstration of the cosmic
no-hair conjecture under the presence of gravitational waves.
They also appear to be the only exact examples of black hole
formation in nonspherical spacetimes which are not asymptotically
flat. Hopefully, these models may serve as tests of various
approximation methods, and as test beds in numerical studies of
more realistic situations in cosmology.

%%%%%%%%%%%%%%%%%%%%%%%%%%%%%%%%%%%%%%%%%%%%%%%%%%%%%%%%%%%%%%%%%%%%%%%%
\section{The boost-rotation symmetric radiative spacetimes}
%%%%%%%%%%%%%%%%%%%%%%%%%%%%%%%%%%%%%%%%%%%%%%%%%%%%%%%%%%%%%%%%%%%%%%%%
In this section we would like to describe briefly the only
explicit solutions available today which are radiative and
represent the fields of finite sources. Needless to say, we
cannot hope to find explicit analytic solutions of the Einstein
equations without imposing a symmetry. A natural first assumption
is axial symmetry, i.e. the existence of a spacelike
rotational Killing vector ${\partial}/{\partial \varphi}$. However,
it appears hopeless to search for a radiative solution with only one
symmetry. We are now not interested in colliding plane waves
since these do not represent finite sources; we wish our spacetime
to be as ``asymptotically flat as possible''. The unique role
of the boost-rotation symmetric spacetimes is exhibited by a
theorem, formulated precisely and proved for the vacuum case
with hypersurface orthogonal Killing vectors in \cite{am},
and generalized to electrovacuum spacetimes with Killing 
vectors which need not be hypersurface orthogonal in \cite{BPa} 
(see also references therein). 
This theorem roughly states that  in {\it axially} symmetric, locally
asymptotically flat spacetimes (in the sense that a null infinity
satisfying Penrose's requirements exists, but it need not
necessarily exist globally), the only {\it additional} symmetry
that does not exclude radiation is the {\it boost} symmetry.

In Minkowski spacetime the boost Killing vector has the form
\begin{equation}
\label{Equ95}
{\cal \zeta}_{boost} = z \frac {\partial}{\partial t}+ t \frac
{\partial}{\partial z},
\end{equation}
so that orbits of symmetry to which the Killing vector is tangent
are hyperbolas $z^2 - t^2 = B = \const,$ $x,$ $y = \const$. Orbits with
$B>0$ are timelike; they can represent worldlines of uniformly
accelerated particles in special relativity. Imagine, for
example, a charged particle, axially symmetric about the $z$-axis,
moving with a uniform acceleration along this axis. The
electromagnetic field produced by such a source will have 
boost-rotation symmetry.

Figure 12 shows two particles uniformly
accelerated in opposite directions along the $z$-axis.
In the space diagram (left), the ``string'' connecting the particles
is also indicated. In the spacetime diagram, the particles'
worldlines are shown in bold. Thinner hyperbolas represent the orbits of
the boost Killing vector (\ref{Equ95}) in the regions $t^2>z^2$ where it
is spacelike. In Figure 13
the corresponding compactified diagram indicates that null
infinity cannot be smooth everywhere since it contains four
singular points in which particles' worldlines ``start'' and ``end''.
Notice that in electromagnetism the presence of {\it two}
particles, one moving along $z>0$, the other along $z<0$, makes
the field symmetric also with respect to inversion $z \rightarrow
- z$. The electromagnetic field can be shown to be analytic
everywhere, except for the places where the particles occur.
These two particles move independently of each other, since
their worldlines are divided by two null hypersurfaces $z = t, z
= -t$. This is analogous to the boost-rotation symmetric
spacetimes in general relativity that we are now going to discuss.

Specific examples of solutions
representing ``uniformly accelerated particles'' have been analyzed
for 35 years, starting with the first solutions of this type obtained
by Bonnor and Swaminarayan \cite{ar}, and Israel and Khan \cite{IKH}. In a curved spacetime the ``uniform acceleration'' is
understood with respect to a fictitious Minkowski background, and
the ``particles'' mean singularities or black holes. For a more
extensive description of the history of these specific solutions
discovered before 1985, see \cite{an}. From a unified point of view,
boost-rotation symmetric spacetimes (with hypersurface orthogonal 
Killing vectors) were defined and treated
geometrically in \cite{ao}. We refer to this detailed work for rigorous
definitions and theorems. Here we shall only sketch some of the
general properties and some applications of these spacetimes.

$\!$The metric of a general boost-rotation symmetric spacetime in
``Cartesian-type'' coordinates  $\{t,x,y,z\}$ reads:
\begin{eqnarray}
\label{Equ96}
ds^2&=&~~\frac{1}{x^2+y^2}
\left[(e^{\lambda}x^2+e^{-\mu}y^2)dx^2 +
2xy(e^{\lambda}-e^{-\mu})dxdy
\right] +
\nonumber \\
& &+\frac{1}{x^2+y^2}
(e^{\lambda}y^2+e^{-\mu}x^2)dy^2 +
\frac{1}{z^2-t^2}(e^{\lambda}z^2-e^{\mu}t^2)dz^2-
\nonumber \\
&&-\frac{1}{z^2-t^2}\left[
2zt(e^{\lambda}-e^{\mu})dzdt
+ (e^{\mu}z^2-e^{\lambda}t^2)dt^2 \right]\;,
\end{eqnarray}
where $\mu$ and $\lambda$ are functions of $\rho^2 = x^2 + y^2$ and
$z^2 - t^2$. As a consequence of the vacuum Einstein equations,
the function $\mu$ must satisfy an equation of the form which is
identical to the flat-space wave equation;
and function $\lambda$ is determined in terms of $\mu$ by
quadrature. Now it can easily be seen that the metric (\ref{Equ96}) admits
axial and boost Killing vectors which have exactly the same form
as in Minkowski space, i.e. the axial Killing vector
$\partial/{\partial\varphi}$ and the boost Killing vector (\ref{Equ95}). In
fact, the whole structure of group orbits in boost-rotation
symmetric {\it curved} spacetimes outside the sources (or
singularities) is the same as the structure of the orbits
generated by the axial and boost Killing vectors in Minkowski
space. In particular, the boost Killing vector (\ref{Equ95}) is timelike in
the region $z^2>t^2$. The invariance of a metric (or of any other
field) in a time-direction (determined in a coordinate-free
manner by a timelike Killing vector) means stationarity, and of course, we
could hardly expect to find radiative properties there.
Intuitively, the existence of a timelike Killing vector in the
region $z^2>t^2$ is understandable because there
(generalized) uniformly accelerated reference frames can be
introduced in which sources are at rest, and the fields are time
independent.

\begin{figure}
\centering
\includegraphics[width=.71\textwidth]{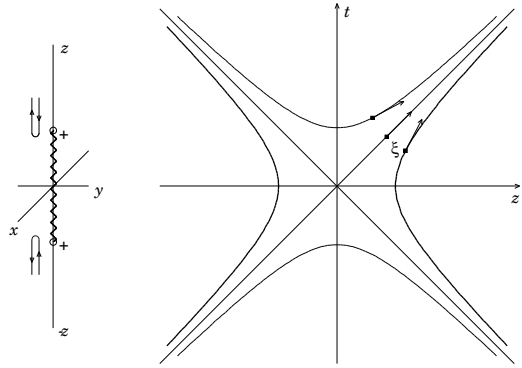}
\label{Figure 12}
\caption{Two particles uniformly accelerated in opposite directions.
The orbits of the boost Killing vector (thinner hyperbolas) are spacelike
in the region $t^2>z^2$.}
\end{figure}

However, in the other ``half'' of the spacetime, $t^2>z^2$, the boost
Killing vector (\ref{Equ95}) is spacelike (see the lines representing
orbits of the boost Killing vector in Fig. 12). Hence in this
region the metric (\ref{Equ96}) is nonstationary. Here we expect to
discover radiative properties. Indeed, it can be shown that for
$t^2>z^2+\rho^2$ the metric (\ref{Equ96}) can {\it locally} be transformed
into the metric of Einstein-Rosen cylindrical waves.
Although locally in the whole region $t^2>z^2$ the metric (\ref{Equ96}) can
be transformed into a radiative metric, the global properties of
the boost-rotation symmetric solutions are quite different from
those of cylindrical waves. Again, we have to refer to the work
\cite{ao}
for a detailed analysis. Let us only say that the boost-rotation
symmetric solutions, if properly defined -- with appropriate
boundary conditions on functions $\lambda$ and $\mu$ -- always
admit asymptotically flat null infinity $\cal J$ at least
locally. Starting with arbitrary solutions $\lambda$ and $\mu$,
and adding suitable constants to both $\lambda$ and $\mu$
(Einstein's equations are then still satisfied), we can always
guarantee that even {\it global} $\cal J$ exists in the sense that
it admits smooth spherical sections.
For the special type of solutions for $\lambda$ and $\mu$,
{\it complete} $\cal J$ satisfies Penrose's
requirements, except for four points in which the sources ``start'' and
``end'' (cf. Fig. 13). In all cases one finds that the
gravitational field in smooth regions of the null infinity is
radiative \cite{BPa,ap}. In particular, the leading term of the Riemann
curvature tensor, proportional to $r^{-1}$ (where $r^2 = \rho^2 +
z^2)$, is nonvanishing and has the same algebraic structure as
the Riemann tensor of plane waves. This is fully analogous to
the asymptotic properties of radiative electromagnetic fields
outside finite sources. Recently, general forms of the news functions 
have been obtained for electrovacuum spacetimes with boost-rotation symmetry
and with Killing vectors which need not be hypersurface orthogonal \cite{BPa}.

\begin{figure}
\centering
\includegraphics[width=.71\textwidth]{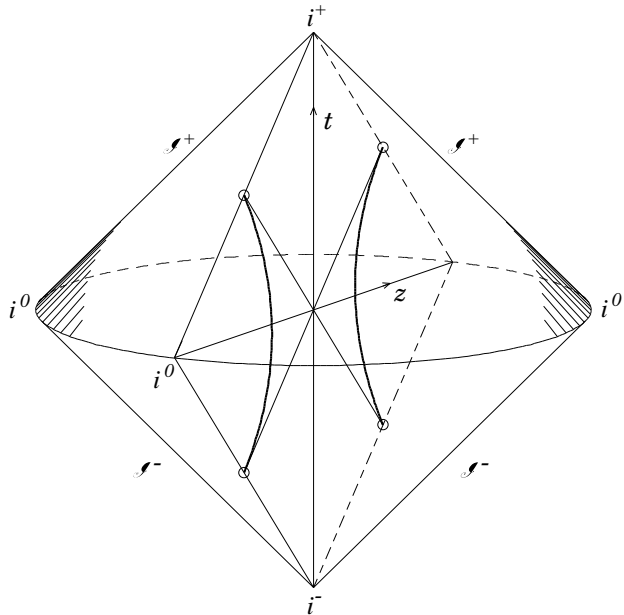}
\label{Figure 13}
\caption{The Penrose compactified diagram of a boost-rotation symmetric spacetime. Null infinity can admit smooth sections.}
\end{figure}

It is well known that in general relativity the ``causes'' of motion
are always incorporated in the theory -- in contrast to
electrodynamics where they need not even be describable by
Maxwell's theory. In a general case of the boost-rotation
symmetric solutions there exist nodal (conical)
singularities of the metric distributed along the $z$-axis which can
be considered as ``strings'', and cause particles to
accelerate. They reveal themselves also at $\cal J$. However, the
distribution of nodes can always be arranged in such a manner
that $\cal J$ admits smooth regular sections as mentioned above.

In exceptional cases, when $\cal J$ is regular except for
four points, either the particles are ``self-accelerating'' due to
their ``inner'' multipole structure, which has to include a negative
mass; or there are more particles distributed along $z>0$ (and
symmetrically along $z<0$) with the signs and the magnitudes of
their masses and accelerations chosen appropriately. (For the
concept of a negative mass in general relativity, and the first
discussion of a ``chasing'' pair of a positive and a negative mass
particle, see classical papers by Bondi, and Bonnor and
Swaminarayan \cite{ar}.) An infinite number of different analytic
solutions representing self-accelerating particles was
constructed explicitly \cite{as}. Although a negative mass cannot
be bought easily in
the shop (as Bondi liked to say), these solutions are the only
exact solutions of Einstein's equations available today for which
one can find such quantities of physical interest as radiation
patterns (angular distribution of gravitational radiation), or
total radiation powers \cite{an}. From a mathematical point of view,
these solutions represent the only known spacetimes in which
{\it arbitrarily strong} (boost-rotation symmetric) {\it initial
data} can be chosen on a hyperboloidal hypersurface in the region
$t^2>z^2$, which will lead to a complete smooth null infinity
and a regular timelike future infinity. With these specific
examples one thus does not have to require weak-field
initial data as one has to in the work of Friedrich, 
and Christodoulou and Klainerman, mentioned at the end of Section 1.3.

\vskip -0.1pt
The boost-rotation symmetric radiative spacetimes can be used as
test beds for approximation methods or numerical relativity.
Bi\v{c}\'{a}k, Reilly and Winicour \cite{at} found the explicit
boost-rotation symmetric ``initial null cone solution'', which
solves initial hypersurface and evolution equations in
``radiative'' coordinates employed in the null cone version of
numerical relativity. This solution has been used for checking and
improving numerical codes for computing gravitational radiation
from more realistic sources; a new solution of this type has
also been found \cite{au}. Recently, the specific boost-rotation
symmetric spacetimes constructed in \cite{as} were used as test beds in the standard
version of numerical relativity based on spacelike hypersurfaces \cite{AGS}.

\vskip -0.1pt
There exist ``generalized'' boost-rotation symmetric spacetimes
which are {\it not} asymptotically flat, but are of considerable
physical interest. They describe accelerated particles in
asymptotically ``uniform'' external fields. One can construct such
solutions from asymptotically flat boost-rotation symmetric
solutions for the pairs of accelerated particles by a limiting
procedure, in which one member of the pair is ``removed'' to
infinity, and its mass parameter is simultaneously increased
\cite{av}.
Since the resulting spacetimes are not asymptotically flat, their
radiative properties are not easy to analyze. Only if the
external field is weak will there exist regions in which the
spacetimes are approximately flat; and here their radiative
properties might be investigated. So far no systematic analysis
of these spacetime has been carried out. Nevertheless, they appear to
offer the best rigorous examples of the motion of relativistic
objects. No nodal singularities or negative masses are necessary
to cause an acceleration.

\vskip -0.1pt
As an eloquent example of such a spacetime consider a 
charged (Reissner-Nordstr\"{o}m) black hole with mass
$M$ and charge $Q$, immersed in an electric field ``uniform at
infinity'', characterized by the field-strength parameter $E$. An
exact solution of the Einstein-Maxwell equations exists which
describes this situation \cite{ERN}. It goes over into an approximate
solution obtained by perturbing the charged black hole spacetime
by a weak external electric field which is uniform at infinity
\cite{aw}. One of the results of the analysis of this solution
is very simple:
a charged black hole in an electric field starts to accelerate
according to Newton's second law, $Ma = QE$, where all the
quantities can be determined -- and in principle measured -- in an
approximately flat region of the spacetime from the asymptotic
form of the metric. Recall T.~S.~Eliot again: ``There is only the fight
to recover what has been lost / And found and lost again
and again.''

These types of generalized boost-rotation symmetric
spacetimes (``generalized {\it C}-metrics'') have been used by Hawking,
Horowitz, Ross, and others \cite{ax} in the context of quantum
gravity -- to describe production of black hole pairs in strong
background fields.

Recently, we have studied the spinning {\it C}-metric  discovered by 
Pleba\'nski and Demia\'nski \cite{PDE}. Transformations can be found which
bring this metric into the canonical form of spacetimes
with boost-rotation symmetry \cite{BPD}. The metric represents 
two uniformly {\it accelerated, rotating} black holes, 
either connected by conical singularity, or with conical singularities 
extending from each of them to infinity.
The spacetime is radiative. No other spacetime of this type,
with two Killing vectors which are not hypersurface orthogonal,
is available in an explicit form.

%%%%%%%%%%%%%%%%%%%%%%%%%%%%%%%%%%%%%%%
%%%%%%%%\input cosmo.tex
%%%%%%%%%%%%%%%%%%%%%%%%%%%%%%%%%%%%%%%
\section{The cosmological models}
%%%%%%%%%%%%%%%%%%%%%%%%%%%%%%%%%%%%%%%
In light of Karl Popper's belief that ``all science is cosmology'', it 
appears unnecessary to justify the choice of solutions for this last section. 
As in the whole article, these will be primarily vacuum solutions. On 
the other hand, in light of the light coming from about $10^{11}$ galaxies, each  
with about $10^{11}$ stars, it may seem weird to consider {\it vacuum} models 
of the Universe. Indeed, it has become part of the present-day culture that 
spatially homogeneous and isotropic, expanding Friedmann-Robertson-Walker (FRW) 
models, filled with uniformly distributed matter, correspond well to basic 
observational data. In order to achieve a more precise correspondence, it 
appears sufficient to consider just perturbations of these ``standard 
cosmological models''. To explain some ``improbable'' features of these models such as 
their isotropy and homogeneity, one finds an escape in inflationary scenarios. 
These views of a ``practical cosmologist'' are, for example, embodied in one of the 
most comprehensive recent treatise on physical cosmology by Peebles \cite{PEE}. 
	
Theoretical (or mathematical) cosmologists, however, point out that more 
general cosmological models exist which differ significantly from a FRW model at 
early times, approach the FRW model very closely for a certain epoch, and may 
diverge from it again in the future. Clearly, the FRW universes represent only a 
very  special class of viable cosmological models, though the simplest and 
most suitable for interpretations of ``fuzzy'' cosmological observational data. 

Simple exact solutions play a significant role in the evolution of more 
general models, either as asymptotic or intermediate states. By an ``intermediate 
state'' one means the situation when the universe enters and remains in a 
small neighbourhood of a saddle equilibrium point. A simple example is the 
Lema\^{\i}tre matter-filled, homogeneous and isotropic model with a 
nonvanishing cosmological constant (see e.g. \cite{PEE}), which expands from a 
dense state (the big bang, or ``primeval atom'' in Lema\^{\i}tre's 1927 
terminology), passes through a quasistatic epoch in which all parameters are 
close to those of the static Einstein universe (cf. Section 1.2), and then the 
universe expands again. An ``asymptotic state'' means close either to an initial 
big bang (or possibly a final big crunch) singularity, or the situation at late 
times in forever expanding universes. It is easy to see that at late 
times in indefinitely expanding universes the matter density decreases, and 
vacuum solutions may become important. However, as we shall discuss below, 
vacuum models play an important role also close to a singularity, when the matter 
terms in Einstein's equations are negligible compared to the ``velocity 
terms'' (given by the rate of change of scale factors) or to the curvature terms 
(characterizing the curvature of spacelike hypersurfaces). In particular, 
the pioneering (and still controversial) work started at the end of 
the 1950s by Lifshitz and Khalatnikov, and developed later on by Belinsky, 
Khalatnikov and Lifshitz, has shown that the fact that the presence 
of matter does not influence the qualitative 
behaviour of a cosmological model near a singularity has a very general 
significance (see \cite{BKL1} and \cite{BKL2} for the main 
original references, and \cite{LL} for a brief review).  

In gaining an intuition in the analysis of general 
cosmological singularities, the class of spatially homogeneous 
anisotropic cosmological models have played a crucial role.
These so called {\it Bianchi models} admit a simply transitive 3-dimensional 
homogeneity group. Among the Bianchi vacuum models 
there are special exact explicit solutions, in particular the Kasner and the 
Bianchi type II solutions, which exhibit some aspects of general cosmological 
singularities. The Bianchi models have also had an impact on other issues in 
general relativity and cosmology. 

Much work, notably in recent years, has been devoted to the class of both 
vacuum and matter-filled cosmological solutions which are homogeneous only
on 2-dimensional spacelike orbits. Thus they  depend on time and on one spatial 
variable, and can be used to study spatial inhomogeneities as density 
fluctuations or gravitational waves. The vacuum cosmological models with two 
spacelike Killing vectors, sometimes called the {\it Gowdy models},\footnote 
{In fact, by Gowdy models, one more often means only the cases 
with closed group orbits, with two commuting spacelike    
othogonally-transitive Killing vectors (the surface elements
orthogonal to the group orbits are surface-forming).} 
are interpreted as gravitational waves in an expanding (or 
contracting) universe with compact spatial sections. We shall discuss these two 
classes separately.

\subsection{Spatially homogeneous cosmologies}

The simplest solutions, the Minkowski, de Sitter, and anti de Sitter 
spacetimes, which have also been used in cosmological contexts (cf. Section 1.3), 
are 4-{\it dimensionally homogeneous}. As noted in Section 8.1, the vacuum plane 
waves (see equations (\ref{Equ51}), (\ref{Equ53}), (\ref{Equ61})) are 
also homogeneous spacetimes; and since they can 
be suitably sliced by spacelike hypersurfaces with expanding normal congruence, 
they can become asymptotic states in homogeneous expanding cosmologies. There 
exist several important non-vacuum homogeneous spacetimes, for example, the 
Einstein static universe (cf. Section 1.2), and G\"{o}del's stationary, rotating 
universe (see e.g. \cite{KSH,GOE}), famous for the first 
demonstration that Einstein's equations with a physically permissible matter 
source are compatible with the existence of closed timelike lines, i.e. with 
the violation of causality.

\vskip -0.3pt
Here we shall consider models in which the symmetry group does not make 
spacetime a homogeneous space, but in which each event in spacetime 
is contained in a spatial hypersurface that is homogeneous. 
The standard FRW models represent a 
special case of such models (they admit, in addition, an isotropy group $SO(3)$ 
at each point). The general spatially homogeneous solutions comprise of the {\it 
Kantowski-Sachs universes} and a much wider class of Bianchi models. By 
definition, the Bianchi models admit a simply transitive 3-dimensional 
homogeneity group $G_3$. There exist special ``locally rotationally symmetric'' 
(LRS) Bianchi models which admit a 4-dimensional isometry group $G_4$ acting on 
homogeneous spacelike hypersurfaces, but these groups have a simply transitive 
subgroup $G_3$. In contrast to this, Kantowski-Sachs spacetimes admit $G_4$ 
(acting on homogeneous spacelike hypersurfaces) which does {\it not} have any 
simply transitive subgroup $G_3$; it contains a multiply transitive $G_3$ acting 
on 2-dimensional surfaces of constant curvature, $G_4 = \RRe \times SO(3)$. 
A special case of the vacuum Kantowski-Sachs universe is represented by the 
Schwarzschild metric inside the horizon (with $t$ and $r$ interchanged). There has 
been a continuing interest in the Kantowski-Sachs models since their discovery in 
1966 \cite{KASA}, to which, as the authors acknowledge, J. Ehlers contributed by 
his advice. Some of these models had already appeared in the PhD thesis of Kip Thorne 
in 1965 (see also \cite{KIP} for magnetic Kantowski-Sachs models). Here, 
however, we just refer the reader to \cite{RYS,MAC1} for their 
classical description, to \cite{RY} for a canonical and quantum treatment, and 
to \cite{NOO} for the latest discussion of the Kantowski-Sachs 
quantum cosmologies.

\vskip -0.3pt
Although the 3-dimensional Lie groups which are simply transitive on homogenous 
3-spaces were classified by Bianchi in 1897, the importance of Bianchi's work 
for constructing vacuum cosmological models was only discovered by Taub in 
1951 \cite{Tau}, when the Taub space (cf. Section 7) was first given.
It is less known that at approximately the same time, if not 
earlier, the first explicit spatially homogeneous expanding and rotating 
cosmological models with matter (of the Bianchi type IX) were constructed by 
G\"{o}del,\footnote{G\"{o}del's profound ideas and results in cosmology, and 
their influence on later developments have been discussed in depth by G. Ellis 
in his lecture at the G\"{o}del '96 conference in Brno, Czech Republic, where 
G\"{o}del was born in 1906 (78 years before G\"{o}del, Ernst Mach was born in a 
place which today belongs to Brno). In the extended written version of Ellis' 
talk \cite{GOE} it is indicated that G\"{o}del's work also initiated the 
investigation of Taub. This may well be true with G\"{o}del's paper on 
the stationary rotating universe, but Taub's paper on Bianchi models was received 
by the Annals of Mathematics on May 15, 1959, i.e. before G\"{o}del's lecture 
on expanding and rotating models at the Congress of Mathematics took place.} who 
first presented his results at the International Congress of Mathematics held at 
Cambridge (Mass.) from August 30 till September 5, 1950. 

An exposition of Bianchi models has been given in a number of places: in the account on 
relativistic cosmology by Heckmann and Sch\"{u}cking \cite{HES} (complementing the 
chapter on exact solutions by Ehlers and Kundt \cite{EK}), in the monographs 
of Ryan and Shepley \cite{RYS}, and Zel'dovich and Novikov \cite{ZN2}, in 
several comprehensive surveys by MacCallum (see e.g. \cite{MAC1} and 
\cite{MAC2} for his latest review containing a number of references), most recently, in 
the book on the dynamical system approach in cosmology (in the Bianchi 
models in particular) edited by Wainwright and Ellis \cite{WE}; and, first but 
not least, in the classics of Landau and Lifshitz \cite{LL}. The Hamiltonian 
approach initiated by Misner \cite{MSR} in 1968, and used in, amongst other things, the 
construction of various minisuperspace models in quantum gravity, has been 
reviewed by Ryan \cite{Ry}; for more recent accounts, see several contributions 
to Misner's Festschrift \cite{MIF}. An interesting framework which unifies 
the Hamiltonian approach to the solutions which admit homogeneous hypersurfaces 
either spacelike (as Bianchi models) or timelike (as static spherical, or 
stationary cylindrical models) was recently developed by Uggla, Jantzen and 
Rosquist in \cite{UJR} (with 115 references on many exact solutions). Herewith 
we shall only briefly introduce the Bianchi models, note their special role 
in understanding the character of an initial cosmological singularity, and 
mention some of the most recent developments not covered by the reviews cited above.

The line element of the Bianchi models can be expressed in the form  
\begin{equation}
 ds^2 = - dt^2 ~+ ~g_{ab}(t)~\omega^a~\omega^b,
\end{equation}
where the time-independent 1-forms $\omega^a$ ($=E^a_\alpha dx^\alpha$), 
$a=1,2,3$, are dual to time-independent\footnote{The gravitational degrees of 
freedom are associated with the component (scalar) functions $g_{ab}(t)$ -- 
the so called metric approach. Alternatively, in the orthonormal frame approach, one 
chooses $g_{ab}(t) = \delta_{ab} $ and describes the evolution by time-dependent 
forms $\omega^a$. In still another approach one employs the automorphism of the 
symmetry group to simplify the spatial metric $g_{ab}$ (see \cite{MAC2,WE} 
for more details).} spatial frame vectors ${\bf E}_a$ (often an arbitrary 
time-variable $\tilde t$ is introduced by $dt=N(\tilde t)~d \tilde t$, $N$ being the 
usual lapse function). Both $\omega^a$ and ${\bf E}_a$ are group-invariant, commuting 
with the three Killing fields which generate the homogeneity group. They satisfy the 
relations 
\begin{eqnarray}
d \omega^a   &=&  ~-{1\over 2}C^a_{bc}~\omega^b \wedge \omega^c , \\
\left[ {\bf E}_a , {\bf E}_b \right]  &=& C^c_{ab} {\bf E}^c,
\end{eqnarray}
 where $d$ is the exterior derivative and $C^a_{bc}$ are the structure constants 
of the Lie algebra of the homogeneity group. The models are classified according 
to the possible distinct sets of the structure constants. They are first divided 
into two classes: in class A the trace $C^a_{ba}=0$, and in class B, $C^a_{ba}\neq 
0$. In class A one can choose $C^a_{bc} = n^{(a)} \epsilon_{abc}$ (no summation 
over $a$), and classify various symmetry types by parameters $n^{(a)}$ with values 
$0,\pm 1$. In class B, in addition to  $n^{(a)}$, one needs the value of a 
constant scalar $h$ (related to $C^a_{ba}$) to characterize types VI${}_h$ and 
VII${}_h$ (see e.g. \cite{WE}). 

The simplest models are the Bianchi I cosmologies in class A with 
$n^{(a)}=0$, i.e. $C^a_{bc}=0$, so that all three Killing vectors (the group 
generators) commute. They contain the standard Einstein-de Sitter model with 
flat spatial hypersurfaces (curvature index $k = 0$). In the vacuum case, all 
Bianchi I models are given by the well-known 1-parameter family of {\it Kasner 
metrics} (found in 1921 by E. Kasner and in 1933 by G. Lema\^{\i}tre without 
considering the Bianchi groups) 
\begin{equation}
\label{Equx99}
 ds^2 = - dt^2 ~+ ~t^{2p_1} dx^2~+ ~t^{2p_2} dy^2~+ ~t^{2p_3} dz^2,
\end{equation}
 where 
\begin{equation}
\label{Equx99a}
p_1+p_2+p_3=1,~p_1^2+p_2^2+p_3^2=1.
\end{equation}
These metrics were first used to investigate various effects in anisotropic 
cosmological models. For example, in contrast to standard FRW models with 
``point-like'' initial singularities, the Kasner metrics can permit the so called 
{\it ``cigar''} and {\it ``pancake'' singularities}. To be more specific, consider the 
congruence of timelike lines with unit tangent vectors $n^\alpha$ orthogonal to 
constant time hypersurfaces, and define the expansion tensor 
$\theta_{\alpha\beta}$ by 
$\theta_{\alpha\beta} = \sigma_{\alpha\beta} + {1\over 3}\theta h_{\alpha\beta}$,
where $h_{\alpha\beta} = g_{\alpha\beta} + n_\alpha n_\beta$ is a 
projection tensor, 
$\sigma_{\alpha\beta} = n_{(\alpha;\beta)} - {1\over 3}\theta h_{\alpha\beta}$
is the shear, and $\theta = \theta^{~\alpha}_{\alpha}$. Determining the three spatial 
eigenvectors of $\theta_{\alpha\beta}$ with the corresponding 
eigenvalues $\theta_{i}$ ($i=1,2,3$), one can define the scale factors 
$l_i$ by the relation $\theta_{i} = (dl_i/dt)/l_i$ , and the Hubble scalar 
$H = {1\over 3}(\theta_1+\theta_2+\theta_3)$. In the FRW models,  
all $l_i \rightarrow 0$ at the big bang singularity. In the Kasner models at 
$t  \rightarrow 0$ one finds that either two of the 
$l_i$ go to zero, whereas the third unboundedly increases 
(a cigar); or one of the $l_i$ tends to zero, while the other two approach a finite 
value (pancake). Also there is the {\it ``barrel'' singularity} in which 
the two of the $l_i$ go to zero, and the third approaches a finite value. There 
is an open question as to whether some other possibilities exist \cite{WE}. Even in 
the perfect fluid Kasner model, the approach to the singularity is 
{\it ``velocity-dominated''} -- the ``vacuum terms'' given by the rates of change of the 
scale factors dominate the ``matter terms'' (curvature terms vanish since the 
Kasner models are spatially flat). 

The general vacuum Bianchi type II cosmologies (with one $n^{(a)} = +1$, 
and the other two vanishing), discovered by Taub in \cite{Tau}, contain 
two free parameters: 
\begin{equation}
\label{Equx100}
 ds^2 = - A^2 dt^2 ~+ ~A^{-2}~t^{2p_1} (dx+4 p_1 b z\; dy)^2~+~ 
 A^2 (t^{2p_2}\; dy^2~+ ~t^{2p_3}\; dz^2),
\end{equation}
where 
\begin{equation}
A^2 = 1+ b^2 t^{4 p_1},~
p_1+p_2+p_3=1,~p_1^2+p_2^2+p_3^2=1.
\end{equation}
If we put the parameter $b=0$, the metrics (\ref{Equx100}) become 
the Kasner solutions (\ref{Equx99}). Near the big bang the general Bianchi type II 
solution is asymptotic to a Kasner model. In the future it is asymptotic again to a 
Kasner model, but with different values of parameters $p_i$ (see e.g. \cite{WE}).
This fact will be important in the following.

The general Bianchi type V vacuum solutions are also known -- these are given by the  
1-parameter family of Joseph solutions \cite{WE}. The type V models are 
the simplest metrics in class B (with all $n^{(a)} = 0$ but 
$C^a_{bc} = 2 a^{}_{[b} \delta^a_{c]}$, $a_b = \const$), and are the simplest Bianchi 
models which contain the standard FRW open universes ($k=-1$). The Joseph solutions 
are asymptotic to the specific Kasner solution in the past, and tend to the {\it 
``isotropic Milne model''} in the future. This is intuitively understandable since 
open FRW models, as they expand indefinitely into the future with matter density 
decreasing, also approach the Milne model. As is well known, the Milne model 
is just an empty flat (Minkowski) spacetime in coordinates adapted to  
homogeneous spacelike hypersurfaces (the ``mass hyperboloids''), with expanding 
normals (see e.g. \cite{PEE}): 
\begin{equation}
 ds^2 = -  d\tau^2 ~+ ~\tau^2 \left[
(1+\rho^2)^{-1} d\rho^2 + \rho^2 (d\theta^2+\sin^2\theta~d\varphi^2)\right],
\end{equation}
with $\tau = t (1-u^2)^{1/2}$, $\rho = u (1-u^2)$, $ u = r/t <1 $,
where $t,r,\theta,\varphi$
are standard Minkowski (spherical) coordinates. 
Because of its significance as an asymptotic solution and its simplicity, 
the Milne model has been used frequently in pedagogical expositions of relativistic 
cosmology (see e.g. \cite{PEE,Ri}) as well as in cosmological perturbation 
theory and quantization (see e.g. \cite{TTS} and references 
therein). The Milne universe is also an asymptotic state of other Bianchi models 
such as, for example, the intriguing {\it Lukash vacuum} type VII${}_h$ {\it solution} 
\cite{LUK}, which can be interpreted as two monochromatic, circularly polarized 
waves of time-dependent amplitude travelling in opposite directions on a FRW 
background, with flat or negative curvature spacelike sections. As was noticed 
earlier, some indefinitely expanding Bianchi models approach the homogeneous plane wave 
solutions. Barrow and Sonoda \cite{BASO} studied the future asymptotic 
behaviour of the known Bianchi solutions in detail by using nonlinear 
stability techniques; and in \cite{WE} dynamical system methods were used.

From the late 1960s onwards the greatest amount of work was probably 
devoted to the Bianchi type IX vacuum models, baptized {\it the Mixmaster 
universe}\footnote {The name comes from the fact that, in contrast to a 
standard FRW model, which has a horizon preventing the equalization of possible 
initial inhomogeneities over large scales, the horizon in a type IX universe is 
absent, so that mixing is in principle possible. However, as was shown e.g. in 
\cite{ZN2}, ``repeated circumnavigations of the universe by light are impossible in 
the Mixmaster model''.} by Misner \cite{MSR}. Type IX models are the most 
general class A models with all parameters $n^{(a)} = +1$. They are the only Bianchi 
universes which recollapse. If a perfect fluid is permitted as the matter source, the 
non-vacuum type IX solutions contain the closed FRW models ($k=+1$) with space sections 
having spherical topology. As a Bianchi I space admits a group isomorphic with 
translations in a 3-dimensional Euclidean space, the group of type IX spaces is 
isomorphic to the group of rotations. None of the pairs of three Killing vectors 
commute. A general Bianchi IX vacuum solution is not known, but a particular 
solution is available: the Taub-NUT spacetime, or rather, its spatially 
homogeneous anisotropic region -- the Taub universe (see Section 7.2). This fact 
was, for example, employed in an attempt to understand the limitations of the 
minisuperspace methods of quantum gravity: by reducing the degrees of freedom to 
a general Mixmaster universe and then further to the Taub universe one can 
see what such restrictions imply \cite{KURY}.

\vskip -0.2pt
The dynamics of general Bianchi cosmologies -- and of the Mixmaster models 
in particular -- close to the big bang singularity has been approached with 
essentially three methods \cite{WE}: (i) piecewise approximation methods, (ii) 
Hamiltonian methods, and (iii) dynamical system methods. In the first method, 
used primarily by Russian cosmologists (cf. \cite{BKL1,BKL2}), the 
evolution is considered to be a sequence of periods in which certain terms in 
the Einstein equations dominate whereas other terms can be neglected. The 
Hamiltonian methods appeared first in the ``Mixmaster paper'' by Misner 
\cite{MSR}, were reviewed by Ryan \cite{Ry}, and more recently by Uggla in \cite{WE}. 
With the Hamiltonian (canonical) approaches, minisuperspace methods entered 
general relativity (cf. Section 9.2 on midisuperspace for cylindrical waves). 
In this approach, infinitely many degrees of freedom are reduced to a finite number: the state of 
the universe is described by a ``particle'' moving inside and reflecting 
instantaneously from the moving potential walls, which approximate the 
time-dependent potentials in the Hamiltonian. In the third method one employs the fact 
that Einstein's equations in the case of homogeneous cosmologies can be put into 
the form of an autonomous system of first-order (ordinary) differential 
equations on a finite dimensional space $\RRe^n$. This is of the form 
$d{\bf x}/dt = {\bf f}({\bf x})$, with ${\bf x} \in \RRe^n$ 
representing a state of the model (for 
example, the suitably normalized components of the shear $\sigma$, the Hubble scalar 
$\theta$, and parameters related to $n^{(a)}$,  can serve as the ``components'' of ${\bf x}$).
A study of the orbits ${\bf x}(t)$ indicates the 
behaviour of the model. Dynamical system methods are the focus of the book 
\cite{WE}. They also are the main tools of the monograph \cite{BOG}. 

\vskip -0.2pt
In the case of the Bianchi IX models (either vacuum or with perfect 
fluid), all three methods imply (though do not supply a rigorous proof) that 
an approach to the past big bang singularity is composed of an infinite 
sequence of intervals, in each of which the universe behaves approximately as 
a specific Kasner model (\ref{Equx99}).
The transition ``regimes'' between two different 
subsequent Kasner epochs, in which the contraction proceeds along subsequently 
different axes, is approximately described by Bianchi type II vacuum solutions 
(\ref{Equx100}). This famous and enigmatic {\it ``oscillatory approach to the 
singularity''} (or ``Mixmaster behaviour'') has rightly
entered the classical literature (cf. e.g. \cite{MTW,Wa,LL}).
It indicates that the big bang singularity (and, similarly, a 
singularity formed during a gravitational collapse) can be much more complicated 
than the ``point-like'' singularity in the standard FRW models. This oscillatory 
character has been suggested not only by the qualitative methods mentioned above, but 
also by extensive numerical work (see e.g. \cite{WE,DET}). So far, however, it 
has resisted a rigorous proof. 

In the ``standard'' picture of the Mixmaster model it is supposed that the 
evolution of the Bianchi type IX universe near the singularity can be 
approximated by a mapping of the so called {\it Kasner circle} onto itself. This 
is the unit circle in the $\Sigma_+\Sigma_-$ plane, where $\Sigma_\pm = \sigma_\pm/H$
describes the anisotropy in the 
Hubble flow (cf. e.g. Fig. 6.2 in \cite{WE}). Each point on the circle 
corresponds to a specific Kasner solution with given fixed values of parameters 
$p_i$ satisfying the conditions (\ref{Equx99a}).
There are three exceptional points on 
the circle -- those at which one of the $p_i=+1$, and the other two vanish. From each 
non-exceptional point $P_1$ on the Kasner circle there leads a 1-dimensional unstable 
orbit given by the vacuum Bianchi II solution (\ref{Equx100}), which joins $P_1$ to 
another point $P_2$ on the circle, then $P_2$  is mapped to $P_3$ , etc. This ``Kasner map'' in 
the terminology of \cite{WE}, called frequently also the BKL 
(Belinsky-Khalatnikov-Lifshitz) map, describes subsequent changes of Kasner epochs during 
the oscillatory approach to a singularity. Recent rigorous results of Rendall 
\cite{REN} show that for any {\it finite} sequence generated by the BKL map, 
there exists a vacuum Bianchi type IX solution which reproduces the sequence 
with any required accuracy.\footnote{A. Rendall (private communication) 
reports that the main points of the BKL picture for homogeneous universes have 
been rigorously confirmed in a recent work of H. Ringstr\"{o}m (to be 
published).} 

The vacuum Bianchi IX models have been extensively analyzed in the context 
of deterministic chaos and their stochasticity, attracting the interest of leading 
experts in these fields \cite{DET,SIN}. Above all, it is the numerical work which strongly 
suggests that it is impossible to make long-time predictions of the evolution of the
system from the initial data, which is the most significant property of a
chaotic system. 

Most recently, interest in the Bianchi cosmologies with (homogeneous) 
magnetic and scalar fields has been revived. Following his previous work with 
Wainwright and Kerr on magnetic Bianchi VI${}_0$ cosmologies \cite{LWK}, 
LeBlanc \cite{LBL1} has shown that even in Bianchi I cosmologies one finds  
an oscillatory approach towards the initial singularity if a magnetic field in a 
general direction is present. (The points on the Kasner circle are now joined 
by Rosen magneto-vacuum solutions.) Hence, Mixmaster-like oscillations occur due 
to the magnetic field degrees of freedom, even in the absence of an anisotropic 
spatial curvature (present in the vacuum type IX models) -- the result 
anticipated by Jantzen \cite{JAT} in his detailed work on Hamiltonian methods 
for Bianchi cosmologies with magnetic and scalar fields. Similar conclusions 
have also been arrived at in \cite{LBL2} for magnetic Bianchi II cosmologies. 
(LeBlanc's papers contain some new exact magnetic Bianchi solutions and a 
number of references to previous work.) Interestingly, in contrast 
to the magnetic field, scalar fields in general 
{\it suppress} the Mixmaster oscillations when approaching the initial 
singularity \cite{BEKA,BERG}.

\vskip -0.2pt
The theory of spatially homogeneous, anisotropic models is an elegant, 
intriguing branch of mathematical physics. It has played an important role in 
general relativity. The classical monograph of Zel'dovich and Novikov 
\cite{ZN2}, or the new volume of 
Wainwright and Ellis \cite{WE} analyze in detail 
the possible observational relevance of these models: they point out spacetimes 
close to FRW cosmologies (at least during an epoch of finite duration) which are 
compatible with observational data. For the most recent work on Bianchi VII${}_h$ 
cosmologies which are potentially compatible with the highly isotropic microwave 
background radiation, see \cite{WCE} (and references therein). Nevertheless, 
the present status is such that, in contrast to for example the Kerr solution, which 
is becoming an increasingly strong attractor for practical 
astrophysicists (cf. Section 4.3), the anisotropic models have not really 
entered (astro)physical cosmology so far. 
Peebles, for example, briefly comments in \cite{PEE}: ``The homogeneous 
anisotropic solutions allowed by general relativity are a very useful tool for 
the study of departures from the Robertson-Walker line element. As a realistic 
model for our Universe, however, these solutions seem to be of limited interest, 
for they require very special initial conditions: if the physics of the early 
universe allowed appreciable shear, why would it not also allow appreciable 
inhomogeneities?'' 

\vskip -0.2pt
An immediate reaction, of course, would be to point out that the FRW models 
require still more ``special initial conditions''. However, there 
appears to be a deeper reason why the oscillatory approach towards a singularity 
may be of fundamental importance. Belinsky, Khalatnikov and Lifshitz 
\cite{BKL1,BKL2} employed their piecewise approximation method, 
and concluded 30 years ago that a singularity in a  {\it general, inhomogeneous} 
cosmological model is spacelike and locally oscillatory: i.e. in their 
scenario, the evolution at different spatial point decouples. At each spatial 
point the universe approaches the singularity as a distinct Mixmaster universe. 
This view, often criticized by purists, appears now to be gaining an increasing number 
of converts, even among the most rigorous of relativists. As mentioned above, the 
homogeneous magnetic Bianchi type VI${}_0$ models, investigated by LeBlanc et al., 
show Mixmaster behaviour. The Bianchi VI${}_0$  models have, as do all Bianchi models, 
three Killing vectors, but two of them commute. The models can thus be 
generalized by relaxing the symmetry connected with the third Killing vector; one can so 
obtain effectively the inhomogeneous (in one dimension) Gowdy-type spacetimes. 
Weaver, Isenberg and Berger \cite{WIB}, following this idea of Rendall, analyzed 
these models numerically, and discovered that the Mixmaster behaviour is reached 
at different spatial points. The numerical evidence for an oscillatory 
singularity in a generic vacuum $U(1)$ symmetric cosmologies with the spatial 
topology of a 3-torus has been found still more recently by Berger and Moncrief 
\cite{BEM}.

Before turning to the Gowdy models, a last word on the ``oscillatory 
approach towards singularity''. I heard E. M. Lifshitz giving a talk on this 
issue a couple of times, with Ya. B. Zel'dovich in the audience. In discussions 
after the talk, Zel'dovich, who appreciated much this work (its detailed 
description is included in \cite{ZN2}), could not resist pointing out that the 
number of oscillations and Kasner epochs will be very limited (to only about 
ten) because of quantum effects which arise when some scale of a model is 
smaller than the Planck length $l_{Pl} \sim 10^{-33}{\rm cm}.$ 
This, however, seems to make the scenario 
still more intriguing. If this is confirmed rigorously within classical relativity, 
how will a future quantum gravity modify this picture? 

\subsection{Inhomogeneous cosmologies}

Among all of the known vacuum inhomogeneous models, the {\it Gowdy solutions} \cite{GOW} 
have undoubtedly played the most distinct role. They belong to the class of solutions with 
two commuting spacelike Killing vectors. Within a cosmological context, they form 
a subclass of a wider class of $G_2$ {\it cosmologies} -- as are now commonly denoted  
models which admit an Abelian group $G_2$ of isometries with orbits being spacelike 
2-surfaces. A 2-surface with a 2-parameter isometry group must be a space of 
constant curvature, and since neither a 2-sphere nor a 2-hyperboloid possess 
2-parameter subgroups, it must be intrinsically flat. 
If the 2-surface is an Euclidean plane or 
a cylinder, then one speaks about planar or cylindrical universes. Gowdy universes 
are compact -- the group orbits are 2-tori $T^2$.

The metrics with two spacelike Killing vectors are often called the 
generalized Einstein-Rosen metrics as, for example, by Carmeli, Charach and 
Malin \cite{CCM} in their comprehensive survey of inhomogeneous cosmological 
models of this type. In dimensionless coordinates ($t,z,x^1,x^2$), the line element can be 
written as ($A,B = 1,2$) 
\begin{equation}
\label{Equx102}
ds^2 / L^2 ~=~ e^F (-dt^2 + dz^2) + \gamma_{AB} dx^A dx^B,
\end{equation}
where $L$ is a constant length, $F$ and $\gamma_{AB}$  depend on $t$ and $z$ 
only, and thus the spacelike Killing vectors are
$
{}^{(1)}\xi^\alpha= (0,0,1,0), {}^{(2)}\xi^\alpha= (0,0,0,1). 
$

The local behaviour of the solutions of this form is described by the 
gradient of the ``volume element'' of the group orbits 
$W = \left( |\det(\gamma_{AB})|\right)^{1/2}$. Classical 
cylindrical Einstein-Rosen waves (cf. Section 9) are obtained if $W_{,\alpha}$ is 
globally spacelike. In Gowdy models, $W_{,\alpha}$ varies from one region to
another.\footnote {The same is true in the boost-rotation symmetric spacetimes 
considered in Section 11: the part $t^2 > z^2$ of the spacetimes, where the boost Killing 
vector is spacelike, can be divided into four different regions, in two of which 
vector $W_{,\alpha}$ is spacelike, and in the other two timelike -- see \cite{ao} for 
details.} 

Considering for simplicity the polarized Gowdy models (when the Killing 
vectors are hypersurface orthogonal), the metric (\ref{Equx102}) can be written in  
diagonal form (cf. equations (\ref{Equ62}), (\ref{Equ63}), and 
(\ref{Equ65}), (\ref{Equ66}) in the analogous cases of 
plane and cylindrical waves)
\begin{equation}
 ds^2/L^2 ~=~ e^{-2U} \left[  
e^{2\gamma} (-dt^2+dz^2) + W^2 dy^2
\right] +e^{2U} dx^2,
\end{equation}
 in which $U(t,z)$ and $\gamma(t,z)$ satisfy wavelike dynamical equations and constraints 
following from the vacuum Einstein equations; the function $W(t,z)$, 
which determines the volume element of the group orbit, can be cast into a standard 
form which depends on the topology of $t=$~constant spacelike hypersurfaces $\Sigma$. 

As mentioned above, in Gowdy models one assumes these hypersurfaces to be 
compact. Gowdy \cite{GOW} has shown that 
$\Sigma$ can topologically be (i) a 3-torus $T^3 = S^1 \otimes S^1 \otimes S^1$
and $W=t$ (except for the trivial 
case when spacetime is identified as a Minkowski space), (ii) a 3-handle (or 
hypertorus, or ``closed wormhole'') $S^1 \otimes S^2$ with $W=\sin z \sin t$, 
or (iii) a 3-sphere $S^3$, again with $W=\sin z \sin t$.
(For some subtle cases not covered by Gowdy, see \cite{PCL}.)
As the form of $W$ suggests, in the case of a
$T^3$ topology, the universe starts with a big bang singularity at $t=0$ and then expands 
indefinitely, whereas in the other two cases it starts with a big bang at $t=0$, 
expands to some maximal volume, and then recollapses to a ``big crunch'' 
singularity at $t=\pi$. One can determine exact solutions for metric functions in 
all three cases in terms of Bessel functions \cite{GOW1}. Hence, for the first 
time {\it cosmological models closed by gravitational waves} were constructed. 
Charach found Gowdy universes with some special electromagnetic fields 
\cite{CHA}, and other generalized Gowdy models were obtained. We refer to the 
detailed survey \cite{CCM} for more information, including the work on 
canonical and quantum treatments of these models, done at the beginning of the 1970s 
by Berger and Misner, and for extensive references. 

Let us only add a few remarks on some more recent developments in which the 
Gowdy models have played a role. Gowdy-type models have been used to study the 
{\it propagation and collision of gravitational waves 
with toroidal wavefronts} (as mentioned earlier, 
2-tori $T^2$ are the group orbits in the Gowdy cosmologies) in the FRW closed 
universes with a stiff fluid \cite{BIJG}. In the standard Gowdy spacetimes it 
is assumed that the ``twists'' associated with the isometry group on $T^2$ vanish. In 
\cite{BCIM} the generalized Gowdy models without this assumption are considered, 
and their global time existence is proved. 

As both interesting and non-trivial models, the Gowdy spacetimes have recently  
attracted the attention of mathematical and numerical relativists with an 
increasing intensity, as indicated already at the end of the previous section. 
Chru\'{s}ciel, Isenberg and Moncrief \cite{GIM} proved that Gowdy spacetimes developed from a 
dense subset in the initial data set cannot be extended past their singularities, 
i.e. in ``most'' Gowdy models the strong cosmic censorship is satisfied.

On cosmic censorship and spacetime singularities, especially in the 
context of compact cosmologies, we refer to a review by Moncrief \cite{VMO}, 
based on his lecture in the GR14 conference in Florence in 1995. The review shows 
clearly how intuition gained from such solutions as the Gowdy models or the Taub-NUT 
spaces, when combined with new mathematical ideas and techniques, can produce 
rigorous results with a generality out of reach until recently. To such 
results belongs also the very recent work of Kichenassamy and Rendall \cite{KIR} 
on the sufficiently general class of solutions (containing the maximum 
number of arbitrary functions) representing unpolarized Gowdy spacetimes. The new 
mathematical technique, developed by Kichenassamy \cite{KIM}, the so called 
Fuchsian algorithm, enables one to construct singular (and nonsingular) 
solutions of partial differential equations with a large number of arbitrary 
functions, and thus provide a description of singularities. Applying the Fuchsian 
algorithm to Einstein's equations for Gowdy spacetimes with topology $T^3$, 
Kichenassamy and Rendall have proved that general solutions behave at the (past) 
singularity in a Kasner-like manner, i.e. they are asymptotically velocity 
dominated with a diverging Kretschmann (curvature) invariant. One needs an 
additional magnetic field not aligned with the two Killing vectors of the Gowdy 
unpolarized spacetimes in order to get a general oscillatory (Mixmaster) 
approach to a singularity, as shown by the numerical calculations \cite{WIB} 
mentioned at the end of the previous section. 

\vskip -0.3pt
Much of the work on exact inhomogeneous vacuum cosmological models has been 
related to ``large perturbations'' of Bianchi universes. In \cite{AZF} the 
authors confined attention to ``plane wave'' solutions propagating over Bianchi 
backgrounds of types I-VII. They found universes which are highly inhomogeneous 
and ``chaotic'' at early times, but are transformed into clearly ``recognizable'' 
gravitational waves at late times. 

\vskip -0.3pt
Other types of metrics can be considered as exact {\it ``gravitational 
solitons''} propagating on a cosmological background. These are usually obtained 
by applying the inverse scattering or ``soliton'' technique of Belinsky and 
Zakharov \cite{BEZ} to particular solutions of Einstein's equations as ``seeds''. 
For example, Carr and Verdaguer \cite{CAV} found gravisolitons by applying the 
technique to the homogeneous Kasner seed. Similarly to previous work \cite{AZF},
their models are very inhomogeneous at early times, but evolve towards 
homogeneity in a wavelike manner at late times. 

\vskip -0.3pt
More recently, Belinsky \cite{BEL}, by applying a two-soliton inverse 
scattering technique to a Bianchi type VI${}_0$ solution as a seed, constructed 
an intriguing solution which he christened as a {\it ``gravitational breather''}, in 
analogy with the Gordon breather in the soliton theory of the sine-Gordon equation. 
Gravisolitons and antigravisolitons, characterized by an opposite topological 
charge, can be heuristically introduced and shown to have an attractive 
interaction. The breather is a bound state of the gravisoliton and 
antigravisoliton. Belinsky suggests that a time oscillating breather 
exists; but a later discussion \cite{KORD} indicates that the oscillations quickly decay.
Alekseev, by employing his generalization of the inverse scattering method to the 
Einstein-Maxwell theory, obtained exact electrovacuum solutions generalizing Belinsky's breather
(see his review \cite{ALE}, containing a general introduction on exact solutions).

Verdaguer \cite{t} prepared a very complete review of solitonic 
solutions admitting two spacelike Killing vector fields, with the main emphasis on 
cosmological models. Among various aspects of such solutions, he has 
noted the role of the Bel-Robinson superenergy tensor in the interpretation of 
cosmological metrics. This tensor and its higher-order generalizations  has also  been 
significantly used in estimates in the proofs of long-time existence 
theorems \cite{CHK,VMO}. Recently, differential conservation laws for 
large perturbations of gravitational field with respect to a given 
curved background have been fomulated \cite{JOKB}, which found an application 
in solving equations for cosmological perturbations corresponding to
topological defects \cite{TUN}. They should bring more light also on 
various solitonic models in cosmology.

%%%%%%%%%%%%%%%%%%%%%%%%%%%%%%%%%%%%%%%
\section{Concluding remarks}
%%%%%%%%%%%%%%%%%%%%%%%%%%%%%%%%%%%%%%%

It is hoped that the preceding pages have helped to elucidate at least one issue: that in such a
complicated nonlinear theory as general relativity, it is not possible to ask relevant
questions of a general character without finding and thoroughly analyzing specific exact 
solutions of its field equations. The role of some of the solutions in our understanding of 
gravity and the universe has been so many-sided that 
to exhibit this role properly on even more than a hundred pages is not really feasible ...

        Although we have concentrated on only (electro)vacuum solutions, 
there remains a number of such solutions that have 
also played some role in various contexts, but, owing to the absence 
of additional space and time, or the presence of the author's ignorance, have not been 
discussed. Tomimatsu-Sato solutions and their generalizations, static plane and cylindrical 
metrics, and some algebraically special solutions are examples.

        In his review of exact solutions, Ehlers \cite{EHLO} wrote 35 years ago that ``it 
seems desirable to construct material sources for vacuum solutions'', and 30 years later Bonnor 
\cite{Bo}, in his review, expressed a similar view. In the above we have noted 
only some of the thin disk sources of static and 
stationary spacetimes in Section 6. To find physically reasonable 
material sources for many of the known vacuum solutions remains a difficult open task. In 
order to make solutions of Einstein's equations with the right-hand side more tractable, one 
is often tempted to sacrifice realism and consider materials, again using Bondi's phraseology, 
which are not easy to buy in the shops. Nevertheless, there are solutions representing 
spacetimes filled with matter which would certainly belong in a more complete discussion of 
the role of exact solutions.

          For example, one of the simplest, the spherically symmetric Schwarzschild interior 
solution with an incompressible fluid as matter source,
modelling ``a star of uniform density'', gives 
surprisingly good estimates of an upper bound on the masses of 
neutron stars; on a more general level, 
it supplies an instructive example of relativistic hydrostatics \cite{MTW}. Many other 
spherical perfect fluid solutions are listed in \cite{KSH}. The proof of a very plausible 
fact that any equilibrium, isolated stellar model which is 
nonrotating must be spherically symmetric, was finally completed in \cite{BESI} and \cite{LIM}. Physically more adequate spherically symmetric static solutions with collisionless 
matter described by the Boltzmann (Vlasov) equation have been studied \cite{ZN} 
(yielding, for example, arbitrarily large 
central redshifts); and some of their aspects have been recently reviewed from a 
rigorous, mathematical point of view \cite{RFL}. Going over to the description of matter in 
terms of physical fields, we should mention the first spherically symmetric regular 
solutions of the Einstein-Yang-Mills equations (``non-Abelian solitons'' discovered by Bartnik 
and McKinnon \cite{BMK} in 1988), and non-Abelian black holes with ``hair'', which were found soon 
afterwards. They stimulated a remarkable activity in the search for models in which gravity is 
coupled with Yang-Mills, Higgs, and Skyrmion fields. Very recently these solutions have been 
surveyed in detail in the review by Volkov and Gal'tsov \cite{VG}.

\vskip -0.3pt
       The role of the standard FRW cosmological models on the development of relativity 
and cosmology can hardly be overemphasized. As for two more recent examples of this 
influence let us just recall that the existence of cosmological
horizons in these models was one of 
the crucial points which inspired the birth of inflationary cosmology (see e.g. \cite{PEE}); 
and the very smooth character of the initial singularity has led Penrose \cite{ROG} to 
formulate his Weyl curvature hypothesis, related to a still unclear concept of gravitational 
entropy. Homogeneous but anisotropic Bianchi models filled with perfect fluid are extensively 
analyzed in \cite{WE}. Very recent studies of Bianchi models with collisionless matter 
\cite{RET} reveal how the matter content can qualitatively alter the character of the model.
 
\vskip -0.3pt
A number of Bianchi models approach self-similar solutions. Perfect fluid solutions 
admitting a homothetic vector, which in this case implies  both geometrical and physical 
self-similarity, have been reviewed most recently by Carr and Coley \cite{CAC}. In their 
review various astrophysical and cosmological applications of such solutions are also 
discussed. Self-similar solutions have played a crucial role in the critical phenomena in 
gravitational collapse. Since their discovery by Choptuik in 1993, they have attracted much 
effort, which has revealed quite unexpected facts. In \cite{CAC} these phenomena are 
analyzed briefly. For a more comprehensive review, see \cite{GUN}. 

\vskip -0.3pt
        Self-similar, spherically symmetric solutions have been very relevant in  
constructing examples of the formation of naked singularities in gravitational collapse 
(see \cite{CAC} for a brief summary and references). In particular, the inhomogeneous, 
spherically symmetric Lema\^{\i}tre-Bondi-Tolman universes containing dust have been employed in 
this context. Solutions with null dust should be mentioned as well, especially the 
spherically symmetric Vaidya solutions: imploding spherical null-dust models have been 
constructed in which naked singularities arise at their centre (see \cite{Jo} for summary and 
references).

        The Lema\^{\i}tre-Bondi-Tolman models are the most frequently analyzed inhomogeneous 
cosmological models which contain the standard FRW dust models as special cases (see e.g. 
\cite{PEE,Jo}). In his recent book Krasi\'{n}ski \cite{KRA} has compiled and discussed 
most if not all of these exact inhomogeneous cosmological solutions found so far which can be 
viewed as ``exact perturbations'' of the FRW models. 

Many solutions known already still wait for their role to be uncovered. The role of many 
others may forever remain just in their ``being''. However, even if new solutions of a ``Kerr-like 
significance'' will not be obtained in the near future, we believe that one should not cease 
in embarking upon journeys for finding them, and perhaps even more importantly, for revealing new roles of 
solutions already known. The roads may not be easy, but with todays equipment like Maple or 
Mathematica, the speed is increasing. Is there another so explicit way of how to learn more about 
the rich possibilities embodied in Einstein's field equations?

        The most remarkable figure of Czech symbolism, Otokar B\v rezina (1868-1929) has 
consoling words for those who do not meet the ``Kerr-type'' metric on the road: ``Nothing is 
lost in the world of the spirit; even a stone thrown away may find its place in the hands of a 
builder, and a house in flames may save the life of someone who has lost his way...''.   
\vskip 6mm

\noindent
{\bf  Acknowledgements}

\vskip 4mm

\noindent
Interaction with J\"{u}rgen Ehlers has been important for me over the years:
Thanks to my regular visits to his group, 
which started seven years before the hardly penetrable barrier between Prague and the West 
disappeared, I have been in contact with ``what is going on'' much more than I could have been 
at home. Many discussions with J\"{u}rgen, collaboration and frequent discussions with Bernd 
Schmidt, and with other members of Munich$\rightarrow$Potsdam$\rightarrow$Golm relativity
 group are fondly recalled and appreciated.

For helpful comments on various parts of the manuscript I am grateful to 
Bobby Beig, Jerry Griffiths, Petr H\'aj\'\i\v cek, Karel Kucha\v r, 
Malcolm MacCallum, Alan Rendall and Bernd Schmidt. For discussions 
and help with references I thank also Piotr Chru\'sciel, Andy Fabian, Joseph Katz, Jorma Louko, 
Donald Lynden-Bell, Reinhard Meinel, Vince Moncrief, Gernot Neugebauer, Martin Rees, Carlo Ungarelli, 
Marsha Weaver, and my Prague colleagues. Peter Williams kindly corrected my worst Czechisms. 
My many thanks  go to  Eva Kotal\'\i kov\'a for her patience 
and skill in technical help with the long manuscript.
Very special thanks to Tom\'a\v s Ledvinka: he prepared all the figures 
and provided long-standing technical help and admirable speed, without which the manuscript 
would certainly not have been finished in the required time and form. 
Support from the Albert Einstein Institute and from the grant 
No. GA\v CR 202/99/0261 of the Czech Republic is gratefully acknowledged.

%\newpage

%INDEX%%%%%%%%%%%%%%%%%%%%%%%%%%%%%%%%%%%
%\clearpage
%\addcontentsline{toc}{section}{Index}
%\flushbottom
%\printindex
%%%%%%%%%%%%%%%%%%%%%%%%%%%%%%%%%%%%%%%


\begin{thebibliography}{288}
\addcontentsline{toc}{section}{References}

%\input bibord.tex
% 1. 
\bibitem{FEM} Feynman, R. (1992) The Character of Physical Law,
Penguin books edition, with Introduction by Paul Davies; the
original edition published in 1965

% 2. 
\bibitem{HHT} Hartle, J. B., Hawking, S. W. (1983) Wave function
of the Universe, Phys. Rev. {\bf D28}, 2960. For more recent
developments, see Page, D. N. (1991) Minisuperspaces with
conformally and minimally coupled scalar fields, J. Math. Phys.
{\bf 32}, 3427, and references therein

% 3. 
\bibitem{KP} Kucha\v{r}, K. V. (1994) private communication based on
unpublished calculations. See also Peleg, Y. (1995) The spectrum
of quantum dust black holes, Phys. Lett. {\bf B356}, 462

% 4. 
\bibitem{CHDR} Chandrasekhar, S. (1987) Ellipsoidal Figures of
Equilibrium, Dover paperback edition, Dover Publ., Mineola, N.
Y.

% 5. 
\bibitem{TAS} Tassoul, J.-L. (1978) Theory of Rotating Stars,
Princeton University Press, Princeton, N. J.

% 6. 
\bibitem{BT} Binney, J., Tremaine, S. (1987) Galactic Dynamics,
Princeton University Press, Princeton.
The idea first appeared in the work of Kuzmin, 
G. G. (1956) Astr. Zh. {\bf 33}, 27

% 7. 
\bibitem{TI} Taniguchi, K. (1999) Irrotational and Incompressible
Binary Systems in the First post-Newtonian Approximation of
General Relativity, Progr. Theor. Phys. {\bf 101}, 283. For an
extensive review, see Taniguchi, K. (1999) Ellipsoidal Figures of
Equilibrium in the First post-Newtonian Approximation of General
Relativity, Thesis, Department of Physics, Kyoto University

% 8. 
\bibitem{ABL} Ablowitz, M. J., Clarkson, P. A. (1991) Solitons,
Nonlinear Evolution Equations and Inverse Scattering, London
Mathematical Society, Lecture Notes in Mathematics {\bf 149},
Cambridge University Press, Cambridge

% 9. 
\bibitem{MAW} Mason, L. J., Woodhouse, N. M. J. (1996)
Integrability, Self-Duality, and Twistor Theory, Clarendon Press,
Oxford

% 10. 
\bibitem{ATI} Atiyah, M. (1998) Roger Penrose -- A Personal
Appreciation, in The Geometric Universe: Science, Geometry, and
the work of Roger Penrose, eds. S. A. Hugget, L. J. Mason, K. P.
Tod, S. T. Tsou and N. M. J. Woodhouse, Oxford University Press,
Oxford

% 11. 
\bibitem{GEB} Bi\v{c}\'ak, J. (1989) Einstein's Prague articles
on gravitation, in Proceedings of the 5th M. Grossmann Meeting on
General Relativity, eds. D. G. Blair and M. J. Buckingham, World
Scientific, Singapore. A more detailed technical account is given
in Bi\v{c}\'ak, J. (1979) Einstein's route to the general theory
of relativity (in Czech), \v{C}s. \v{c}as. fyz. {\bf A29}, 222

% 12. 
\bibitem{EAB} Einstein, A. (1912) Relativity and Gravitation.
Reply to a Comment by M. Abraham (in German), Ann. der Physik
{\bf 38}, 1059

% 13. 
\bibitem{EG} Einstein, A., Grossmann, M. (1913) Outline of a
Generalized Theory of Relativity and of a Theory of Gravitation
(in German), Teubner, Leipzig; reprinted in Zeits. f. Math. und
Physik {\bf 62}, 225

% 14. 
\bibitem{EG1} Einstein, A., Grossmann, M. (1914) Covariance
Properties of the Field Equations of the Theory of Gravitation
Based on the Generalized Theory of Relativity (in German), Zeits.
f. Math. und Physik {\bf 63}, 215

% 15. 
\bibitem{Pai} Pais, A. (1982) `Subtle is the Lord...' -- The
Science and the Life of Albert Einstein, Clarendon Press, Oxford

% 16. 
\bibitem{EFE} Einstein, A. (1915) The Field Equations of
Gravitation (in German), K\"onig. Preuss. Akad. Wiss. (Berlin)
Sitzungsberichte, 844

% 17. 
\bibitem{CRS} Corry, L., Renn, J. and Stachel, J. (1997)
Belated Decision in the Hilbert-Einstein Priority Dispute,
Science {\bf 278}, 1270



% 18. 
\bibitem{MTW} Misner, C., Thorne, K. S. and Wheeler, J. A. (1973)
Gravitation, W. H. Freeman and Co., San Francisco

% 19. 
\bibitem{Wa} Wald, R. M. (1984) General Relativity, The
University of Chicago Press, Chicago

% 20. 
\bibitem{ECO} Einstein, A. (1917) Cosmological Considerations in
the General Theory of Relativity (in German), K\"onig. Preuss.
Akad. Wiss. (Berlin) Sitzungsberichte, 142

% 21. 
\bibitem{PRG} Prosser, V., Folta, J. eds. (1991) Ernst Mach and
the Development of Physics, Charles University -- Karolinum,
Prague

% 22. 
\bibitem{TUB} Barbour, J., Pfister, H. eds. (1995) Mach's
Principle: From Newton's Bucket to Quantum Gravity, Birkh\"auser,
Boston-Basel-Berlin

% 23. 
\bibitem{DJJ} Lynden-Bell, D., Katz, J. and Bi\v{c}\'ak J. (1995)
Mach's principle from the relativistic constraint equations, Mon.
Not. Roy. Astron. Soc. {\bf 272}, 150; Errata: Mon. Not. Astron.
Soc. {\bf 277}, 1600

% 24. 
\bibitem{HORA} Ho\v{r}ava, P. (1999) M theory as a holographic
field theory, Phys. Rev. {\bf D59}, 046004

% 25. 
\bibitem{dS} De Sitter, W. (1917) On Einstein's Theory of
Gravitation, and its Astronomical Consequences, Part 3, Mon. Not.
Roy. Astron. Soc. {\bf 78}, 3; see also references therein

% 26. 
\bibitem{HE} Hawking, S. W., Ellis, G. F. R. (1973) The large
scale structure of space-time, Cambridge University Press,
Cambridge

% 27. 
\bibitem{Pen} Penrose, R. (1968) Structure of Space-Time, in
Batelle Rencontres (1967 Lectures in Mathematics and Physics),
eds. C. M. DeWitt and J. A. Wheeler, W. A. Benjamin, New York


% 28. 
\bibitem{PEE} Peebles, P. J. E. (1993) Principles of Physical
Cosmology, Princeton University Press, Princeton

% 29. 
\bibitem{BLG} Bertotti, B., Balbinot, R., Bergia, S. and Messina,
A. eds. (1990) Modern Cosmology in Retrospect, Cambridge University
Press, Cambridge. See especially the contributions by J.
Barbour, J. D. North, G. F. R. Ellis, and W. C. Seitter and H.
W. Duerbeck

% 30. 
\bibitem{DIN} d'Inverno, R. (1992) Introducing Einstein's
Relativity, Clarendon Press, Oxford

% 31. 
\bibitem{GHO} Geroch, R., Horowitz, G. T. (1979) Global structure
of spacetimes, in General Relativity, An Einstein Centenary
Survey, eds. S. W. Hawking and W. Israel, Cambridge University
Press, Cambridge

% 32. 
\bibitem{Jo} Joshi, P. S. (1993) Global Aspects in Gravitation
and Cosmology, Oxford University Press, Oxford

% 33. 
\bibitem{HJS} Schmidt, H. J. (1993) On the de Sitter space-time --
the geometric foundation of inflationary cosmology, Fortschr. d.
Physik {\bf 41}, 179

% 34. 
\bibitem{GRN} Eriksen, E., Gr{\o}n, O. (1995) The de Sitter universe
models, Int. J. Mod. Phys. {\bf 4}, 115

% 35. 
\bibitem{BOU} Bousso, R. (1998) Proliferation of de Sitter space,
Phys. Rev. {\bf D58}, 083511;
see also
Bousso, R. (1999) Quantum global structure of de Sitter space,
Phys. Rev. {\bf D60}, 063503

% 36. 
\bibitem{MDC} Maldacena, J. (1998) The large $N$ limit of
superconformal field theories and supergravity, Adv. Theor.
Math. Phys. {\bf 2}, 231

% 37. 
\bibitem{BLB} Balasubramanian, V., Kraus, P. and Lawrence, B.
(1999) Bulk versus boundary dynamics in anti-de Sitter spacetime,
Phys. Rev. {\bf D59}, 046003

% 38. 
\bibitem{VEN} Veneziano, G. (1991) Scale factor duality for
classical and quantum string, Phys. Lett. {\bf B265}, 287;
Gasperini, M., Veneziano, G. (1993) Pre-big bang in string
cosmology, Astropart. Phys. {\bf 1}, 317. For the most recent
review, in which also some answers to the critism of the
pre-big-bang scenario and possible observational tests can be
found, see Veneziano, G. (1999) Inflating, warming up, and
probing the pre-bangian universe, hep-th/9902097

% 39. 
\bibitem{CHK} Christodoulou, D., Klainerman, S. (1994) The Global
Nonlinear Stability of the Minkowski Spacetime, Princeton University
Press, Princeton

% 40. 
\bibitem{KVB} Bi\v{c}\'ak, J. (1997) Radiative spacetimes: Exact
approaches, in Relativistic Gravitation and Gravitational
Radiation (Proceedings of the Les Houches School of Physics),
eds. J.-A. Marck and J.-P. Lasota, Cambridge University Press,
Cambridge

% 41. 
\bibitem{HF1} Friedrich, H. (1986) On the existence of
n-geodesically complete or future complete solutions of
Einstein's field equations with smooth asymptotic structure,
Commun. Math. Phys. {\bf 107}, 587

% 42. 
\bibitem{HF2} Friedrich, H. (1995) Einstein equations and
conformal structure: existence of anti-de Sitter-type
space-times, J. Geom. Phys. {\bf 17}, 125

% 43. 
\bibitem{HF3} Friedrich, H. (1998) Einstein's Equation and
Geometric Asymptotics, in Gravitation and Relativity: At the turn
of the Millenium (Proceedings of the GR-15 conference), eds. N.
Dadhich and J. Narlikar, Inter-University Centre for Astronomy and Astrophysics Press, Pune

% 44. 
\bibitem{MOL} M{\o}ller, C. (1972) The theory of Relativity,
Second Edition, Clarendon Press, Oxford

% 45. 
\bibitem{SG} Synge, J. L. (1960) Relativity: The General Theory,
North-Holland, Amsterdam

% 46. 
\bibitem{EPS} Ehlers, J., Pirani, F. A. E. and Schild, A. (1972)
The geometry of free-fall and light propagation, in General
Relativity, Papers in Honor of J. L. Synge, ed. L. O.
O'Raifeartaigh, Oxford University Press, London

% 47. 
\bibitem{MAS} Majer, U., Schmidt, H.-J. eds. (1994) Semantical
Aspects of Spacetime Theories, BI-Wissenschaftsverlag, Mannheim,
Leipzig, Wien

% 48. 
\bibitem{MER} Misner, C. (1969) Gravitational Collapse, in
Brandeis Summer Institute 1968, Astrophysics and General
Relativity, eds. M. S. Chr\'etien, S. Deser and J. Goldstein,
Gordon and Breach, New York

% 49. 
\bibitem{HP} H\'{a}j\'{\i}\v{c}ek, P. (1999) Choice of gauge in
quantum gravity, in Proc. of the 19th Texas symposium on
relativistic astrophysics, Paris 1998, to be published;
gr-qc/9903089


% 50. 
\bibitem{ECH} Ehlers, J. (1981) Christoffel's Work on the
Equivalence Problem for Riemannian Spaces and Its Importance for
Modern Field Theories of Physics, in E. B. Christoffel: The
Influence of His Work on Mathematics and the Physical Sciences,
eds. P. L. Butzer, F. Feh\'er, Birkh\"auser Verlag, Basel


% 51. 
\bibitem{KL} Karlhede, A. (1980) A review of the geometrical
equivalence of metrics in general relativity, Gen. Rel. Grav.
{\bf 12}, 693
 
% 52. 
\bibitem{PRM} Paiva, F. M., Rebou\accent24cas, M. J. and MacCallum, M.
A. H. (1993) On limits of spacetimes -- a coordinate-free
approach, Class. Quantum Grav. {\bf 10}, 1165

% 53. 
\bibitem{EK} Ehlers, J., Kundt, K. (1962) Exact Solutions of the
Gravitational Field Equations, in Gravitation: an introduction to
current research, ed. L. Witten, J. Wiley\&Sons, New York

% 54. 
\bibitem{EHD} Ehlers, J. (1957)
Konstruktionen und Charakterisierungen von L\"osungen der Einsteinschen
Gravitationsfeldgleichungen, Dissertation, Hamburg



% 55. 
\bibitem{EHRO} Ehlers, J. (1962) Transformations of static
exterior solutions of Einstein's gravitational field equations
into different solutions by means of conformal mappings, in Les
Th\'eories Relativistes de la Gravitation, eds. M. A.
Lichnerowicz, M. A. Tonnelat, CNRS, Paris


% 56. 
\bibitem{EHLO} Ehlers, J. (1965) Exact solutions, in
International Conference on Relativistic Theories of Gravitation,
Vol. II, London (mimeographed)

% 57. 
\bibitem{JEK} Jordan, P., Ehlers, J. and Kundt, W. (1960) Strenge
L\"{o}sungen der Feldgleichungen der Allgemeinen
Relativit\"{a}tstheorie, Akad. Wiss. Lit. Mainz, Abh. Math.
Naturwiss. Kl., Nr. 2

% 58. 
\bibitem{JEK2} Jordan, P., Ehlers, J. and Sachs, R. K. (1961)
Beitr\"age zur Theorie der reinen Gravitationsstrahlung, Akad.
Wiss. Lit. Mainz, Abh. Math. Naturwiss. Kl., Nr. 1

% 59. 
\bibitem{CHS} Chandrasekhar, S. (1986) The Aesthetic Base of the
General Theory of Relativity. The Karl Schwarzschild lecture,
reprinted in  Chandrasekhar, S. (1989) Truth and Beauty,
Aesthetics and Motivations in Science, The University of Chicago
Press, Chicago

% 60. 
\bibitem{SCH} Chandrasekhar, S. (1975) Shakespeare, Newton, and
Beethoven or Patterns of Creativity. The Nora and Edward Ryerson
Lecture, reprinted in Chandrasekhar, S. (1989) Truth and Beauty,
Aesthetics and Motivations in Science, The University of Chicago
Press, Chicago

% 61. 
\bibitem{KSH} Kramer, D., Stephani, H., Herlt, E. and MacCallum,
M. A. H. (1980) Exact solutions of Einstein's field equations,
Cambridge University Press, Cambridge

% 62. 
\bibitem{ROPE} Penrose, R. (1999) private communication; see the
paper which will appear in special issue of Class. Quantum
Gravity celebrating the anniversary of the Institute of Physics

% 63. 
\bibitem{AI} Einstein, A. (1950) Physics and Reality, in Out of
My Later Years, Philosophical Library, New York. Originally
published in the Journal of the Franklin Institute {\bf 221}, No.
3; March, 1936

% 64. 
\bibitem{Bo} Bonnor, W. B. (1992) Physical Interpretation of
Vacuum Solutions of Einstein's Equations. Part I.
Time-independent solutions, Gen. Rel. Grav. {\bf 24}, 551

% 65. 
\bibitem{BoGM} Bonnor, W. B., Griffiths, J. B. and MacCallum, M.
A. H. (1994) Physical Interpretation of Vacuum Solutions of
Einstein's Equations. Part II. Time-dependent solutions,
Gen. Rel. Grav. {\bf 26}, 687

% 66. 
\bibitem{e}Bondi, H., van der Burg, M. G. J. and Metzner, A. W. K.
(1962) Gravitational Waves in General Relativity. VII. 
Waves from Axi-symmetric Isolated Systems,
Proc. Roy. Soc. Lond. A {\bf 269}, 21

% 67. 
\bibitem{JE} Ehlers, J. (1973) Survey of General Relativity
Theory, in Relativity, Astrophysics and Cosmology, ed. W. Israel,
D. Reidel, Dordrecht

% 68. 
\bibitem{Kun} K\"{u}nzle, H. P. (1967) Construction of
singularity-free spherically symmetric space-time manifolds,
Proc. Roy. Soc. Lond. {\bf A297}, 244

% 69. 
\bibitem{Schm} Schmidt, B. G. (1967) Isometry groups with
surface-orthogonal trajectories, Zeits. f. Naturfor. {\bf 22a},
1351

% 70. 
\bibitem{IS} Israel, W. (1987) Dark stars: the evolution of an
idea, in 300 years of gravitation, eds. S. W. Hawking and W.
Israel, Cambridge University Press, Cambridge

% 71. 
\bibitem{CW} Ciufolini, I., Wheeler, J. A. (1995) Gravitation and
Inertia, Princeton University Press, Princeton

% 72. 
\bibitem{Will} Will, C. M. (1996) The Confrontation between
General Relativity and Experiment: A 1995 Update, in General
Relativity (Proceedings of the 46th Scottish Universities
Summer School in Physics), eds. G. S. Hall and J. R. Pulham,
Institute of Physics Publ., Bristol

% 73. 
\bibitem{SE} Schneider, P., Ehlers, J. and Falco, E. E. (1992)
Gravitational Lenses, Springer-Verlag, Berlin

% 74. 
\bibitem{Haw} Hawking, S. W. (1973) The Event Horizon, in Black
Holes (Les Houches 1972), eds. C. DeWitt and B. S. DeWitt, Gordon and
Breach, New York-London-Paris

% 75. 
\bibitem{TP} Thorne, K. S., Price, R. H. and MacDonald, D. A.
(1986) Black Holes: The Membrane Paradigm, Yale University Press,
New Haven

% 76. 
\bibitem{FN} Frolov, V., Novikov, I. (1998) Physics of Black
Holes, Kluwer, Dordrecht

% 77. 
\bibitem{Cla} Clarke, C. J. S. (1993) The Analysis of Space-Time
Singularieties, Cambridge University Press, Cambridge

% 78. 
\bibitem{BO} Boyer, R. H. (1969) Geodesic Killing orbits and bifurcate Killing horizons, Proc.
Roy. Soc. (London) {\bf A311}, 245

% 79. 
\bibitem{CA} Carter, B. (1972) Black Hole Equilibrium States,  in
Black Holes (Les Houches 1972), eds. C. De Witt and B. S. De
Witt, Gordon and Breach, New York-London-Paris

% 80. 
\bibitem{PC} Chru\'sciel, P. T. (1996) Uniqueness of stationary,
electro-vacuum black holes revisited, Helv. Phys. Acta {\bf 69}, 529

% 81. 
\bibitem{HEU} Heusler, M. (1996) Black Hole Uniqueness Theorems, Cambridge University
Press, Cambridge

% 82. 
\bibitem{RW} Wald, R. M. (1994) Quantum Field Theory in Curved Spacetime and Black
Hole Thermodynamics, The University of Chicago Press, Chicago

% 83. 
\bibitem{RAW} R\'acz, I., Wald R. M. (1996) Global extensions of
spacetimes describing asymptotic final states of black holes, Class.
Quantum Grav. {\bf 13}, 539

% 84. 
\bibitem{RPT} Penrose, R. (1980) On Schwarzschild Causality  -- A Problem for
``Lorentz Covariant'' General Relativity, in Essays in General Relativity, eds. F. J. Tipler,
Academic Press, New York

% 85. 
\bibitem{WB} Weinberg, S., Gravitation and Cosmology (1972) J.
Wiley, New York (see in particular Ch. 6, part 9)

% 86. 
\bibitem{Ro} Zel'dovich, Ya. B., Grishchuk, L. P. (1988) The
general theory of relativity is correct!, Sov. Phys. Usp. {\bf
31}, 666.
This very pedagogical paper contains a number of references on
the field-theoretical approach to gravity

% 87. 
\bibitem{JRG} Ehlers, J. (1998) General Relativity as Tool for
Astrophysics, in Relativistic Astrophysics, eds. H. Riffert et al.,
Vieweg, Braunschweig/Wiesbaden

% 88. 
\bibitem{Re} Rees, M. (1998) Astrophysical Evidence for Black
Holes, in Black Holes and Relativistic Stars, ed. R. M. Wald, The
University of Chicago Press, Chicago

% 89. 
\bibitem{Na} Menou, K., Quataert, E. and Narayan, R. (1998)
Astrophysical Evidence for Black Hole Event Horizons, in
Gravitation and Relativity: At the turn of the Millennium
(Proceedings of the GR-15 Conference), eds. N. Dadhich and J.
Narlikar, Inter-University Centre for Astronomy and Astrophysics
Press, Pune; also astro-ph/9712015

% 90. 
\bibitem{Ca} Carr, B. J. (1996) Black Holes in Cosmology and
Astrophysics, in General Relativity (Proceedings of the 46th
Scottish Universities Summer School in Physics), eds. G. S. Hall
and J. R. Pulham, Institute of Physics Publishing, London

% 91. 
\bibitem{Chan} Chandrasekhar, S. (1984) The Mathematical Theory
of Black Holes, Clarendon Press, Oxford

% 92. 
\bibitem{Abr} Abramowicz, M. A. (1993) Inertial forces in general
relativity, in The Renaissance of General Relativity and
Cosmology, eds. G. Ellis, A. Lanza and J. Miller, Cambridge
University Press, Cambridge

% 93. 
\bibitem{Se} Semer\'{a}k, O. (1998) Rotospheres in Stationary
Axisymmetric Spacetimes, Ann. Phys. (N.Y.) {\bf 263}, 133; see
also 69 references quoted therein

% 94. 
\bibitem{Fey} Feynman, R. P., Morinigo, F. B., Wagner W. G.
(1995) Feynman lectures on gravitation, Addison-Wesley Publ. Co.,
Reading, Mass.

% 95. 
\bibitem{TeSh} Shapiro, S. L., Teukolsky, S. A. (1983) Black
Holes, White Dwarfs, and Neutron Stars, J. Wiley, New York

% 96. 
\bibitem{Fr} Frank, J., King, A. and Raine, D. (1992) Accretion
Power in Astrophysics, 2nd edition, Cambridge University Press,
Cambridge

% 97. 
\bibitem{KST} Thorne, K. S. (1998) Probing Black Holes and
Relativistic Stars with Gravitational Waves, in Black Holes and
Relativistic Stars, ed. R. M. Wald, The University of Chicago
Press, Chicago. See also lectures by E. Seidel, J. Pullin, and E.
Flanagan, in Gravitation and Relativity: At the turn of the Millennium
(Proceedings of the GR-15 Conference), eds. N. Dadhich and J.
Narlikar, Inter-University Centre for Astronomy and Astrophysics
Press, Pune

% 98. 
\bibitem{Pull} Pullin, J. (1998) Colliding Black Holes: Analytic
Insights, in Gravitation and Relativity: At the turn of the Millennium
(Proceedings of the GR-15 Conference), eds. N. Dadhich and J.
Narlikar, Inter-University Centre for Astronomy and Astrophysics
Press, Pune

% 99. 
\bibitem{GB} Graves, J. C., Brill, D. R. (1960) Oscillatory
character of Reissner-Nordstr\"{o}m metric for an ideal charged
wormhole, Phys. Rev. {\bf 120}, 1507

% 100. 
\bibitem{Bou} Boulware, D. G. (1973) Naked Singularities, Thin
Shells, and the Reissner-Nordstr\"om Metric, Phys. Rev. {\bf D8},
2363

% 101. 
\bibitem{ZN} Zel'dovich, Ya. B., Novikov, I. D. (1971)
Relativistic Astrophysics, Volume 1: Stars and Relativity, The
University of Chicago Press, Chicago

% 102. 
\bibitem{ROG} Penrose, R. (1979) Singularities and
time-asymmetry, in General Relativity, An Einstein Centenary
Survey, eds. S. W. Hawking and W. Israel, Cambridge University
Press, Cambridge

% 103. 
\bibitem{Ori} Burko, L., Ori, A. (1997) Introduction to the
internal structure of black holes, in Internal Structure of Black
Holes and Spacetime Singularities, eds. L. Burko and A. Ori,
Inst. Phys. Publ., Bristol, and The Israel Physical Society,
Jerusalem

% 104. 
\bibitem{BiDv} Bi\v{c}\'ak, J., Dvo\v{r}\'ak, L. (1980)
Stationary electromagnetic fields around black holes III. General
solutions and the fields of current loops near the
Reissner-Nordstr\"om black hole, Phys. Rev. {\bf D22}, 2933

% 105. 
\bibitem{Mo} Moncrief, V. (1975) Gauge-invariant perturbations of
Reissner-Nordstr\"om black holes, Phys. Rev. {\bf D12}, 1526; see
also references therein

% 106. 
\bibitem{JiBi}  Bi\v{c}\'ak, J. (1979) On the theories of the
interacting perturbations of the Reissner-Nordstr\"om black hole,
Czechosl. J. Phys. {\bf B29}, 945

% 107. 
\bibitem{Bca}  Bi\v{c}\'ak, J. (1972) Gravitational collapse with
charge and small asymmetries, I: Scalar perturbations, Gen. Rel.
Grav. {\bf 3}, 331

% 108. 
\bibitem{Pri} Price, R. H. (1972) Nonspherical perturbations of
relativistic gravitational collapse, I: Scalar and gravitational
perturbations, Phys. Rev. {\bf D5}, 2419

% 109. 
\bibitem{Pri1} Price, R. H. (1972)  Nonspherical perturbations of
relativistic gravitational collapse, II: Integer-spin,
zero-rest-mass fields, Phys. Rev. {\bf D5}, 2439

% 110. 
\bibitem{Bca1}  Bi\v{c}\'ak, J. (1980) Gravitational collapse
with charge and small asymmetries, II: Interacting
electromagnetic and gravitational perturbations, Gen. Rel. Grav.
{\bf 12}, 195

% 111. 
\bibitem{POI} Poisson, E., Israel, W. (1990) Internal structure
of black holes, Phys. Rev. {\bf D41}, 1796

% 112. 
\bibitem{BOVA} Bonnor, W. B., Vaidya, P. C. (1970) Spherically
Symmetric Radiation of Charge in Einstein-Maxwell Theory, Gen.
Rel. Grav. {\bf 1}, 127

% 113. 
\bibitem{Chamb} Chambers, C. M. (1997) The Cauchy horizon in
black hole-de Sitter spacetimes, in Internal Structure of Black
Holes and Spacetime Singularities, eds. L. Burko and A. Ori,
Inst. Phys. Publ. Bristol, and The Israel Physical Society,
Jerusalem

% 114. 
\bibitem{PCH} Penrose, R. (1998) The Question of Cosmic Censorship, in
Black Holes and Relativistic Stars, ed. R. M. Wald, The
University of Chicago Press, Chicago

% 115. 
\bibitem{BMM} Brady, P. R., Moss, I. G. and Myers, R. C. (1998)
Cosmic Censorship: As Strong As Ever, Phys. Rev. Lett. {\bf 80},
3432

% 116. 
\bibitem{VH} Huben\'y, V. E. (1999) Overcharging a Black Hole and
Cosmic Censorship, Phys. Rev. {\bf D59}, 064013

% 117. 
\bibitem{BIK} Bi\v{c}\'ak, J. (1977) Stationary interacting
fields around an extreme Reissner-Nordstr\"om black hole, Phys.
Lett. {\bf 64A}, 279. See also the review Bi\v{c}\'ak, J. (1982),
Perturbations of the Reissner-Nordstr\"om black hole, in the
Proceedings of the Second Marcel Grossmann Meeting on General
Relativity, ed. R. Ruffini, North-Holland, Amsterdam, and
references therein

% 118. 
\bibitem{Gib} H\'aj\'{\i}\v{c}ek, P. (1981) Quantum wormholes
(I.) Choice of the classical solution, Nucl. Phys. {\bf B185},
254

% 119. 
\bibitem{PACH} Aichelburg, P. C., G\"uven, R. (1983) Remarks on
the linearized superhair, Phys. Rev. {\bf D27}, 456; and references
therein

% 120. 
\bibitem{SSE} Schwarz, J. H., Seiberg, N. (1999) String theory,
supersymmetry, unification, and all that, Rev. Mod. Phys. {\bf
71}, S112

% 121. 
\bibitem{Carl} Carlip, S. (1995) The (2+1)-dimensional black
hole, Class. Quantum Grav. {\bf 12}, 2853

% 122. 
\bibitem{MPe} Myers, R. C., Perry, M. J. (1986) Black holes in
higher dimensional space-times, Ann. Phys. (N.Y.) {\bf 172}, 304

% 123. 
\bibitem{GiHo} Gibbons, G. W., Horowitz, G. T. and Townsend, P.
K. (1995) Higher-dimensional resolution of dilatonic black-hole
singularities, Class. Quantum Grav. {\bf 12}, 297

% 124. 
\bibitem{HoTe} Horowitz, G. T., Teukolsky, S. A. (1999) Black
holes, Rev. Mod. Phys. {\bf 71}, S180

% 125. 
\bibitem{WD} Wald, R. M. (1998) Black Holes and Thermodynamics,
in Black Holes and Relativistic Stars, ed. R. M. Wald, The
University of Chicago Press, Chicago

% 126. 
\bibitem{HORO} Horowitz, G. T. (1998) Quantum States of Black
Holes, in Black Holes and Relativistic Stars, ed. R. M. Wald, the
University of Chicago Press, Chicago

% 127. 
\bibitem{SKE} Skenderis, K. (1999) Black holes and branes in
string theory, hep-th/9901050

% 128. 
\bibitem{AA} Ashtekhar, A., Baez, J., Corichi, A. and Krasnov, K.
(1998) Quantum Geometry and Black Hole Entropy, Phys. Rev. Lett.
{\bf 80}, 904

% 129. 
\bibitem{YO} Youm, D. (1999) 
Black holes and solitons in string theory, 
Physics Reports {\bf 316}, Nos. 1-3, 1

% 130. 
\bibitem{CEG} Chamblin, A., Emparan, R. and Gibbons, G. W. (1998)
Superconducting p-branes and extremal black holes, Phys. Rev.
{\bf D58}, 084009

% 131. 
\bibitem{ERN} Ernst, F. J. (1976) Removal of the nodal
singularity of the C-metric, J. Math. Phys. {\bf 17}, 54; see
also Ernst, F. J., Wild, W. J. (1976) Kerr black holes in a
magnetic universe, J. Math. Phys. {\bf 17}, 182

% 132. 
\bibitem{KVO} Karas, V., Vokrouhlick\'y, D. (1991) 
On interpretation of the magnetized Kerr-Newman black hole,
J. Math. Phys. {\bf 32}, 714

% 133. 
\bibitem{KER} Kerr, R. P. (1963) Gravitational field of a
spinning mass as an example of algebraically special metrics,
Phys. Rev. Lett. {\bf 11}, 237

% 134. 
\bibitem{STW} Stewart, J., Walker, M. (1973) Black holes: the
outside story, in Springer tracts in modern physics, Vol. {\bf
69}, Springer-Verlag, Berlin

% 135. 
\bibitem{TRN} Thorne, K. S. (1980) Multipole expansions of
gravitational radiation, Rev. Mod. Phys. {\bf 52}, 299

% 136. 
\bibitem{HN} Hansen, R. O. (1974) Multipole moments of stationary
space-times, J. Math. Phys. {\bf 15}, 46

% 137. 
\bibitem{BeS} Beig, R., Simon, W. (1981) On the multipole
expansion for stationary space-times, Proc. Roy. Soc. Lond. {\bf
A376}, 333

% 138. 
\bibitem{FEC} de Felice, F., Clarke, C. J. S. (1990) Relativity
on curved manifolds, Cambridge University Press, Cambridge

% 139. 
\bibitem{LL} Landau, L. D., Lifshitz, E. M. (1962) The Classical
Theory of Fields, Pergamon Press, Oxford


% 140. 
\bibitem{ONE} O'Neill, B. (1994) The Geometry of Kerr Black
Holes, A. K. Peters, Wellesley

% 141. 
\bibitem{KLB} Katz, J., Lynden-Bell, D. and Bi\v{c}\'ak, J.
(1998) Instantaneous inertial frames but retarded
electromagnetism in rotating relativistic collapse, Class.
Quantum Grav. {\bf 15}, 3177

% 142. 
\bibitem{OLS} Semer\'ak, O. (1996) Photon escape cones in the
Kerr field, Helv. Phys. Acta {\bf 69}, 69

% 143. 
\bibitem{BST} Bi\v{c}\'ak, J., Stuchl\'{\i}k, Z. (1976) The fall
of the shell of dust onto a rotating black hole, Mon. Not. Roy.
Astron. Soc. {\bf 175}, 381

% 144. 
\bibitem{BISH} Bi\v{c}\'ak, J., Semer\'ak, O. and Hadrava, P.
(1993) Collimation effects of the Kerr field, Mon. Not. Roy.
Astron. Soc. {\bf 263}, 545

% 145. 
\bibitem{KN} Newman, E. T., Couch, E., Chinnapared, K., Exton, A.,
Prakash, A. and Torrence, R. (1965) Metric of a rotating charged
mass, J. Math. Phys. {\bf 6}, 918

% 146. 
\bibitem{GAT} Garfinkle, D., Traschen, J. (1990) Gyromagnetic
ratio of a black hole, Phys. Rev. {\bf D42}, 419

% 147. 
\bibitem{BAR} Bardeen, J. M. (1973) Timelike and Null Geodesics
in the Kerr Metric, in Black Holes, eds. C. DeWitt and B. S.
DeWitt, Gordon and Breach, New York

% 148. 
\bibitem{Rin} Rindler, W. (1997) The case against space dragging,
Phys. Lett. {\bf A233}, 25

% 149. 
\bibitem{JT} Jantzen, R. T., Carini, P. and Bini, D. (1992) The
Many Faces of Gravitoelectromagnetism, Ann. Phys. (N.Y.) {\bf 215},
1; see also the review (1999) 
The Inertial Forces / Test Particle Motion Game, 
in the Proceedings of the 8th M.
Grossmann Meeting on General Relativity, ed. T. Piran, World
Scientific, Singapore

% 150. 
\bibitem{KaVo} Karas, V., Vokrouhlick\'{y}, D. (1994)
Relativistic precession of the orbit of a star near a
supermassive rotating black hole, Astrophys.
J. {\bf 422}, 208

% 151. 
\bibitem{Bl} Blandford, R. D., Znajek, R. L. (1977)
Electromagnetic extraction of energy from Kerr black holes, Mon.
Not. Roy. Astron. Soc. {\bf 179}, 433. See also Blandford, R.
(1987) Astrophysical black holes, in 300 years of gravitation,
eds. S. W. Hawking and W. Israel, Cambridge University Press,
Cambridge

% 152. 
\bibitem{BiJ}Bi\v{c}\'ak, J., Jani\v{s}, V. (1985) Magnetic
fluxes across black holes, Mon. Not. Roy. Astron. Soc. {\bf 212},
899

% 153. 
\bibitem{Puns} Punsly, B., Coroniti, F. V. (1990) Relativistic
winds from pulsar and black hole magnetospheres, Astrophys. J.
{\bf 350}, 518. See also Punsly, B. (1998) High-energy gamma-ray
emission from galactic Kerr-Newman black holes. The central
engine, Astrophys. J. {\bf 498}, 640, and references therein

% 154. 
\bibitem{MAB} Abramowicz, M. (1998) private communication

% 155. 
\bibitem{Mir} Mirabel, I. F., Rodr\'{\i}guez, L. F. (1998)
Microquasars in our Galaxy, Nature {\bf 392}, 673

% 156. 
\bibitem{FuMa} Futterman, J. A. H., Handler, F. A. and Matzner,
R. A. (1988) Scattering from black holes, Cambridge University
Press, Cambridge

% 157. 
\bibitem{BDV} Bi\v{c}\'ak, J., Dvo\v{r}\'ak, L. (1976) Stationary
electromagnetic fields around black holes II. General solutions
and the fields of some special sources near a Kerr black hole,
Gen. Rel. Grav. {\bf 7}, 959

% 158. 
\bibitem{SNA} Sasaki, M., Nakamura, T. (1990) Gravitational
Radiation from an Extreme Kerr Black Hole, Gen. Rel. Grav. {\bf
22}, 1551; and references therein

% 159. 
\bibitem{KrP} Krivan, W., Price, R. H. (1999) Formation of a
rotating Black Hole from a Close-Limit Head-On Collision, Phys.
Rev. Lett. {\bf 82}, 1358

% 160. 
\bibitem{CLO} Campanelli, M., Lousto, C. O. (1999) Second order
gauge invariant gravitational perturbations of a Kerr black hole,
Phys. Rev. {\bf D59}, 124022




% 161. 
\bibitem{Fa}Fabian, A. C. (1999) Emission lines: signatures of
relativistic rotation, in Theory of Accretion Disks, eds. M.
Abramowicz, G. Bj\"{o}rnson, J. Pringle, Cambridge University
Press, Cambridge

% 162. 
\bibitem{Ips} Ipser, J. R. (1998) Low-Frequency Oscillations of
Relativistic Accretion Disks, in Relativistic Astrophysics, eds.
H. Riffert et al., Vieweg, Braunschweig, Wiesbaden

% 163. 
\bibitem{BiPo} Bi\v{c}\'{a}k, J., Podolsk\'{y}, J. (1997) The
global structure of Robinson-Trautman radiative space-times with
cosmological constant, Phys. Rev. {\bf D55}, 1985


% 164. 
\bibitem{HH} Hartle, J. B., Hawking, S. W. (1972) Solutions of
the Einstein-Maxwell equations with many black holes, Commun.
Math. Phys. {\bf 26}, 87

% 165. 
\bibitem{HS} Heusler, M. (1997) 
On the Uniqueness of the Papapetrou-Majumdar metric,
Class. Quantum Grav. {\bf 14}, L129

% 166. 
\bibitem{CI} Chru\'sciel, P. T. (1999)
Towards the classification of static electro-vacuum
space-times containing an asymptotically flat
spacelike hypersurface with compact interior,
Class. Quantum Grav. {\bf 16}, 689.
See also Chru\'sciel's very general result 
for the vacuum case in the preceding paper:
The classification of static vacuum space-times
containing an asymptotically flat spacelike
hypersurface with compact interior,
Class. Quantum Grav. {\bf 16}, 661

% 167. 
\bibitem{KrNe} Kramer, D., Neugebauer, G. (1984) 
B\"{a}cklund Transformations in General Relativity, in
Solutions of Einstein's Equations: Techniques and Results, eds. C.
Hoenselaers and W. Dietz, Lecture Notes in Physics 205, Springer-Verlag,
Berlin

% 168. 
\bibitem{BiHo} Bi\v{c}\'ak, J., Hoenselaers, C. (1985) Two equal
Kerr-Newman sources in stationary equilibrium, Phys. Rev. {\bf D31}, 2476

% 169. 
\bibitem{Wein} Weinstein, G. (1996) N-black hole stationary and axially symmetric solutions of the Einstein/Maxwell equations,
Comm. Part. Diff. Eqs. {\bf 21}, 1389

% 170. 
\bibitem{DiHo} Dietz, W., Hoenselaers, C. (1982)
Stationary System of Two Masses Kept Apart
by Their Gravitational Spin-Spin Interaction,
Phys. Rev. Lett. {\bf 48}, 778; see also Dietz,~W.
(1984) HKX-Transformations: Some Results, in Solutions of 
Einstein's Equations: Techniques and Results, eds.
C. Hoenselaers and W. Dietz, Lecture Notes in Physics 205,
Springer-Verlag, Berlin

% 171. 
\bibitem{KaTr} Kastor, D., Traschen, J. (1993) Cosmological
multi-black-hole solutions, Phys. Rev. {\bf D47}, 5370

% 172. 
\bibitem{BHK} Brill, D. R., Horowitz, G. T., Kastor, D. and
Traschen, J. (1994) Testing cosmic censorship with black hole
collisions, Phys. Rev. {\bf D49}, 840

% 173. 
\bibitem{WLC} Welch, D. L. (1995) Smoothness of the horizons of multi-black-hole
solutions, Phys. Rev. {\bf D52}, 985

% 174. 
\bibitem{BHA} Brill, D. R., Hayward, S. A. (1994) Global structure of a black hole cosmos and its extremes, Class. Quantum Grav. {\bf 11}, 359
% 175. 
\bibitem{INSH} Ida, D., Nakao, K., Siino, M. and Hayward, S. A.
(1998) Hoop conjecture for colliding black holes, Phys. Rev. {\bf
D58}, 121501

% 176. 
\bibitem{SS} Scott, S. M., Szekeres, P. (1986) The Curzon
singularity I: spatial section, Gen. Rel. Grav. {\bf 18}, 557;
The Curzon singularity II: global picture, Gen. Rel. Grav. {\bf
18}, 571

% 177. 
\bibitem{BLK} Bi\v c\'ak, J., Lynden-Bell, D. and Katz, J.
(1993) Relativistic disks as sources of static vacuum spacetimes,
Phys. Rev. {\bf D47}, 4334

% 178. 
\bibitem{BLP} Bi\v c\'ak, J., Lynden-Bell, D. and Pichon, C.
(1993) Relativistic discs and flat galaxy models,
Mon. Not. Roy. Astron. Soc. {\bf 265}, 26

% 179. 
\bibitem{EZ} Evans, N. W., de Zeeuw, P. T. (1992)
Potential-density pairs for flat galaxies,
Mon. Not. Roy. Astron. Soc. {\bf 257}, 152

% 180. 
\bibitem{Chr} Chru\'{s}ciel, P., MacCallum, M. A. H. and
Singleton, P. B. (1995) Gravitational waves in general
relativity XIV. Bondi expansions and the `polyhomogeneity' of
${\cal {J}}$, Phil. Trans. Roy. Soc. Lond. {\bf A350}, 113

% 181. 
\bibitem{SZZ} Semer\'ak, O., Zellerin, T. and \v{Z}\'a\v{c}ek, M.
(1999) The structure of superposed Weyl fields, Mon. Not. Roy.
Astron. Soc., {\bf 308}, 691 and 705

% 182. 
\bibitem{LEL} Lemos, J. P. S., Letelier, P. S. (1994) 
Exact general relativistic thin disks around black holes, 
Phys. Rev. {\bf D49}, 5135

% 183. 
\bibitem{GOL} Gonz\'alez, G. A., Letelier, P. S. (1999) 
Relativistic Static Thin Disks with Radial Stress Support,
Class. Quantum Grav. {\bf 16}, 479

% 184. 
\bibitem{PLE} Letelier, P. S. (1999) 
Exact General Relativistic Disks with Magnetic Fields, 
gr-qc/9907050

% 185. 
\bibitem{Kr}  Krasi\' nski, A. (1978) Sources of the Kerr metric,
Ann. Phys. (N.Y.) {\bf 112}, 22

% 186. 
\bibitem{Mc}  McManus, D. (1991) A toroidal source for the Kerr
black hole geometry, Class. Quantum Grav. {\bf 8}, 863


% 187. 
\bibitem{BW} Bardeen, J. M., Wagoner, R. V. (1971) Relativistic
disks. I. Uniform rotation, Astrophys. J. {\bf 167}, 359


% 188. 
\bibitem{BiLe} Bi\v c\' ak, J., Ledvinka, T. (1993)
Relativistic Disks as Sources of the Kerr Metric, Phys. Rev.
Lett. {\bf 71}, 1669. See also (1993) Sources for stationary
axisymmetric gravitational fields, Max-Planck-Institute 
for Astrophysics, Green report MPA 726, Munich


% 189. 
\bibitem{PL} Pichon, C., Lynden-Bell, D. (1996) New sources for
Kerr and other metrics: rotating relativistic discs with pressure
support, Mon. Not. Roy. Astron. Soc. {\bf 280}, 1007


% 190. 
\bibitem{Bar} Barrab\` es, C., Israel, W. (1991) Thin shells in
general relativity and cosmology: the lightlike limit, Phys. Rev.
{\bf D43}, 1129

% 191. 
\bibitem{LeB} Ledvinka, T., Bi\v c\' ak, J. and \v{Z}ofka, M.
(1999) Relativistic disks as sources of Kerr-Newman fields, in
Proc. 8th M. Grossmann Meeting on General Relativity, ed. T.
Piran, World Sci., Singapore


% 192. 
\bibitem{GN1} Neugebauer, G., Meinel, R. (1995) General
Relativistic Gravitational Fields of a Rigidly Rotating Disk of Dust:
Solution in Terms of Ultraelliptic Functions, Phys. Rev. Lett.
{\bf 75}, 3046

% 193. 
\bibitem{GN2} Neugebauer, G., Kleinw\"achter, A. and Meinel, R.
(1996) Relativistically rotating dust, Helv. Phys. Acta {\bf 69},
472

% 194. 
\bibitem{Me} Meinel, R. (1998) The rigidly rotating disk of dust
and its black hole limit, in Proc. of the Second Mexican School
on Gravitation and Mathematical Physics, eds. A. Garcia et al.,
Science Network Publishing, Konstanz, gr-qc/9703077

% 195. 
\bibitem{COR} Breitenlohner, P., Forg\'acs, P. and Maison, D. (1995)
Gravitating Monopole Solutions II, Nucl. Phys. {\bf 442B}, 126

% 196. 
\bibitem{Mi} Misner, Ch. (1967) Taub-NUT Space as a
Counterexample to Almost Anything, in Relativity Theory and
Astrophysics 1, Lectures in Applied Mathematics, Vol. 8, ed. J.
Ehlers, American Math. Society, Providence, R. I.

% 197. 
\bibitem{Tau} Taub, A. H. (1951) Empty space-times admitting a
three parameter group of motions, Ann. Math. {\bf 53}, 472

% 198. 
\bibitem{NTU} Newman, E., Tamburino, L. and Unti, T. (1963)
Empty-space generalization of the Schwarzschild metric, J. Math.
Phys. {\bf 4}, 915

% 199. 
\bibitem{LBNZ} Lynden-Bell, D., Nouri-Zonoz, M. (1998) Classical monopoles: Newton, NUT
space, gravomagnetic lensing, and atomic spectra, Rev. Mod. Phys.
{\bf 70}, 427

% 200. 
\bibitem{Ger} Geroch, R. (1971) A method for generating solutions of Einstein's equations,
J. Math. Phys. {\bf 12}, 918 and J. Math. Phys. {\bf 13}, 394

% 201. 
\bibitem{ETN} Ehlers, J. (1997) Examples of Newtonian limits of
relativistic spacetimes, Class. Quantum Grav. {\bf 14}, A119

% 202. 
\bibitem{JAW} Wheeler, J. A. (1980) The Beam and Stay of the Taub
Universe, in Essays in General Relativity, eds. F. J. Tipler,
Academic Press, New York


% 203. 
\bibitem{Haj} H\'{a}j\'{\i}\v{c}ek, P. (1971) Extension of the
Taub and NUT spaces and extensions of their tangent bundles,
Commun. Math. Phys. {\bf 17}, 109; Bifurcate spacetimes, J.
Math. Phys. {\bf 12}, 157; Causality in non-Hausdorff
spacetimes, Commun. Math. Phys. {\bf 21}, 75

% 204. 
\bibitem{KT} Thorne, K. S. (1993) Misner Space as a Prototype for
Almost Any Pathology, in Directions in General Relativity, Vol.
1, eds. B. L. Hu, M. P. Ryan and C. V. Vishveshwara, Cambridge
University Press, Cambridge

% 205. 
\bibitem{GiMa} Gibbons, G. W., Manton, N. S. (1986) Classical and
Quantum Dynamics of BPS monopoles, Nuclear Physics {\bf B274}, 183

% 206. 
\bibitem{KrB} Kraan T. C., Baal P. (1998) Exact T-duality between calorons
and Taub -- NUT spaces, INLO-PUB-4/98, hep-th/9802049

% 207. 
\bibitem{BiPy} Bi\v{c}\'ak, J., Podolsk\'y, J. (1999)
Gravitational waves in vacuum spacetimes with cosmological
constant. I. Classification and geometrical properties of
non-twisting type $N$ solutions. II. Deviation of geodesics and
interpretation of non-twisting type $N$ solutions, J. Math. Phys.
{\bf 44}, 4495 and 4506

% 208. 
\bibitem{AiB} Aichelburg, P. C., Balasin, H. (1996) Symmetries of
pp-waves with distributional profile, Class. Quantum Grav. {\bf
13}, 723

% 209. 
\bibitem{AiB2}  Aichelburg, P. C., Balasin, H. (1997) Generalized
symmetries of impulsive gravitational waves, Class. Quantum
Grav. {\bf 14}, A31

% 210. 
\bibitem{AS} Aichelburg, P. C., Sexl, R. U. (1971) On the
gravitational field of a massless particle, Gen. Rel. Grav. {\bf
2}, 303

% 211. 
\bibitem{Prs} Penrose, R. (1972) The geometry of impulsive
gravitational waves, in General Relativity, Papers in Honour of J.
L. Synge, ed. L. O'Raifeartaigh, Clarendon Press, Oxford

% 212. 
\bibitem{Gr} Griffiths, J. B. (1991) Colliding Plane Waves in
General Relativity, Clarendon Press, Oxford

% 213. 
\bibitem{ae} Bondi, H., Pirani, F. A. E. and Robinson, I. (1959)
Gravitational waves in general relativity. III. Exact plane
waves, Proc. Roy.
Soc. Lond. A {\bf 251}, 519

% 214. 
\bibitem{Ri} Rindler, W. (1977) Essential Relativity (2nd
edition), Springer, New York-Berlin


% 215. 
\bibitem{RPE} Penrose, R. (1965) A remarkable property of plane
waves in general relativity, Rev. Mod. Phys. {\bf 37}, 215

% 216. 
\bibitem{LoSa} Lousto, C. O., S\'anchez, N. (1989) The
ultrarelativistic limit of the Kerr-Newman geometry and particle
scattering at the Planck scale, Phys. Lett. {\bf B232}, 462

% 217. 
\bibitem{FePe} Ferrari, V., Pendenza, P. (1990) Boosting the Kerr
Metric, Gen. Rel. Grav. {\bf 22}, 1105

% 218. 
\bibitem{BaNa} Balasin, H., Nachbagauer, H. (1995) The
ultrarelativistic Kerr-geometry and its energy-momentum tensor,
Class. Quantum Grav. {\bf 12}, 707

% 219. 
\bibitem{PoGr1} Podolsk\'y, J., Griffiths, J. B. (1998) Boosted
static multipole particles as sources of impulsive gravitational
waves, Phys. Rev. {\bf D58}, 124024

% 220. 
\bibitem{HoTa} Hotta, M., Tanaka, M. (1993) Shock-wave geometry
with non-vanishing cosmological constant, Class. Quantum
Grav. {\bf 10}, 307

% 221. 
\bibitem{PoGr2} Podolsk\'y, J., Griffiths, J. B. (1997) Impulsive
gravitational waves generated by null particles in de Sitter and
anti-de Sitter backgrounds, Phys. Rev. {\bf D56}, 4756

% 222. 
\bibitem{PE} D'Eath, P. D. (1996) Black Holes: Gravitational
Interactions, Clarendon Press, Oxford

% 223. 
\bibitem{Hoo} 't Hooft, G. (1987) Graviton dominance in
ultra-high-energy scattering, Phys. Lett. {\bf B198}, 61

% 224. 
\bibitem{Fab} Fabbrichesi, M., Pettorino, R., Veneziano, G. and
Vilkovisky, G. A. (1994) Planckian energy scattering and surface
terms in the gravitational action, Nucl. Phys. {\bf B419}, 147

% 225. 
\bibitem{Kus1} Kunzinger, M., Steinbauer, R. (1999) A note on the
Penrose junction conditions, Class. Quantum Grav. {\bf 16}, 1255

% 226. 
\bibitem{KuS} Kunzinger, M., Steinbauer, R. (1999) A rigorous
solution concept for geodesic and geodesic deviation equations in
impulsive gravitational waves, J. Math. Phys. {\bf 40}, 1479

% 227. 
\bibitem{PoVe} Podolsk\'y, J., Vesel\'y, K. (1998) Chaotic motion
in pp-wave spacetimes, Class. Quantum Grav. {\bf 15}, 3505

% 228. 
\bibitem{Per} Levin, O., Peres, A. (1994) Quantum field theory
with null-fronted metrics, Phys. Rev. {\bf D50}, 7421

% 229. 
\bibitem{AKl} Klim\v{c}\'{\i}k, C. (1991) Gravitational waves as
string vacua I, II, Czechosl. J. Phys. {\bf 41}, 697 (see also
references therein)

% 230. 
\bibitem{Gi} Gibbons, G. W. (1999) Two loop and all loop finite
4-metrics, Class. Quantum Grav. {\bf 16}, L 71

% 231. 
\bibitem{BiPr} Bi\v{c}\'{a}k, J., Pravda, V. (1998) Curvature
invariants in type $N$ spacetimes, Class. Quantum Grav. {\bf 15},
1539

% 232. 
\bibitem{PRD} Pravda, V. (1999) 
Curvature invariants in type-{\it III} spacetimes, 
Class. Quantum Grav. {\bf 16}, 3321

% 233. 
\bibitem{KhPe} Khan, K. A., Penrose, R. (1971) Scattering of two
impulsive gravitational plane waves, Nature {\bf 229}, 185

% 234. 
\bibitem{Se1}Szekeres, P. (1970) Colliding gravitational waves,
Nature {\bf 228}, 1183

% 235. 
\bibitem{Se2} Szekeres, P. (1972) Colliding plane gravitational
waves, J. Math. Phys. {\bf 13}, 286

% 236. 
\bibitem{Yu1} Yurtsever, U. (1988) Structure of the singularities
produced by colliding plane waves, Phys. Rev. {\bf D38}, 1706

% 237. 
\bibitem{HET} Hauser, I., Ernst, F. J. (1989)
Initial value problem for colliding gravitational waves -- I/II,
J. Math. Phys. {\bf 30}, 872 and 2322;
(1990) and (1991) Initial value problem for colliding
gravitational waves -- III/IV, J. Math. Phys. 
{\bf 31}, 871 and {\bf 32}, 198;

% 238. 
\bibitem{HET1} Hauser, I., Ernst, F. J. (1999)
Group structure of the solution manifold of the hyperbolic 
Ernst equation -- general study of the
subject and detailed elaboration of mathematical proofs,
216 pages, gr-qc/9903104

% 239. 
\bibitem{NH} Nutku, Y., Halil, M. (1977) Colliding impulsive
gravitational waves, Phys. Rev. Lett. {\bf 39}, 1379

% 240. 
\bibitem{MT} Matzner, R., Tipler, F. J. (1984) Methaphysics of
colliding self-gravitating plane waves, Phys. Rev. {\bf D29}, 1575

% 241. 
\bibitem{ChF} Chandrasekhar, S., Ferrari, V. (1984) On the
Nutku-Halil solution for colliding impulsive gravitational waves,
Proc. Roy. Soc. Lond. {\bf A396}, 55

% 242. 
\bibitem{ChX2} Chandrasekhar, S., Xanthopoulos, B. C. (1986) A
new type of singularity created by colliding gravitational waves,
Proc. Roy. Soc. Lond. {\bf A408}, 175

% 243. 
\bibitem{ChX1} Chandrasekhar, S., Xanthopoulos, B. C. (1985) On
colliding waves in the Einstein-Maxwell theory, Proc. Roy. Soc.
Lond. {\bf A398}, 223

% 244. 
\bibitem{r}Bi\v{c}\'{a}k, J. (1989) Exact radiative space-times,
in {Proceedings of the fifth
Marcel Grossmann Meeting on General Relativity}, eds. D. Blair 
and M. J. Buckingham, World Scientific, Singapore


% 245. 
\bibitem{Y1} Yurtsever, U. (1987) Instability of Killing-Cauchy
horizons in plane-symmetric spacetimes, Phys. Rev. {\bf D36},
1662


% 246. 
\bibitem{Y2} Yurtsever, U. (1988) Singularities in the
collisions of almost-plane gravitational waves, Phys. Rev. {\bf
D38}, 1731


% 247. 
\bibitem{CHR} Chandrasekhar, S. (1986) Cylindrical waves in
general relativity, Proc. Roy. Soc. Lond. {\bf A408}, 209

% 248. 
\bibitem{ER} Einstein, A., Rosen, N. (1937) On Gravitational
Waves, J. Franklin Inst. {\bf 223}, 43

% 249. 
\bibitem{Be} Beck, G. (1925) Zur Theorie bin\"{a}rer
Gravitationsfelder, Z. Phys. {\bf 33}, 713

% 250. 
\bibitem{St} Stachel, J. (1966) Cylindrical Gravitational News,
J. Math. Phys. {\bf 7}, 1321

% 251. 
\bibitem{RI} d'Inverno, R. (1997) Combining Cauchy and
characteristic codes in numerical relativity, 
in Relativistic Gravitation and Gravitational Radiation 
(Proceedings of the Les Houches School of Physics),
eds. J.-A. Marck and J.-P. Lasota, Cambridge University Press,
Cambridge

% 252. 
\bibitem{Pi} Piran, T., Safier, P. N. and Katz, J. (1986)
Cylindrical gravitational waves with two degrees of freedom: An
exact solution, Phys. Rev. {\bf D34}, 331

% 253. 
\bibitem{KiT} Thorne, K. S. (1965) C-energy, Phys. Rev. {\bf
B138}, 251

% 254. 
\bibitem{GVe} Garriga, J., Verdaguer, E. (1987) Cosmic strings
and Einstein-Rosen waves, Phys. Rev. {\bf D36}, 2250

% 255. 
\bibitem{X} Xanthopoulos, B. C. (1987) Cosmic strings coupled
with gravitational and electromagnetic waves, Phys. Rev. {\bf
D35}, 3713

% 256. 
\bibitem{ChV} Chandrasekhar, S., Ferrari, V. (1987) On the
dispersion of cylindrical impulsive gravitational waves, Proc.
Roy. Soc. Lond. {\bf A412}, 75

% 257. 
\bibitem{To} Tod, K. P. (1990) Penrose's quasi-local mass and
cylindrically symmetric spacetimes,
Class. Quantum Grav. {\bf 7}, 2237

% 258. 
\bibitem{BER} Berger, B. K., Chru\'{s}ciel, P. T. and Moncrief,
V. (1995) On ``Asymptotically Flat'' Space-Times with
$G_2$-Invariant Cauchy Surfaces, Ann. Phys. (N.Y.) {\bf 237}, 322

% 259. 
\bibitem{Ku} Kucha\v{r}, K. V. (1971) Canonical quantization of
cylindrical gravitational waves, Phys. Rev. {\bf D4}, 955

% 260. 
\bibitem{AP} Ashtekar, A., Pierri, M. (1996) Probing quantum
gravity through exactly soluble midisuperspaces 1, J. Math.
Phys. {\bf 37}, 6250

% 261. 
\bibitem{KOS} Korotkin, D., Samtleben, H. (1998) 
Canonical Quantization of Cylindrical Gravitational 
Waves with Two Polarizations, 
Phys. Rev. Lett. {\bf 80}, 14

% 262. 
\bibitem{ABS1} Ashtekar, A., Bi\v{c}\'{a}k, J. and Schmidt, B. G.
(1997) Asymptotic structure of symmetry-reduced general
relativity, Phys. Rev. {\bf D55}, 669

% 263. 
\bibitem{ABS2}  Ashtekar, A., Bi\v{c}\'{a}k, J. and Schmidt, B. G.
(1997) Behaviour of Einstein-Rosen waves at null infinity, Phys.
Rev. {\bf D55}, 687

% 264. 
\bibitem{PEN} Penrose, R. (1963) Asymptotic properties of fields
and space-times, Phys. Rev. Lett. {\bf 10}, 66; (1965) Zero
rest-mass fields including gravitation: asymptotic behaviour,
Proc. Roy. Soc. Lond. {\bf A284}, 159

% 265. 
\bibitem{CG} Ehlers, J., Friedrich, H. eds. (1994) in
Canonical Gravity: From Classical to Quantum, Springer-Verlag,
Berlin-Heidelberg

% 266. 
\bibitem{Ry} Ryan, M. (1972) Hamiltonian Cosmology,
Springer-Verlag, Berlin

% 267. 
\bibitem{MC} MacCallum, M. A. H. (1975) Quantum Cosmological
Models, in Quantum Gravity, eds. C. J. Isham, R. Penrose and D.
W. Sciama, Clarendon Press, Oxford

% 268. 
\bibitem{HL} Halliwell, J. J. (1991) Introductory Lectures on
Quantum Cosmology, in Quantum Cosmology and Baby Universes, eds.
S. Coleman, J. Hartle, T. Piran and S. Weinberg, World
Scientific, Singapore

% 269. 
\bibitem{HLL} Halliwell, J. J. (1990) A Bibliography of Papers on
Quantum Cosmology, Int. J. Mod. Phys. {\bf A5}, 2473

% 270. 
\bibitem{KK} Kucha\v{r}, K. V. (1973) Canonical Quantization of
Gravity, in Relativity, Astrophysics and Cosmology, ed. W.
Israel, Reidel, Dordrecht

% 271. 
\bibitem{KU} Kucha\v{r}, K. V. (1992) Time and Interpretations of
Quantum Gravity, in Proceedings of the 4th Canadian Conference
on General Relativity and Relativistic Astrophysics, eds. G.
Kunstatter, D. Vincent and J. Williams, World Scientific,
Singapore

% 272. 
\bibitem{KVK} Kucha\v{r}, K. V. (1994) Geometrodynamics of
Schwarzschild black holes, Phys. Rev. {\bf D50}, 3961

% 273. 
\bibitem{RTo} Romano, J. D., Torre, C. G. (1996) Internal Time
Formalism for Spacetimes with Two Killing Vectors, Phys. Rev.
{\bf D53}, 5634. See also Torre, C. G. (1998) Midi-superspace
Models of Canonical Quantum Gravity, gr-qc/9806122



% 274. 
\bibitem{LW} Louko, J., Whiting, B. F. and Friedman, J. L. (1998)
Hamiltonian spacetime dynamics with a spherical null-dust shell,
Phys. Rev. {\bf D57}, 2279

% 275. 
\bibitem{GM} Griffiths, J. B., Miccicho, S. (1997) The
Weber-Wheeler-Bonnor pulse and phase shifts in gravitational
soliton interactions, Phys. Lett. {\bf A233}, 37

% 276. 
\bibitem{PiR} Piran, T., Safier, P. N. and Stark, R. F. (1985)
General numerical solution of cylindrical gravitational waves,
Phys. Rev. {\bf D32}, 3101

% 277. 
\bibitem{Wil} Wilson, J. P. (1997) Distributional curvature of
time dependent cosmic strings, Class. Quantum Grav. {\bf 14},
3337

% 278. 
\bibitem{BiS}  Bi\v{c}\'{a}k, J., Schmidt, B. G. (1989) On the
asymptotic structure of axisymmetric radiative spacetimes, Class.
Quantum Grav. {\bf 6}, 1547

% 279. 
\bibitem{BPa} Bi\v{c}\'{a}k, J., Pravdov\'{a}, A. (1998)
Symmetries of asymptotically flat electro\-vacu\-um spacetimes and
radiation, J. Math. Phys. {\bf 39}, 6011

% 280. 
\bibitem{BAP} Bi\v{c}\'{a}k, J., Pravdov\'{a}, A. (1999)
Axisymmetric electrovacuum spacetimes with a translational
Killing vector at null infinity, Class. Quantum Grav. {\bf 16}, 2023
% 281. 
\bibitem{aj} Robinson, I., Trautman, A. (1962) Some spherical
gravitational waves in general relativity, Proc. Roy. Soc. Lond.
{\bf A265}, 463 ; see also \cite{KSH}

% 282. 
\bibitem{ak}Chru\'{s}ciel, P. T. (1992)
On the global structure of Robinson-Trautman space-times,
Proc. Roy. Soc. Lond. A {\bf 436}, 299;
Chru\'{s}ciel, P. T., Singleton, D. B. (1992) 
Non-Smoothness of Event Horizons of Robinson-Trautman Black Holes,
Commun. Math. Phys. {\bf 147}, 137, and references therein

% 283. 
\bibitem{al} Bi\v{c}\'{a}k, J., Podolsk\'{y}, J. (1995) Cosmic
no-hair conjecture and black-hole formation: An exact model with
gravitational radiation, Phys. Rev.
{\bf D52}, 887


% 284. 
\bibitem{am}Bi\v{c}\'{a}k, J., Schmidt, B. G. (1984)
Isometries compatible with gravitational radiation,
J. Math. Phys. {\bf 25}, 600

% 285. 
\bibitem{ar} Bonnor, W. B., Swaminarayan, N. S. (1964) An exact solution for
uniformly accelerated particles in general relativity,
Zeit. f. Phys. {\bf 177}, 240.
See also the original paper on negative mass in general
relativity by Bondi, H. (1957) Rev. Mod. Phys. {\bf 29}, 423

% 286. 
\bibitem{IKH} Israel, W., Khan, K. A. (1964) Collinear particles
and Bondi dipoles in general relativity, Nuov. Cim. {\bf 33}, 331

% 287. 
\bibitem{an}Bi\v{c}\'{a}k J. (1985) On exact radiative solutions
representing finite sources, in Galaxies, axisymmetric systems
and relativity (Essays presented to W. B. Bonnor on his 65th
birthday), ed. M. A. H. MacCallum, Cambridge University Press,
Cambridge

% 288. 
\bibitem{ao} Bi\v{c}\'{a}k, J., Schmidt, B. G. (1989)
Asymptotically flat radiative space-times with boost-rotation
symmetry: the general structure, Phys. Rev.  {\bf D40}, 1827

% 289. 
\bibitem{ap}Bi\v{c}\'{a}k J. (1987) Radiative properties of
spacetimes with the axial and boost symmetries, in 
Gravitation and Geometry (A volume in honour of Ivor Robinson),
eds. W. Rindler and A. Trautman, Bibliopolis, Naples

% 290. 
\bibitem{as} Bi\v{c}\'{a}k, J., Hoenselaers, C. and Schmidt, B. G.
(1983) The solutions of the Einstein equations for uniformly
accelerated particles without nodal singularities II.
Self-accelerating particles, Proc. Roy. Soc. Lond. {\bf A390}, 411

% 291. 
\bibitem{at} Bi\v{c}\'{a}k, J., Reilly, P. and Winicour, J.
(1988) Boost rotation symmetric gravitational null cone data,
Gen. Rel. Grav. {\bf 20}, 171

% 292. 
\bibitem{au}G\'{o}mez R., Papadopoulos P. and Winicour J.
(1994) J. Math. Phys. {\bf 35}, 4184 

% 293. 
\bibitem{AGS} Alcubierre, M., Gundlach, C. and Siebel, F. (1997)
Integration of geodesics as a test bed for comparing exact 
and numerically generated spacetimes,
in Abstracts of Plenary Lectures and Contributed Papers (GR15),
Inter-University Centre for Astronomy and Astrophysics Press, Pune
% 294. 
\bibitem{av} Bi\v{c}\'{a}k, J., Hoenselaers, C. and Schmidt B.G.,
(1983) The solutions of the Einstein equations for uniformly
accelerated particles without nodal singularities I. Freely
falling particles in external fields,
Proc. Roy. Soc. Lond. {\bf A390}, 397

% 295. 
\bibitem{aw} Bi\v{c}\'{a}k, J. (1980) The motion of a charged
black hole in an electromagnetic field, Proc. Roy. Soc. Lond. {\bf
A371}, 429

% 296. 
\bibitem{ax} Hawking, S. W., Horowitz, G. T. and Ross, S. F. (1995)
Entropy, area, and black hole pairs, Phys. Rev. {\bf D51}, 4302;
Mann, R. B., Ross, S. F. (1995) Cosmological production of
charged black hole pairs, Phys. Rev. {\bf D52}, 2254;
Hawking, S. W., Ross, S. F. (1995) Pair production of black
holes on cosmic strings,  Phys. Rev. Lett. {\bf 75}, 3382

% 297. 
\bibitem{PDE} Pleba\'{n}ski, J., Demia\'{n}ski, M. (1976)
Rotating, charged and uniformly accelerating mass in general relativity,
Ann. Phys. (N.Y.) {\bf 98}, 98

% 298. 
\bibitem{BPD} Bi\v{c}\'ak, J., Pravda, V. (1999) 
Spinning C-metric: radiative spacetime with accelerating, 
rotating black holes, 
Phys. Rev. {\bf D60}, 044004

% 299. 
\bibitem{BKL1} Belinsky, V. A., Khalatnikov, I. M. and Lifshitz, E. M. (1970) Oscillatory approach to a singular point in the relativistic cosmology, Adv. in Phys. {\bf 19}, 525

% 300. 
\bibitem{BKL2} Belinsky, V. A., Khalatnikov, I. M. and Lifshitz, E. M. (1982) A general solution of the Einstein equations with a time singularity, Adv. in Phys. {\bf 31}, 639

% 301. 
\bibitem{GOE} Ellis, G. F. R. (1996) Contributions of K. G\"{o}del to Relativity and Cosmology, in G\"{o}del '96: Logical Foundations of Mathematics, Computer Science and Physics -- Kurt G\"{o}del's Legacy, 
ed. P. H\'ajek, Springer-Verlag, Berlin-Heidelberg; see also preprint 1996/7 of the Dept. of Math. and Appl. Math., University of Cape Town

% 302. 
\bibitem{KASA} Kantowski, R., Sachs, R. K. (1966) Some Spatially Homogenous Anisotropic Relativistic Cosmological Models, J. Math. Phys. {\bf 7}, 443

% 303. 
\bibitem{KIP} Thorne, K. S. (1967) Primordial element formation, primordial magnetic fields, and the isotropy of the universe, Astrophys. J. {\bf 148}, 51

% 304. 
\bibitem{RYS} Ryan, M. P., Shepley, L. C. (1975) Homogeneous Relativistic Cosmologies, Princeton University Press, Princeton

% 305. 
\bibitem{MAC1} MacCallum, M. A. H. (1979) Anisotropic and inhomogeneous relativistic cosmologies, in General Relativity (An Einstein Centenary Survey), eds. S. W. Hawking and W. Israel, Cambridge University Press, Cambridge

% 306. 
\bibitem{RY} Obreg\'on, O., Ryan, M. P. (1998) Quantum Planck size black hole states without a horizon, Modern Phys. Lett. A {\bf 13}, 3251; see also references therein

% 307. 
\bibitem{NOO} Nojiri, S., Obreg\'on, O., Odintsov, S. D. and Osetrin, K. E. (1999)\\
(Non)singular Kantowski-Sachs universe from quantum spherically reduced matter, Phys. Rev. {\bf D60}, 024008

% 308. 
\bibitem{HES} Heckmann, O., Sch\"{u}cking, E. (1962) Relativistic Cosmology, in Gravitation: an introduction to current research, ed. L. Witten, J. Wiley and Sons, New York

% 309. 
\bibitem{ZN2} Zel'dovich, Ya. B., Novikov, I. D. (1983) Relativistic Astrophysics, Volume 2: The Structure and Evolution of the Universe,
The University of Chicago Press, Chicago

% 310. 
\bibitem{MAC2} MacCallum, M. A. H. (1994) Relativistic cosmologies, in Deterministic Chaos in General Relativity, eds. D. Hobill, A. Burd and A. Coley, Plenum Press, New York

% 311. 
\bibitem{WE} Wainwright, J., Ellis, G. F. R. eds. (1997) Dynamical Systems in Cosmology, Cambridge University Press, Cambridge

% 312. 
\bibitem{MSR} Misner, C. W. (1969) Mixmaster universe, Phys. Rev. Lett. {\bf 22}, 1071

% 313. 
\bibitem{MIF} Hu, B. L., Ryan, M. P. and Vishveshwara, C. V. eds. (1993) Directions in General Relativity, Vol. 1 (Papers in honor of Charles Misner), Cambridge University Press, Cambridge

% 314. 
\bibitem{UJR} Uggla, C., Jantzen, R. T. and Rosquist, K. (1995) Exact hypersurface-homogeneous solutions in cosmology and astrophysics, Phys. Rev. {\bf D51}, 5522

% 315. 
\bibitem{TTS} Tanaka, T., Sasaki, M. (1997) Quantized gravitational waves in the Milne universe, Phys. Rev. {\bf D55}, 6061

% 316. 
\bibitem{LUK} Lukash, V. N. (1975) Gravitational waves that conserve the homogeneity of space, Sov. Phys. JETP {\bf 40}, 792

% 317. 
\bibitem{BASO} Barrow, J. D., Sonoda, D. H. (1986) Asymptotic stability of Bianchi type universes, Physics Reports {\bf 139}, 1

% 318. 
\bibitem{KURY} Kucha\v r, K. V., Ryan, M. P. (1989) Is minisuperspace 
quantization valid?: Taub in Mixmaster, Phys. Rev. {\bf D40}, 3982.
The approach was first used in 
Kucha\v r, K. V., Ryan, M. P. (1986)
Can Minisuperspace Quantization be Justified?,
in Gravitational Collapse and Relativity,
eds. H. Sato and T. Nakamura, World Scientific, Singapore

% 319. 
\bibitem{BOG} Bogoyavlenski, O. I. (1985) Methods in the Qualitative Theory of Dynamical Systems in Astrophysics and Gas Dynamics, Springer-Verlag, Berlin

% 320. 
\bibitem{DET} Hobill, D., Burd, A. and Coley, A. eds. (1994) Deterministic Chaos in General Relativity, Plenum Press, New York

% 321. 
\bibitem{REN} Rendall, A. (1997) Global dynamics of the Mixmaster model, 
Class. Quantum Grav. {\bf 14}, 2341

% 322. 
\bibitem{SIN} Khalatnikov, I. M., Lifshitz, E. M., Khamin, K. M., Shehur, L. N. and Sinai, Ya. G. (1985) On the Stochasticity in Relativistic Cosmology, J. of Statistical Phys. {\bf 38}, 97

% 323. 
\bibitem{LWK} LeBlanc, V. G., Kerr, D. and Wainwright, J. (1995)
Asymptotic states of magnetic Bianchi VI${}_0$ cosmologies,
Class. Quantum Grav. {\bf 12}, 513

% 324. 
\bibitem{LBL1} LeBlanc, V. G. (1977) Asymptotic states of magnetic Bianchi I cosmologies, Class. Quantum Grav. {\bf 14}, 2281

% 325. 
\bibitem{JAT} Jantzen, R. T. (1986) Finite-dimensional Einstein-Maxwell-scalar field system, 
Phys. Rev. {\bf D33}, 2121

% 326. 
\bibitem{LBL2} LeBlanc, V. G. (1998) Bianchi II magnetic cosmologies, 
Class. Quantum Grav. {\bf 15}, 1607

% 327. 
\bibitem{BEKA} Belinsky, V. A., Khalatnikov, I. M. (1973) Effect of scalar and vector fields on 
the nature of the cosmological singularity, Soviet Physics JETP {\bf 36}, 591

% 328. 
\bibitem{BERG} Berger, B. K. (1999) Influence of scalar fields on the approach to a cosmological singularity, gr-qc/9907083

% 329. 
\bibitem{WCE} Wainwright, J., Coley, A. A., Ellis, G. F. R. and Hancock, M. (1998) On the isotropy of 
the Universe: do Bianchi VII${}_h$ cosmologies isotropize? Class. Quantum Grav. {\bf 15}, 331

% 330. 
\bibitem{WIB} Weaver, M., Isenberg, J. and Berger, B. K. (1998) Mixmaster Behavior in 
Inomogeneous Cosmological Spacetimes, Phys. Rev. Lett. {\bf 80}, 2984

% 331. 
\bibitem{BEM} Berger, B. K., Moncrief, V. (1998) Evidence for an oscillatory 
singularity in generic $U(1)$ cosmologies on $T^3 \times R$, 
Phys. Rev. {\bf D58}, 064023

% 332. 
\bibitem{GOW} Gowdy, R. H. (1971) Gravitational Waves in Closed Universes, 
Phys. Rev. Lett. {\bf 27}, 826; Gowdy, R. H. (1974) 
Vacuum Spacetimes with Two-Parameter Spacelike Isometry Groups and 
Compact Invariant Hypersurfaces: 
Topologies and Boundary Conditions, Ann. Phys. (N.Y.) {\bf 83}, 203

% 333. 
\bibitem{CCM} Carmeli, M., Charach, Ch. and Malin, S. (1981) 
Survey of cosmological models with gravitational scalar and 
electromagnetic waves, Physics Reports {\bf 76}, 79

\bibitem{PCL} Chru\'sciel, P. T. (1990) On Space-Times with $U(1)\times U(1)$
Symmetric Compact Cauchy Surfaces,
Ann. Phys. (N. Y.) {\bf 202}, 100

% 335. 
\bibitem{GOW1} Gowdy, R. H. (1975) Closed gravitational-wave universes: 
Analytic solutions with two-parameter symmetry, J. Math. Phys. {\bf 16}, 224

% 336. 
\bibitem{CHA} Charach, Ch. (1979) Electromagnetic Gowdy universe, 
Phys. Rev. {\bf D19}, 3516

% 337. 
\bibitem{BIJG} Bi\v c\'ak, J.,  Griffiths, J. B. (1996) Gravitational Waves Propagating into Friedmann-Robertson-Walker Universes, Ann. Phys. (N.Y) {\bf 252}, 180

% 338. 
\bibitem{BCIM} Berger, B. K., Chru\'sciel, P. T., Isenberg, J. and Moncrief, V. (1997) 
Global Foliations of Vacuum Spacetimes with $T^2$ Isometry, Ann. Phys. (N.Y.) {\bf 260}, 117

% 339. 
\bibitem{GIM} Chru\'sciel, P. T., Isenberg, J. and Moncrief, V. (1990) 
Strong cosmic censorship in polarized Gowdy spacetimes, Class. Quantum Grav. {\bf 7}, 1671

% 340. 
\bibitem{VMO} Moncrief, V. (1997) 
Spacetime Singularities and Cosmic Censorship, in Proc. of the 14th 
International Conference on General Relativity and Gravitation, eds. M. Francaviglia, G. Longhi, L. Lusanna and E. Sorace, World Scientific, Singapore

% 341. 
\bibitem{KIR} Kichenassamy, S., Rendall, A. D. (1998) Analytic description of singularities in 
Gowdy spacetimes, Class. Quantum Grav. {\bf 15}, 1339

% 342. 
\bibitem{KIM} Kichenassamy, S. (1996) Nonlinear Wave Equations, Marcel Dekker Publ. New York

% 343. 
\bibitem{AZF} Adams, P. J., Hellings, R. W., Zimmermann, R. L., Farhoosh, H., 
Levine, D. I. and Zeldich, S. (1982) Inhomogeneous cosmology: 
gravitational radiation in Bianchi backgrounds, Astrophys. J. {\bf 253}, 1

% 344. 
\bibitem{BEZ} Belinsky, V., Zakharov, V. (1978) Integration of the Einstein 
equations by means of the inverse scattering problem technique and 
construction of exact soliton solutions, Sov. Phys. JETP {\bf 48}, 985

% 345. 
\bibitem{CAV} Carr, B. J., Verdaguer, E. (1983) Soliton solutions and 
cosmological gravitational waves, Phys. Rev. {\bf D28}, 2995

% 346. 
\bibitem{BEL} Belinsky, V. (1991) Gravitational breather and topological 
properties of gravi\-solitons, Phys. Rev. {\bf D44}, 3109

% 347. 
\bibitem{KORD} Kordas, P. (1993) Properties of the gravibreather, Phys. Rev. {\bf D48}, 5013

% 348. 
\bibitem{ALE} Alekseev, G. A. (1988) Exact solutions in the general theory of relativity, 
Proceedings of the Steklov Institute of Mathematics, Issue 3, p. 215

% 349. 
\bibitem{t} Verdaguer, E. (1993) Soliton solutions in spacetimes
with spacelike Killing fields, Physics Reports {\bf 229}, 1

% 350. 
\bibitem{JOKB} Katz, J., Bi\v c\'ak, J. and Lynden-Bell, D. (1997) 
Relativistic conservation laws and integral constraints 
for large cosmological perturbations, 
Phys. Rev. {\bf D55}, 5957

% 351. 
\bibitem{TUN} Uzan, J. P., Deruelle, M. and Turok, N. (1998)
Conservation laws  and cosmological perturbations in curved universes,
Phys. Rev. {\bf D57}, 7192

% 352. 
\bibitem{BESI} Beig, R., Simon, W. (1992) On the Uniqueness of Static Perfect-Fluid 
Solutions in General Relativity, Commun. Math. Phys. {\bf 144}, 373

% 353. 
\bibitem{LIM} Lindblom, L., Masood-ul-Alam (1994) On the Spherical Symmetry of 
Static Stellar Models, Commun. Math. Phys. {\bf 162}, 123

% 354. 
\bibitem{RFL} Rendall, A. (1997) Solutions of the Einstein equations with matter, in 
Proc. of the 14th International Conference on General Relativity and 
Gravitation, eds. M. Francaviglia, G. Longhi, L. Lusanna and E. Sorace, 
World Scientific, Singapore

% 355. 
\bibitem{BMK} Bartnik, R., McKinnon, J. (1988) Particlelike Solutions of the 
Einstein-Yang-Mills Equations, Phys. Rev. Lett. {\bf 61}, 141

% 356. 
\bibitem{VG} Volkov, M. S., Gal'tsov, D. V. (1999) Gravitating Non-Abelian Solitons 
and Black Holes with Yang-Mills Fields, Physics Reports {\bf 319}, 1 

% 357. 
\bibitem{RET} Rendall, A. D., Tod, K. P. (1999) Dynamics of spatially 
homogeneous solutions of the Einstein-Vlasov equations which are 
locally rotationally symmetric, Class. Quantum Grav. {\bf 16}, 1705

% 358. 
\bibitem{CAC} Carr, B. J., Coley, A. A. (1999) Self-similarity in general relativity, 
Class. Quantum Grav. {\bf 16}, R 31

% 359. 
\bibitem{GUN} Gundlach, C. (1998) Critical Phenomena in Gravitational 
Collapse, Adv. Theor. Math. Phys. {\bf 2}, 1

% 360. 
\bibitem{KRA} Krasi\'{n}ski, A. (1997) Inhomogeneous Cosmological Models, 
Cambridge University Press, Cambridge 
 


\end{thebibliography}
\end{document}